\documentclass[10pt,b5paper,twoside]{book}

\usepackage{dsfont}
\usepackage{multicol}
\usepackage{ThALV}
\usepackage[utf8]{inputenc}
\usepackage[T1]{fontenc}
\usepackage{amsmath, amssymb, amsthm, mathabx}
\usepackage{graphicx}
\usepackage[usenames,dvipsnames]{xcolor}
\usepackage{hyperref}
\usepackage{subcaption}
\usepackage{braket}
\usepackage{makecell}
\hypersetup{
    linktocpage,
    colorlinks   = true,
    citecolor    = RoyalBlue, 
    linkcolor    = RoyalBlue,
    urlcolor     = RoyalBlue  
}

\usepackage{todonotes}
\usepackage{lipsum}
\usepackage{marginnote}

\usepackage{geometry}
\usepackage{tikz}
\usepackage{lipsum, afterpage}
\usepackage{placeins}
\usepackage{float}
\usepackage{mathtools}
\DeclareUnicodeCharacter{0308}{cy}

\theoremstyle{definition}
\newtheorem{definition}{Definition}[section]

\newtheorem{theorem}{Theorem}[section]
\DeclareMathOperator{\He}{He}
\DeclareMathOperator{\expect}{\mathbb{E}}
\theoremstyle{remark}

\usepackage{enumitem}

\usepackage{fancyhdr} 
\fancypagestyle{my-fancy}{
    \fancyhf{}
    \fancyhead[LE]{\fontsize{10}{10}\selectfont\nouppercase\thepage}
    \fancyhead[RE]{\fontsize{10}{10}\selectfont\nouppercase\leftmark} 
    \fancyhead[RO]{\fontsize{10}{10}\selectfont\nouppercase\thepage} 
    \fancyhead[LO]{\fontsize{10}{10}\selectfont\nouppercase\rightmark} 
    \fancyfoot[CE,CO]{} 
    \fancyfoot[LE,RO]{} 
}
\fancypagestyle{plain}{%
  \fancyhf{}%
}
\usepackage{titlesec} 
\titleformat{\chapter}[display]
{\bfseries\Large\center}
{\MakeUppercase{\chaptertitlename} \Huge\thechapter}
{2ex}
{\titlerule\vspace{1ex}}
[\vspace{1ex}\titlerule]

\usepackage{minitoc}
\dominitoc[c]
\nomtcrule

\usepackage{caption}
\DeclareMathOperator*{\argmax}{arg\,max}

\DeclareMathOperator*{\Tr}{Tr}

\newcommand{\bigO}{\mathcal{O}}

\usepackage{tcolorbox}
\tcbuselibrary{breakable}
\DeclareRobustCommand{\details}[2][gray!10]{%
\begin{tcolorbox}[   
        breakable,
        left=0pt,
        right=0pt,
        top=0pt,
        bottom=0pt,
        colback=#1,
        colframe=#1,
        width=\dimexpr\textwidth\relax, 
        enlarge left by=0mm,
        boxsep=5pt,
        arc=0pt,outer arc=0pt,
        left skip = 0pt,
        right skip = 0pt,
        before upper={\parindent15pt\noindent},
        ]
        #2
\end{tcolorbox}
}

\DeclareMathOperator{\id}{id}
\DeclareMathOperator{\diag}{diag}

\usepackage[backend=biber,
            style=trad-alpha,
            backref=true]
            {biblatex}
\usepackage{url}
\begin{document}

\frontmatter
\maketitle

\nocite{*}
\chapter*{Abstract}
Deep Neural Networks (DNNs) are at the forefront of a technological revolution, displaying capabilities that frequently match or surpass human expertise in a diverse array of tasks. From image recognition to complex decision-making processes, DNNs are reshaping industries and scientific research. Despite their extensive application, the internal mechanisms of these algorithms remain largely opaque, often drawing comparisons to ``black boxes''. While we can experimentally enhance their performance, achieving a deep and comprehensive understanding of their internal processes remains a formidable challenge.

For the field of Statistical Mechanics, dealing with computer science problems is not a new challenge. This thesis is situated at this crossroads, embracing an interdisciplinary approach that leverages physics-based methods to deepen our understanding of DNNs.

The core of this thesis is developed through three distinct but interconnected research approaches. 
The first is a data-averaged one, which we use to establish an asymptotic bound on the generalization performance of DNNs. This result not only greatly improves on a classical bound of Statistical Learning Theory, but also illustrates how a physics-based perspective can transcend traditional analytical limitations. 
Notably, our result only depends on the last layer size rather than on the total number of network parameters, highlighting how, in deep networks, information is processed differently through different layers.

The second approach takes a data-dependent perspective, focusing on the behavior of DNNs in a specific thermodynamic limit beyond the well-known infinite-width limit. This line of action involves a detailed analysis of the network's forward dynamics, allowing for a more precise statistical description of the inner workings of these algorithms in a realistic regime where most real-world DNNs operate. 
This advance allows us to obtain (i) a closed formula for the generalization error associated to a regression task in a finite-width one-hidden layer network; 
(ii) an approximate expression of the partition function for deep architectures and (iii) a link between deep neural networks in the thermodynamic limit and Student's t-processes.

Finally, the thesis adopts a task-explicit approach, which involves a (preliminary) examination of how DNNs interact with and adapt to the structure of a simple and controlled dataset. This analysis aims to discern whether DNNs are truly capable of resonating with the structure of the dataset forming an internal representation of its features, as opposed to merely memorizing it. This aspect of the research could be useful for understanding when DNNs are forced to learn the structure of the data, instead of simply memorizing it.

In summary, this thesis wants to be a journey into one of the most intriguing and impactful areas of modern technology. A journey that, exploiting the fruitful dialogue between Statistical Physics and Machine Learning, wants to contribute to shed light on the inner behavior of Deep Neural Networks.
This line of research, of which this thesis is a very small part, has the potential to influence not just the field of Deep Neural Networks but also the myriad fields where these systems are applied, hopefully paving the way for more interpretable and transparent Artificial Intelligence.


\tableofcontents

\mainmatter

\chapter{Introduction}
\begin{refsection}
\section{Motivation}

Over the past decade, Deep Neural Networks (DNNs) have sparked a revolution across various domains, from scientific research to industries. 
Their unparalleled success is not solely rooted in their ability to learn intricate patterns from datasets; rather, it lies in their capacity to apply this knowledge to entirely novel, unseen data points, achieving performance comparable to human capabilities. 
This capacity, known as generalization, forms the bedrock of DNNs' practical applicability in real-world scenarios.

Let me show you a few examples of how DNNs are reshaping the world through their remarkable generalization abilities:

\begin{itemize}
    \item Self-driving cars: DNNs are integral to autonomous vehicles, enabling them to identify diverse elements on the road, from other vehicles and pedestrians to traffic signs. Moreover, they possess the capability to predict the behaviors of other drivers and pedestrians \cite{8014794,dong2023applications}. This adaptability to new situations is imperative for ensuring the safe operation of self-driving cars in real-world environments.
    \item Medical imaging: DNNs assume a pivotal role in medical imaging for disease diagnosis. They can discern cancer cells in mammograms \cite{nasser2023deep}, detect Alzheimer's disease in brain scans \cite{THANGAVEL2023105215}, and even predict the risk of future ailments like heart disease or stroke \cite{garcia2023heart}. Their ability to generalize across a spectrum of patient profiles is essential for enhancing the precision of medical diagnoses and treatments.
    \item Fraud detection: DNNs are instrumental in fraud detection \cite{udeze2022application,kim2019fraud,zhou2018state,singla2022class}, facilitating the identification of fraudulent transactions, encompassing credit card fraud, insurance fraud, and money laundering. Their agility in adapting to emerging forms of fraudulent activity is critical for safeguarding businesses and consumers against financial losses.
    \item Natural language processing: DNNs find utility in a range of natural language processing (NLP) tasks, such as machine translation between languages \cite{johnson2017googl,carl2019machine}, summarizing lengthy documents \cite{yao2015automatic}, and responding to questions based on textual content \cite{singh2023chat}. Their capacity to generalize across various languages and text types is indispensable for the daily use of these algorithms. 
\end{itemize}

And, as DNN technology continues to progress, we can expect to see even more innovative and groundbreaking applications in the future.
Examples include the utilization of DNNs in drug development \cite{chen2023dnn}, the creation of more efficient and personalized educational programs \cite{tang2022personalized}, and the enhancement of the safety and reliability of our infrastructure \cite{ajayi2020deep}.

At this point, it is not so strange to contemplate the possibility that, in the coming years, Deep Neural Networks (DNNs) might assist or even replace humans in various decision-making tasks. 
This raises some problems, like for example,  from an ethical perspective, although this aspect falls outside the scope of this thesis. 
Our focus is on the technical problems: in fact, nowadays, we tend to employ these algorithms as "black boxes," akin to magical entities that can address a wide array of problems.  

To be frank, would you trust an algorithm to drive your car if you didn't comprehend how it functions? 
How can you place faith in an algorithm when you're unsure about its impartiality, fairness, predictability, and resilience against adversarial attacks? 
No one would want a car accident caused by AI misinterpreting an unusual asphalt pattern as a car.
The risk can be mitigated by gaining a deep understanding of how these algorithms operate, specifically how they represent dataset features within DNNs and employ these representations to accomplish tasks. At the moment, however, we lack a suitable theoretical framework that describes how these algorithms learn and generalize.
Developing a machine learning framework is a challenging task because many factors interact in complex ways. 
For example, the choice of architecture for a given task depends on the nature of the task itself\footnote{For example, convolutional feed feedforward neural networks (CNNs) are well-suited for image classification because of their translational invariance and unidirectional information flow, while recurrent neural networks (RNNs) are well-suited for speech recognition because of their ability to store memory of past signals.}, which in turn is closely linked to the nature and structure of the dataset\footnote{Task can also change the form of the dataset itself, think about the tokenization of a text in the context of natural language model (NLM).}. 
However, the dataset structure also interacts with the architecture itself\footnote{ Training a machine learning model is often modeled as a high-dimensional risk minimization problem, where the dataset is a constraint. 
Therefore, the geometric structure of the dataset plays a fundamental role in the learning dynamics.}.
So, we can say that \textit{the complex interplay of the many components of deep neural networks makes modeling them a challenging task.}

In this thesis, I have adopted a widely used approach from the Statistical Mechanics of Complex Systems to investigate these systems.
My aim was to reduce their complexity by exploring simplified cases or special regimes, with the goal of deriving theoretical insights that could enhance our understanding of the internal dynamics of Deep Neural Networks.
This approach opens the door to a wealth of powerful techniques rooted in Statistical Physics, which can be readily applied to DNNs. 
One notable advantage is that such an approach diverges from the common theoretical framework for these systems known as Statistical Learning Theory (SLT), which primarily concerns worst-case scenarios. In contrast, Statistical Mechanics focuses on the average-case scenarios. 
Consequently, we can glean a larger quantity of information from typical behaviors, making this perspective not only more informative but also more practical for analyzing DNNs.

In this manuscript, I addressed three distinct questions, each with its unique approach.
In Chapter 2, I explored the generalization properties of DNNs, aiming for a broad perspective.
Chapter 3 delves into the transformations a dataset undergoes within a DNN, particularly in a specific proportional regime (where the number of training examples scales with the number of neurons per layer).
Finally, Chapter 4 examines the significance of inner representations and their impact on the generalization performance of DNNs.

\section{Organization and main results}

Despite the \textit{fil rouge} (i.e. common theme) that threads through the entirety of the manuscript, this thesis is composed of four distinct chapters thoughtfully crafted to stand on their own, facilitating independent reading. This structure not only provides a cohesive narrative but also allows for each section to be understood in isolation, making it suitable for readers who may have an interest in specific topics covered within the thesis.

The initial \textbf{Chapter 1} offers a concise overview of Machine Learning with a more detailed focus on neural networks, the subject of this thesis. We also provide an introduction to the principles of the Statistical Mechanics of Deep Learning. 
Subsequent chapters methodically introduce and detail novel results achieved together with my collaborators, in the context of supervised learning. 
These results are presented in a pedagogical manner, introducing topics broadly before navigating into detailed exploration. 
To streamline the discussion, non-essential technical details are highlighted in gray boxes; feel free to bypass these if you're not keen on delving into technical minutiae.
The primary focus is on the Statistical Mechanics of Deep Neural Networks, which is however approached differently in each chapter.

\textbf{Chapter 2} focuses on the generalization capabilities of DNNs. 
Although the problem has been explored in various ways within the Statistical Learning Theory (SLT) setting, classic results often rest on worst-case analysis and uniform distributions of the model (in practice, training is not accounted for).
These methods, while theoretically robust, yield results of limited practical utility by inclusively accounting for numerous pathological situations. 
My group and I adopted an alternative \textit{data-averaged approach}, grounded in the quenched average, which enabled us to restrict the class of models considered, exclusively focusing on networks with optimized final layers.
Our results establish a new bound on generalization performance, achieving an improvement of several orders of magnitude over prior SLT results, particularly in the asymptotic regime.

\textbf{Chapter 3} investigates the behavior of Deep Neural Networks (DNNs) in an asymptotic scenario, characterized by certain ``dimensions'' of the networks increasing towards infinity. This scenario, within the framework of Statistical Mechanics, is commonly referred to as the thermodynamic limit.
It is a widely accepted fact that as the number of neurons per layer approaches infinity, while the training set size remains fixed, a DNN can be approximated by a Gaussian Process (GP).
This revelation is crucial as it connects the behavior of our algorithms to a well-established mathematical framework with an extensive body of literature.
However, this principle carries a notable limitation. 
In practical, state-of-the-art deep network scenarios, the number of neurons in any given layer and the size of the training set are usually comparable. 
This reality complicates the direct application of GP theory to depict the behavior of contemporary algorithms.
Building on this, through a \textit{data-dependent approach}, my collaborators and I have devised a framework capable of delineating the statistical mechanics of DNNs in this proportional regime, aiming to forge a pathway for studying real-world algorithms. 
Additionally, this approach has unearthed an intriguing dualism between DNNs in the proportional regime and the Student's t Process.

\textbf{Chapter 4} delves into the formation of hidden representations within DNNs. In the literature, it's widely acknowledged that Convolutional Neural Networks (CNNs) evolve to form filters, which the network then uses to identify features within data points.
This machine-interpretable data representation appears to be a pivotal aspect of ``feature learning'', a regime wherein the algorithm develops an internal representation of the most representative dataset features and exhibits high generalization capabilities.
Building on this premise, we devised a straightforward \textit{task-explicit} framework for a fully connected DNN.  
This was done to scrutinize the conditions under which the algorithm could generate a hidden representation and to explore how such representations influence generalization performance.

\printbibliography[heading=subbibnumbered, title=Chapter bibliography]
\clearpage
\end{refsection}

\chapter{Machine Learning \& StatMech}
\begin{refsection}
\label{ch.1MLSM}
Today, humanity boasts unprecedented computational capabilities. Yet, despite relentless efforts, numerous contemporary challenges lack optimal solutions. Consequently, much of this immense computational power often gets channeled into sub-optimal algorithms. This reality underscores the significant growth and importance of the Machine Learning (ML) field in recent years. At its core, machine learning embodies the science and the art of helping machines derive their algorithm to address a specific problem.
Among all the techniques of ML, Deep Neural Networks (DNNs) are likely the most known algorithm in the field.  
In this chapter, I will usher you into the enchanting world of Machine Learning, with a substantial focus on Deep Neural Networks (DNNs). Additionally, I will provide a succinct introduction to the Statistical Mechanics of Learning.

\vspace{0.5cm}
\minitoc
\vfill
\newpage

\section{Machine Learning: a qualitative overview}

In this section, I will provide a concise, qualitative introduction to the field of Machine Learning. 
After briefly discussing the most common techniques and paradigms, our focus will shift to Deep Neural Networks. 
This subset of Machine Learning algorithms, which are the primary subject of this thesis, differs significantly from others. 
In fact, unlike traditional Machine Learning algorithms that necessitate a preliminary process of feature engineering to tailor them to a specific problem, Deep Neural Networks are capable of extracting the necessary information directly from raw datasets. 
A more comprehensive and detailed treatment of the field can be found in the works of Bishop \cite{bishop2006pattern}, Murphy \cite{murphy2012machine}, Hastie et al. \cite{hastie2009elements}, and Goodfellow et al. \cite{goodfellow2016deep}.

\subsection{One algorithm to rule them all}\label{1MLSM:one}

Imagine being faced with an incredibly intricate and lengthy task: categorizing every book ever written by its genre.  
Initially, you'd immerse yourself in each book, reading it from cover to cover to pinpoint its category.  
As you progress through volumes, patterns emerge. Words like ``Elf'', ``Dwarf'', or ``Hobbit'' hint at a fantasy theme, while mentions of figures such as Gaius Julius Caesar, Hannibal, or Attila typically signify a historical context. 
With these insights, you soon realize that sometimes just a chapter—or even a few pages—might suffice for categorization.
Given the staggering number of books in existence\footnote{over 129 million and counting!}, it's tempting to automate this categorization process. 
Drafting the list of indicative words to scan each book for them might seem like a clever shortcut. Congratulations, you've now classified Dante Alighieri's ``Divina Commedia'' as a veritable historical account! 
While this might offer a chuckle in hindsight, it underscores the limitations of algorithms that lean too heavily on simplicity.

Rethinking your strategy, you opt for a more nuanced approach: designing an algorithm that learns from examples.
By pairing each book you've read with its genre, you curate a dataset.
This serves as the training ground for your algorithm, which progressively hones its ability to identify patterns.
Unlike the initial simplistic method, this refined algorithm recognizes that isolated keywords won't be enough; understanding intricate structures and contexts is the main point.
This newfound knowledge forms the bedrock for the ability to \textit{\textbf{generalize}}, reflecting an intuition that was hard to encapsulate in your initial algorithm. 
Indeed, whereas the first algorithm mechanically segregated genres based solely on predefined, rigid rules, the second algorithm, guided by the dataset, constructs its own more complex and nuanced rules. 
These evolved rules, inherently more adaptable, stand a far better chance of accurately categorizing the myriad of books that exist.

Now, picture applying this dynamic, learning-driven methodology to a spectrum of problems where creating the perfect algorithm seems impossible. You make your first step in the realm of Machine Learning!

To give you a flavor of what machine learning can do, here you have some examples of the most common tasks.

\paragraph{Regression}
Regression involves predicting a continuous output value based on one or more input features. Essentially, it is about estimating the relationships among variables. This technique is crucial in various fields such as Economics, Meteorology, Engineering, and many others. Key machine learning algorithms that are commonly used in this task include Linear Regression, Polynomial Regression, Ridge Regression, Decision Trees, Random Forests, and Neural Networks. These algorithms aim to model the underlying trend in data so that they can predict numerical values for new, unseen data points. 
The theoretical analysis of Neural Networks carried on in this thesis work is primarily focused on regression problems.

\paragraph{Classification}
Classification entails predicting the categorical label of a given input. 
Put simply, it involves categorizing a specific input into one of several predefined groups or classes. 
This technique is pivotal in numerous fields such as Medical Diagnosis, Handwriting Recognition, Text Categorization, Anomaly Detection, and Filtering. 
Several machine learning algorithms excel in this domain, including Logistic Regression, Decision Trees, Random Forests, Support Vector Machines (SVMs), K-Nearest Neighbors, and Neural Networks.

\paragraph{Writing}

Text generation is the art of crafting coherent and contextually relevant textual content through machine learning. 
This intricate process involves generating sequences of words or characters that authentically replicate the style and structure of human-written prose. 
While chatbots are a quintessential application, text generation's versatility extends to content creation, poetry, scriptwriting, and even the drafting of PhD theses. 
Predominant models in this arena encompass Recurrent Neural Networks (RNNs), Long Short-Term Memory (LSTM) networks, and state-of-the-art Transformers like Generative Pre-trained Transformer (GPT) and Bidirectional Encoder Representations from Transformers (BERT).

\paragraph{Drawing}

Drawing and artistic expression, once solely the realm of human creativity, now they are now being enhanced and mirrored by machine learning algorithms.
Central to this progression is the aim to produce visuals that align with human aesthetic sensibilities. 
Uses span from aiding digital art creation and completing sketches automatically to crafting entirely original art pieces. 
Prominent machine learning models in this field include Convolutional Neural Networks (CNNs), Generative Adversarial Networks (GANs), and Variational Autoencoders (VAEs).

\paragraph{Automatization}
Automation, empowered by machine learning algorithms, is revolutionizing various industries by performing tasks with increased efficiency and reducing the necessity for human intervention. 
In the realm of automation, the core objective is to enhance systems to perform repetitive tasks or make decisions automatically, based on data input and learned patterns. 
Applications are vast, encompassing robotic process automation, predictive maintenance, automated data analysis, and even autonomous vehicles. 
Notable machine learning algorithms pivotal for automation include Decision Trees, Neural Networks, Reinforcement Learning (RL), and clustering algorithms like K-Means. 
\bigskip

A note of caution is imperative when discussing machine learning, especially when it comes to Neural Networks.
Contrary to what some may believe,  these algorithms are far from infallible or all-knowing.  
For them to operate optimally, they require rigorous design, thorough tuning, and meticulous optimization. 
Unfortunately, much of the fine-tuning in this field often relies heavily on a trial-and-error methodology for algorithm refinement. 
This iterative approach, while often necessary given the complexity of the problems at hand, can inadvertently open doors for unanticipated errors or biases to creep in. 
These potential pitfalls aren't just technical challenges; they have ethical and practical implications, especially when algorithms make decisions that affect human lives. 
This very conundrum has been the catalyst for the recent surge in advancements within Machine Learning Theory. 
Researchers in this field are working to deepen their understanding, with a particular focus on forging a theoretical framework wherein the inner workings of these algorithms are not only explicit but also comprehensively elucidated. The ultimate goal is to make these algorithms not just robust and resilient, but also crystal clear in how they work. This will boost their reliability and ensures they're safe to use.



\subsection{Machine Learning Paradigms}

Consider the actions of the human brain when attempting to catch a mosquito\footnote{A challenge I've just faced!}.  In an instant, it analyzes visual input, anticipates the mosquito's flight path, and orchestrates a coordinated body response to snag that tiny menace. It is remarkable how seamlessly the brain executes simultaneously complex and different tasks. Furthermore, it acquired this skill independently, without external aid. In your early years of life, your brain learns to identify a mosquito by processing numerous examples of this creature. It comprehends how to predict future motions using preceding frames of an action and masters body movement through countless trials and errors. Truly, the human brain is an astounding machine!

Regrettably, in the domain of Machine Learning, we are not yet able to construct a single algorithm capable of simultaneously performing such varied tasks. This limitation arises partly due to different tasks requiring diverse datasets and distinct interaction methods for learning\footnote{Computational power and energy supply can also be limiting factors: the human brain can perform the equivalent of $10^{18}$ mathematical operations per second using just 20 watts.}.

These variances have led to the development of different approaches, termed learning paradigms. Although numerous paradigms exist, allow me to introduce just the three principal ones.

\paragraph{Supervised Learning}

Imagine a teacher guiding a student through a math problem, where the student attempts to solve it, and the teacher provides the correct answer afterward.
Over time, with enough practice and corrections, the student starts making fewer mistakes.
Similarly, in supervised learning, the algorithm (or ``student'') learns from the training data (or ``lessons'') you provide, continually adjusting its predictions based on the actual outcomes (or ``correct answers'').  
The training dataset typically consists of input data paired with the correct outputs. 
The goal of these algorithms is to leverage this data to develop a mathematical function that, when presented with new inputs, can predict the corresponding outputs. 
This is accomplished by iteratively tweaking and fine-tuning this function, with the aim of minimizing errors, measured by the so-called loss function. 
Supervised learning algorithms have a large spectrum of uses, even in our daily lives, often unbeknownst to us. 
For instance, when we check our emails, these algorithms work behind the scenes, analyzing incoming messages and filtering them as spam or not based on learned patterns from labeled data. 
Similarly, in the realm of financial forecasting, they sift through historical stock data, learning patterns, and making predictions for future prices. 
The healthcare system also leverages supervised learning to enhance diagnostic accuracy, where algorithms analyze patient data and images, helping professionals diagnose diseases by drawing upon past instances where inputs (symptoms or image data) and outputs (diagnoses) were explicitly provided during training. 
Furthermore, e-commerce platforms employ these algorithms to curate personalized shopping experiences, predicting your next potential purchase by learning from your past browsing and purchase history. 
In other words, supervised learning has found applications where algorithms need to make predictive decisions based on previously learned examples.

\paragraph{Unsupervised Learning}
Imagine a curious child handed a box of various shaped blocks without any instructions. 
Over time, the child might begin grouping the blocks based on similarities, like color, size, or shape, drawing conclusions based on patterns they observe. 
In the realm of unsupervised learning, algorithms behave similarly, diving into datasets without any explicit instructions or labeled responses. 
Instead of being ``told'' the right answer, they seek to unearth hidden structures and patterns within the data. 
Everyday applications are abundant.
In marketing, unsupervised learning aids businesses in customer segmentation, identifying clusters of customers with similar purchasing behaviors, even when no initial categorization exists.
In social media platforms, unsupervised learning might detect emerging trends or topics by clustering similar posts or discussions. 
In genetics, it can be employed to identify groups of genes with similar expression patterns. 
Without needing explicit guidance, unsupervised learning algorithms venture into vast data terrains, identifying patterns and connections that might elude the human eye, enhancing our understanding, and offering novel insights.

\paragraph{Reinforced Learning}
Imagine a player in a video game navigating through various levels, where each action leads to a consequence, either earning points or facing setbacks.
As the player progresses, they refine their strategies based on the feedback received, whether it's leveling up or being defeated.
Similarly, in reinforcement learning, the algorithm (or ``player'') interacts with an environment and takes actions that yield either rewards or penalties. 
The aim is not just to make correct decisions but to maximize the cumulative reward over time. 
These algorithms essentially ``play'' through countless scenarios, fine-tuning their strategies based on the feedback received, encapsulated as the reward signal.
Reinforcement learning is used in an extensive number of real-world scenarios. 
For example, in robotics, these algorithms guide robots to adapt to new terrains and challenges, learning optimal paths and actions through trial and error. 
In the world of finance, reinforcement learning aids in optimizing trading strategies by evaluating the long-term benefits of trading decisions. 
In personalized content recommendation, like on streaming platforms, the algorithms adapt to user preferences over time, optimizing suggestions based on the user's past interactions and feedback.
Moreover, adding a touch of nuance, within video games themselves, machines aid players in crafting innovative strategies, even for time-honored games such as Chess or Go.
In essence, reinforcement learning thrives in environments where decision-making is sequential, outcomes have delayed repercussions, and the aim is to optimize for long-term rewards.

\subsection{Neural Network: from the Neuron to the Brain}
In the dynamic landscape of machine learning, few tools have made as significant an impact as Neural Networks. 
These computational models, over the past few years, have not just risen to prominence, but have also outshone many other techniques, setting benchmarks and opening up avenues previously thought impossible. 
They have ingrained themselves as central components in an array of industries, from finance and healthcare to entertainment and transportation.
As implied by their name, Neural Networks draw inspiration from certain aspects of the brain, a parallel that has also influenced the evolution of their architecture, which will be explored in this section. A deeper dive into the technical aspects will be reserved for the following section.  


\subsubsection{Examples of architectures}
In this section, we aim to traverse the journey of neural networks from their conceptual genesis to their contemporary relevance. 
We will talk about the most important architectures that have marked milestones in the evolution of the field, emphasizing the foundational principles that enable their functionality.

\paragraph{Perceptron}
The Perceptron, introduced by Frank Rosenblatt in 1957 \cite{rosenblatt1958perceptron}, is one of the earliest and simplest forms of neural networks and holds an important place in the history of artificial intelligence and machine learning. 
Initially conceived as a machine, the fundamental idea behind it is to mimic the basic function of a neuron in the human brain.
This algorithm takes multidimensional binary inputs, processes them, and produces a single binary output. 
This ``processing'' is akin to a neuron firing: if the combined inputs surpass a certain threshold, the outputs will be ``1'' (or ``fires''), and otherwise, it outputs will be ``0''.
Each input has an associated weight, and these weights determine the importance or influence of each input on the final output. 
To determine its output, the Perceptron multiplies each input by its weight, sums the products, and then passes this sum through a step function, often called an activation function. 
During the learning process, these weights are adjusted to optimize the accuracy. 
One of the main limitations of the original instance is its inability to handle non-linear problems.
It works well for problems where data can be separated into categories by a straight line, termed linearly separable problems. 
However, for more complex, non-linearly separable problems, this algorithm will fail.
Though the Perceptron may seem simplistic in light of contemporary advancements, its inception marked a transformative moment in machine learning. 
Its foundational principles served as cornerstones, paving the way for the sophisticated neural network architectures that dominate the landscape today.

\paragraph{Support Vector Machine}

The Support Vector Machine (SVM) \cite{cortes1995support}, developed in the 1990s, is a significant milestone in the realm of machine learning, particularly known for its prowess in classification tasks.
Contrary to simpler linear classifiers, SVM does not merely seek a decision boundary; it aims to find the optimal hyperplane that distinctly classifies data points of different classes with the maximum possible margin. 
This hyperplane is determined by a subset of data points termed ``support vectors'', which essentially lie closest to the boundary of opposing classes. 
By maximizing the margin between the support vectors of two classes, SVM ensures a better generalization capability in unseen data. 
In cases where the data is not linearly separable, SVM leverages kernel tricks—functions that transform the data into a higher dimension where it becomes linearly separable. 
This versatility allows SVM to tackle complex, non-linear problems.
While SVMs might seem overshadowed by the rise of deep learning techniques in recent times, their core concept of maximizing margins between classes has deeply influenced various domains of machine learning, reinforcing SVM's fundamental role in the field's evolution.

\paragraph{Deep Linear Network}
The Deep Linear Network (DLN) or Multi-Layer Linear Perceptron (MLLP) is characterized by its multi-layered structure, typically consisting of an input layer, several hidden layers, and an output layer. While it fundamentally employs linear transformations, similar to the basic Perceptron, the DLN is distinguished by its layered architecture. In theoretical contexts, the DLN is particularly appreciated for its ability to model multi-layered yet solvable networks.

\paragraph{Fully-Connected Deep Neural Network}

The Fully-Connected Deep Neural Network (DNN) \cite{rumelhart1986learning} represents a significant advancement beyond the basic Perceptron and MLLP, introducing a more complex and versatile architecture that has greatly enhanced the capabilities of neural networks. 
Similar to the DLN, DNNs consist of multiple layers of neurons. 
However, a crucial difference lies in the inclusion of non-linear functions, such as sigmoid or ReLU, applied to the outputs of neurons in the hidden layers. These non-linearities allow DNNs to learn and represent more complex patterns and relationships within the dataset. 
The introduction of DNNs was a crucial step in machine learning, highlighting the potential of layered neural architectures in addressing complex computational problems.
Their success has significantly propelled the proliferation, development, and application of Neural Networks, establishing them as an essential tool in Computer Science. 
Despite the advent of more advanced deep learning architectures, DNNs continue to be a foundational element, forming the cornerstone of many neural network innovations. This thesis delves into the theoretical aspects of these influential structures.

\paragraph{Convolutional Neural Network}

The Convolutional Neural Network (CNN) \cite{fukushima1980neocognitron} is a specialized type of deep neural network that is especially proficient in handling grid-like data structures, such as images. 
Its design draws inspiration from the human visual system, particularly the structure and function of the optic nerve, which processes visual information from the retina before it reaches the brain. 
Unlike traditional fully connected networks, CNNs exploit spatial hierarchies in the data through the use of convolutional layers, which apply filters to local features, enabling the network to recognize patterns ranging from simple edges to complex structures.
As data progresses through the network, pooling layers reduce its spatial dimensions, focusing on the most crucial information while discarding redundancies. 
This design not only aids in reducing computational load but also helps in achieving translational invariance, allowing the model to recognize features regardless of their position in the input, mirroring how our eyes and optic nerve process visual stimuli.
Another significant advantage of CNNs is weight sharing, which reduces the number of parameters, enabling them to efficiently process high-dimensional data.
CNN's groundbreaking capacity to directly learn from image data without the need for manual feature extraction has revolutionized fields like computer vision, leading to significant advancements in image classification, object detection, and more.
Its introduction underscored the power of tailoring neural network architectures to specific data characteristics, opening doors to a myriad of applications and further innovations in deep learning.

\paragraph{Recurrent Neural Network}
The Recurrent Neural Network (RNN) \cite{williams1989learning} represents a significant advancement in the realm of neural architectures, specifically tailored to handle sequential data. 
Unlike traditional feedforward neural networks, RNNs are characterized by their ability to maintain a form of ``memory'' through internal loops, enabling them to process information from previous inputs and use it to process the next ones in the sequence. 
This unique structure makes them particularly adept at tasks involving time series data, natural language processing, speech recognition, and other applications where temporal dynamics and context from earlier inputs are crucial. 
The essence of the RNN lies in its capacity to remember patterns over time, allowing it to recognize dependencies in sequences and make informed predictions based on prior information. 
This capacity to process and predict sequences has solidified its position as a fundamental tool in modern machine learning toolkits, addressing challenges that traditional neural networks could not tackle. 
An important evolution of these algorithms is the Long Short Time Networks, that have the capability  to retain information for long periods. This feature makes them extraordinarily effective for applications such as language modeling and translation, speech recognition, and anomaly detection in time series data. By maintaining a more stable gradient during training, they can learn from and remember extensive sequences of data, far surpassing the capabilities of traditional RNNs. The architectural advancements embodied in LSTM networks thus represent a significant step forward in the evolution of neural network technologies, making it possible to delve deeper into complex sequential data and extract meaningful patterns that were previously inaccessible.

\paragraph{Attention layer \& Transformers}
The Attention mechanism, introduced \cite{vaswani2017attention} within the domain of neural networks, has revolutionized various tasks, especially in the context of natural language processing and machine translation. Unlike traditional architectures that process input data uniformly, the Attention layer allows models to focus on specific parts of the input, dynamically determining which segments are more relevant to the task at hand. 
This mimics the human cognitive process of paying ``attention'' to particular details while overlooking others. By enabling this selective focus, models equipped with Attention layers have achieved remarkable improvements in efficiency and accuracy, addressing long-standing challenges like handling long sequences and managing dependencies between distant words or features. 
Today, the concept of Attention has become central in the design of state-of-the-art neural architectures, most notably in models like the Transformer.
This architecture, introduced in the paper ``Attention is All You Need'' by Vaswani et al. in 2017, represents a significant shift in the world of neural network-based models.
In fact, Transformers entirely rely on attention mechanisms to draw global dependencies between input and output, allowing them to process input data in parallel, vastly improving efficiency and scalability. 
The architecture splits its attention mechanism into what is termed ``self-attention'', allowing each component in the input sequence to focus on different parts of the sequence, thus capturing a diverse range of relationships. 
Since its introduction, the Transformer model has become the foundation for several state-of-the-art models in tasks ranging from machine translation to text generation, most notably models like BERT, GPT, and T5. The Transformer's ability to handle complex patterns and relationships in data has solidified its place as a cornerstone in modern deep-learning techniques.

\bigskip

\section{Neural Network: an Introduction}
As highlighted in Section \ref{1MLSM:one}, the development of a comprehensive and robust theory for Neural Networks is not just a scholarly pursuit but a crucial step towards understanding their inner workings, optimizing their performance and ensuring their safe utilization. 
However, the quest for such a theory has proven elusive, prompting much debate and research. 
In the initial portion of this section, we will navigate through the fundamental components that make up Neural Networks. 
These elements form the bedrock upon which any theory must be constructed. 
Subsequently, we will address the prevailing questions that permeate this domain. 
We will describe some of the challenges faced by the community and also highlight the diverse strategies and approaches researchers have adopted to answer them. 
Through this dual exploration, we aim to offer a panoramic view of the current landscape and the continued efforts to refine our understanding of Neural Networks.

\subsection{The four main ingredients}
The outstanding performance of Neural Networks (NN) is largely attributed to the intricate interplay of their core components. In this section, I will provide a propaedeutic exploration, emphasizing the mathematical aspect of these essential components. For clarity purposes, I will consider only the supervised setting - the focus of this Thesis - but all the concepts can be easily generalized with some caution.

\subsubsection{Dataset \& Task}

Machine Learning algorithms intrinsically hinge on their datasets to unearth patterns, making the dataset a vital ingredient in the learning process. 
In supervised learning, the focus on this Thesis, the dataset comprises pairs of input samples $x^\mu$  and their corresponding labels $y^\mu$, which signify the desired prediction or the outcome of a specific task.
Input samples typically are elements from a metric space, with  $\mathcal{R}^N$ being a prevalent choice. 
Labels, versatile in nature, might manifest as real numbers or vectors (with dimension $N^{out}$), or even categorical variables, leading us to talk about regression in the former instance and classification in the latter.
A standard practice in the world of machine learning is to partition the dataset. The larger part is typically called the training set. 
This set is essential, serving as the primary source from which the algorithm learns. 
The smaller counterpart, named the test set, is utilized to gauge the algorithm's ability to generalize its learning to similar but unseen data.
Interestingly, from a mathematical standpoint, one might observe that the structure of datasets hasn't been the spotlight of extensive research. 
However, there are some detailed studies of how different structural (and geometrical) aspects of a dataset influence the learning and the predictive capabilities of neural network models \cite{NEURIPS2022_c23f3852,PhysRevE.102.032119,erba2020random}. 

\details{
\begin{center}
    \textbf{A simple example: MNIST \& digits recognition.}
\end{center}
In these ``A simple example'' \textit{details} sections we will present a simplified picture of a real-world application of neural networks. 
These examples not only aid in grasping the four main ingredients of these algorithms but also serve as a foundation for the topics addressed in the upcoming chapter.

We turn our attention to one of the most renowned datasets in this field: MNIST \cite{lecun1998gradient}, in Fig. \ref{MNISTfig}.
This dataset comprises $60,000$ images of single-digit handwritten numbers, ranging from $0$ to $9$.
Each image is represented as a $28x28$ matrix, whose entries represent the image grayscale. 
Each pixel quantifies the ink intensity left by the pen, with values ranging from $0$ (indicating no ink) to $255$ (representing maximum ink intensity). 
While each image initially consists of this $28\times 28$ grid, it's common practice to convert each data point into a $728$-dimensional vector.
MNIST is frequently employed in a supervised learning context, as each data point (a single image) is associated with a label that corresponds to the digit written in the image.

While the MNIST dataset resides in a $728$-dimensional space, it becomes increasingly apparent that a significant portion of these dimensions holds little to no valuable information.
Consider, for instance, the border regions of the images, which consistently remain free of ink due to the central framing of the handwritten digits. 
Researchers have conducted analyses that reveal the intrinsic dimensionality of this dataset, representing the minimum number of essential variables needed for a concise representation, to be less than 15 \cite{PopeZAGG21}. 

This intriguing observation highlights a crucial characteristic of MNIST—its inherent simplicity.
In fact, even the most straightforward neural architectures can efficiently learn from this dataset, making it an ideal and widely accepted testing ground within the field of machine learning and computer vision.
Its ease of use allows researchers to focus on the development and evaluation of novel algorithms and techniques, ultimately advancing the broader understanding of deep learning.
As a result, MNIST (and the comparison with other more complicated, but well-defined, datasets) has played a pivotal role in benchmarking models, fostering innovation, and serving as a foundational resource for the machine learning community. 
Its simplicity, paradoxically, has made it an invaluable asset for tackling more complex challenges in the realm of artificial intelligence.

In the next \textit{simple example} sections, we will see how to build a neural network that accurately recognizes digits in the MNIST dataset.
We represent our dataset as a collection $\mathcal{T}$ of $P$ pairs, where each pair consists of a datapoint $x^\mu \in \mathbb{R}^{728}$ and its corresponding label $y^\mu \in \mathbb{R}$, with $\mu$ ranging from 1 to $P$.}

\afterpage{
\begin{figure}
    \centering
    \includegraphics[width=0.8\textwidth]{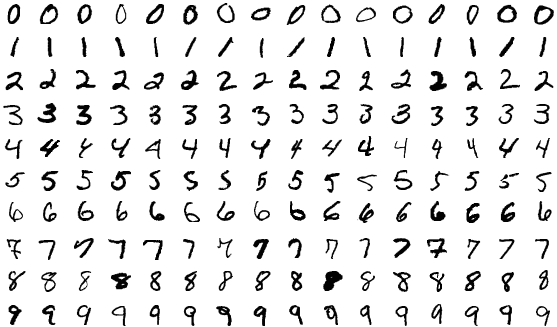}
    \caption{\textbf{The MNIST dataset.} 
    This dataset contains $60,000$ images, each depicting a single handwritten digit ranging from $0$ to $9$. Every image is formatted as a $28\times 28$ matrix, where each matrix element corresponds to a grayscale value in the image. The values of each pixel vary from $0$, denoting the absence of ink, to $255$, indicating the highest intensity of ink. Image taken from \url{https://en.wikipedia.org/wiki/MNIST_database}.} 
    \label{MNISTfig}
\end{figure}
\clearpage
}

\subsubsection{Architecture}
The architecture of the algorithm emerges as the second main ingredient in the creation of a Neural Network. 
As previously explored, there's a vast spectrum of potential designs when constructing these algorithms.
While they may appear divergent — especially when examined mathematically — they fundamentally are a variational ansatz of a high-dimensional function. 
At their core, these algorithms take input and channel it through a series of (possibly non-linear) transformations. 
This transformation process, governed by the architecture, is intricately modulated by the variational parameters of our ansatz, more commonly referred to as weights. 
To elaborate, these algorithms process input data and methodically subject it to a succession of linear and non-linear operations. 
Each network node or computational entity usually termed a \textit{neuron}, calculates the weighted sum (determined by the weights themselves) of its inputs, then employs a non-linear function, known as the \textit{activation function}, to generate its output.
In this Thesis, we focus on feedforward networks, where the input is processed sequentially through a series of layers. In each given layer, every neuron combines linearly (according to the relevant weights) the outputs of the previous layer neurons. This linear combination, called preactivation is then filtered through a nonlinear activation function and then transmitted to the next layer. The first network layer (called the input layer) directly inputs the data, while the last one provides the network output.
The simplest kind of linear combination used to transfer the information from one layer to the next is global coupling (each neuron output is transmitted to all neurons in the next layer), giving rise to the so-called fully connected (FC) neural networks. However, other schemes are possible, such as Convolutional Neural Networks (CNNs) - typically used to process images - where, in order to preserve the spatial nature of the input, different neurons are only connected to different regions of the previous layer.
The number of neurons in the network obviously influences its capability to store and process information. In feedforward networks neurons are organized in L layers (the network depth), each composed of $N_i$ neurons (the width), with $i=1,\ldots,L$  
The kind of linear interactions between individual neurons, their number and organization and the kind of nonlinear activation function used collectively constitute the architecture of a neural network and are typical subjects of study in this field.
\details{
\begin{center}
    \textbf{A simple example: Fully Connected 1 Hidden Layer Neural Network.}
\end{center}
For our \textit{simple example}, we consider a basic neural network architecture. While Convolutional Neural Networks (CNNs) are the preferred choice for image-related tasks (as our dataset MNIST), for pedagogical reasons, we will opt for a fully connected (FC) neural network with one hidden layer (1HL).

This neural network comprises three layers of neurons: the input layer with a fixed number of neurons ($728$ in our case), the second layer known as the hidden layer, responsible for linear and non-linear transformations, and finally, the output layer. Although a one-dimensional output may seem intuitive, we must remember that we are dealing with a classification task. Hence, our output layer will consist of a vector with a dimensionality equal to the number of classes (which is $10$ in this context).
This neural network can be written as a variational function:
\begin{equation}
f_{\text{DNN}}(x;\mathcal{W};\mathcal{B})_i = \frac{1}{\sqrt{N_{\text{hid}}}}\sum_j^{N_{\text{hid}}} v_{i,j}\; \sigma\left(\sum_k^{N_{\text{in}}}\frac{W_{j,k} x_k}{\sqrt{N_{\text{in}}}} + b_j\right) + b_i\;.
\end{equation}
Here, $x$ is the input, $\mathcal{W} = \{v, W\}$ represents the weights of the first layer\footnote{Between the input and the hidden layer.} and the second layer\footnote{Between the hidden layer and the output.} respectively, while $\mathcal{B}={b}$ are the biases. $\sigma(*)$ denotes the activation function, and $N_{\text{in}}$ and $N_{\text{hid}}$ correspond to the number of neurons in the input and hidden layers, respectively.
Note that normalizing the preactivation by the network width is a common procedure. The square-root choice of this optimization, via the Central Limit Theorem, implies a zero-average, random distribution of the weights.

Initially, the weights are chosen out of some zero mean distribution (typically Gaussian), and our objective will be to optimize the weights for improve the prediction accuracy. 
Despite its simplicity, this model is among the basic architectures for deep neural networks, yet it can achieve remarkably accurate results for our task (i.e. learning to recognize the MNIST digits). 
This phenomenon stems from the mathematical insight that a fully connected single hidden layer deep neural network (FC-1HL-DNN) serves as a Universal Approximator \cite{hornik1991approximation} and its approximation performance depends on the number of neurons in the hidden layer.

For a more complete understanding, and given its central role in this Thesis, let me introduce also neural networks with more than one hidden layer. As discussed in the previous section, in a multi-layer architecture, the linear and non-linear transformations are iteratively applied across each layer. For instance, a Deep Neural Network (DNN) with \( L \) layers can be expressed as follows:
\begin{equation}
    \begin{split}
        f_{\text{DNN}}(x;\mathcal{W};\mathcal{B})_j &= h^{(L)}(x)_j  \\
        z^{(\ell)}_j(x)&=\sigma\left(h^{(\ell-1)}_j(x)\right) \;,\\
        h^{(\ell)}(x)_j &= \sum_k^{N_{\ell}}\frac{W^{(\ell)}_{j,k} z^{(\ell)}_k}{\sqrt{N_{\ell}}} + b_j^{(\ell)}\;,\\
        z^{(0)}_k(x) &=x_k\;,
    \end{split}
\end{equation}

where \( h^{(\ell)}_j \) are the preactivations of the \( \ell \)-th layer, \( \sigma \) is the activation function, and \( N_\ell \) is the number of neurons in the \( \ell \)-th layer.
}

\subsubsection{Loss Function}

To enhance the performance of these algorithms, one must first quantify the discrepancy between the current predictions (the network output) and the actual input label. 
This necessitates the formulation of a function that gauges the average error the algorithm commits over the training set. 
This metric, widely referred to as the \textit{Loss Function}, is our third ingredient. 
While a lot of Loss Functions exist, their core objective remains the same: to evaluate the performance of the Deep Neural Network (DNN) in its present state and to guide the training in adjusting the weights for improved predictions.

\details{
\begin{center}
    \textbf{A simple example: Cross Entropy Loss.}
\end{center}
Now we want to define a function that quantifies the disparity between the predicted values and the actual labels generated by our Deep Neural Network (DNN). 
In the context of our classification problem, where the output involves multiple classes (in this case, with the number of classes $C = 10$), this scenario typically involves two fundamental steps: normalization of the output using a procedure known as SoftMax, followed by the application of the Cross-Entropy function as a distance metric.

The SoftMax function, denoted as $s(h)$, transforms the raw model outputs ($h_i$) into a probability distribution. It does this by exponentiating each output component $h_i$ and then normalizing it by the sum of exponentiated values across all classes:
\begin{equation}
    s(h)_i = \text{SoftMax}(h_i) = \frac{e^{h_i}}{\sum_j^C e^{h_j}}\;.
\end{equation}
The result is a set of probabilities, where each value represents the likelihood of the input belonging to a specific class.

Subsequently, the Cross-Entropy function, denoted as \( l(y, x) \), measures the dissimilarity between the predicted probabilities, \( s(x)_i \), and the true labels, \( y_i \), which are expressed as one-hot vectors (where all elements are ``0'' except for the element corresponding to the correct category, which is ``1'').
It does this by computing the negative logarithm of the predicted probabilities for the correct class labels and summing these values across all classes:
\begin{equation}
    l(y, h) = -\sum_i^C y_i \log\left(s(h)_i\right)\;.
\end{equation}
This ``distance metric'' effectively measures how much the model's predictions deviate from the actual labels for a single example. In essence, it captures the divergence between the predicted probability distribution and the true distribution represented by the labels, serving as a crucial component in training and evaluating the performance of our neural network.

As we already pointed out, the theoretical analysis carried on in this thesis work is primarily focused on regression problems.  For this setting, the most commonly used loss function is the Mean Squared Error (MSE). This function calculates the squared norm of the distance between the current prediction of our DNN and the target value we aim to predict. The MSE is defined as:
\begin{equation}
    l(y, h(x)) = (y - h(x))^2\,,
\end{equation}
where \( h(x) \) represents the scalar output of the network associated with the input $x$, and the scalar \( y \) is the target value (the ``input label'') we seek to predict. 
}

\subsubsection{Optimization algorithm}
The final ingredient essential to the functionality of Deep Neural Networks (DNNs) is the optimization algorithm that let the DNN  ``learn'' the dataset.
Optimization is achieved by adjusting and fine-tuning the network's weights, based on feedback from the loss function and contingent upon the training set. 
It should drive the DNN towards a solution that minimizes the Loss Function, thereby giving predictive power to the network. 
Stochastic Gradient Descent (SGD) emerges as one of the most iconic and effective of these algorithms, celebrated for its simplicity and efficacy. Nevertheless, the realm of optimization is diverse, hosting a wide array of strategies, each with its unique flair and characteristics \cite{bottou2010large,robbins1951stochastic,kingma2014adam}.

\details{
\begin{center}
    \textbf{A simple example: Gradient Descent.}
\end{center}
Now, we aim to update the weights to minimize the loss function. While various approaches and techniques exist for this task, we've opted for the simplest one: Gradient Descent. This optimization algorithm comprises three main components.
Firstly, it computes the average value of the loss function summed over the entire training dataset.
\begin{equation}
    \mathcal{L}(\mathcal{W},\mathcal{B}) = \left\langle\;  l\left(y^\mu,f_{\text{DNN}}(x^\mu;\mathcal{W},\mathcal{B} )\right)\; \right\rangle_{\mathcal{T}} = \frac{1}{P} \sum_\mu^P l\left(y^\mu,f_{\text{DNN}}(x^\mu; \mathcal{W},\mathcal{B})\right)\;.
\end{equation}
Following that, the algorithm computes the gradients of the loss function with respect to the weights and biases. To achieve this, it employs a widely used approach known as Backpropagation \cite{Linnainmaa1976taylor}, which simplifies the derivative calculation in feedforward networks, fundamentally by applying the chain rule. 
After that, it updates the weights and the bias in the (negative) direction of the gradient to reduce and eventually minimize the loss.
\begin{equation}
    \begin{split}
        \mathcal{W}(t+1) &= \mathcal{W}(t) - \gamma \nabla_\mathcal{W} \mathcal{L}(\mathcal{W},\mathcal{B})\;,\\
        \mathcal{B}(t+1) &= \mathcal{B}(t) - \gamma \nabla_\mathcal{B} \mathcal{L}(\mathcal{W},\mathcal{B})\;.\\
    \end{split}
\end{equation}
Here, the symbol $\gamma$ represents a fixed hyperparameter that controls the size of the step, ensuring that the updates are neither too large (avoiding overshooting jumping over the minima) nor too small (avoiding slow convergence).

This procedure is typically repeated a large number of times, often referred to as epochs, until the optimization problem converges. However, Gradient Descent has a significant limitation. 
Due to its deterministic nature and the complex landscape of the function we aim to minimize, there is no guarantee that it will converge to the global minimum of the loss function. 
In practice, it could become trapped in a local minimum (typical problems are non-convex and present several different local minimal with several suboptimal ones).
That's why stochastic variants are often preferred over purely deterministic Gradient Descent algorithms. In practice, adding some effective noise, allows the gradient descent dynamics to escape local minima that otherwise trap fully deterministic dynamics. 
A common way to introduce stochasticity is to consider the gradient of the loss computed on a single, randomly chosen, datapoint and to adjust the weight accordingly. In each training epoch, this is repeated over P randomly chosen datapoint. This random sequential ordering introduces an effective noise in the training dynamics. 
A computationally more efficient implementation, \textit{batch gradient descent}, amounts to dividing the entire training dataset into smaller subsets (batches) and iteratively updating the model's parameters using the gradient of the loss function calculated on randomly chosen batches.
This method effectively combines the computational efficiency of batch processing with the stochastic nature of single-instance updates, enhancing the overall robustness and speed of the model's training process.\\
Now that we have the four main ingredients of our ``simple example'', combining all these elements allows us to easily achieve an accuracy of about 90\%.
}


\subsection{FAQ}
With all the essential components and the operational flow of these algorithms in place, we are now ready to delve into some fundamental questions.
I present to you some of the most pressing theoretical problems, accompanied by pertinent state-of-the-art references. However, bear in mind that these questions are essentially open ones, that, for the time being, remain largely unresolved.

\subsubsection{Generalization capacity \& Overfitting}
Imagine you are trying to fit a line through a given set of points on a graph to predict a trend. 
Using a simple straight line (which can be represented by two parameters: slope and y-intercept) might capture the general direction of the points, giving you a decent understanding of the overall trend.
However, let's say you decide to use a polynomial regression, which has more parameters allowing it to create curves. 
You choose a high-degree polynomial that can have many wiggles and turns. 
With a large enough degree (and an equally large number of parameters to fit) this polynomial might pass through every single point on your graph, making it a ``perfect'' fit for your data.
The problem? When you introduce a new data point not in your training set, this polynomial, because of its many parameters and wiggles, might predict a value that is way off the mark. This is because the polynomial is too adapted to the training data and not flexible enough for new data.
In machine learning, this scenario exemplifies ``overfitting''. 
The model with many parameters has fit the training data exceptionally well, down to its noise and outliers, but might not perform well on new, unseen data. 
Essentially, it has become too specialized in the training set and lost its ability to generalize, even when applied to more data points from the same source.
In traditional machine learning algorithms, heavy regularization (i.e. technique that penalizes the model complexity) is often essential to mitigate overfitting. There are various types of regularization techniques, with the most common being Lasso regularization \cite{tibshirani1996regression} and Ridge regularization \cite{hoerl1970ridge}.
Yet, this is not the case for deep neural networks (DNNs). Surprisingly, despite their vast capacity — characterized by a large number of neurons, which can range from millions to billions and a vastly larger number of weights — DNNs do not overfit as easily as one might expect.
This intriguing characteristic forms the foundation of their success. 
In practice, it is often possible to design a DNN with an enormous number of neurons and parameters to optimize, placing it in the so-called overparameterized regime, without experiencing overfitting. 
Given its central role in the efficacy of these algorithms, this phenomenon has been extensively investigated from both theoretical  \cite{neyshabur2017pac, bartlett2017spectrally, arora2018stronger} and technical perspectives \cite{ioffe2015batch, he2015delving, wan2013regularization, srivastava2014dropout, goodfellow2014explaining, keskar2016large, arpit2017closer, wager2013dropout}.
Undoubtedly, one of the most seminal contributions in this domain is the paper by Zhang et al. \cite{zhang2016understanding}. Their experiments, which involved training on datasets with completely random labels and still achieving zero training error, underscored a crucial point: our traditional frameworks for understanding generalization, such as the classic bias-variance tradeoff and Occam's Razor, might not fully capture the nuances of deep learning \cite{yang2020rethinking}. You can find a more complete description of this problem in Chapter \ref{ch.2AB}.
Moreover, the research hinted at the possibility that optimization methods, particularly stochastic gradient descent, might inherently introduce regularization effects when training DNNs, offering another layer of explanation to the observed phenomena. Despite the notable efforts, the (lack of) overfitting problem remains poorly understood.

\subsubsection{Feature Learning \& Inner Representation}
Deep Neural Networks (DNNs) are renowned for their ability to learn hierarchical features through their layered structures. 
Early layers typically detect rudimentary patterns, such as edges or textures in image data. In contrast, deeper layers delve into more abstract concepts, resulting in a comprehensive and sophisticated interpretation of the data.
Yet, this intricate process, termed ``feature learning'' in machine learning, is laden with complexities. 
Key questions emerge: How exactly do these networks extract these features? 
At what stage or depth within the DNN does this extraction become effective? 
Furthermore, the criteria by which networks discern and prioritize certain features over others remain elusive. 
Indeed, understanding how these networks internally encapsulate features is pivotal, as it has significant implications for the network's capacity to generalize across diverse data.
This aspect of deep learning has piqued the curiosity of many researchers. One of the groundbreaking contributions in this area is the work by Zeiler and Fergus \cite{zeiler2014Visualizing}, which introduced visualization techniques to decipher the inner workings of convolutional networks. This work \cite{olah2017feature} on feature visualization provides insights into what neural networks ``see'' at each layer. Other studies, such as \cite{radford2015unsupervised}, explore the latent space of generative models to understand the richness of these inner representations.
This and other works \cite{bau2017network, bengio2013representation, goodfellow2016deep, mahendran2015understanding, yosinski2015understanding, mallat2016understanding, zhou2015cnnlocalization} illuminated how networks progressively build up complexity and abstraction through layers.
As with generalization and overfitting, the realms of feature learning and inner representation in DNNs continue to be areas of intense study and exploration, offering both challenges and opportunities for further breakthroughs.

\subsubsection{Structure and Goodness of Minimun}
Optimizing the large number of DNN parameters on typical tasks is not a simple convex problem, characterized by a single well-defined end easily found optimal configuration. 
Typically, it is rather a complex non-convex problem, driven by a loss landscape (the hypersurface defined by the loss function values in parameter space) with a complex structure and a large number of local minima.
DNNs explore these intricate loss landscapes in search of the optimal model for a given dataset.
This journey culminates in the network reaching a minimum on the loss surface, which ideally translates to superior performance on similar unseen data. Yet, not every minimum is equally beneficial. 
The structure and quality of these minima have evolved into central themes of deep learning research.
In fact, given the vast configuration space that DNNs operate within, a series of questions arise: How do DNNs consistently converge to effective minima, and what defines the ``goodnes'' of a minimum? 
Do DNNs always converge to global minima, or do they often find only local minima? 
How do architectural attributes, such as depth and width, shape the found minimum's quality?
It's noteworthy that some local minima may lead to better generalization than others, even if their training accuracies are similar. 
The intricacies of the loss landscape—factors like the sharpness or flatness of a minimum—play a pivotal role in determining a model's robustness and its ability to generalize.
This domain has magnetized extensive research, driven by the implications it holds for training stability and model elucidation. Noteworthy is the work by Keskar et al. \cite{keskar2016large}, which bridges the relationship between minima sharpness and generalization.
Other works, such as \cite{hochreiter1997flat}, emphasize the importance of flat minima. Concurrently, studies like that of Li et al. \cite{li2017visualizing} harness visualization tools to elucidate the optimization landscapes confronting neural networks.

\bigskip
While some questions dominate the discourse in the field, the study of neural networks offers a vast expanse of intriguing avenues. 
For instance, the choice of optimization algorithm impacts performance, especially in terms of speed of convergence to the optimal solutions and thus computational efficiency. Although Stochastic Gradient Descent (SGD) has been the mainstay for a decade, emerging algorithms \cite{zhu1997lbfgs, liu1989lbfgs, dozat2016nadam, kingma2014adam, hinton2012rmsprop} are showcasing superior results. Current research investigates how different algorithms influence weight dynamics and the nature of minima achieved.
Likewise, the choice of non-linearity functions \cite{klambauer2017self, clevert2016fast, gers1999learning, nair2010rectified, lecun1998gradient} influences the loss landscape structure, calling for deeper explorations into their distinct implications.
Regularization techniques are pivotal in shaping network dynamics. 
Tools like Dropout \cite{srivastava2014dropout} -  where some backward and forward connections are randomly and temporarily removed until the next training epoch - and Weight Decay \cite{krogh1992simple}-  a form of L2 regularization that modifies the loss function to penalize large weights - have demonstrably improved the overall performance.
Yet, the interplay between these techniques and the loss landscape structure remains a fertile ground for investigation \cite{gal2016dropout, huang2017understanding, huang2017understanding, goodfellow2016deep, bengio2003regularization}. 
A better understanding of this relationship has the potential to help in developing more robust and adaptive neural architectures.

Moreover, ever since the advent of CNNs, the significance of symmetries in the evolution and efficacy of neural networks has been evident \cite{weiler2018symmetries, basri2020symmetry, scellier2016symmetry,}. 
Harnessing insights from symmetries could illuminate the dynamics of learning trajectories (i.e. the evolution during training of the network weights) and ultimately enhance pattern recognition capabilities.
Navigating the intricate paths of Deep Network Theory, it becomes clear that we're not just working with computational models. 
We're engaging with complex systems where every facet, be it an algorithm, a regularization strategy, or a design symmetry, contributes to the grand scheme of machine learning. 
It is thus natural to investigate DNNs with the array of tools developed by physicists to study complex systems.
The quest to decipher these elements offers the promise of not just enhanced algorithms but a profound insight into the essence of automated learning itself.

\section{Statistical Mechanics of Learning}
In this section, we will provide a concise introduction to the fascinating field of Statistical Mechanics of Learning. We will begin with a brief overview to lay the foundation, followed by an exploration of the common approach borrowed from statistical physics to tackle the challenges of deep learning. Our guidance in this endeavor will be drawn from one of the seminal books in this domain, ``Statistical Mechanics of Learning'' by Engel and Van den Broeck \cite{EngelVanDenBroeck}. In the final section of this chapter, we will present some of the key findings that have emerged from this interdisciplinary field.

\subsection{Why?}

Statistical Mechanics has a rich history of dealing with complex systems and modeling their behavior.
It typically begins with the intricate interactions among numerous constituent elements in a physical system and aims to unveil the macroscopic laws that govern the system's behavior.
This foundational approach has endowed Statistical Mechanics with remarkable versatility, allowing it to often transcend disciplinary boundaries.
It has ventured into various domains, including the structure of matter itself (exploring phenomena from ferromagnetism to superconductivity), biology (from the structure of proteins to the behavior of cellular motors) \cite{zaman2009statistical, bialek2012biophysics}, ecology (from the dynamics of bird flocks and the rule of entire ecosystems) \cite{fort2013statistical,}, soft matter (analyzing liquid crystals and self-repairing systems) \cite{kleman2003soft}, economics (in the form of econophysics) \cite{savoiu2013econophysics}, and countless other scientific fields.
The underlying principle of its success is straightforward: constructing a model entails capturing the large-scale essential features of a problem while abstracting away from the intricate microscopic details, thereby revealing the fundamental processes that govern the macroscopic system.

In the context of deep learning, one of the most pressing challenges has been the apparent ``black box'' nature of neural networks.
Understanding the underlying mechanisms driving these algorithms has been elusive.
However, Statistical Mechanics has emerged as a promising avenue to model and study these neural network black boxes.
As early as the mid-1980s, it became evident that applying the principles of Statistical Mechanics could provide a fresh and functional perspective on the study of neural networks.
Key contributions, such as the Hopfield model for neural networks and the work by the late Elizabeth Gardner on neural network capacity, marked pivotal milestones in this journey.

Over the past four decades, the field of Statistical Mechanics of Learning has expanded significantly.
It has taken on diverse challenges, each contributing to the development of models and frameworks aimed at deciphering the intricate behaviors of learning algorithms.
The overarching goal has been to lift the veil on these algorithms, moving beyond their ``passive'' use as mere black boxes solely engineered by trial and error empirical attempts.
Through the tools of Statistical Mechanics, researchers have endeavored to gain deeper insights into the underlying processes, dynamics, and optimization landscapes that govern the behavior of neural networks. 
In doing so, they have brought us closer to the day when these powerful learning algorithms are no longer shrouded in mystery but are understood and harnessed with precision.

\subsection{Common approach}\label{subsec:1CommApp}

In this section, we explore the fundamental concepts and methodologies employed in the field of Statistical Mechanics of Learning.
It's important to recognize that while the specific approach may vary based on the problem at hand, the foundational concept often commences with the calculation of the partition function, denoted as  $Z$.
In equilibrium systems, by meticulously accounting for all possible configurations of particles or variables and weighing them according to their associated energies, the partition function encapsulates a wealth of information about a system's microscopic constituents and their interactions, effectively bridging the gap between the microscopic world and macroscopic observables. 
Moreover, the partition function, or better its logarithm, defines the method by which ensemble averages of observables are computed, allowing researchers to gain insights into the macroscopic behavior of systems composed of many elements and make predictions about their properties, phase transitions, equilibrium states, and the behavior of specific essential thermodynamic quantities such as free energy, entropy, and specific heat.

Consider a system composed of a large number $N$ of components, each represented by a state \( s_i \). Let \( s = \{s_i\}_i^N \) denote the state of the entire system. The Hamiltonian of the system, denoted as \( H \), assigns an energy \( H(s) \) to a given state \( s \).

In the canonical ensemble, where the energy can fluctuate but the temperature \( T \) is fixed, the partition function \( Z \) is defined by the sum (or integral) over all possible configurations of the system's components, weighted exponentially by a factor derived from the system's energy, known as the Boltzmann weights:
\begin{equation}
    Z = \sum_{s} e^{-\beta H(s)} \;,
\end{equation}
where $\beta = 1/k_B T$ and \( k_B \) is the Boltzmann constant.
From this, we can derive the probability distribution of the states:
\begin{equation}
    \mathcal{P}(S) = \frac{1}{Z} e^{-\beta H(S)}\,.
\end{equation}

A key concept linking the microscopic and macroscopic worlds is the Free Energy, defined as:
\begin{equation}
    F = -k_B T \log Z \;,
\end{equation}
which coincidence with the thermodynamic Helmholtz Free Energy:
\begin{equation}
    F = U - TS \;,
\end{equation}
where \( U \) is the Internal Energy, and \( S \) is the microcanonical Entropy.

Various observables of interest, such as the average energy \( \langle E \rangle \) and the average magnetization M, can be directly obtained from the logarithm of the partition function (see the next \textit{details section}), which constitute the link to the thermodynamic functions of state and observables. 

In the canonical ensemble, the microscopic energy $H(s)$ is not constant but can fluctuate around its mean value. 
However, it can be shown that in the thermodynamic limit (\( N \to \infty \)), its fluctuations are negligible.
Consider, for instance, the probability $\mathcal{P}(E)$ of observing a given energy $H(s) = E$. It is given by the marginalization of the probability distribution over all the states with energy $E$:
\begin{equation}
    \mathcal{P}(E) = \sum_{S:H(S)=E} \mathcal{P}(S) = \frac{1}{Z} \sum_{S:H(S)=E} e^{-\beta H(S)} = \frac{e^{-\beta E}}{Z} \sum_{S:H(S)=E} = \frac{e^{-\beta E} \Omega(E)}{Z}\,,
\end{equation}
where $\Omega(E)$ is the phase space volume occupaied by microstates of energy $E$.
The corresponding Boltzmann entropy is $S(E) = k_B \log \Omega(E)$, so that we can rewrite:
\begin{equation}
    \mathcal{P}(E) = \frac{e^{-\beta (E - TS)}}{Z}  = \frac{e^{-\beta F(E)}}{Z} =  \frac{e^{-\beta N f(E)}}{Z}\;,
\end{equation}
where $F(E)$ is the (extensive) Free Energy function evaluated at energy $E$, and $f(E) = F(E)/N$ is the (intensive) Free Energy density.

For finite N and T, the largest probability will be given by the absolute minimum of the Free Energy $F(\bar{E})$, but different energy states have still a non-zero probability of being observed.
For $N \to \infty$ or $T \to 0$, however, probability \textbf{fully concentrates} on the energy $\bar{E}$ that minimizes the Free Energy density $f(E)$:
\begin{equation}
    \frac{\mathcal{P}(E)}{\mathcal{P}(\bar{E})} = e^{-\beta N [f(E) - f(\bar{E})]} \xrightarrow{\beta N \to \infty} \delta(E - \bar{E}) \;.
\end{equation}

In this limit $F(\bar{E})$ coincides with the thermodynamic Free Energy $F = U - TS$, and all microscopic configurations with energy $E \neq \bar{E}$ are completely negligible.
In this situation, we have that the partition function $Z= e^{-\beta F}$  only depends on the microstates such that $H(S) = \bar{E}$. It can be further shown that the relative fluctuations of the energy $H(S)$ vanish in the thermodynamic limit:
\begin{equation}
    \frac{\langle H^2 \rangle - \langle H \rangle^2}{\langle H \rangle^2} \xrightarrow{N \to \infty} 0 \;.
\end{equation}
In this scenario, we also find that \( U = \bar{E} = \langle H \rangle \) and all other thermodynamical quantities may be derived from the free energy \cite{Huangbook}.

\details{ 

\begin{center}
    \textbf{The Ising model.}
\end{center}

Let's momentarily step away from deep neural networks and revisit one of the most renowned systems in statistical mechanics—the Ising model. 
This system comprises interacting spins arranged on a lattice. 
Our goal is to identify the most probable configurations, around which the probability distribution concentrates in the thermodynamic limit (where the number of spin goes to infinity). We aim to measure the average values of specific observables associated with these states, while also taking into account the influence of temperature.
Mathematically, the partition function ($Z$) is the cornerstone of this endeavor. It is defined as the summation (or integration) of all possible configurations of the system's spin configurations. Each configuration is weighted by a factor derived from the system's energy, which can be expressed in terms of a Hamiltonian:
\begin{equation}
    H(\{s_{i}\}) = -J\sum_{i}^{N-1} \sigma_i \sigma_{i+1} - h\sum_i^{N} \sigma_i\;.
\end{equation}
In this expression, $\sigma_i = \pm 1$ represents the spins, $J$ denotes the spin interactions, and $h$ represents an external field. 
The Hamiltonian provided here pertains to the one-dimensional Ising model.
While there exist various methods to derive the partition function, it ultimately takes the form:
\begin{equation}
    Z= \prod_i^N \sum_{\{\sigma_i= \pm 1\}} e^{-\beta H(\{\sigma_{i}\})} =  \prod_i^N \sum_{\{\sigma_i= \pm 1\}} e^{-\beta \left(-J \sigma_i \sigma_{i+1} - \frac{h}{2}(\sigma_i+\sigma_{i+1}) \right)} \;.
\end{equation}
Commonly, this expression must undergo modifications to become practically useful. 
However, the effort expended in this endeavor is well justified, as it opens the door to an array of valuable insights. 
By effectively manipulating this expression, we can calculate the average values of observables with relative ease. To illustrate, let's focus on computing the total energy of the system:
\begin{equation}
    \langle E \rangle = \sum_{s} E(s) \mathcal{P}(s) = \frac{1}{Z}\sum_{s} E(s) e^{-\beta E(s)} = -\frac{\partial \log Z}{\partial \beta}\;,
\end{equation}
where \( s \) denotes the state of the system.
Similarly, we can consider consider the magnetization:
\begin{equation}
    M = \left\langle \sum_i^N \sigma_i \right\rangle = -\frac{\partial \log Z}{\partial h}\;,
\end{equation}
Furthermore, the use of partition functions equips researchers with a variety of techniques for understanding finite systems and extends the scope of investigation to systems composed of an infinite number of components. 
This broader perspective enables the exploration of the behavior of systems with a vast number of interacting elements. 
It allows for an in-depth analysis of collective properties, phase transitions, and emergent phenomena that arise in infinitely complex systems.
}

This foundational concept appears to adapt seamlessly to the context of deep learning. Indeed, it is not difficult to draw an analogy between the Loss Function in learning algorithms and the energy of a physical system, with the weights acting as the system's components (i.e. the spins in the previous \textit{details} section). This analogy highlights the scope of the Statistical Mechanics of Learning: in the thermodynamic limit (which needs to be accurately defined), the probability distribution of the state concentrates around the minimum of the Loss Function. This is indeed the target of algorithmic optimization, the set of DNN parameters that minimize the loss.
By computing the partition functions:
\begin{equation}
    Z(\mathbf{x}) = \int \mathcal{DW} \; e^{-\beta \mathcal{L}(\mathcal{W},\mathbf{x})}  = \int \mathcal{DW}\; e^{-\frac{\beta}{P} \sum_\mu^P l\left(y^\mu, f_{\text{DNN}}(x_\mu,\mathcal{W})\right) }\;,
    \label{1:Zintegral}
\end{equation}
where the integral is carried on over all the DNN weights $\mathcal{W}$ (we assumed zero biases for simplicity), $\mathcal{L}$ is the loss function, and $\mathbf{x}$ is the training dataset, we obtain a result which mainly depends on the Loss function minimum. 
In the limit $\beta \to \infty$, it exclusively depends on the configuration of the weight corresponding to the Loss minimum.
Relevant observables, such as the training and the generalization error, can be derived from the logarithm of the partition function. In the limit $\beta \to \infty$ they will describe the properties of fully optimized (trained) DNNs.
Remarkably, this perspective allows us to study the typical behavior and performance of the optimized DNNs.
This approach has the potential to provide profound insights into the learning behavior of algorithms. It can shed light on the underlying principles that govern their convergence, generalization, and overall performance, offering a deeper understanding of the dynamics at play in complex learning systems.


However, within the domain of DNNs, we encounter a further difficulty w.r.t. simple models such as the Ising one described above.
In this context, two major components come into play. 
Firstly, we have the weights, which serve as direct analogs to the physical variables found in traditional physical systems. 
Secondly, we must consider another component of the system: the training set. 
The (finite) dataset assumes a fundamental role within the loss function, effectively influencing the system's ``energy'' (the Loss function) which depends on it. It follows that also the partition function \ref{1:Zintegral} depends on the given realization of the dataset over which it is optimized. 

Fortunately, this kind of situation is not foreign to Statistical Physics.
In fact, a multitude of techniques has been developed specifically to address and navigate this kind of problem. In the following \textit{details section} we present a classical example.

\details{
\begin{center}
    \textbf{The Edwards-Anderson model.}
\end{center}
Let us consider again the 1D Ising model, but with a slight difference: now the interaction between spins depends on the index of the spins themselves. The new Hamiltonian of the system is:
\begin{equation}
    H(\{\sigma_i\}) = -\sum_i^N J_i \sigma_i \sigma_{i+1}\;,
\end{equation}
where we have set $h=0$ for simplicity and $J_i$ represents the interaction between the $i$-th spin and the subsequent one.
In contrast to the previous case, the interactions between the spins are not uniform. Instead, they are randomly drawn from a probability distribution, typically a Gaussian distribution.

As before, our aim is to calculate the partition function, a quantity defined as the summation (or integration) over all conceivable configurations of the system's variables. 
However, it's evident that the configuration that minimizes the energy, and thus dominates the partition function, may depend on the specific values of the interactions.  
Our partition function itself will depend on the specific realization of the interactions J.
Consequently, if we are unable to know directly the values taken by the interactions J (think of a real disordered ferromagnet, how do you access the microscopic interactions?) and/or if you are interested in a more general result, we must average over all possible configurations of the interaction vector $J$. 
Naively done, this leads us to the following partition function, often referred to as the \textit{``annealed partition function''}:
\begin{equation}
Z = \langle Z \rangle_J = \prod_i^N \int dJ_i \sum_{{\sigma_i= \pm 1}} e^{-\beta \left(-J_i \sigma_i \sigma_{i+1} - \frac{h}{2}(\sigma_i+\sigma_{i+1}) \right)} ;.
\end{equation}
However, upon closer examination, it becomes apparent that this approach has certain inconsistencies. 
Specifically, it treats the variables $\sigma$ and $J$ as if they fluctuate in unison. 
In reality, these two components are fundamentally distinct. 
The spins $\sigma$ represent the physical entities capable of changing during the system's evolution, while the interactions $J$ provide the quenched background against which the spins operate. 
In other words, our goal is to find the spin configurations that dominate the partition function given some ``quenched'' interactions and to later average over different interaction configurations.

To address this issue correctly, we must develop a new method for calculating this average. 
For a second, let our focus shift from the partiction function to a different (physical) object — the Free Energy ($F$), which possesses the property of being ``self-averaging''.
This means that it does not depend on the realization of disorder, provided the considered system is sufficiently large. 
Mathematically, we express this as follows:

\begin{align}
        F_N(J,\beta) &=  -\frac{1}{\beta} \log Z(J) =  -\frac{1}{\beta} \log \left(\prod_i^N  \sum_{\{\sigma_i= \pm 1\}} e^{-\beta \left(-J_i  \sigma_i \sigma_{i+1} - \frac{h}{2}(\sigma_i+\sigma_{i+1}) \right)} \right)\\ 
        & \\
        F(\beta) &= -\lim_{N\to\infty} F_N(J) =\\ 
        &= -\lim_{N\to\infty} \frac{1}{\beta} \int dJ  \; \log \left(\prod_i^N  \sum_{\{\sigma_i= \pm 1\}} e^{-\beta \left(-J_i  \sigma_i \sigma_{i+1} - \frac{h}{2}(\sigma_i+\sigma_{i+1}) \right)} \right)\\
        &= -\lim_{N\to\infty} \frac{1}{\beta}\, \langle\langle \log Z(J)\rangle\rangle_J = F^{(\infty)}(\beta)\,,
\end{align} 
where the double bracket $\langle\langle \cdot \rangle\rangle_J$ represents a quenched average.

In simpler terms, this property underscores a crucial point: when we aim to incorporate the interaction component into our calculations, we can't just directly average the partition function $Z$ over the probability distribution of interactions. 
Instead, what we need to average is the natural logarithm of $Z$. The reason for this choice is that $\log Z$ is an extensive quantity and possesses the desirable property of being self-averaging. Note that from the properly averaged free energy, one can then derive the full thermodynamics of the disordered system.
However, it's important to note that even in the simplest cases, this type of calculation can be extremely challenging. Nevertheless, there are methods and techniques, like the \textit{replica method} \cite*{mezard1987spin} which will be discussed in Chapter \ref{ch.2AB}, that can help us tackle these complexities and make progress in understanding such systems.}

Again, these techniques seem to adapt well to the case of DNNs. In fact, the influence of the dataset on the loss function can be likened to a conformational disorder, akin to the vector \( J \) discussed in the previous \textit{details} sections. For each given dataset, the partition function concentrates around the corresponding minimum of the loss. To obtain a more general, dataset-independent result, we may need therefore to perform a quenched average over the dataset distribution. 

Let's introduce the quenched free energy of our DNN:
\begin{equation}
    F(\beta,P) = -\frac{1}{\beta} \langle\langle \log Z(\{y_\mu,x_\mu\}_\mu^P) \rangle\rangle_{\mathcal{T}}\;,
\end{equation}
where $\mathcal{T} = (\{y_\mu,x_\mu\}_\mu^P)$ represent the distribution of our fixed dataset.
Importantly, this quantity no longer explicitly depends on the dataset itself.
The most interesting observables of these systems are the training error ($\epsilon_t$) and the generalization error ($\epsilon_t$), defined as:
\begin{align}
        \epsilon_t &= P^{-1} \left\langle\left\langle \int \mathcal{DW}\; \mathcal{P}(\mathcal{W})\, H(\mathcal{W}\;) \right\rangle\right\rangle_\mathcal{T}\\
        \epsilon_g &= \left\langle\left\langle \int \mathcal{DW}\; \mathcal{P}(\mathcal{W}) \int dx\,dy\, \mathcal{P}(x,y)\;l(x,y,\mathcal{W}) \right\rangle\right\rangle_\mathcal{T},
\end{align}
where the first represents the average error computed by the neural network on the training set, while the second is the error computed by the neural network over the input-output distribution $\mathcal{P}(x,y)$.
It's worth noting that:
\begin{equation}
\epsilon_t = \frac{1}{P} \frac{\partial \left( \beta F(\beta, P)\right)}{\partial \beta}
\end{equation}

The calculations involved are typically notably challenging, primarily due to the complexities in characterizing the probability distribution of the input-output relationship within the datasets. Moreover, the introduction of a hidden layer further complicates the analytical computation, rendering it infeasible without resorting to approximations. This is a common scenario in the statistical physics of strongly correlated systems. 
A notable approximation employed in this context is the Gaussian Equivalence Principle, as discussed in \cite{goldt2022gaussian}, where the distribution of neuron's output is approximated to a Gaussian. This approximation, along with others, will be explored in more detail in the upcoming sections. Additionally, the use of scaling arguments, as demonstrated in \cite{Seung1992Statistical}, can still yield valuable insights despite these complexities.

In Chapter \ref{ch.2AB}, this approach will be explored in depth, while Chapter \ref{ch.3BIW} will show how the dataset exerts a more significant influence on the behaviour of the weights than what is typically observed in simple conformational disorder.

By drawing from these well-established methodologies, we gain valuable tools to analyze and understand the intricate behavior of DNNs and their interactions with training data.
In the field of Statistical Mechanics of Learning, some of the most crucial observables measure the generalization properties of algorithms. 
Fortunately, the partition function offers convenient access to these important observables.

\details{\begin{center}
    \textbf{A simple example: the Perceptron \& Student-Teacher scenario.}
\end{center}

In this section, we explore a practical scenario that focuses on the foundational element of any Deep Neural Network (DNN), the Perceptron. 
As discussed earlier in this chapter, the Perceptron is an algorithm that accepts multi-dimensional binary inputs, processes them through its internal linear mechanism, and outputs a single binary value.

Mathematically, the Perceptron's function can be expressed as the sign of an affine transformation:
\begin{equation}
f_S(x) = \text{sign}(\mathbf{S}^T \mathbf{x})
\end{equation}
where $\mathbf{T}\in R^N$ are the weights and $\mathbf{x}$, also an N-dimensional vector, is an input data point.

Next, we need to choose a dataset for our example. In this case, we aim to construct our own dataset. To do this, we select a function, referred to as the \textit{Teacher}, and use it to generate the labels of our dataset.

For this example, we employ another perceptron as the \textit{Teacher}. The function of the \textit{Teacher} perceptron is defined as:
\begin{equation}
f_T(x) = \text{sign}(\mathbf{T}^T \mathbf{x})
\end{equation}
where the teacher vector $\mathbf{T}\in R^N$

It's important to note that in this example the norm of vectors $S$ and $T$ do not impact the classification process. A standard practice in this context involves normalizing the vectors of both the Student and the Teacher, as well as the input vector. This normalization can be represented as:
\begin{equation}
|\mathbf{S}|^2 = N, \quad |\mathbf{T}|^2 = N, \quad |\mathbf{x}|^2 = N,
\end{equation}
where each vector lies on the surface of an N-dimensional sphere with a radius of $\sqrt{N}$.

To assess the similarities and differences in classifications made by the teacher perceptron $f_T$ and the student perceptron $f_S$, we project the input examples onto the plane that is defined by the coupling vectors of the teacher and the student.

It becomes evident that the projections located in the region formed by the intersection of these two vectors represent inputs classified differently by the teacher and the student. When the inputs are randomly selected, the probability of a classification disagreement, which equates to the generalization error $\epsilon_g$, is the same as the probability of a projection landing within this intersecting region.

Therefore, we can deduce that the generalization error is given by:
\begin{equation}
\epsilon_g = \frac{\theta}{\pi},
\end{equation}
where $\theta$ denotes the angle between the Teacher and the Student vectors.

It is convenient introduce the so-called \textit{teacher-student overlap}, defined as:
\begin{equation}
    R_S=\frac{S^TT}{N}.
\end{equation}
Since we fixed the lengths of the vectors equal to $\sqrt{N}$, $R_S$ is nothing but the cosine of the angle $\theta$, and the generalization error can be written as:
\begin{equation}
    \epsilon_g = \frac{\arccos(R)}{\pi}.
\end{equation}

It is beneficial to introduce the concept known as \textit{teacher-student overlap}, defined by the equation:
\begin{equation}
R_S = \frac{\mathbf{S}^T \mathbf{T}}{N}.
\end{equation}
Given that we have set the lengths of the vectors to $\sqrt{N}$, $R_S$ simply represents the cosine of the angle $\theta$. Consequently, the generalization error can be reformulated as:
\begin{equation}
\epsilon_g = \frac{\arccos(R_S)}{\pi}.
\end{equation}

Now, let us select a dataset and consider all the Student vectors that predict the same outcomes as the Teacher for the entire dataset. The collection of these vectors denoted as $\mathcal{V}$, is referred to as the \textit{version space}. All the Student vectors within this space are termed \textit{compatible}. Essentially, this space encompasses potential students that are perfectly optimized over the dataset (i.e., those with a test error of $\epsilon_t = 0$).

Our focus now shifts to the average generalization error of a well-performing Student vector $S$ in the $\mathcal{V}$ space. In other words, we aim to understand the typical performance of a Student that is compatible or optimized. This approach is historically known as \textit{zero temperature Gibbs Learning}.

In the context of Gibbs learning, the generalization error diminishes as the size of the training set increases. This reduction is due to the exclusion of more Student vectors that are deemed incompatible with the examples, leading to a gradual reduction in the size of the version space $\mathcal{V}$. If we can quantify the likelihood of a Student vector remaining viable upon the introduction of a new example, we could estimate the average trajectory of the generalization error as training progresses.

Grouping the couplings into classes based on their overlap with the teacher allows us to accurately calculate the probability of a Student with a specific overlap $\hat{R}$ producing the same output as the teacher on a randomly chosen input. This probability is given by:
\begin{equation}
\mathcal{P}(f_S(x)=f_T(x) \mid R_s = \hat{R}) = 1 - \epsilon_g.
\end{equation}
Utilizing this, we can determine the average volume of compatible students with a generalization error $\epsilon$ after presenting $P$ training examples:
\begin{equation}
\Omega_P(\epsilon) = \Omega_0(\epsilon) \left(1-\epsilon\right)^P,
\end{equation}
where $\Omega_0(\epsilon)$ represents the initial volume of Students with overlap $\hat{R}$ before training begins.

In the thermodynamic limit, where the dimensionality of the input approaches infinity ($N \to \infty$), the volumes $\Omega_0(\epsilon)$ and $\Omega_P(\epsilon)$ can be approximated as follows:
\begin{align}
\Omega_0(\epsilon) &= \int dS , \delta(S^2-N), \delta\left( \frac{S^T}{N} - \cos(\pi\epsilon) \right) \sim \\
&\sim \exp\left(\frac{N}{2}[1 + \log(2\pi) + \log\sin(2\pi\epsilon)]\right)\, , \\
\Omega_P(\epsilon) &\sim \exp\left(\frac{N}{2}[1 + \log(2\pi) + \log\sin(2\pi\epsilon) + 2\alpha\log(1-\epsilon)]\right)\, ,
\end{align}
where $\alpha = P/N$.

Given the exponential scaling of $\Omega_P(\epsilon)$ with the large factor $N$, it is reasonable to assume that a randomly chosen Student vector from the version space will, with high probability and for large $N$, have a value of $\epsilon$ that maximizes $\Omega_P(\epsilon)$. This leads us to:
\begin{equation}
\epsilon(\alpha) = \argmax\left[\frac{1}{2}\log\sin^2(\pi\epsilon)+\alpha\log(1-\epsilon)\right]
\end{equation}
While this result seems theoretically sound, a discrepancy arises when compared with experimental findings. The issue here is the use of an \textit{annealed} approximation, where we averaged over both the examples and the Student $S$. In the context of Gibbs learning, this implies that both quantities are adapted to minimize the error.

However, our goal is actually to minimize the error by adapting the Student to a fixed set of examples. This approach is the essence of the \textit{Quenched} averaging, which will be explored in detail in the following chapter (Chapter \ref{ch.2AB}).
}

Note that the approach outlined above only deals with the ``static'' properties of the trained DNNs, a situation that exactly mirrors equilibrium thermodynamics.
The dynamical learning process, that is, the way by which DNNs explore the loss landscape to find optimal solutions, is clearly beyond the scope of equilibrium statistical mechanics. 
Several lines of research attempt to address this problem going beyond the static description with techniques from dynamical system theory \cite{NARENDRA1992109,doi:10.1137/20M131727X}, non-equilibrium statistical physics \cite{PhysRevE.98.062120, Mignacco_2021} and statistical field theory \cite{helias2020statistical}.


\subsection{Some notable results}
Trying to list all the important results in this field is quite a task since there are so many of them, making it tough to pin down a clear list of the top works.
So, I've decided to break down the task into the major subfields and aim to give a sneak peek into the literature in each area. 
This way, it's a bit easier to dig into the key contributions without getting lost in a vast ocean of information.

\paragraph{Storage capacity of the Perceptron}
Historically, one of the first questions that captivated the field is: How many patterns can a network store?
Let me rephrase it, quoting Elizabeth Gardner and Bernard Derrida. Consider a set of P patterns, each pattern $\mu$ consisting of N input bits $S_i^\mu = \pm 1 $ for $1 \leq i \leq N $ and one output bit $R^\mu = \pm 1$. 
Is it possible to find a Boolean function F of N variables such that 
\begin{equation}
    R^\mu = F(S^\mu_1,S^\mu_2,\dots,S^\mu_N)
\end{equation}
is satisfied for each pattern $\mu$? In general, this is always possible, as every function $F$ of $N$ variables is fully determined by $2N$ bits corresponding to its $2^N$ possible inputs.
The question becomes more complex if we restrict the class of functions to those that can be represented by a neural network.
Initially, the Hopfield model was taken into account, and the typical approach was to study the capacity with fixed rules for the pattern \cite{amit1985spin,amit1987statistical,kohonen2012self,personnaz1985information,kanter1987associative,toulouse1986spin,mezard1986solvable, PhysRevLett.55.1530, Parisi_1986}.
It was only in 1988 that a solution for random patterns was discovered \cite{gardner1988optimal}, and in that same year, the theory was applied to the Perceptron \cite{gardner1989three}, finding the maximal fraction of patterns $\alpha = P/N$ that could be stored by a perceptron of size N. For this simple model, the maximal capacity is $\alpha = 2$.
Starting from this approach, dubbed as the Gardner volume in honor of its developer, the maximal capacity has been explored for a myriad of different systems \cite{Aubin_2019}.

\paragraph{Loss Landscape}

Deep neural networks learn by optimizing a high-dimensional, non-convex loss function, usually employing a stochastic gradient descent (SGD) approach. Remarkably, this learning process often identifies effective minimizers without becoming trapped in local critical points. Additionally, these minimizers tend to resist overfitting. The ability of such complex networks, comprising millions of adjustable connections, to consistently achieve these two feats remains a profound and unresolved question.
One of the approaches suggested to tackle this problem is to study in depth the landscape of the loss function. Empirical studies consistently demonstrate a notable correlation between the ``flatness'' of minima reached by optimization algorithms and their associated generalization performance \cite{DBLP:journals/corr/KeskarMNST16, DBLP:journals/corr/abs-1912-02178,dziugaite2019entropysgd}. This suggests that neural network loss functions possess extensive flat minima regions that are both easily reachable by optimization algorithms and conducive to good generalization \cite{draxler2018essentially, li2018visualizing,huang2020understanding,verpoort2020archetypal,}. Intriguingly, such minima are observed even with randomized labels or across varied datasets, pointing to their resilience as an inherent network property \cite{zhang2021understanding}.
This viewpoint gains further credibility from contemporary research employing statistical physics techniques \cite{baldassi2020shaping, PhysRevLett.123.170602, PhysRevResearch.3.033290}. In simpler, nonconvex neural network models, a plethora of sub-optimal minima coexists with a select set of broader, flat minima, colloquially referred to as ``high local entropy minima''. These latter minima are particularly noteworthy for their near-optimal generalization prowess \cite{PhysRevLett.115.128101}. Such findings are grounded in large-deviation methods, emphasizing minima encircled by a substantial count of other minima within a predefined proximity. Numerical analyses further underscore these insights, revealing that even basic algorithms, without specifically aiming at, can effectively access these wide flat minima, avoiding the prevalent suboptimal ones \cite{baldassi2016unreasonable}.
Recent developments have increased our understanding of these minima and how affect the performance of DNNs \cite{PhysRevLett.127.278301,annesi2023starshaped}.

\paragraph{Influence of data structure}

In Statistical Physics of Machine Learning, many results have historically been based on the limiting assumption that the inputs of a training set are independent, identically distributed random variables, without any relation to their labels. 
Such assumptions, although convenient, often don't align with the complexities of real-world data. 
In recent times, researchers have shifted their focus, aiming to understand the implications of more nuanced and realistic data generation models on existing theoretical frameworks.
Research into the linear classification of perceptual manifolds has led to initial quantitative evaluations of deep neural networks' capability to classify such complex object manifolds \cite{PhysRevX.8.031003, PhysRevE.93.060301, cohen2020separability}.
Within the context of Hopfield models, it's been proposed \cite{PhysRevE.95.022117} that hierarchical structures with hidden layers inherently arise when training patterns manifest as superpositions of specific random features. This phenomenon aligns with patterns commonly observed in real-world data \cite{PhysRevX.8.021023, PhysRevE.98.012315}.
Furthermore, thorough investigations have provided precise insights into generalization errors utilizing the replica method. These studies center on two distinct synthetic data scenarios: one highlighting random features \cite{gerace2020generalisation}, and the other anchored in the hidden manifold model \cite{goldt2020modeling}.
Recently, a compelling model that does not require the replica approach was introduced \cite{petrini2023deep}. This new model not only presents a fresh perspective but also potentially expands the scope of understanding including the convolutional neural network.

\paragraph{Statistical Physics of DNNs}
Building upon the initial inquiry regarding pattern storage, the trajectory toward more generic frameworks has led to the development of a statistical mechanics approach for neural networks. Throughout the final decade of the previous century, this framework was meticulously crafted from a theoretical standpoint, as shown in the preceding section \cite{58339, PhysRevLett.65.1683}. 
Initially, the exploration was confined to simpler models, such as the perceptron, the committee machine\cite{NIPS1992_5d44ee6f}, and the support vector machine (SVM) \cite{dietrich1999statistical}. 
These models served as a foundational springboard, enabling a deeper understanding and analysis of neural networks within a statistical mechanical context.
Moreover, in recent years, the applications of this powerful framework expanded to encapsulate more complex systems. 
A significant milestone was achieved with the advent of a Gaussian theory for the Random Feature Model \cite{loureiro2021learning}. This development paved the way for the analytical examination of single hidden layer neural networks (1HL NNs), enriching the theoretical landscape and providing new avenues for exploration \cite{canatar2021spectral}.
The momentum of these advancements didn't wane; rather, it propelled the field further into uncharted territories. 
Recently, the complex case of two hidden layer neural networks (2HL NNs) was also integrated into this robust theoretical framework \cite{mei2018mean,sirignano2020mean,chizat2018global}. 

Chapter \ref{ch.2AB} will go more into the details of this expanding subfield.

\paragraph{Kernel methods}

A significant breakthrough in recent research emerged from the exploration of neural networks in the infinite-width limit. In this context, while the number of training data points, denoted as $P$, remains fixed, the dimensions or sizes of the hidden layers expand indefinitely. A pivotal observation here was the equivalence between such deep models and Gaussian processes (GPs). This finding, supported by a series of studies \cite{Neal, NIPS1996_ae5e3ce4,g.2018gaussian, LeeGaussian, garriga-alonso2018deep, novak2019bayesian, JacotNTK, ChizatLazy, lee2019wide}, bridged the gap between the realms of deep learning and kernel methods \cite{cortes1995support}. It also facilitated a deeper understanding of this regime from the perspective of statistical physics \cite{pmlr-v119-bordelon20a, canatar2021, PhysRevLett.82.2975}.
However, while the infinite-width limit offers intriguing insights, it doesn't encapsulate the full of deep learning potential and neither the description of many practical DNN architectures. Many in the field concur that a holistic theory of deep learning should venture beyond this limit \cite{Seleznova2022, Vyas2022, antognini2019finite, yaida2020nonGauss, hanin2023random, zavatone2021prior}. This assertion stems from the reality of practical neural networks. Real-world neural networks often operate in a different regime where the number of training samples typically surpasses the size or width of the most expansive layer in the network.

In Chapter \ref{ch.3BIW}, we will delve deeper into this issue and discuss its implications and applications.

\paragraph{Beyond the Gaussian Realm}
The vast majority of results garnered thus far, in one way or another, have been obtained under the assumptions of Gaussianity. From the Gaussian Equivalence Principle \cite{goldt2022gaussian} to Gaussian Processes \cite{Lee2017DeepNN}, the Gaussian realm has largely shaped the lens through which this field has been viewed. However, it's apparent that both the dataset and the behaviors of DNNs can only be approximated to a Gaussian. It's only recently that the community has begun to venture beyond this paradigm, unearthing intriguing findings regarding how the moments of probability distributions—beyond the second moment, which is the last non-zero moment in a Normal distribution—affect the behavior and performance of our algorithms \cite{pmlr-v202-refinetti23a, ingrosso2022data,redman2023sparsity,szekely2023learning}.

In Chapter \ref{ch.4RTS}, we will explore further this emerging field, discussing some work (yet unpublished) where we describe an example of non-Gaussian behavior of DNNs.

\paragraph{Feature Learning}
 
In the domain of fully connected deep neural networks (DNNs), understanding the relationship between their capability to generalize effectively and their ability to learn and internalize the features of a dataset—a process known as ``feature learning''—is a subject of active research and discussion. Insights have been gained using a Statistical Mechanics approach. For instance, it has been suggested that feature learning in fully connected DNNs might lead to the development of sparse weight matrices \cite{petrini2023learning}. Additionally, the behavior in more complex architectures, like convolutional neural networks (CNNs), has been explored, yielding significant insights as reported in Refs. \cite{seroussi2023separation,naveh2021self, aiudi2023local}. Another intriguing area of research is the investigation of the grokking effect \cite{humayun2024deep, rubin2023droplets}.

In Chapter \ref{ch.4RTS}, we will delve deeper into this topic, discussing some yet unpublished work where, in a controlled setting, we study the ability of DNNs to learn features.

\clearpage
\printbibliography[heading=subbibnumbered, title=Chapter bibliography]
\end{refsection}

\chapter{Data-Averaged approach: Asymptotic bound}
\begin{refsection}
\label{ch.2AB}
Commonly, there's a belief that models with numerous adjustable parameters are prone to overfitting their training data, which would typically result in poorer generalization performance on novel, unseen data sets. 
However, empirical evidence shows that this assumption does not hold for general Deep Neural Networks (DNNs). 
Statistical Learning Theory (SLT) is acknowledged as a solid theoretical probabilistic framework for assessing the generalization capabilities of machine learning models \cite{Vapnik1999_2, Bousquet2004}. 
SLT specifically provides reliable predictions about potential declines in a model's performance when it encounters new data. 
These predictions are grounded in assessments of model capacity or complexity, using metrics like the Vapnik-Chervonenkis dimension \cite{vapnik1999} or Rademacher complexity \cite{shalev2014understanding}. 
Yet, these predictions often do not guarantee the effectiveness of modern, highly overparameterized models on test data. 
In this Chapter, we aim to delve into the causes of this mismatch and examine how statistical mechanics might offer improved approaches to refining these generalization predictions, as further discussed in \cite{PhysRevE.105.064309}.

\vspace{0.5cm}
\minitoc
\vfill
\newpage

\section{Statistical Learning Theory \& Generalization Bound}\label{sec2:1SLT}
As we noted in the first chapter, selecting an architecture for a deep neural network is essentially choosing a specific form for a high-dimensional parametric function. 
A fundamental initial question arises: How does this choice impact the generalization performance of our algorithms?

Statistical Learning Theory (SLT) sits at the intersection of computer science and statistics, providing a theoretical framework to address this question. In this section, we will delve into the intricacies of SLT, introducing some of its key concepts, such as ``capacity'', and discussing the principal findings related to the problem of generalization.

\subsection{Capacity (or complexity, expressive power, and richness)}
Let's start with an example. Consider a regression estimation problem.
Suppose we have $m$ empirical observations,
\begin{equation}
(x_1,y_1),\dots , (x_m,y_m) \in \mathcal{X} \times \mathcal{Y}
\end{equation}
where, for simplicity, we assume $\mathcal{X}=\mathcal{Y}=\mathbb{R}$. Figure \ref{fig2:2ab-1} illustrates a plot of this dataset, alongside two potential functional relationships that might explain the data. 
A physicist analyzing these data points might argue that it's improbable for the measurements to align almost perfectly on a straight line by mere coincidence, preferring to attribute the residuals to measurement errors rather than an incorrect model. 
But, can we define how a straight line exemplifies simplicity, and why this simplicity might suggest it is more closely aligned with an underlying true dependency?

\afterpage{
\begin{figure}[t]
    \centering
    \includegraphics[width = 0.95\textwidth]{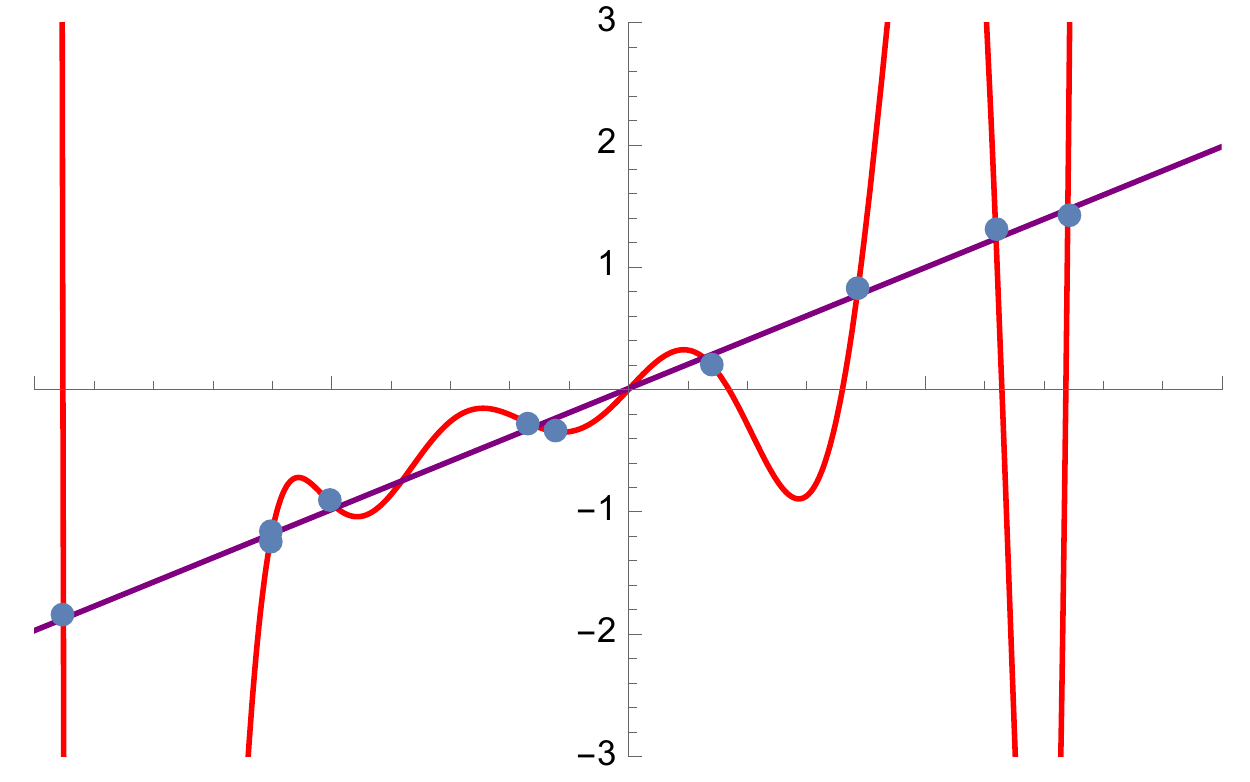}
    \caption{}
    \label{fig2:2ab-1}
    \caption*{\textbf{A set of data and two possible relationships between the variables.} The data points are represented by the blue dots, and the two possible relationships are represented by the red line and the purple line. The red line is a linear fit, while the purple line is a high-degree polynomial fit. The linear fit is simpler than the polynomial fit, but it also captures the underlying trend in the data while the polynomial fit captures the noise.}
\end{figure}
\clearpage
}

This question has long preoccupied researchers in the field of learning. In classical statistics, it's examined through the lens of the \textit{bias-variance dilemma}. 
If we were to compute a linear fit for every dataset we encounter, then every functional relationship we identify would invariably be linear. 
However, this outcome wouldn't arise from the data itself; it would be a \textit{bias} introduced by our methodology. 
Conversely, if we fit an arbitrarily high-degree polynomial to any given dataset, we could always achieve a perfect fit, but this approach would likely lead to overfitting, where the model captures not only the underlying trend but also the noise inherent in the data.
This results in a high \textit{variance}, meaning the model becomes overly sensitive to the specific details and anomalies of the particular dataset, reducing its ability to generalize and perform well on new, unseen data. 
Therefore, the challenge lies in finding a balance between the simplicity of a linear model, which may underfit complex data, and the flexibility of a high-degree polynomial, which risks overfitting, to accurately capture the true underlying dependency in the data.

This \textit{bias-variance dilemma} can be rephrased as a problem in terms of \textit{capacity}. 
In statistical learning theory, the \textit{capacity} refers to the ability of a given set of functions, the hypothesis space, to explain or model data. 
These functions are essentially different hypotheses or theories about the underlying patterns in the data. Imagine an observer (like a scientist or a machine learning algorithm) who has access to a variety of such functions. 
The observer collects empirical data and then selects one function from this set that best explains the data observed so far.
The key idea behind the capacity is that quantify the explanatory power of this set of functions. A set with a larger capacity can model a wider range of data patterns, but there is a trade-off: the larger the capacity, the more data you need to confidently choose a function that will not only explain past observations but will also reliably predict future ones. 
In other words, with a high-capacity set, there's a greater risk of overfitting the current data unless you have a sufficiently large dataset.

This underscores the need to develop a capacity metric, which is crucial for predicting the generalization capabilities of a hypothesis space. In the past 50 years, several approaches for defining such metrics have emerged. In the following subsections, we will explore the most significant among these and examine their role in establishing upper bounds for the generalization performance of machine learning algorithms.

\subsection{Vapnik-Chervonenkis dimension}

In this section, we will introduce some elements that are taken from \cite{vapnik1999}.

Let's consider a fixed set of samples, denoted as $\{x^\mu\}_{\mu=1}^P\in\mathcal{X}$, and a set of predictor functions, $f_w:\mathcal{X}\to\mathcal{Y}$, parametrized by the parameter set $w$, which constitute our hypothesis space $\mathcal{H}$.
For every possible classification labeling of the samples, $\{y^\mu\}_\mu^P\in\mathcal{Y}$, we can investigate whether there exists at least one choice of parameter values, $w^*$, such that:
\begin{equation}
    f_{w^*}(x^\mu) = y^\mu \quad \text{for all} \, \mu = 1, \dots, P.
\end{equation}
This means that the predictor function can accurately reproduce any given labeling of the sample set.
A hypothesis space with a more expressive architecture will have a greater number of labelings it can reproduce.
However, this definition currently lacks precision, as it relies on the specific set of samples $\{x^\mu\}_{\mu=1}^P$; a more precise definition will be provided shortly.

To refine our definition, it's necessary to introduce the concepts of \textit{dichotomies}, \textit{shattering}, and being \textit{in general position}.
\begin{definition}[Dichotomies]
    Let $\mathcal{H}$ be a hypothesis space and $S=\{x^\mu\}_{\mu=1}^P$ a set of $P$ elements. The dichotomies generated by $\mathcal{H}$ on the points in $S$ are defined as:
    \begin{equation}
        \mathcal{H}(\{x^\mu\}_{\mu=1}^P) = \{ (f_w(x_1), \dots , f_w(x_P)) \,|\, f_w \in \mathcal{H}, \, x_i \in S \}
    \end{equation}
\end{definition}
In other words, dichotomies refer to the various ways the hypothesis space can classify the elements of set $S$.  Essentially, for each element in the set $S$, the hypothesis space provides a corresponding output or label, resulting in a collection of outputs for the entire set. This collection represents a dichotomy. When considering all possible functions $f_w$ in the hypothesis space $\mathcal{H}$, we are essentially exploring all the different labeling patterns (or dichotomies) that $\mathcal{H}$ can produce for the set $S$.
\begin{definition}[Shattering]
    A hypothesis space $\mathcal{H}$ is said to \textit{shatter} a set $S$ if it can generate all possible dichotomies on that set.
\end{definition}
The term ``shattering'' describes a situation where a hypothesis space $\mathcal{H}$ is so versatile or powerful that it can produce every possible dichotomy for a given set $S$. In other words, no matter how you want to classify or label the elements of $S$, the hypothesis space $\mathcal{H}$ has a function that can do exactly that. Shattering is an indication of the expressiveness or richness of the hypothesis space: the more sets a space can shatter, the more complex or flexible it is considered to be.
\begin{definition}[General Position]
    A set of points in a $d$-dimensional affine space is said to be in \textit{general position} if, for any $k$ points in the set (where $k = 2, 3, \dots, d+1$), no subset of these $k$ points lies entirely within a $(k-2)$-dimensional subspace.
\end{definition}
This definition ensures that the points are not aligned or arranged in a way that is ``too regular'' or ``predictable''. In simpler terms, points in \textit{general position} do not all lie on the same line, plane, or any lower-dimensional subspace, providing a more diverse and spread-out arrangement in the space.

These definitions enable us to introduce a critical measure of capacity, known as the \textbf{Vapnik-Chervonenkis dimension}. 
\begin{definition}[VC Dimension]
    The Vapnik-Chervonenkis (VC) dimension of a hypothesis set $\mathcal{H}$, denoted as $d_{VC}(\mathcal{H})$, is the maximum cardinality $P$ for which $\mathcal{H}$ can shatter a set of $P$ points in general position.
\end{definition}
This measure is particularly insightful as it clearly defines what it quantifies. Importantly, the VC dimension is a measure of worst-case expressiveness. This means it looks at the most complex patterns or hardest examples that the model can fit or shatter. In other words, it's not about how well the model performs on average or on typical data; it's about understanding the maximum capacity of the model to adapt to any possible scenario, no matter how challenging.

\details{
\begin{center}
    \textbf{VC dimension examples}
\end{center}
Let me give some practical examples of the VC dimension for simple hypothesis space:
\begin{itemize}[rightmargin=+.1in,leftmargin=+.2in]
    \item $\mathcal{H}$ is a set of constant classifiers (with no parameters); Its VC dimension is 0 since it cannot shatter even a single point. 
    \item $\mathcal{H}$ is a set of single-parametric threshold classifiers on real numbers; i.e., for a certain threshold $\theta$ , the classifier $f_{\theta}$ returns 1 if the input number is larger than $\theta$  and 0 otherwise. The VC dimension of this hypothesis space is 1 because it can shatter a single point but it cannot shatter all the sets with two points. For every set of two numbers, if the smaller is labeled 1, then the larger must also be labeled 1, so not all labelings are possible.
    \item $\mathcal{H}$ is a set of single-parametric interval classifiers on real numbers; i.e., for a certain parameter $\theta$, the classifier $f_{\theta}$ returns 1 if the input number is in the interval $[\theta,\theta + 4]$  and 0 otherwise. The VC dimension of this hypothesis space is 2 because it can shatter some sets of two points, but cannot shatter any set of three points. For every set of three numbers, if the smallest and the largest are labeled 1, then the middle one must also be labeled 1, so not all labelings are possible.
    \item $\mathcal{H}$ is a set of straight lines as a classification model on points in a two-dimensional plane (this is the model used by a perceptron). The line should separate positive data points from negative data points. There exist sets of 3 points that can indeed be shattered using this model (any 3 points that are not collinear can be shattered). However, no set of 4 points can be shattered: by Radon's theorem, any four points can be partitioned into two subsets with intersecting convex hulls, so it is not possible to separate one of these two subsets from the other. Thus, the VC dimension of this particular classifier is 3. It is important to remember that while one can choose any arrangement of points, the arrangement of those points cannot change when attempting to shatter for some label assignment.
    \item $\mathcal{H}$ is a set of single-parametric sine classifiers, i.e., for a certain parameter 
    $\theta$ the classifier $f_{\theta}$ returns 1 if the input $x$ has $\sin(\theta x) > 0 $  and 0 otherwise. The VC dimension of this set is infinite since it can shatter any finite subset of the set $\{2^{-m}| m \in \mathbb{N}\}$.
    \item $\mathcal{H}$ is a set of NN classifiers with $L$ layers, $\theta \in \mathbb{R}^{N_{tot}}$, and a piecewise linear activation function, i.e for a certain choice of parameters $\theta$ the classifier $f_{\theta}$ returns 1 if the input $x$ has $f(\theta; x) > 0 $  and 0 otherwise. The VC dimension of this set is $\bigO(L\cdot N_{tot})$ as shown in \cite{bartlett2019nearly}.
\end{itemize}
}

Turning back to the main focus of this Thesis, calculating the VC (Vapnik-Chervonenkis) dimension for Deep Neural Networks (DNNs) is typically not straightforward due to the non-linear interactions between the (millions of) parameters and data points. However, a commonly accepted approximation is as follows:

\begin{equation}
    d_{VC}(DNNs) \approx L \cdot N_{tot}
\end{equation}
where \( N_{tot} \) represents the total number of parameters, and \( L \) is the number of layers. This formula, which was developed for DNNs with piecewise linear activation functions \cite{bartlett2019nearly}, indicates that DNNs have a very large capacity. It's important to note, however, that this approach does $\textbf{not}$ take into account the optimization of weights that is, it measures the capacity of the general hypothesis space and does not focus on any trained subspace.

Now, we have a measure of the capacity of our hypothesis space, and we aim to use it to gain insights into the generalization performance of our machine learning model. Before delving into the main result, let's more formally reintroduce the quantity of interest.

We begin by defining the \textit{Training Error}, which is the error our model incurs on the dataset used during training.

\begin{definition}[Empirical Risk or Training Error]
    Given a dataset $\mathcal{T}=\{x^\mu,y^\mu\}$, let $f_{w^*}$ be a prediction function from our hypothesis space $\mathcal{H}$ and let $\mathcal{L}:\mathcal{H}\to\mathbb{R}$ be the \textit{Loss Function}, a measure of the discrepancy between our hypothesis's predictions and the actual labels. The \textit{Empirical Risk} (or \textit{Training Error}) is defined as:
    \begin{equation}
        \epsilon_t(f_{w^*},\mathcal{T},\mathcal{L})=\frac{1}{P}\sum_{\mu=1}^{P}\mathcal{L}\left(f_{w^*}(x^\mu),y^\mu\right)\,.
    \end{equation}    
\end{definition}

As discussed in the previous chapter, the entire process of neural network optimization can be viewed as an optimization of the parameters $w$ in order to minimize the Empirical Risk (ER). However, ER is not the most critical quantity in this context. Indeed, as illustrated at the beginning of this chapter, a model with zero ER might fail to capture the real patterns in the dataset and perform poorly on similar but unseen data. 
Hence, the most crucial concept in Statistical Learning Theory (SLT) is the \textit{Risk} or \textit{Generalization Error}.

\begin{definition}[Risk or Generalization Error]
    Consider $\mathcal{P}(x,y)$ as the underlying distribution of our dataset, $f_{w^*}$ as a prediction function from our hypothesis space $\mathcal{H}$, and $\mathcal{L}$ as the \textit{Loss Function}. The Risk (or Generalization Error) is defined as: 
    \begin{equation}
        \epsilon_g(f_{w^*},\mathcal{P}(x,y),\mathcal{L}) = \int \mathcal{L}\left(f_{w^*}(x),y\right)\,d\mathcal{P}(x,y)
    \end{equation}
\end{definition}

This definition, as logical as it may seem, presents a significant challenge: in most cases, the probability distribution $\mathcal{P}(x,y)$ is unknown, making the Risk nearly impossible to calculate directly.

Another central quantity is the \textit{Generalisation Gap}, defined as the difference between the \textit{Generalisation Error} and the \textit{Training Error}:
\begin{definition}[Generalization Gap]
    Given a dataset $\mathcal{T}=\{x^\mu,y^\mu\}$, and consider $\mathcal{P}(x,y)$ the underlying distribution of our dataset, $f_{w^*}$ as a prediction function from our hypothesis space $\mathcal{H}$, and $\mathcal{L}$ as the \textit{Loss Function}. The Generalization Gap is defined as: 
    \begin{equation}
        \Delta \epsilon(f_{w^*},\mathcal{P}(x,y),\mathcal{L}) = \epsilon_g(f_{w^*},\mathcal{P}(x,y),\mathcal{L}) - \epsilon_t(f_{w^*},\mathcal{T},\mathcal{L})
    \end{equation}
\end{definition}

Given this formal definition of our key quantities, we can express a theorem that imposes an upper bound to the generalization performance. 

\begin{theorem}[VC Generalization Bound, \cite{Vapnik1999_2}]
    Let $\mathcal{H}$ be a hypothesis space with VC dimension $d_{VC}$. Assume that the loss function $\mathcal{L}$ is bounded by $M$. Then, for any $\delta > 0$, with probability at least $1 - \delta$ over the choice of a training set of size $P$, the following bound holds for all $f_w \in \mathcal{H}$:
    \begin{equation}
        \Delta \epsilon = \epsilon_g(f_w) - \epsilon_t(f_w) \leq M \sqrt{\frac{d_{VC}(\log(2P) -\log(d_{VC}) + 1) - \log(\delta/4)}{P}}
    \end{equation}
    where $\epsilon_g(f_w)$ is the generalization error and $\epsilon_t(f_w)$ is the training error of hypothesis $f_w$.
\end{theorem}

This result appears quite remarkable: for a given class of hypotheses, we can calculate the worst possible performance (at least probabilistically). For many years, this bound played a crucial role in the theoretical analysis of many machine learning algorithms. Unfortunately, with the advent of Deep Neural Networks (DNNs) \cite{Goodfellow-et-al-2016}, this bound has become somewhat less relevant in certain contexts.

Take, for example, the use of a well-known architecture like VGG18 \cite{simonyan2014} for classifying the MNIST dataset \cite{MNIST}. The hypothesis space in this scenario has a VC dimension that is roughly proportional to the total number of parameters of the architecture, which, for VGG18, amounts to 45 million. Given that the MNIST dataset has $P = 600,000$, this approximation suggests that, according to the VC theorem, $\Delta\epsilon \leq \sqrt{45 \times 10^6/60 \times 10^3} \approx 30$. However, this bound is not particularly informative, considering that we expect $\Delta\epsilon$ to be significantly smaller, ideally less than one.
Also, a notable limitation of this bound is that the concept of VC dimension is inherently linked to binary classification problems, thus limiting its applicability in the broader context of machine learning tasks.

\subsection{Rademacher Complexity}
In this section, we will focus on a different setup, that include also real-valued functions in the hypothesis class.
To begin, let us define the \textit{Rademacher complexity of a set}:

\begin{definition}[Rademacher complexity of a set]
    Given a set $A \subseteq \mathbb{R}^d$, the Rademacher complexity of $A$ is defined as follows:
    \begin{equation}
        \text{Rad}(A) = \underset{\sigma}{\mathbb{E}}\left[\sup_{a \in A}\frac{1}{P}\sum_{\mu=1}^P \sigma_\mu a_\mu\right]
    \end{equation}
    where $\sigma = (\sigma_1, \dots, \sigma_P)$ represents a vector of i.i.d. Rademacher variables, which are independent and uniformly distributed random variables taking values in $\{-1, +1\}$.
\end{definition}

When considering a hypothesis space, it is pertinent to define the \textit{Empirical Rademacher Complexity}. 
This concept is crucial for understanding the behavior of function classes with respect to specific samples.
\begin{definition}[Empirical Rademacher complexity]
    Let $\mathcal{H}$ be a family of functions mapping from $\mathbb{R}^d$ to $\mathbb{R}$. The empirical Rademacher complexity of $\mathcal{H}$ for a sample $S = \{x_\mu\}_{\mu=1}^P$ is defined as:
    \begin{equation}
        \text{Rad}_S(\mathcal{H}) = \underset{\sigma}{\mathbb{E}}\left[\sup_{f_w \in \mathcal{H}}\frac{1}{P}\sum_{\mu=1}^P \sigma_\mu f_w(x_\mu)\right]
    \end{equation}
    where $\sigma = (\sigma_1, \dots, \sigma_P)$ is a vector of i.i.d. Rademacher variables, which are independent, uniformly distributed random variables taking values in $\{-1, +1\}$.
\end{definition}

It is noteworthy that this quantity can be reinterpreted as a measure over a set, analogous to the measure of dichotomies. Specifically, we can express $\text{Rad}_S(\mathcal{H})$ as $\text{Rad}(\mathcal{H} \circ S)$, where $\mathcal{H} \circ S$ denotes the function composition defined by $\mathcal{H} \circ S := \{(f_w(x_{1}), \ldots, f_w(x_{P})) \mid f_w \in \mathcal{H}\}$.

To extend our analysis beyond finite datasets and include all possible data patterns, we define the \textit{Rademacher complexity}. This measure is crucial for understanding the behavior of function classes in the context of potentially infinite datasets.

\begin{definition}[Rademacher complexity]
    Let $\mathcal{P}$ be a probability distribution over $\mathbb{R}^d$. The Rademacher complexity of a function class $\mathcal{H}$, with respect to $\mathcal{P}$ and for a sample size of $P$, is defined as:
    \begin{equation}
        \text{Rad}_{\mathcal{P},P}(\mathcal{H}) := \underset{S \sim \mathcal{P}^P}{\mathbb{E}} \left[\text{Rad}_{S}(\mathcal{H})\right]
    \end{equation}
\end{definition}

This definition enables the evaluation of the complexity of $\mathcal{H}$ in a comprehensive manner, considering the distribution $\mathcal{P}$ over the entire space $\mathbb{R}^d$.

Intuitively, Rademacher complexity is applied to a function class to measure their ability to classify points drawn from a probability space under arbitrary labelings. When the function class is rich enough, it contains functions that can appropriately adapt for each arrangement of labels, simulated by the random draw of $\sigma _{i}$ under the expectation, so that this quantity in the sum is maximized.

\details{
    \begin{center}
        \textbf{Rademacher Complexity Example}
    \end{center}
    Consider a hypothesis space defined as $\mathcal{H} = \{f_w(x) = w^T x \mid \lVert w \rVert_2 \leq \omega \}$. Assume our dataset is uniformly distributed such that $\mathcal{P}(x)$ is uniform for $\lVert x \rVert_2 \leq B$.
    By definition, the Rademacher complexity is given by:
    \begin{equation}
        \text{Rad}_\mathcal{P}(\mathcal{H}) := \underset{S\sim \mathcal{P}^{P}}{\mathbb{E}} \underset{\sigma}{\mathbb{E}} \left[ \sup_{w, \lVert w \rVert \leq \omega }\frac{1}{P}  \left\lvert\sum_\mu^P \sigma_\mu w^T x_\mu \right\rvert \right].
    \end{equation} 
    Applying the Cauchy-Schwarz inequality, we derive:
    \begin{equation}
        \text{Rad}_\mathcal{P}(\mathcal{H}) \leq \frac{1}{P} \underset{S\sim \mathcal{P}^{P}}{\mathbb{E}} \underset{\sigma}{\mathbb{E}} \left[ \sup_{w, \lVert w \rVert \leq \omega } \lVert w \rVert_2 \left\lVert \sum_\mu^P \sigma_\mu x_\mu \right\rVert_2 \right].
    \end{equation}
    Maximizing this and applying a normal transformation, we obtain:
    \begin{equation}
        \text{Rad}_\mathcal{P}(\mathcal{H}) \leq \frac{\omega}{P} \sqrt{  \underset{\sigma}{\mathbb{E}} \left[  \sum_{\mu,\nu}^P \sigma_\mu \sigma_\nu \langle x_\mu, x_\nu \rangle_{\mathcal{P}^{P}} \right]}.
    \end{equation}
    Utilizing Jensen's inequality, we further deduce:
    \begin{equation}
        \text{Rad}_\mathcal{P}(\mathcal{H}) \leq \frac{\omega}{P} \sqrt{     \sum_{\mu,\nu}^P \underset{\sigma}{\mathbb{E}}\left[\sigma_\mu \sigma_\nu\right] \langle x_\mu, x_\nu \rangle_{\mathcal{P}^{P}} }.
    \end{equation}
    Given that $\underset{\sigma}{\mathbb{E}}\left[\sigma_\mu \sigma_\nu\right] = \delta_{\mu,\nu}$, we ultimately conclude:
    \begin{equation}
        \text{Rad}_\mathcal{P}(\mathcal{H}) \leq \frac{\omega B}{\sqrt{P}}.
    \end{equation}

    This result indicates that the ability of our function class to fit random noise is influenced by several factors. Specifically, the norm \( \omega \) of the weight vector \( w \) and the norm \( B \) of the input vector \( x \) both play a role in increasing this capability. A larger \( \omega \) expands the range of possible weight vectors, thereby increasing the likelihood of fitting noise. Similarly, a larger \( B \) broadens the range of input values that the model can handle, which also contributes to an increased complexity.
    
    On the other hand, the number of examples \( P \) that the model needs to fit has the opposite effect. As \( P \) grows, the task of fitting all examples becomes more challenging, effectively reducing the model's capacity to fit random noise. This is because with more data points, the model must find a general pattern that applies across all examples, rather than fitting individual or random variations.
    
    In essence, while larger values of \( \omega \) and \( B \) enhance the model's flexibility and its potential to fit a wide variety of patterns (including noise), a larger dataset size \( P \) imposes a constraint on this flexibility, steering the model towards capturing more general, underlying trends in the data rather than fitting to noise.
}

Similar to the VC dimension, when we have a capacity measure, it can be utilized to establish a formal bound on the generalization performance of a given hypothesis space made of real-valued functions, such as a Deep Neural Network (DNN). The Rademacher complexity provides a framework for this purpose:
\begin{theorem}[Rademacher complexity bound]
    Let $X \subseteq \mathbb{R}^d$ and $Y \subseteq \mathbb{R}$, with $Z = X \times Y$. Consider a hypothesis space $\mathcal{H}$ and define $\mathcal{L}(\mathcal{H}) = \{L_{f_w} \mid f_w \in \mathcal{H}\}$, where $\mathcal{L}_h : Z \to \mathbb{R}$ represents loss functions. For any $\delta > 0$, with probability at least $1 - \delta$ over the choice of a training set of size $P$, the following bound holds for all $f_w \in \mathcal{H}$:
    \begin{equation}
        \Delta \epsilon = \epsilon_g(f_w) - \epsilon_t(f_w) \leq 2 \text{Rad}(\mathcal{H}\circ S) - 4\sqrt{\frac{2\log(4/\delta)}{P}}
    \end{equation}
    where $\epsilon_g(f_w)$ denotes the generalization error and $\epsilon_t(f_w)$ the training error of the hypothesis $f_w$.
\end{theorem}
It is noteworthy that this bound can be related to the VC dimension one using the \textit{Growth funcion}, as showed in \cite{Bousquet2004}. For example, one can show that exists a constant $C$, such that any class of $\{0,1\}$-indicator functions with $d_{VC} = d$ has Rademacher complexity upper-bounded by $C\sqrt{d/P}$. This suggests that for DNNs, the Rademacher complexity can be estimated as $\sqrt{N_{tot}/P}$

While the Rademacher Complexity appears to offer a more profound and well structured theoretical approach compared to the VC Dimension, it, much like its predecessor, falls short in providing informative bounds when applied to modern Deep Neural Networks (DNNs).

\subsection{Great limitation of SLT and a New Hope}
Originally developed for simpler models, traditional theoretical tools developed by the Statistical Learning Theory (SLT), like the VC dimension and Rademacher complexity bounds, face significant challenges when applied to contemporary neural architectures.
The intricate design of these networks, marked by deep layers and a vast number of parameters, often leads to traditional theoretical bounds being overly conservative or disconnected from the actual performance metrics of modern-day DNNs in the overparametrised regime, making them essentially useless.
This situation underscores a critical gap in machine learning theory. Consequently, there is an urgent need for novel theoretical frameworks and metrics that can offer more relevant and practical insights into the generalization abilities of these complex models.

Anchored to its origins (from statistics, functional analysis, and computer science), SLT is characterized by rigorous statistical theorems, typically centered around the most agnostic models of a hypothesis space and worst-case scenarios. In fact, the core objective of SLT theorems is to provide distribution-independent uniform bounds that accurately reflect the difference between generalization and training errors, as demonstrated by the VC bound and Rademacher complexity.

However, the focus on generality and rigor in SLT may limit the practical applicability of these theorems to modern algorithms. This limitation has been addressed by Bottou in \cite{bottou2015making} and further examined in a recent review \cite{belkin2021fit} by Belkin. The practical inadequacies of these theoretical models can be primarily attributed to two factors:
\begin{itemize}
\item Real-world data distributions often exhibit smaller deviations than those anticipated in worst-case scenarios \cite{PhysRevResearch.2.023169, PhysRevLett.125.120601, PhysRevE.102.032119,e23030305, Pastore_2021}.
\item While uniform bounds apply to all functions within a hypothesis space, more accurate bounds could be derived by concentrating on trained functions that obviously perform well on specific and meaningful training sets.
\end{itemize}

These insights suggest a need for a more nuanced approach in Statistical Learning Theory (SLT), balancing mathematical rigor with the practical requirements of modern machine learning algorithms. 
In the following sections, we expand upon these insights to develop a mean-field theory addressing the \textit{generalization gap}, denoted as $\Delta\epsilon$, in deep neural networks. 
Our method begins with the application of statistical physics tools used for disordered systems. 
This approach facilitates the calculation of both generalization and training errors in models with quenched features, which we will introduce shortly. This leads to the derivation of straightforward formulas that apply to a meaningful subspace of the hypothesis space which, we will argue, contains trained networks.
We will finally focus our results on the regime of large datasets and network width, typical in most state-of-the-art deep neural networks, and identify a simple asymptotic upper bound.

\section{Quenched feature}
When analyzing Deep Neural Networks (DNNs), it's crucial to account for two distinct sources of randomness: the weights and the dataset. As previously mentioned (\ref{subsec:1CommApp}), these two elements require different treatments. 
Specifically, the weights are the primary variables over which we aim to perform averages (or optimize in the Gibbs learning scenario). 
On the other hand, the dataset represents a type of ``conformational disorder''.
While, for a general description of the system performance, it's necessary to average over all possible instances of the datasets, it's important to recognize that the dataset influences the system's dynamics without evolving within it. In other words, we can say that it is ``quenched'' from the point of view of weight optimization. 

In this section, we employ and expand upon a framework recently introduced by \cite{pmlr-v119-20a}, based on the concept of the \textit{quenched average}, which appropriately handles both sources of randomness.
Utilizing this framework, along with standard techniques from Statistical Mechanics such as the \textit{replica trick}, enables us to derive analytical formulas for both training and generalization errors.

\subsection{Setting}

We start with a brief description of the supervised learning problem that we will study in the so-called \textit{teacher-student framework} \cite{EngelVanDenBroeck}. 
Let us consider a training set $\mathcal{T}$ made of $P$ independent identically distributed (iid) random observations, $\mathcal{T} = \{(\mathbf{x}^\mu, y^\mu)\}_{\mu=1}^P$. 
Here, $\mathbf{x}^\mu \in \mathbb{R}^D $ are drawn from an input probability distribution $\rho(\mathbf{x})$, and the scalar outputs $y^\mu \in \mathbb{R}$ are provided by a teacher function $f_T$, i.e., $y^\mu = f_T(\mathbf{x}^\mu)$.
Under these assumptions, the \textit{joint input/output probability distribution} can be written as $\rho_{I/O}(\mathbf{x}, y) = \rho(x)\delta(y-f_T(\mathbf{x}))$.

Next, we want to define a student function to represent our algorithm.
The student will be then optimized to reproduce as best as it can the teacher function on the training dataset. 
This function is described in a different fashion, the reason for which will become apparent shortly. 
We conceptualize a general fully connected Deep Neural Network (DNN) as comprising two interconnected parts. 
The first part, embodying the initial $L-1$ layers, functions as a generic (non-linear) feature map $\phi: \mathbb{R}^D \to \mathbb{R}^N$. 
Following this, the final layer is modeled as a simple Perceptron with weights $\mathbf{v}$. 
Therefore, any fully connected DNN can be reformulated as:
\begin{equation}
    f_{\textrm{DNN}}(\mathbf{x}) = \sum_{\alpha}^N v_\alpha \phi(x^\mu)_\alpha = \mathbf{v} \cdot \mathbb{\phi(\mathbf{\mathbf{x}})}
    \label{eq2:student}
\end{equation}
The motivation for this different layer's representation is the fact that we want to take into account only the weights of the last layer during the optimization process. In practice, we assume fixed (and random) features and only optimize the last layer weights. This choice has the technical advantage of keeping the loss landscape of our optimization problem convex.
In fact, it is possible to express the generalization and training errors as a function of the features; as we will discuss in the following, this ingredient is fundamental to provide insight into the generalization gap of fully-trained DNNs.

In the teacher-student setup, the average generalization and training errors are defined as:
\begin{align}
\label{eq2:epsilong}
\epsilon_{\textrm g} &= \left\langle\int d^D \mathbf{x} \; \rho(\mathbf x) \; \mathcal{L}(f_T,f^*_{\textrm{DNN}},\mathbf{x^\mu})\right\rangle_{\mathcal T}\,, \\
\label{epsilont}
  \epsilon_{\textrm t} &=  \left\langle \frac{1}{P}  \mathcal{L}(f_T,f^*_{\textrm{DNN}},\mathbf{x^\mu}) \right\rangle_{\mathcal T}\,,
\end{align}
where $\left\langle \cdot \right\rangle_{\mathcal{T}}$ denotes the average over all possible realizations of a training set of size $P$, and the optimized function $f^*_{\textrm{DNN}}$ corresponds to the choice of vector $\mathbf{v}^*$ that minimizes the loss for a specific instance of the dataset $\mathcal{T}$.
While this approach can be applied to various loss functions \cite{loureiro2021learning, pmlr-v119-20a}, we focus here on quadratic loss and regression problems, which are more straightforward to analyze \cite{canatar2021, pmlr-v139-mel21a, Coolen_2020,hastie2022surprises}:
\begin{equation}
    \mathcal{L}(f_{\textrm{T}}, f_{\textrm{DNN}}, \mathbf{x^\mu}) = \frac{1}{P} \sum_{\mu=1}^P \left[ f_{\textrm{T}} (\mathbf{x^\mu}) - f_{\textrm{DNN}} (\mathbf{x^\mu}) \right]^2.
    \label{eq2:loss}
\end{equation}

In this setting, the calculation of the average generalization and training errors is performed using the well-known replica method, a standard statistical physics technique developed to study disordered systems \cite{mezard1987}.

\subsection{Partition function}
To evaluate the errors, as shown in the previous chapter, one has to introduce the partition function $Z$, using an approach similar to that developed for the Random Feature Model (RFM) \cite{pmlr-v119-20a} and for kernel regression \cite{canatar2021}. Note that here we chose only to integrate out the last layer weights ${\bf v}$.
\begin{equation}
    Z(\mathbf{x}) = \int d \mathbf{v} \, e^{-\frac{\beta}{2}\mathcal{L}(f_T,f_{\textrm{DNN}(\mathbf{v})},\mathbf{x}) - \frac{\beta\lambda}{2} \lVert \mathbf{v}\rVert^2}\,,
    \label{2AB:Z1}
\end{equation}
where $\mathcal{L}_{\textrm{Reg}} = \mathcal{L} + \lambda \lVert \mathbf{v} \rVert^2 $ is the regularized loss function serving as a Hamiltonian in this context, and $\beta$ represents the inverse of an effective temperature, introducing an equilibrium statistical mechanics perspective to describe the training of the output weights.
The term $\lambda \lVert \mathbf{v} \rVert^2$, with $\lambda > 0$,  is an additional regularization term, commonly used in DNNs in order to control the overfitting by penalizing large weights.

As $\beta \to \infty$, the Gibbs measure is fully dominated by the regularized Loss functions's minimum, reflecting the training loss's minimum. 
This scenario is akin to considering a zero-temperature cooled physical system, where the system settles into the state of minimal energy.
As discussed in the previous section the partition function \ref*{2AB:Z1} does depend on the specific choice of the training dataset. 
To meaningfully take averages over different instances of the training dataset we have to take into account the quenched average, that is by averaging the logarithm of the partition function over all the possible realizations of the training set $\mathcal{T}$. 
Physically, this amounts to optimizing first the student weights for any given instance of the dataset T and then averaging the corresponding free energy over all dataset realizations \cite{Gardner_1988}. 

Technically, to perform this quenched average we exploit the following mathematical trick:
\begin{equation}
    \langle \log Z \rangle_{\mathcal{T}} = \lim_{m \to 0} \frac{\langle Z^m\rangle_{\mathcal{T}} -1}{m}\,.
    \label{replica}
\end{equation}
The technique, known as the replica method \cite{mezard1987}, involves initially computing $Z^m$ for an integer number of replica $m$, averaging over the data and later extending this analytically to $m \rightarrow 0$. 

Considering that inputs $\mathbf{x}^\mu$  are independent and identically distributed variables, the integral over the training set can be factorized, leading to:
\begin{equation}
 \langle Z^m \rangle_{\mathcal{T}} =
\int \prod_{a=1}^m    \text{d}^N\mathbf{v}^a \; e^{ - \frac{\beta}{2} \lambda \sum_{a}^m \Vert\mathbf{v}^a\Vert^2 }  \, \left[\, \int\! \text{d}^D\! x\, \rho(\mathbf{x}) e^{- \frac{\beta}{2}\! \sum_a^m (q^a)^2} \,\right]^P\,.
\label{eq2:gauss}
\end{equation}
In this expression, $a$ is a replica index and $q^a$ are auxiliary random variables defined as $q^a \equiv f_{\text{T}}(\mathbf{x}) - \mathbf{v}^a \cdot \pmb{\phi} (\mathbf x)$.

We now consider a key assumption in our calculation. In fact, to advance in our analysis, we observe that each random variable $q^a$ is the sum of $N$ random variables.
In the context of a large last layer size $N$ and input dimension $D$, we approximate their probability distribution with a multivariate Gaussian. 
This Gaussian is characterized by a mean $\mu_q^a ({\mathbf v^a})= \langle q^a \rangle_{\rho}$ and a covariance matrix $Q_{ab}({\mathbf v^a})$.

It is important to note that while this Gaussian approximation is key to our progress, it remains somewhat uncontrolled due to the absence of a formal result confirming its validity. However, similar non-rigorous approximations are commonly employed in the literature on statistical learning. This includes seminal work on support vector machines by Dietrich, Opper, and Sompolinsky \cite{PhysRevLett.82.2975}, recent studies on kernel regression \cite{canatar2021}, and research on the so-called random feature model, where this approximation is known as the \emph{Gaussian equivalence principle} \cite{PhysRevX.10.041044}.
These considerations suggest that the $N$ feature maps $\phi_\alpha(\mathbf{x})$ exhibit only weak mutual correlations, an assumption that is considered reasonable for a broad range of relevant architectures. This is particularly applicable in the thermodynamic limit where $N$ and $D$ are large, but their ratio $N/D$ remains finite \cite{PhysRevX.10.041044}.

The above-mentioned covariance matrix $Q_{ab}({\mathbf v^a})$ plays, therefore, a pivotal role in our results and is defined as:
\begin{equation}
Q_{ab}({\mathbf v^a},{\mathbf v^b}) = \langle q^a q^b \rangle_{\rho} = T +\left(\mathbf{v}^a\right)^T \mathbf{\Phi}\, \mathbf{v}^b - \mathbf{J}^T\cdot (\mathbf{v}^a + \mathbf{v}^b)\,,
\label{eq2:cov}
\end{equation}
where the components $\mathbf{J}$, $\mathbf{\Phi}$, and $T$ (respectively, a vector, a matrix and a scalar) of the covariance matrix have specific interpretations. 
The first two depend on the chosen feature map $\pmb{\phi}$, and represent the teacher-feature and feature-feature overlaps, respectively. The last component, $T$, serves as the trivial predictor \cite{10.1214/20-AOS1990} of the regression problem, providing a natural scale for comparing different learning problems:
\begin{equation}
\begin{split}
    J_\alpha &= \int d^D x \rho(\mathbf{x}) f_{\textrm{T}}(\mathbf{x}) \phi_\alpha (\mathbf{x})\,, \\ 
    \Phi_{\alpha \beta} &= \int d^D x \rho (\mathbf{x}) \phi_{\alpha} (\mathbf{x}) \phi_\beta (\mathbf{x})\,, \\
   T &= \int d^D x \rho (\mathbf{x}) f^2_{\textrm{T}}(\mathbf{x})\,.
\end{split}
\label{eq2:JPhiT}
\end{equation}
with $\alpha,\beta=1,\ldots, N$.

This Gaussian approximation enables us to perform the integration within the square brackets in Eq. \ref{eq2:gauss} leading to:
\begin{equation}
    \begin{split}
        &\int\! \text{d}^D\! x\, \rho(\mathbf{x}) e^{- \frac{\beta}{2}\! \sum_a^m (q^a)^2} \simeq \bigg(\det(\mathbb{I} + \beta \mathbf{Q})\bigg)^{-\frac{1}{2}}\,,
    \end{split}
    \label{eq2:gauss22}
    \end{equation}

Thanks to the convexity of our optimization problem restricted to the last layer, we can work within the framework of the replica symmetric ansatz. Computing the leading order contribution in $m$ we get: 
\begin{equation}
\langle Z^m \rangle_{\mathcal{T}} \approx \int dQ_0 dQ d\hat{Q}_0 d\hat{Q} \,\,e^{-\frac{mP}{2}S_\beta(Q_0, Q, \hat{Q}_0, \hat{Q})}\,,
\label{eq2:Zfin}
\end{equation} 
where $Q_0$ and $Q$ are respectively the diagonal and off-diagonal elements of the replica symmetric covariance matrix while $\hat{Q}_0$ and $\hat{Q}$ are their conjugated variables. From a more physical perspective, $Q_0$ represents the self-overlap within the same replica, whereas $Q$ denotes the overlap between different replicas. This overlap $Q$ remains constant for any choice of different replicas, a condition that arises because we are considering the replica symmetric case. Essentially, the overlap is a measure of the similarity between two configurations \cite*{pedestrian}.
Here, the action $S_\beta(Q_0, Q, \hat{Q}_0, \hat{Q})$ is expressed through a convoluted formula, given in the subsequent \textit{details} section.


\details{
    \begin{center}
        \textbf{Partition Function calculation detail}
    \end{center}

    In this section, starting from Eq. \ref{eq2:gauss}, I will show you the main steps of the calculation. The first operation involves the identity:

    \begin{equation}
        1 = \int \!dQ_{a b} \,\delta\left (Q_{a b} -  \langle q^a q^b \rangle_{\rho} \right) = \int dQ_{a b}\frac{d\hat{Q}_{a b}}{2\pi} \,e^{-i \hat{Q}_{a b} \left (Q_{a b} -  \langle q^a q^b \rangle_{\rho} \right)}
    \label{eq2:ide}
    \end{equation} 
    If we introduce this identity in Eq. \ref{eq2:gauss}, we can deal with the integral in the square brackets: 
    
    \begin{equation}
        \begin{split}
            &\int\! \text{d}^D\! x\, \rho(\mathbf{x}) e^{- \frac{\beta}{2}\! \sum_a^m (q^a)^2} =\\
            &=  \int \prod_a^m dq^a e^{-\frac{\beta}{2}	\sum_a^m (q^a)^2} \int d^D x\; \rho(\mathbf x) \; \delta \left( q_a - \sum_\alpha^{N} v_\alpha^a \phi_\alpha(\mathbf{x})+f_{\textrm T}(\mathbf x) \right)\\
            &\simeq \int \prod_a^m d q^a  \,\frac{e^{-\frac{\beta}{2}\sum_a (q^a)^2 -\frac{1}{2}\sum_{a b} (q^a)^T Q^{-1}_{a b} q^b  }}{\sqrt{(2\pi)^{m} \;\det \mathbf{Q} } } =\\
            &= \bigg(\det(\mathbb{I} + \beta \mathbf{Q})\bigg)^{-\frac{1}{2}} \equiv K(\beta, {\bf Q})^{-\frac{1}{2}}\,.
        \end{split}
        \label{eq2:gauss2}
    \end{equation}
    where we have assumed $\mu_q^a=0$ for simplicity, without loss of generality \cite{canatar2021}.
    Returning to the partition function, we now have:
    \begin{equation}
        \begin{split}
            \langle Z^m \rangle_{\mathcal{T}} = &\int dQ_{a b}d\hat{Q}_{a b}\; K(\beta, Q)^{-\frac{P}{2}} e^{iQ_{a b}\hat{Q}_{a b}} \times\\  &\times \int d^N\mathbf{v}^a e^{-\frac{\beta \lambda}{2}\sum_{a} \Vert\mathbf{v}^a \Vert^2-i\sum_{a\leq b} \hat{Q}_{a b} \left(T +\left(\mathbf{v}^a\right)^T \mathbf{\Phi}\; \mathbf{v}^b - \mathbf{J}^T(\mathbf{v}^a+\mathbf{v}^b) \right)}
        \end{split}
        \label{eq2:Zinit}
    \end{equation}

    Now, our goal is to explicitly perform the integration over the replicated weights $\mathbf{v}^a$ and to evaluate the integrals over $Q_{ab}$ and $\hat{Q}_{ab}$ using the saddle-point method. 
    However, as it occurs in standard spin glass models, a notable challenge arises in executing the delicate limit $m \to 0$, which can potentially lead to the breaking of the so-called replica symmetry in the matrix $Q_{ab}$. Fortunately, in our specific context, the underlying optimization problem is convex. This convexity ensures that the simplest replica symmetric ansatz for the matrix $Q_{ab}$ will yield the exact solution to the saddle-point equations in the $m \to 0$ limit \cite{PhysRevLett.82.2975}.

    Under the assumption of replica symmetry, the matrices $Q_{ab}$ and $\hat{Q}_{ab}$ are structured as follows:
    \begin{equation}
    Q_{a b} = 
    \begin{cases}
      Q_0 \quad a = b \\
      Q\,\,   \quad a \neq b
    \end{cases}
    \quad
    \hat{Q}_{a b} = 
    \begin{cases}
      \hat{Q}_0 \quad a = b \\
      \hat{Q}\,\,   \quad a \neq b\,.
    \end{cases}
    \label{eq2:RSA}
    \end{equation}
    This formulation, exact for convex problems, simplifies the analysis and allows us to proceed with the calculation under the replica symmetric framework.

    With a slight abuse of notation, we can specify our analysis to the replica symmetric ansatz already at the level of Eq. \ref{eq2:Zinit}, thus obtaining:

    \begin{equation}
    \begin{split}
        \langle Z^m \rangle_{\mathcal{T}} =& \int dQ_0\;dQ\;d\hat{Q}_0\;d\hat{Q}\; K^{-\frac{P}{2}}e^{im\left(\hat{Q}_0(Q_0-T)+\frac{m-1}{2}\hat{Q}(Q-T)\right)} \times \\
        &\times \int d^N \mathbf{v}^a \; e^{-\frac{\beta \lambda}{2}\sum_{\alpha,a} (v_\alpha^a)^2- i\left(\hat{Q}_0-\frac{\hat{Q}}{2}\right)\sum_a\sum_{\alpha,\beta}v_\alpha^a \Phi_{\alpha\beta} v_\beta^a} \times \\ &\qquad\quad \;\; \times e^{ -i\frac{\hat{Q}}{2}\sum_{a,b}\sum_{\alpha,\beta}v^a_\alpha\Phi_{\alpha\beta}v_\beta^b + 2i\left(\hat{Q}_0 -(1-m)\frac{\hat{Q}}{2}\right)\sum_a\sum_\alpha J_\alpha v_\alpha^a} \,.
    \end{split}
    \end{equation}
    
    We now turn our attention to the integral over the replicated weights. The initial step involves applying Wick's rotation \cite{Wick} to the order parameters: specifically, transforming $\hat{Q}_0 \to i\hat{Q}_0/2$ and $\hat{Q} \to i\hat{Q}$. Subsequently, we perform a rotation of the weights to diagonalize the matrix $\Phi$. For simplicity, and with a slight abuse of notation, we express the transformed equation as follows:

    \begin{equation}
        \begin{split}
            \langle Z^m \rangle_{\mathcal{T}} =& \int dQ_0\;dQ\;d\hat{Q}_0\;d\hat{Q}\; K^{-\frac{P}{2}}e^{-\frac{m}{2}\left(\hat{Q}_0(Q_0-T)+(m-1)\hat{Q}(Q-T)\right)} \times \\
            &\times \int d^N \mathbf{v}^a \;\text{Jac}(\mathbf{v}) \; e^{-\frac{\beta \lambda}{2}\sum_{\alpha,a} (v_\alpha^a)^2+\frac{1}{2}\left(\hat{Q}_0-\hat{Q}\right)\sum_a \sum_\alpha v^a_\alpha \lambda_\alpha v_\alpha^a} \times \\ &\qquad\quad \;\; \times e^{ +\frac{\hat{Q}}{2}\sum_\alpha \lambda_\alpha  (\sum_{a}v^a_\alpha)^2 - \left(\hat{Q}_0 -(1-m)\frac{\hat{Q}}{2}\right)\sum_a \sum_\alpha \Tilde{J}_\alpha v_\alpha^a} \,,
        \end{split}
    \end{equation}
    In this equation, $\mathbf{JO}=\tilde{\mathbf{J}}$, where $\mathbf{JO}$ represents the matrix of rotation, and $\text{Jac}(v)$ denotes the Jacobian of the transformation. We omit the explicit definition of the Jacobian here, as we intend to revert to the original variables shortly.

    Now, focusing on the integral over the weights, we can use the standard technique known as \textit{Hubbard-Stratonovich trasformation} \cite*{HStra}.
    \begin{equation}
        \begin{split}
            \int \prod_\alpha \frac{dz_\alpha}{\sqrt{(2\pi)^P}} e^{\frac{1}{2}\sum_\alpha z^2_\alpha} &\int d^N \mathbf{v} \;\text{Jac}(\mathbf{v}) \; e^{-\frac{\beta \lambda}{2}\sum_{\alpha,a} (v^a_\alpha)^2+\frac{1}{2}\left(\hat{Q}_0-\hat{Q}\right) \sum_{\alpha,a} v_\alpha^a \lambda_\alpha v_\alpha^a}\\ 
            & \qquad \qquad \qquad e^{+\sum_\alpha \sqrt{\hat{Q} \lambda_\alpha} z_\alpha \sum_{a} v^a_\alpha  - \left(\hat{Q}_0 -(1-m)\frac{\hat{Q}}{2}\right)\sum_{\alpha, a} \Tilde{J}_\alpha v_\alpha^a}  \,.
        \end{split}
    \end{equation}
    At this point is pretty clear that $a$ is a dumb index, so we can rewrite the integral over the weights as:
    \begin{equation}
        \begin{split}
            \left[\int d^N \mathbf{v} \;\text{Jac}(\mathbf{v}) \; e^{-\frac{1}{2}\sum_{\alpha} (\beta\lambda-(\hat{Q}_o)-\hat{Q}) v_\alpha^2+\sum_\alpha \sqrt{\hat{Q} \lambda_\alpha} z_\alpha v_\alpha  - \left(\hat{Q}_0 -(1-m)\frac{\hat{Q}}{2}\right)\sum_{\alpha} \Tilde{J}_\alpha v_\alpha}  \right]^m\,.
        \end{split}
    \end{equation}
    This integral is Gaussian and we can perform it easily, obtaining:
    \begin{equation}
        \prod_\alpha  \left(\frac{2\pi}{\beta\lambda-(\hat{Q}_0-\hat{Q})\lambda_\alpha}\right)^{\frac{m}{2}} \int dz_\alpha e^{-\frac{1}{2}\sum_\alpha z_\alpha^2 + \sum_\alpha \frac{m\left(\sqrt{\hat{Q}\lambda_\alpha}z_\alpha - \left(\hat{Q}_0 -(1-m)\frac{\hat{Q}}{2}\right) \Tilde{J}_\alpha \right)^2}{2(\beta\lambda-(\hat{Q}_0-\hat{Q})\lambda_\alpha)}}\,,
    \end{equation}
    which is itself a Gaussian integral. In the end, performing the integral and rotating back the variable, we obtain: 
    \begin{equation}
        \begin{split}
            \langle Z^m \rangle_{\mathcal{T}} =& \int dQ_0\;dQ\;d\hat{Q}_0\;d\hat{Q}\; K^{-\frac{P}{2}}e^{im\left(\hat{Q}_0(Q_0-T)+\frac{m-1}{2}\hat{Q}(Q-T)\right)} \times\\
            &\qquad (2\pi)^{\frac{m}{2}} e^{-\frac{m-1}{2}\text{Tr}\left[\log\left(\beta\lambda\mathds{1}-(\hat{Q}_0-\hat{Q})\mathbf{\Phi} \right)\right]-\frac{1}{2}\text{Tr}\left[\log\left(\beta\lambda\mathds{1}-(\hat{Q}_0-(1-m)\hat{Q})\mathbf{\Phi}\right)\right]} \\
            &\qquad e^{\frac{m}{2}\left(\hat{Q}_0-(1-m)\hat{Q}\right)^2 \mathbf{J}^T\left(\beta\lambda\mathds{1}-(\hat{Q}_0-(1-m)\hat{Q})\mathbf{\Phi}\right)^{-1}\mathbf{J}}  
        \end{split}
    \end{equation}
    We also note that, under the replica symmetric ansatz, the term $K$ simplifies:
    \begin{equation}
        K = \det(\mathds{1}+\beta \mathbf{Q}) =  (1+\beta(Q_0-Q))^{m-1}(1+\beta(Q_0-(1-m)Q)). 
    \end{equation}

    Now, we want to obtain the leading (linear) order in $m$ as suggested by the recipe of the \textit{replica trick}. By doing that and performing the transformation $\hat{Q}\to P\hat{Q}$ and $\hat{Q}_0\to P\hat{Q}_0$ we obtain a final form for $\langle Z^m \rangle_{\mathcal{T}}$ (Eq. \ref{eq2:Zfin}), where the action amount to:
    \begin{equation}
            \begin{split}
               S_\beta&(Q_0,Q,\hat{Q}_0,\hat{Q})= \; \log(1+\beta(Q_0\!-\!Q)) + \hat{Q}_0(Q_0-T) -\hat{Q}(Q-T) +\\
               & +\frac{1}{P}\text{Tr}\left[\log\left(\beta\lambda\mathds{1}-P(\hat{Q}_0-\hat{Q})\mathbf{\Phi}\right)\right]-\hat{Q}\text{Tr}\left[\mathbf{\Phi}\left(\beta\lambda\mathds{1}-P(\hat{Q}_0-\hat{Q})\mathbf{\Phi}\right)^{-1}\right]+\\
                &-P\left(\hat{Q}_0-\hat{Q}\right)^2 \mathbf{J}^T\left(\beta\lambda\mathds{1}-P(\hat{Q}_0-\hat{Q})\mathbf{\Phi}\right)^{-1}\mathbf{J}+\frac{\beta Q}{1+\beta(Q_0\!-\!Q)}
            \end{split}
        \label{eq2:action}
    \end{equation}

}

To effectively evaluate the integral \ref{eq2:Zfin} in the context of large $P$, we employ the \textbf{saddle-point method}. This method is a powerful technique for approximating certain integrals by focusing on points of maximum curvature, known as saddle points. In our analysis, it is necessary to solve a set of four saddle-point equations. These equations are derived by considering the variation of the action $S$ with respect to each of the order parameters. The solutions to these equations identify the special values of the order parameters that correspond to the action's minimum. This minimum represents a special point that dominates the statistics of the system.

\details{
    \begin{center}
        \textbf{Saddle point calculation}
    \end{center}

    First, we note that differentiating the action $S_\beta$ with respect to $Q$ and $Q_0$ yields explicit expressions for $\hat{Q}$ and $\hat{Q}0$. By setting the derivatives to zero:
\begin{equation}
        \frac{\partial S_\beta}{\partial Q} = 0 \quad \text{and} \quad   \frac{\partial S_\beta}{\partial Q_0} = 0\,,
\end{equation}
we obtain: 
\begin{equation}
\hat{Q} = \frac{\beta^2 Q}{(1+\beta(Q_0-Q))^2} \quad \text{and} \quad  \hat{Q}_0 = \hat{Q} - \frac{\beta}{1+\beta(Q_0-Q)}\,.
\label{SP1}
\end{equation}
Considering the derivative of the action with respect to $\hat{Q}$ and setting it to zero, we find:
\begin{equation}
    \begin{split}
       Q  =
        T  \!+\! P\,\hat{Q}\; \text{Tr}[\mathbf{(\Phi} \tilde{\mathbf{G}}^{-1})^2] 
         \!+\! P \mathbf{J}^T \frac{P\mathbf{\Phi}(\hat{Q}_0 \!-\!\hat{Q})^2 \! \!+\! 2\tilde{\mathbf{G}}(\hat{Q}_0 \!-\!\hat{Q})}{\tilde{\mathbf{G}}^2}\mathbf{J}\,,
    \end{split}
    \label{SP3}
\end{equation}
where $\Tilde{\mathbf{G}}$ is a $N\times N$ matrix defined as $\Tilde{\mathbf{G}} = \beta\lambda\mathds{1}-P(\hat{Q}_0-\hat{Q})\mathbf{\Phi}$. The saddle point equation for $\hat{Q}_0$ then becomes:
\begin{equation}
    \begin{split}
     Q_0 = Q +\text{Tr}\left[\mathbf{\Phi} \tilde{\mathbf{G}}^{-1}\right]\,.
    \end{split}
    \label{SP4}
\end{equation}
Now, let's consider the special combination $\kappa = 1 + \beta (Q_0-Q)$. Using Eq. \eqref{SP4}, we derive:
\begin{equation}
    \begin{split}
        \kappa  = &1 + \beta (Q_0-Q) = 1 + \beta \left( Q +\text{Tr}\left[\mathbf{\Phi} \tilde{\mathbf{G}}^{-1}\right]-Q\right) \\&= 1 +  \beta\;\text{Tr}\left[\mathbf{\Phi} \tilde{\mathbf{G}}^{-1}\right]\,,
    \end{split}
\end{equation}
and the difference $\hat{Q}_0 - \hat{Q}$ (using Eq. \ref{SP1}) can be expressed as:
\begin{equation}
    \begin{split}
        \hat{Q}_0\!-\!\hat{Q} = \hat{Q}\! -\! \frac{\beta}{1\!+\!\beta(Q_0\!-\!Q)} \!-\! \hat{Q} = -\frac{\beta}{1\!+\!\beta(Q_0\!-\!Q)} = -\frac{\beta}{\kappa}\,.
    \end{split}
\end{equation}
Substituting this result into the definition of $\tilde{\mathbf{G}}$, we define the rescaled matrix $\mathbf{G} = \kappa \tilde{\mathbf{G}}/\beta$:
\begin{equation}
    \begin{split}
        \mathbf{G} =  \frac{\kappa}{\beta}\left[\beta\lambda\mathds{1}-P(\hat{Q}_0-\hat{Q})\mathbf{\Phi} \right]= \kappa\lambda\mathds{1}+P\mathbf{\Phi}\,.
    \end{split}
\end{equation}
These steps lead us to a self-consistency equation for $\kappa$::
\begin{equation}
    \begin{split}
        \kappa = 1 + \beta\;\text{Tr} \left[\mathbf{\Phi} \left(\frac{\beta}{\kappa} \mathbf{G}\right)^{-1}\right]= 1 + \kappa \;\text{Tr} \left[\frac{\mathbf{\Phi}}{\kappa\lambda\mathds{1}+P\mathbf{\Phi}}\right]\,.
    \end{split}
    \label{kappa}
\end{equation}
Finally, we can express the solution of the saddle-point equations in a more convenient form (with solutions indicated by an asterisk):
\begin{equation}
    \begin{split}
        &\hat{Q}^*_0 = \hat{Q}^* - \frac{\beta}{\kappa} \qquad \qquad \quad \;\; \hat{Q}^* = \frac{\beta^2Q^*}{\kappa^2}\\
        & Q_0^* = Q^* +\frac{\kappa -1}{\beta} \qquad \qquad Q^* =  \frac{T - P \mathbf{J}^T\frac{2\kappa\lambda+P\mathbf{\Phi}}{\mathbf{G}^2}\mathbf{J}}{1-P\text{Tr}[\mathbf{\Phi}^2 \mathbf{G}^{-2}] }\,.
        \end{split}
        \label{eq2:OP}
\end{equation}
It is important to note that since $\mathbf{G}$ and $\kappa$ are independent of the inverse temperature $\beta$, the solution for the order parameter $Q^*$ is also independent of the temperature. Furthermore, in the limit $\beta \rightarrow \infty$, we find that $Q^* = Q^*_0$, while the conjugated variables diverge, but this does not raise a problem because they are non-physical.

}

\subsection{Train and Generalization error}

We now turn to the macroscopic observable of interest, the generalization $\epsilon_g$ and training $\epsilon_t$ errors, that can be closely linked to the saddle-point replica order parameters. 
To understand this, let's first consider the generalization error. 
Referring to Eqs. \eqref{eq2:epsilong} and \eqref{eq2:student} in the main text, we have:
\begin{equation}
\begin{split}
\label{eq34}
        \epsilon_{\text{g}} &= \int
        d^Dx \rho(\mathbf{x}) \left(f_T(\mathbf{x}) - \mathbf{v}^*\cdot\pmb{\phi}(\mathbf{x}) \right)^2\\
        &= T -2 \mathbf{J}^T\mathbf{v}^* + {\mathbf{v}^*}^T \mathbf{\Phi} \mathbf{v}^*=Q^*\,.
\end{split}
\end{equation}
In the last equality, we utilized Eq. \eqref{eq2:cov} and the replica symmetric ansatz in Eq. \ref{eq2:RSA}.

As $P \to \infty$, it is straightforward to demonstrate that $\kappa \sim 1$ and $\mathbf{G} \sim P \mathbf{\Phi}$, leading the generalization error to converge to the \emph{residual generalization error}:
\begin{equation}
\epsilon_{\text{g}} \rightarrow \epsilon_{\text{g}}^{\textrm{R}} = T - \mathbf{J}^T \mathbf{\Phi}^{-1}\mathbf{J}\,. 
\label{eq2:egr}    
\end{equation}
This residual generalization error represents the optimal performance that the quenched model can achieve on the dataset, assuming complete knowledge of the input/output probability distribution.

This outcome is not unexpected and serves as a first validation of our replica mean field theory. One could also derive it by directly minimizing Eq. \eqref{eq34} with respect to the parameters $\mathbf{v}$. Specifically, $\partial_{\mathbf{v}} \epsilon_{\text g} = - 2 \mathbf{J}^T + 2 \mathbf{\Phi} \mathbf{v}$ leads to $\mathbf{v}^* = \mathbf{J}^T\mathbf{\Phi}^{-1}$, and thus $\epsilon_{\text g}(\mathbf{v}^*) \equiv \epsilon_{\text g}^{\text R} = T - \mathbf{J}^T \mathbf{\Phi}^{-1}\bf{J}$.

Finally, by isolating the residual generalization error in Eq. \eqref{eq2:OP}, we arrive at a concise formula for the generalization error:

\begin{equation}
    \epsilon_{\textrm{g}} = \frac{\epsilon_{\textrm{g}}^{\textrm{R}} + (\kappa \lambda)^2 {\bf J}^T \mathbf{\Phi}^{-1} {\bf G}^{-2}{\bf J}}{1-P \textrm{Tr}\left(\mathbf{\Phi}^2 { \bf G}^{-2}\right)}\,. \\ 
\label{eq2:traintesteq}    
\end{equation}

Moreover, the theory developed thus far also lets us calculate the average training error. The average training error is essentially the average of the loss function defined in Eq. \eqref{eq2:loss}, evaluated at $\lambda = 0$ (up to a factor of $P$). This implies that the training error can be determined by evaluating the action at the saddle point solution and then taking the limit as $\beta \rightarrow \infty$. The equation for this process is:

\begin{equation}
\epsilon_{\text{t}} = \lim_{\beta \to \infty} \frac{1}{\beta}\left. S_\beta (Q^*,Q_0^*,\hat{Q}^*,\hat{Q}_0^*)\right|_{\lambda=0} \,.
\end{equation}

Through straightforward algebraic operations, one can derive an expression for the training error:

\begin{equation}
\epsilon_{\text{t}} = \frac{\epsilon_{\text{g}}}{\kappa}\left(\frac{\kappa-1}{\kappa} - \frac{N}{P} \right) + \frac{\epsilon_{\text{g}}^{\text{R}}}{\kappa}
\label{eq2:gentesteq}   
\end{equation}
This equation provides a direct link between the training error and the other key quantities derived from the replica method, offering a comprehensive view of the learning dynamics in the system.

It is important to note that when the dataset size $P$ becomes infinitely large, $\kappa \rightarrow 1$. In this scenario, it is straightforward to demonstrate that $\epsilon_{\textrm{g}} \rightarrow \epsilon_{\textrm{g}}^{\textrm{R}}$ and $\epsilon_{\textrm{t}} \rightarrow \epsilon_{\textrm{g}}$. These convergences serve as an initial consistency check for the validity of the mean field theory. Furthermore, the self-consistent definition of $\kappa$ and its role in the formulation of the generalization error align with the findings in the recent work on kernel regression by Pehlevan's group \cite{canatar2021}. This alignment is not unexpected, as the general quenched features we use here can be specifically adapted to polynomial, Gaussian, or NTK kernels.

For these specific types of quenched features, Equation \ref{eq2:traintesteq} essentially offers an alternative representation of the generalization error formula presented in \cite{canatar2021}. Moreover, a broader generalization of Equation \ref{eq2:traintesteq} has been recently established in \cite{loureiro2021learning}.

\section{Generalization bound }

In the previous section, we presented explicit formulas for training and generalization errors (see Eq. \ref{eq2:traintesteq} and Eq. \ref{eq2:gentesteq}). While analogous results are known in the literature, our current objective is to apply these formulas to study the asymptotic behavior of generalization performance, particularly in the context of a large number of datapoints \( P \).

A critical measure in Statistical Learning Theory (SLT) for evaluating generalization performance, as discussed earlier in this chapter, is the  \textit{generalization gap}. This is the disparity between the generalization error (\( \epsilon_g \)) and the training error (\( \epsilon_t \)). Utilizing our derived results, the generalization gap can be formulated as:

\begin{equation}
    \Delta\epsilon(\mathcal{W}) = \epsilon_g(\mathcal{W})  - \epsilon_t(\mathcal{W})  = \frac{\epsilon_g(\mathcal{W}) }{\kappa}\left(\frac{\kappa^2 - \kappa + 1}{\kappa} + \frac{N}{P}\right) - \frac{\epsilon^R_g(\mathcal{W}) }{\kappa}\,.
    \label{deltaeg}
\end{equation}
where we use the notation $ \epsilon_{\textrm{g}} \left(\mathcal W\right)$, $\epsilon_{\textrm{g}}^{\textrm{R}} \left(\mathcal W\right)$, $\epsilon_{\textrm{t}} \left(\mathcal W\right)$ to stress that the generalisation and training errors depend on $\mathcal W$ via the teacher-feature ${\bf J}$ and feature-feature $\mathbf{\Phi}$ overlaps, see Eq. \ref*{eq2:JPhiT}.

Let us highlight the crucial aspect of this mean-field theory. Consider a model where the fixed features are represented by a Deep Neural Network (DNN). In a fully connected architecture with a single hidden layer, the implemented function is \( f_{\text{1HL}} (\mathbf{x})= \sum_{\alpha=1}^N v_{\alpha} \sigma \left(W_{\alpha} \cdot \mathbf{x}\right) \), assuming, for simplicity, that all biases are zero. The \( W_\alpha \)'s represent the \( D \)-dimensional vector weights of the hidden layer, and \( \sigma \) is a well-defined activation function. Here, the fixed features are defined as \( \phi^{\text{1HL}}_\alpha (\mathbf{x}, W) =  \sigma \left(W_{\alpha} \cdot \mathbf{x}\right) \).

Expanding our perspective, consider the class of all DNNs with a fully connected final layer. This class encompasses all relevant state-of-the-art architectures. The specific architecture of the initial layers has a negligible impact on this analysis. We denote the dimension of the last layer as \( N_{\text{out}} \) and divide the network weights \( \vartheta \) into \( \vartheta = \{ \mathbf{v}, \mathcal{W}\} \). Here, \( \mathbf{v} \) refers to the \( N_{\text{out}} \)-dimensional weights of the last fully-connected layer, while \( \mathcal{W} \) includes all other weights in the DNN. The feature map \( \phi^{\text{DNN}}_{\alpha} (\mathbf{x}, \mathcal{W}) \) is introduced to represent these fixed features.

The key observation is that our analysis provides the average generalization and training errors for \emph{any} choice of feature map \( \phi \) and \emph{any} choice of weights \( \mathcal{W} \). This means that, for a given configuration \( \bar{\mathcal{W}} \), our theory predicts the corresponding average generalization and training errors, assuming that the weights \( \mathbf{v} \) of the last layer are optimally set to minimize the training loss for a fixed \( \bar{\mathcal{W}} \).

While the Eq. \ref{deltaeg} is comprehensive, it becomes particularly insightful under a specific, yet common\footnote{This limit is satisfied in most state-of-the-art DNNs, even in the overparameterized regime, where $P\ll N_{\text{tot}}$.}, regime. 
If we consider the limit where both \( P \) and \( N \) are large, but with \( N \ll P \), the equation simplifies significantly:
\begin{equation}
    \Delta\epsilon(\mathcal{W}) \simeq 2\epsilon_g^R(\mathcal{W})\frac{N}{P}\,,
    \label{asymptoticgap}
\end{equation} 
In this simplification, we used the fact that in this limit, \( \epsilon_g \to \epsilon_g^R(\mathcal{W})/\kappa \) and \( \kappa \to 1 + N/P \).

Following the previous reasoning, we have that the result in Eq. \eqref{asymptoticgap} holds for each given realization of the weights $\mathcal W$ of the DNN if we assume perfect training over the last layer.
This aligns with the recent conjecture by \cite{loureiro2021learning}.
In particular, this result is valid for a fully trained configuration $\theta^* \equiv \{v^*, \mathcal{W}^*\}$, which is a \textit{local} minimum of the loss function. 
Unfortunately, it is important to note that such a local minimum may vary with the size $P$ of the training set, and accordingly, so does $\epsilon_{\textrm{g}}^{\textrm{R}} (\mathcal{W})$.
However, given that the residual generalization error \eqref{eq2:egr} is positive by definition and bounded by $T$, we deduce that $0 \leq \epsilon_{\textrm{g}}^{\textrm{R}} (\mathcal{W}) / T \leq 1$ for any $\mathcal{W}$. Consequently, this leads to establishing an asymptotic mean field upper bound for the (normalized) generalization performance of a generic DNN, with a number of parameters in the last layer $N_{\text{out}}$ and a training set of size $P$:

\begin{equation}
    \Delta\tilde{\epsilon} \equiv \frac{\Delta\epsilon(\mathcal{W})}{\mathbf{T}}  \leq 2 \,\frac{N_{\text{out}}}{P}\,,
\end{equation}
which is the central result of the Chapter.

Our mean field analysis, though not as rigorously formalized as theorems, sets a more restrictive bound on the generalization gap compared to those derived from traditional statistical learning theory (SLT). 
In fact, we have seen in Section \ref*{sec2:1SLT} that SLT, for example, estimates an upper bound that is approximately proportional to \( \sqrt{N_{\text{tot}}/P} \), where \( N_{\text{tot}} \) represents the \textit{total} number of parameters in the network \cite{bartlett2019nearly}. 
In contrast, our derived upper bound is influenced only by the number of parameters in the final layer of the network.
However, it is important to stress that our bound is obtained under the following assumptions: \(N_{\text{out}} \ll P\) and the existence of an optimum for the last layer weights. While this is a common scenario in practice, it should be noted that this makes our bound less rigorous compared to those derived in Statistical Learning Theory (SLT).
For example, consider a purely noisy training set. In the regime of \(N_{\text{out}} \ll P\), we are beyond the capacity where a perceptron (our last layer) can store the random dataset (\(\alpha = P/N_{\text{out}} \leq 2\)), making it impossible to find the optimal last layer weights. This means that our bound is not applicable in this setting. In fact, if we consider a DNN trained on this random teacher, we will have \(\epsilon_t \simeq 0\) while \(\epsilon_g \simeq 0.5\), resulting in \(\Delta\epsilon \simeq 0.5\) and invalidating our bound if \(N_{\text{out}}/P < 1/4\).
However, for completeness' sake, it must be noted that if we consider a random feature model (random feature plus a perceptron in the last layer), we find that \(\epsilon_t \simeq \epsilon_g \simeq 0.5\) (if \(\alpha > 2\)), leading to \(\Delta\epsilon \simeq 0\), and our bound still holds.

This advancement in determining a tighter, albeit non-rigorous, upper bound for the generalization gap stems from incorporating optimization considerations, particularly in the last layer of the model. While SLT seeks bounds applicable to every possible function within a model class, our method focuses on specific, yet significant, subsets of this class. Specifically, in the case of a DNN with parameters \( \vartheta = \{ \mathbf v, \mathcal W\} \), traditional SLT bounds are applicable across all versions of \( \vartheta \). However, our approach is only valid for all the DNNs for which, given a fixed \( \mathcal W \), the weights \( \mathbf v \) in the final layer are optimized w.r.t.~the training set.

\subsection{Numerical Test}
Although our theoretical framework appears robust, we have undertaken a series of numerical experiments to validate and demonstrate the efficacy of our results. These experiments are crucial in testing the practical applicability and eventual limitations of our proposed upper bound. We employed two distinct types of neural network models for our analysis: toy models and state-of-the-art architectures.

The toy models were selected to cover a broad spectrum of scenarios. By manipulating various parameters and configurations within these simplified models, we could extensively test the boundaries and capabilities of our theoretical predictions. This approach allowed us to explore a wide range of cases.

In parallel, we focused on state-of-the-art architectures to evaluate the relevance of our theory in real-world applications. These complex models, often used in cutting-edge machine learning tasks, provided a more stringent testing ground. By applying our theoretical predictions to these advanced models, we could assess their performance in scenarios that are more reflective of actual use cases. This dual approach, encompassing both toy models and advanced architectures, was instrumental in thoroughly vetting our theory's practicality and effectiveness.


\subsubsection{Toy DNNs with synthetic datasets}

Our test of the generalization upper bound began with a fully connected neural network architecture featuring a single hidden layer and ReLU activation functions. The specifics of this architecture varied in each case studied and will be detailed accordingly as the evaluation progresses. To rigorously test our theoretical predictions, we selected three distinct classes of \textit{teacher} functions, each representing a different level of complexity:

\begin{enumerate}
    \item \textbf{Linear Function}: The simplest case involved a linear function \( f_{\textrm{T}}(\mathbf x) = \mathbf t \cdot \mathbf{x} \), where $\textrm{t} \in \mathcal{R}^D$ is a random uniform vector, that is a random vector with elements from a uniform distribution. This choice allowed us to test the bound in a fundamental scenario, providing a baseline for understanding its behavior in more complex settings.
    \item \textbf{Quadratic Polynomial}: The second scenario involved a quadratic polynomial \( f_{\textrm{T}}(\mathbf x) = \mathbf t \cdot \mathbf{x} + (\mathbf t \cdot \mathbf{x})^2 \), where $\textrm{t} \in \mathcal{R}^D$ is a random uniform vector. This function introduced a non-linear component to the evaluation, enabling us to assess how the bound performs under a modest increase in complexity.
    \item \textbf{Fully-Connected One-Hidden Layer Architecture}: The most complex scenario involved a fully-connected one-hidden layer (1HL) architecture with ReLU activations, described by \( f_{\textrm{T}}(\mathbf x) = \sum_{\alpha =1}^M q_\alpha \textrm{ReLU}\left(\mathbf{S}_\alpha \cdot \mathbf x\right) \), where $q \in  \mathbb{R}^M$ and $\mathbf{S} \in \mathbb{R}^{D\times M}$  are drawn from a normal distribution with zero mean and variance (respectively) $1/M$ and $1/D$. This setup mimic the complexity found in practical applications, providing insights into the bound's performance in scenarios resembling real-world tasks.
\end{enumerate}

These diverse teacher classes were instrumental in creating a comprehensive evaluation framework, ensuring that our analysis spanned from simple to complex models, thereby offering a robust validation of the proposed bound across various levels of neural network architecture complexity.

\paragraph{Lazy Training Regime.} 
We began our analysis by examining the so-called \textit{lazy training regime} \cite{chizat2019lazy}, characterized by training only the last layer of weights, while the preceding layers are maintained at their initial values, denoted as $\bar{\mathcal{W}}$. 
This is exactly the regime in which we derived the generalization and training errors \ref{eq2:gentesteq} and \ref{eq2:traintesteq}. In this regime, we expect the generalization gap to scale with $1/P$. 
This expectation is grounded in the fact that the residual generalization error $\epsilon^R_g(\bar{\mathcal{W}})$ does not depend on $P$, as $\bar{\mathcal{W}}$ remains constant throughout the training process.
\afterpage{
\begin{figure}
    \centering
    \includegraphics[width=0.95\textwidth]{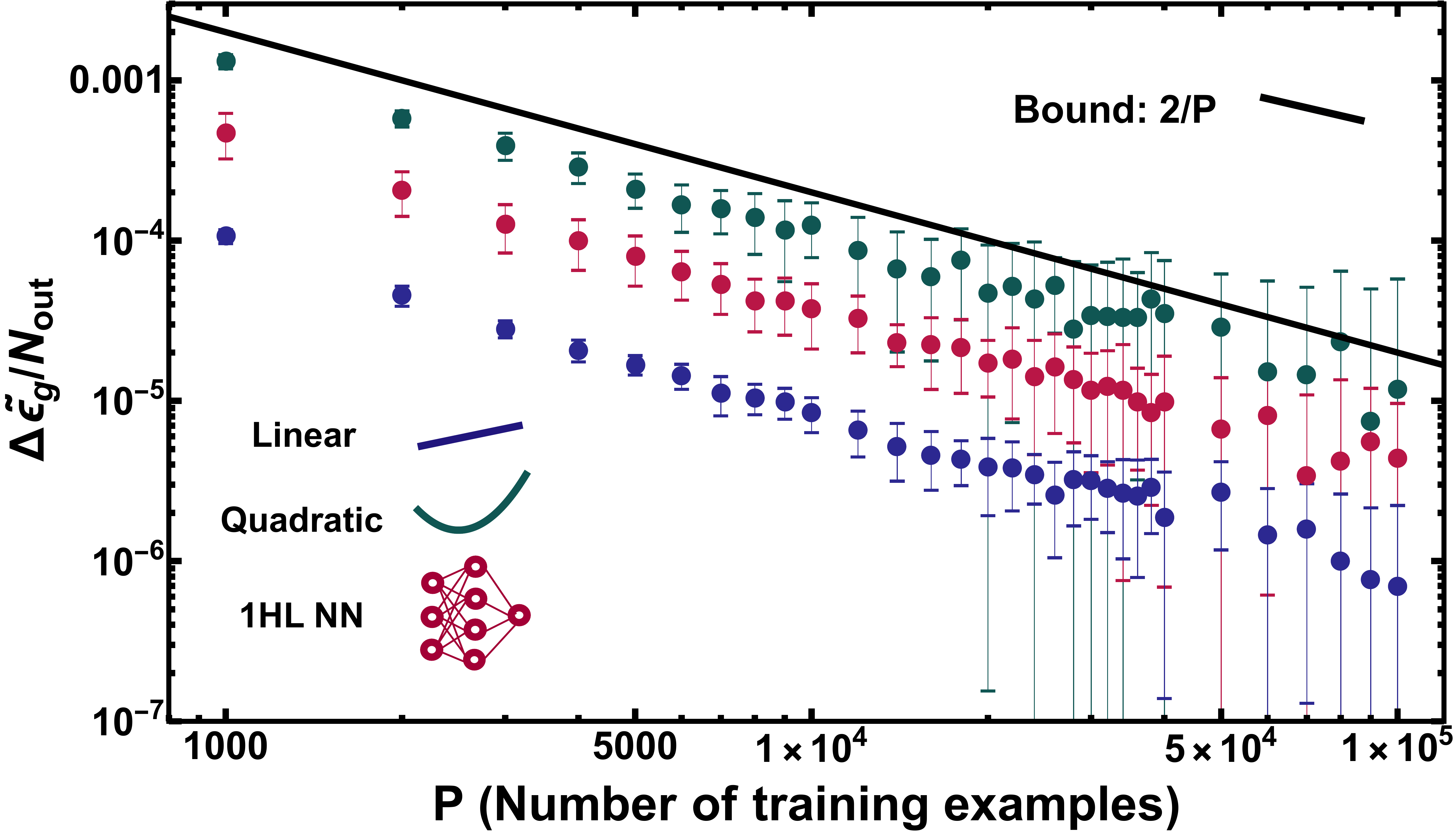}
    \caption{\textbf{Generalisation gap in the lazy-training regime (1).} The behavior of the normalized (rescaled by $N_{\textrm{out}}$) generalization gap as a function of the training dataset $P$, for a one-hidden layer student architecture for the three different classes of teacher outlined in the main text: linear (blue symbols), quadratic (green) and one-hidden layer (red). The solid black line marks our mean field upper bound.
    Data points are the result of an average over $50$ different realizations of the teacher and of the input (of dimension $D = 50$) with $N_{\textrm{out}} = 400$. We observe that the functional form of the rescaled GG $\Delta \tilde{\epsilon}$ is compatible with $y_0/P$ over two decades. }
    \label{fig2:2ab-2}
\end{figure}
\clearpage
}
\afterpage{
\begin{figure}
    \centering
    \includegraphics[width=0.95\textwidth]{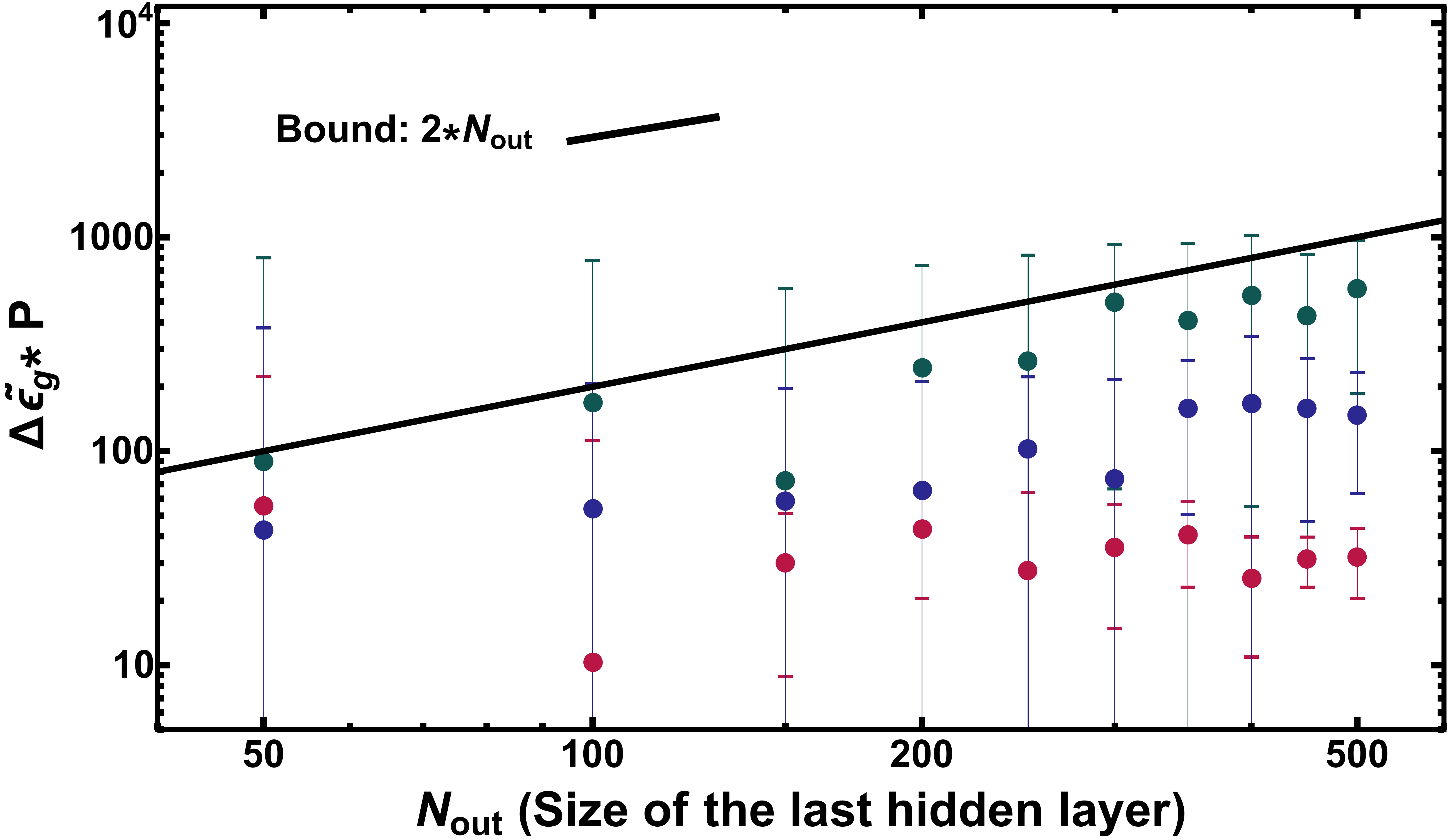}
    \caption{\textbf{Generalisation gap in the lazy-training regime (2).} The behavior of the normalized (multiplied by $P$) generalization gap as a function of the last layer size $N_{\textrm{out}}$, for a one-hidden layer student architecture for the three different classes of teacher outlined in the main text: linear (blue symbols), quadratic (green) and one-hidden layer (red). The solid black line marks our mean field upper bound.
    Data points are the result of an average over $50$ different realizations of the teacher and of the input (of dimension $D = 50$) with $P=2\cdot 10^4$ for the linear teacher, $P=4\cdot 10^4$ for the quadratic one, and $P=8\cdot 10^4$ for the 1HL NN teacher.}
    \label{fig2:2ab-3}
\end{figure}
\clearpage
}

This scale is verified in Fig. \ref{fig2:2ab-2} for the three different teacher classes introduced above. 

On the other hand,  $\epsilon_{\textrm{g}}^{\textrm{R}} \left(\bar{\mathcal W}\right)$ may still retain a dependence on the last layer size $N_{\textrm{out}}$. 
In particular, one may expect that increasing $N_{\textrm{out}}$ will decrease the residual generalization error, as this increases the number of functions available to approximate the target $f_{\textrm{T}} (\mathbf x)$. 
Therefore, for large $P$ and $N_{\textrm{out}}$ we expect $\Delta \tilde{\epsilon}$ to increase {\it at most} linearly as a function of $N_{\textrm{out}}$. 
The numerical behavior of the generalization gap as a function of $N_{\textrm{out}}$ is shown in Fig. \ref{fig2:2ab-3}. Once again, our mean field bound holds, but different scaling with $N_{\textrm{out}}$ can be appreciated. 
In particular, the generalization gap is almost constant for the linear teacher, whereas it has an approximately linear behavior for the quadratic one, reflecting different dependencies of the residual generalization error from the last layer size.
Note also that in both  Fig. \ref{fig2:2ab-2} and \ref{fig2:2ab-3} the generalization gap is systematically lower for the linear teacher case, confirming the intuitive expectation that the linear problem should be the easiest to learn.

\afterpage{
\begin{figure}
    \centering
    \includegraphics[width=0.95\textwidth]{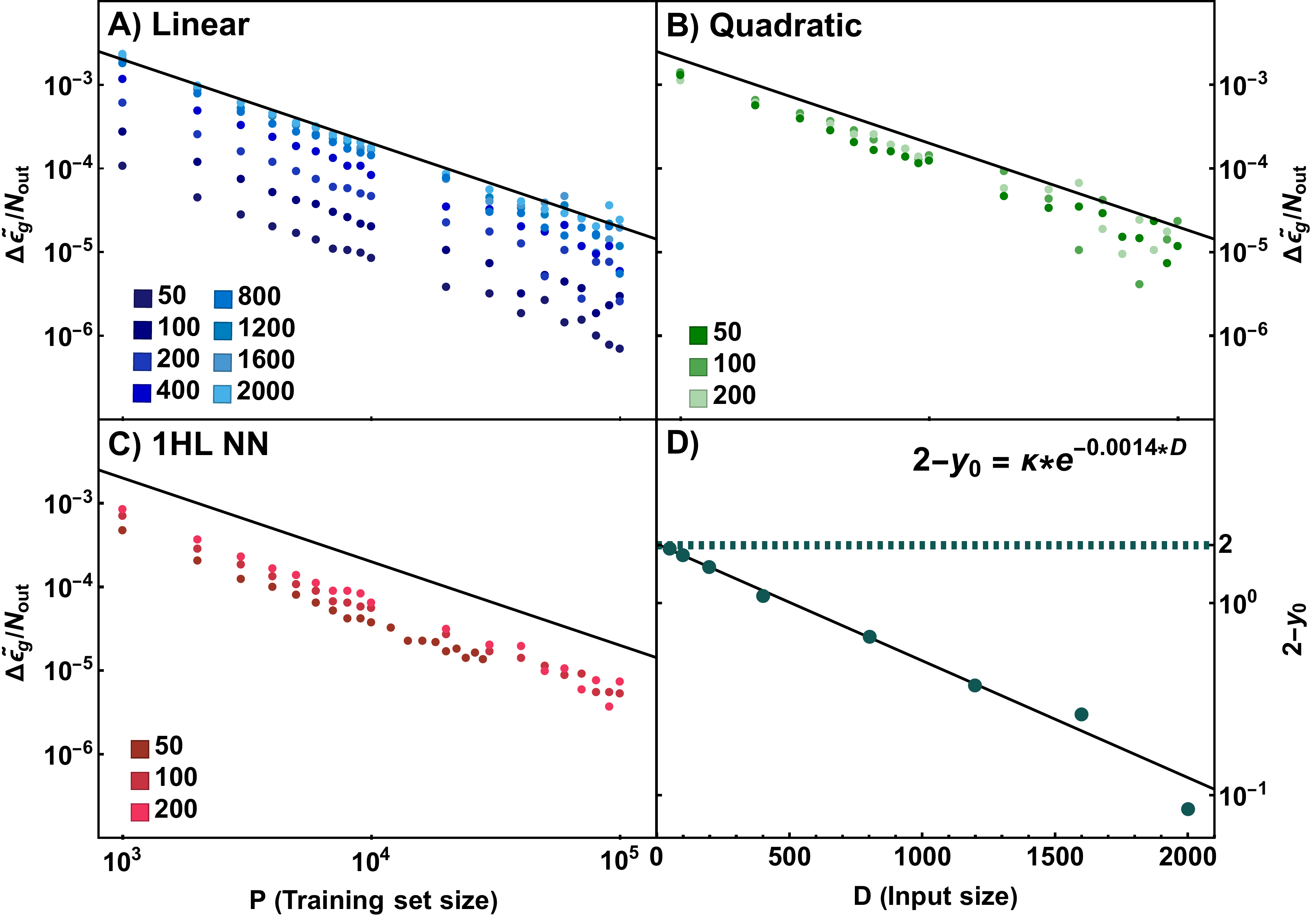}
    \caption{\textbf{Generalization Gap in the Lazy-Training Regime (3).} Panels A, B, C) illustrate the generalization gap of a lazy-training architecture, plotted against the number of training examples for varying input dimensions $D$. These are shown for three teacher models: linear, quadratic, and a 1HL NN, respectively. Throughout these experiments, the size of the hidden layer is maintained at $N_{\text{out}}= 400$. Panel D) Focuses on the linear teacher model across different input dimensions, demonstrating that as $D$ increases, the residual generalization gap rapidly converges to the theoretical bound. This convergence is quantified by the prefactor $y_0$, derived from standard fitting procedures, which exponentially approaches 2 from below as $D\rightarrow \infty$, with a convergence rate of approximately $\gamma \approx 0.0014$.}
    \label{fig2:2ab-4}
\end{figure}
\clearpage
}

It is important to highlight that within our theoretical framework, the input dimension \( D \) influences the outcomes only through the residual generalization error, as detailed in Equation \eqref{eq2:egr}. Intriguingly, as we increase \( D \), the normalized generalization gap appears to asymptotically approach the bound, exhibiting an exponential rate of convergence relative to \( D \). This phenomenon is particularly evident in the case of the linear teacher model, as depicted in Fig. \ref{fig2:2ab-4}. Notably, at this stage, our theoretical understanding does not offer an explanation for this specific trend.

We have also extended our investigation to student architectures with more than one hidden layer, maintaining the lazy training approach. In our theoretical framework, the size of any hidden layer other than the last one influences the behavior of the generalization gap in a complex and non-trivial way. Similar to the observed scaling with input dimension \( D \), the normalized generalization gap as a function of \( N_{\text{hid}} \) in these multi-layer scenarios also tends towards the proposed bound. This phenomenon is illustrated in Fig. \ref{fig2:2ab-5}. Nevertheless, akin to our previous observations, the fundamental reasons behind this phenomenon remain unclear and are beyond the scope of our analysis that does not delve into the internal structure of the features map.

\afterpage{
\begin{figure}
    \centering
    \includegraphics[width=0.95\textwidth]{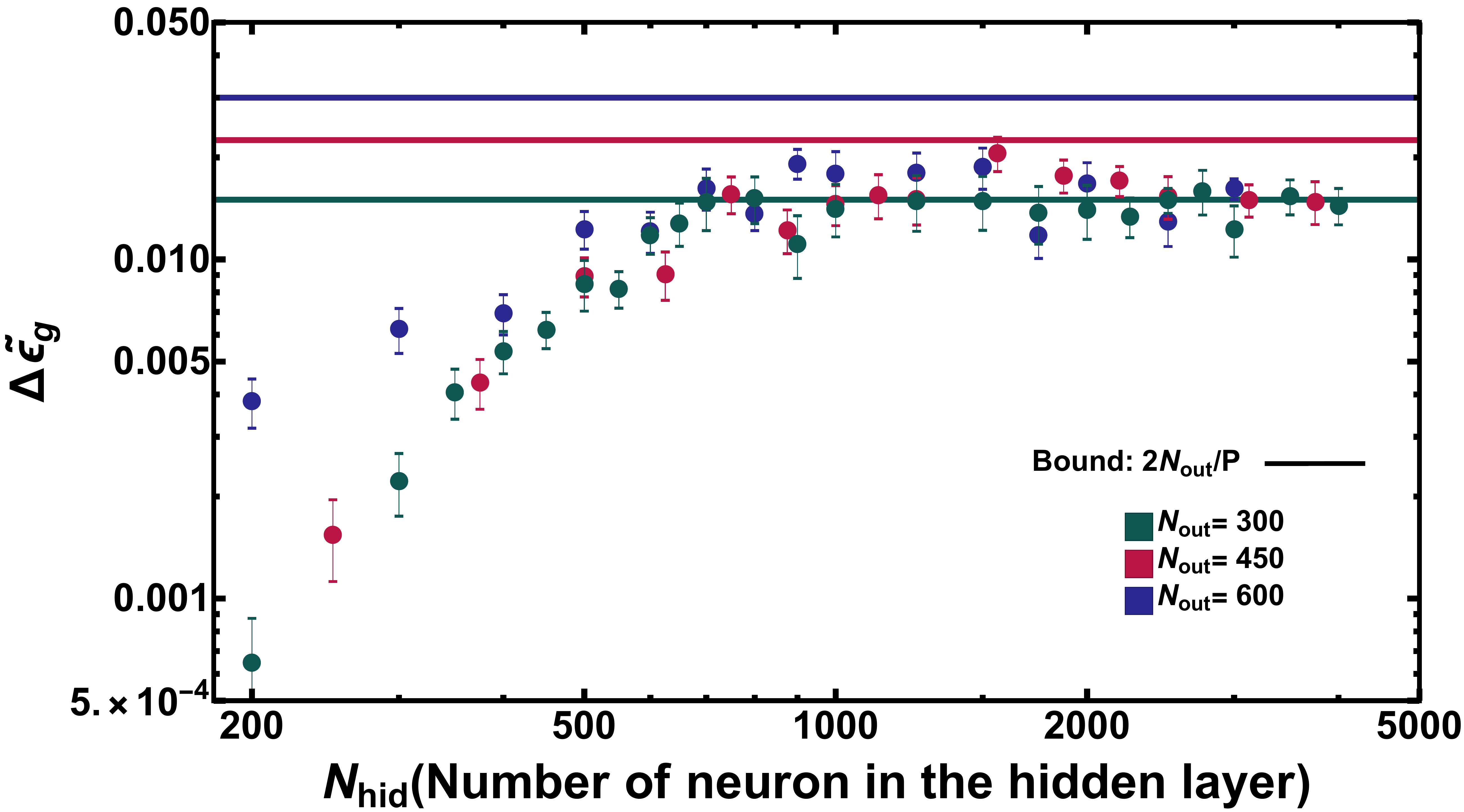}
    \caption{\textbf{Generalization Gap in the Lazy-Training Regime (4).} This figure illustrates the behavior of the normalized generalization gap as a function of the hidden layer size \( N_{\textrm{out}} \), observed in two-hidden layer student architectures for a different number of neurons in the last layer. The context is set for a linear teacher model. Solid lines represent the upper bound as predicted by our mean-field analysis. Data points shown are the average outcomes from \( 10 \) different realizations of both the teacher and the input (with dimension \( D = 50 \)) at a fixed training set size of \( P = 4\cdot 10^4 \).}
    \label{fig2:2ab-5}
\end{figure}
\clearpage
}

\paragraph{Toys DNNs in the fully-trained Regime.} 

We now turn our attention to fully-trained Deep Neural Networks (DNNs). In this context, the weights $\mathcal W$ are trained, and the residual generalization error $\epsilon_{\textrm{g}}^{\textrm{R}} \left(\mathcal W\right)$ potentially varies with the size of the training set $P$.

Consider a scenario where it exists a specific configuration of weights $\mathcal W^\dagger$  such that the residual generalization error $\epsilon_{\textrm{g}}^{\textrm{R}} \left(\mathcal W^\dagger\right)$ is equal to zero. This implies that the DNN is capable of perfectly learning the teacher function. Intuitively, as the dataset size $P$ increases, it is expected that the DNN will more likely find weight configurations $\mathcal W$ that approximate the optimal configuration $\mathcal W^\dagger$ more closely. Consequently, the residual generalization error should decrease, approaching zero as a function of $P$, particularly when $P$ is sufficiently large. This also suggests that the generalization gap will diminish at a rate faster than $1/P$.

Conversely, in situations where the DNN cannot learn the target function effectively, a non-zero residual generalization error $\epsilon_{\textrm{g}}^{\textrm{R}} \left(\mathcal W\right) = \hat{\epsilon}^{\textrm{R}}$ persists even as $P$ approaches infinity. In such cases, the generalization gap's scaling of $1/P$ is expected to be restored asymptotically.

Figure \ref{fig2:2ab-7} presents numerical simulations for fully-trained toys DNNs. These results align with the theoretical considerations previously discussed. Notably, in contrast to the lazy training regime, we do not observe a straightforward $1/P$ scaling of the generalization gap as shown in Figure \ref{fig2:2ab-2}. Instead, the generalization gap curves exhibit two distinct learning phases as a function of the training set size $P$. Remarkably, the generalization gap systematically falls below the mean field bound when $P$ exceeds approximately $10^4$. Entering the second learning stage, which occurs around this size, the gap appears to diminish towards zero as rapidly as $1/P$. This trend is consistent across both one and two hidden layer architectures, as well as various synthetic datasets, indicating a potential constant residual generalization error in this advanced learning phase.

In Figure \ref{fig2:2ab-8}, we consider the generalization performance in relation to the increasing width of the last layer $N_{\textrm{out}}$. These outcomes are fully consistent with our predictions: the established bound is maintained, and a linear or sub-linear decline in generalization performance is consistently observed across the different student and teacher architectures.

\afterpage{
\begin{figure}
    \centering
    \includegraphics[width=0.9\textwidth]{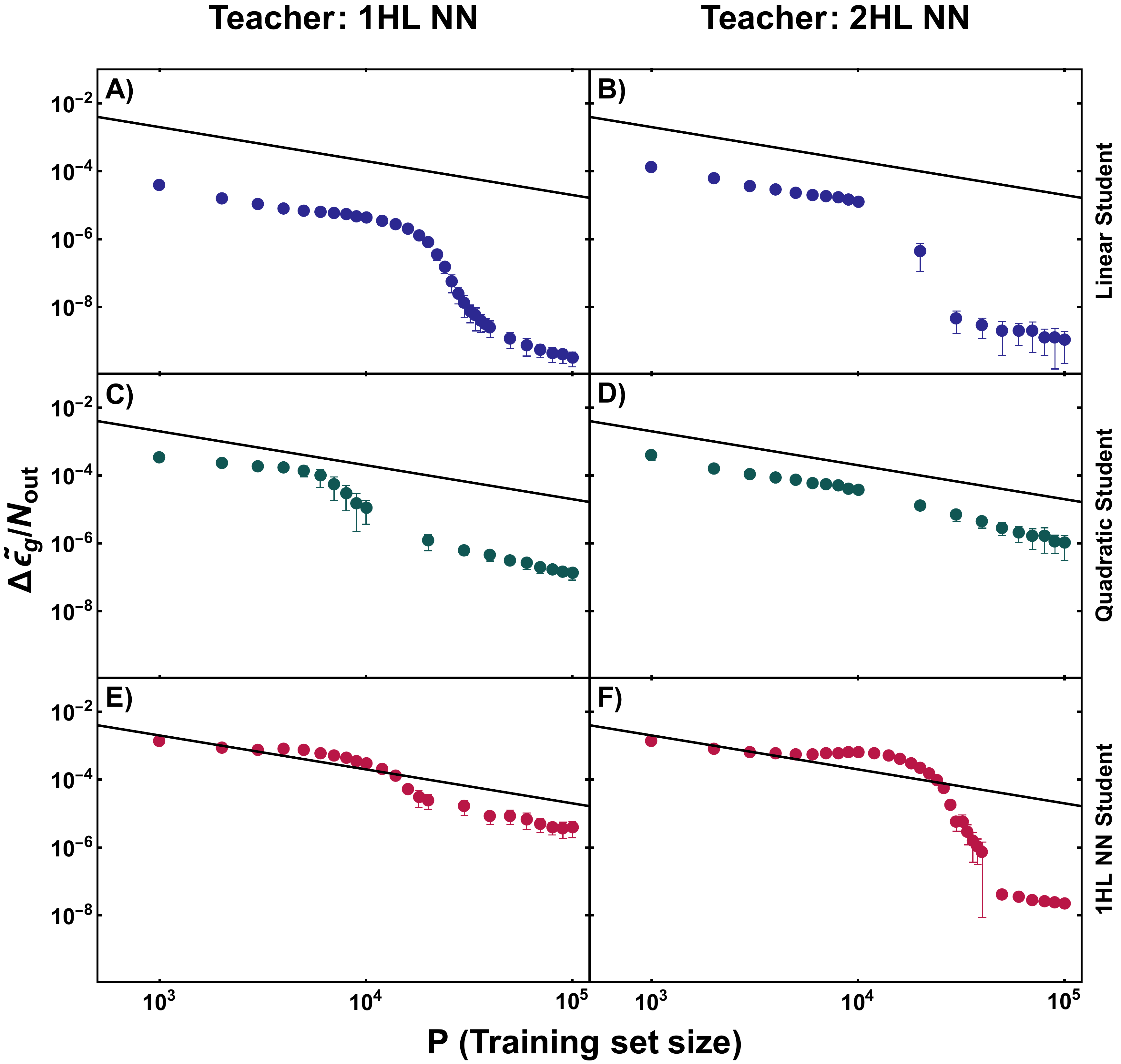}
    \caption{\textbf{Generalization Gap in the Fully-Trained Regime (1).} The normalized generalization gap (rescaled by $N_{\textrm{out}}$) for neural networks with one hidden layer (1HL NN, shown on the left-hand side in panels A, C, and E) and two hidden layers (2HL NN, on the right-hand side in panels B, D, and F). The generalization gap is plotted as a function of the training set size $P$ for the three different teacher classes discussed in the main text. The panels from top to bottom represent different teacher architectures: linear (blue symbols in panels A and B), quadratic (green symbols in panels C and D), and 1HL teacher architectures (red symbols in panels E and F). The solid black line in each panel indicates the mean field upper bound. Data points are the result of an average over $50$ realizations of the teacher and of the input ($D=50$) with $N_{\textrm{out}} = 100$.  Error bars in both panels correspond to one standard error.}
    \label{fig2:2ab-7}
\end{figure}
}

\afterpage{
\begin{figure}
    \centering
    \includegraphics[width=0.9\textwidth]{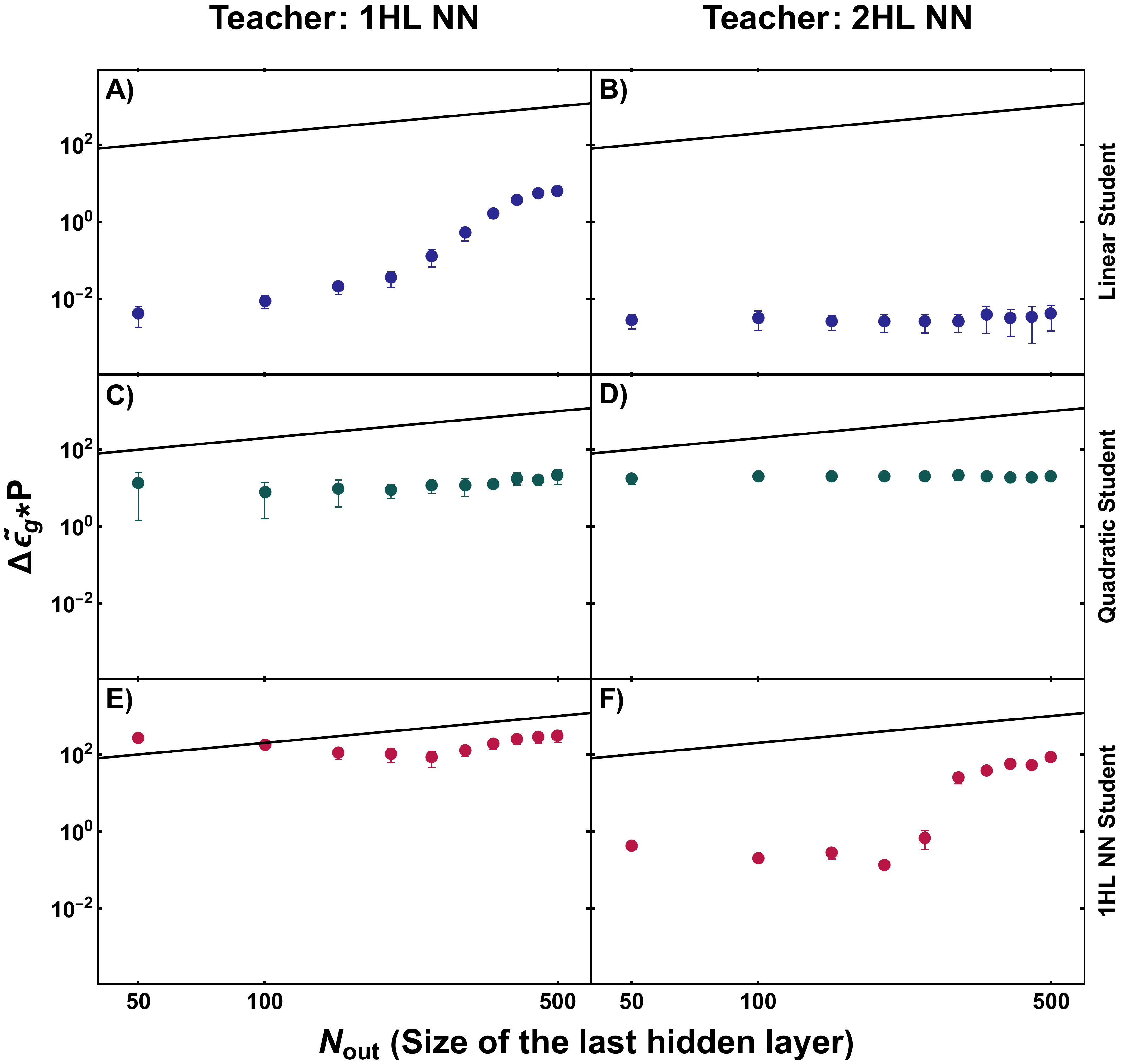}
    \caption{\textbf{Generalization Gap in the Fully-Trained Regime (2).} The figure displays the normalized generalization gap (rescaled by $N_{\textrm{out}}$) for neural networks with one hidden layer (1HL NN, depicted on the left-hand side in panels A, C, and E) and two hidden layers (2HL NN, shown on the right-hand side in panels B, D, and F). The generalization gap is plotted against the number of neurons in the last layer, $N_{\textrm{out}}$, for three distinct teacher classes as discussed in the main text. Sequentially from top to bottom, the panels represent various teacher architectures: linear (with blue symbols in panels A and B), quadratic (green symbols in panels C and D), and 1HL teacher architectures (red symbols in panels E and F). Each panel features a solid black line that marks the mean field upper bound.
    The data presented is averaged over 20 different realizations of teacher and input configurations, with the input dimension set at $D=50$ and the training set size at $P=4\cdot 10^4$. Error bars shown in both sets of panels correspond to one standard error.}
    \label{fig2:2ab-8}
\end{figure}
}

\subsubsection{State-of-the-art architecture with real-world datasets}
As an additional and more challenging test, we have applied it to the generalization gap observed when training three different state-of-the-art convolutional neural network architectures—ResNet18, DenseNet121, and VGG-11—on the MNIST dataset of handwritten digits \cite{MNIST}. It is important to note that while the MNIST problem is fundamentally a classification task, our theory was initially developed for regression tasks. To align with our theoretical framework, we adapted the learning problem into a regression format. In this setup, each digit vector $\mathbf x$ is associated with an output that is the integer number, ranging from $0$ to $9$, corresponding to its class. Consequently, instead of using standard accuracy (i.e., the fraction of correctly classified digits) to measure performance, we employed the mean square deviation between the network's output and the class index.

The results of these simulations are summarized in Fig. \ref{fig2:2AB-9}. Further details about the learning protocols for these DNNs and the previously mentioned toy models are provided in the subsequent \textit{detail} section. Notably, our theoretical bound holds true even for these advanced architectures when trained on a dataset that is significantly relevant in the field of computer vision. Furthermore, it is interesting to observe that in the regime we examined, the generalization gap closes at a rate faster than $1/P$.

\afterpage{
\begin{figure}
    \centering
    \includegraphics[width=0.95\textwidth]{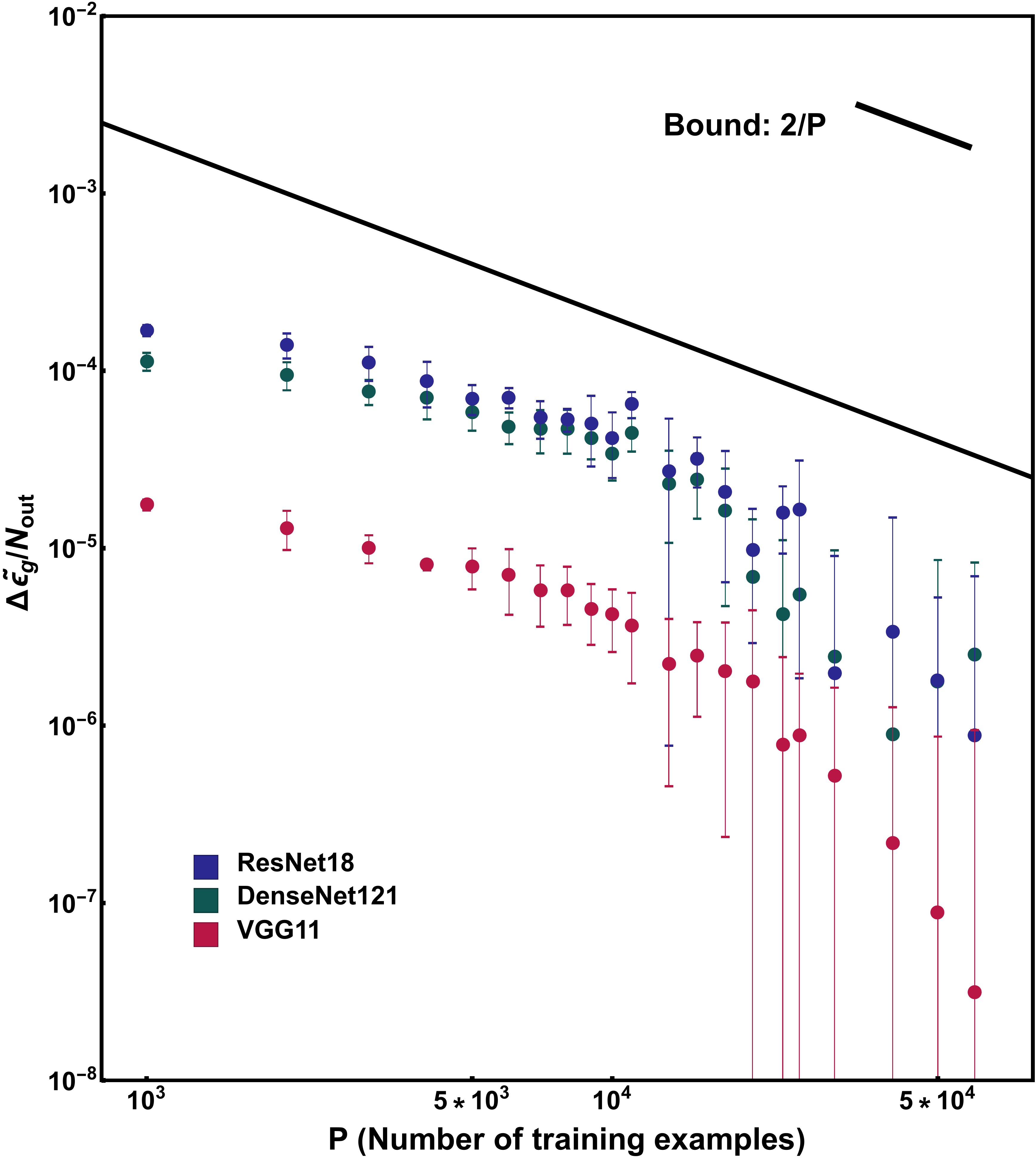}
    \caption{\textbf{Generalization Gap of State-of-the-Art CNNs.} This figure illustrates the behavior of the normalized generalization gap (rescaled by $N_{\textrm{out}}$) as a function of the training dataset size $P$, for three different state-of-the-art convolutional neural networks (CNNs): ResNet18 (blue symbols), DenseNet121 (green), and VGG18 (red). The solid black line indicates our mean field upper bound. Rescaling the generalization gap by $N_{\textrm{out}}$ allows for a more effective comparison between architectures with varying sizes of the last layer, as is the case here. The data shown is averaged over three different initial conditions for the architecture weights, and the error bars represent one standard error.}
    \label{fig2:2AB-9}
\end{figure}

}

\details{
\begin{center}
\textbf{Details of Numerical Experiments}
\end{center}

In this section, we give a detailed report of all our numerical procedures. The code to replicate our experiments can be found at \textit{https://github.com/rosalba-p/Generalisation$\_$DNN}. 

\begin{center}
    \textbf{\textit{Teacher-student architectures}}
\end{center}

\subparagraph*{Student architectures.} We considered six types of student architectures: two toy networks with one- and two-hidden layers (size of the second hidden layer $N_{\rm hid} = 200$) and three state-of-the-art convolutional ones (ResNet18, DenseNet121 and VGG11). The toy architectures have fully connected layers and ReLu activation functions at every layer but the last; the convolutional networks are the standard PyTorch \cite{PyTorch} models modified to yield a scalar output suitable for regression through a last fully connected linear layer with parameters $\mathbf{v}$, instead of the $LogSoftMax$ that is employed for classification. All these architectures have several convolutional layers before a last fully connected one that counts respectively $N_{\text{out}} =512,1024,4096$ hidden units. The total number of trainable parameters in the first two networks (weights and biases) is approximately 10 million, while vgg11 counts 10 times as many.
\subparagraph*{Teacher architectures and inputs for toy DNNs.} Each linear or quadratic teachers is defined by a random uniform vector $\mathbf{t} \in \mathbb{R}^D$ of unitary norm. 1HL teachers are defined by parameters $q_\alpha \in  \mathbb{R} $ and $\mathbf{S}_{\alpha} \in \mathbb{R}^D$  ($\alpha=1,\ldots,M=200$). They are drawn from a normal distribution with zero mean and variance (respectively) $1/M$ and $1/D$. 
Inputs $\mathbf{x} \in \mathbb{R}^D$ are also drawn from normal distribution with zero mean and unit variance.

\begin{center}
    \textbf{\textit{Learning and generalization}}
\end{center}

\subparagraph*{Learning algorithm.} All architectures in our study were trained using the Adam optimizer \cite{Adam_2015}, coupled with a small weight decay ($w_d = 10^{-5}$). The learning rate $\alpha$ was set to $10^{-3}$ for the toy architectures and $10^{-4}$ for the convolutional networks. The weight decay $w_d$ is linked to the regularization parameter $\lambda$ through the relation $w_d = \lambda \alpha$. We have confirmed that our results remain qualitatively consistent when varying $\lambda$ within a reasonable range (up to $\lambda \leq 10^{-1}$).

The loss function employed for regression is the standard mean squared error (MSE), defined as:
\begin{equation}
\text{MSE} = \frac{1}{P} \sum_{\mu=1}^{P} \left( f_{\text{T}}(x^\mu) - f_{\text{S}}(x^\mu) \right)^2
\end{equation}
In the case of the MNIST dataset, we treat the labels as integers, with $f_{\text{S}}(x^\mu)$ taking values in the set $[0,1,2,3,4,5,6,7,8,9]$. The MSE loss for MNIST is computed in the same manner as for the other datasets.

\subparagraph*{Estimation of the Generalization Gap.} To estimate the generalization gap, we first ensured that the training process reached convergence, meaning that both the training and test losses had stabilized at a plateau. For synthetic datasets, achieving this state required a substantial number of training epochs (approximately $3 \cdot 10^4$), whereas the MNIST dataset reached convergence more rapidly (within 100 epochs). During one epoch, the entire dataset is fed to the network, divided into minibatches only when full batch learning is impossible due to memory limitations. Once a steady state is observed where the training loss remains consistent with the test loss, we calculate the generalization gap as the average over the last 100 epochs (or 50 epochs for MNIST). The size of the training set used for testing, denoted as $P_{\text{test}}$, is $10^4$ in all cases.

Figure \ref{fig2:2AB-10} presents several plots of training and test loss versus the number of epochs. These plots demonstrate that for different teacher-student pairs, a plateau in the test loss is consistently reached. Although some fluctuations may still be visible in the training loss, these oscillations are sufficiently minor to not significantly impact the test loss and, consequently, the generalization gap.

\begin{center}
    \textbf{\textit{Trivial predictor}}
\end{center}

To ensure meaningful comparisons across various teacher/student pairings, it's crucial to normalize the loss by its natural scale.
This is achieved by referencing the trivial predictor $T$, as defined in Eq. \eqref{eq2:JPhiT}. 
By adopting this normalization strategy, we guarantee that both the training and test losses are consistently of order $O(1)$ for a randomly initialized architecture (i.e., at epoch 0). It's important to note that $T$ is a characteristic of the dataset, and therefore, its computation varies accordingly.

\subparagraph*{MNIST.}
The MNIST dataset is inherently designed for classification tasks, featuring 10 distinct classes corresponding to the first ten integers. Given this discrete label structure, the integral typically used to describe the trivial predictor in our framework is replaced by a summation over these classes. Another simplification arises from the balanced nature of the MNIST dataset: its $P = 6 \cdot 10^{4}$ training examples are evenly distributed across the 10 classes. This uniform distribution allows for a straightforward analytical computation of the trivial predictor:
\begin{equation}
T = \frac{1}{P} \sum_{n = 0}^{9} \frac{P}{\text{\# classes}} n^2 = \frac{1}{\text{\# classes}} \sum_{n = 0}^{9} n^2 = 28.5
\end{equation}

\subparagraph*{Synthetic datasets.}
For the other teacher models we utilized, an analytical computation of the trivial predictor $T$ is not feasible. Therefore, we resort to a numerical estimation using the following approximation:
\begin{equation}
T \approx \frac{1}{P_{\text{norm}}}\sum_{\mu=1}^{P_{\text{norm}}} f^2_{\text{T}}(x^\mu)
\end{equation}
In practice, this involves drawing an independent dataset comprising $P_{\text{norm}}$ samples $\{ x^\mu \}_{\mu = 1}^{
P_{\text{norm}}}$, and then calculating the average of the squared true labels $f^2_{\text{T}}(x^\mu)$.

We chose $P_{\text{norm}}=10^{6}$ as a balanced compromise between computational time and the reliability of the estimate. The stability of $T$ as $P_{\text{norm}}$ increases is showed in Figure \ref{fig2:2AB-11}.
}

\afterpage{
\begin{figure}[t]
    \centering
    \includegraphics[width = 0.75\textwidth]{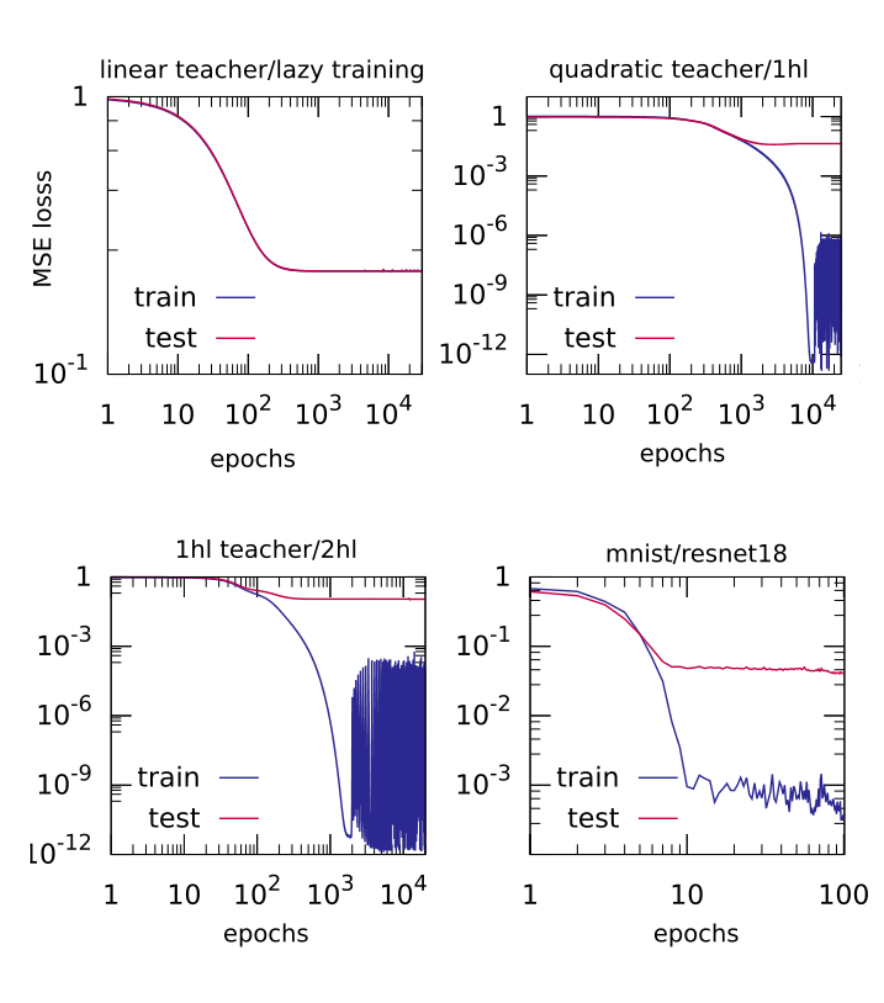}
    \caption{Train (blue) and test (magenta) loss of different teacher/student tasks as a function of the training epochs. The test loss reaches a plateau even if the training loss is noisy, due to the different order of magnitude reached by the two. It is worth remarking that the losses are normalized with the trivial predictor and therefore at epoch 1 are $O(1)$.}
    \label{fig2:2AB-10}
\end{figure}
\clearpage
}

\afterpage{
\begin{figure}[t]
    \centering
    \includegraphics[width = 0.95\textwidth]{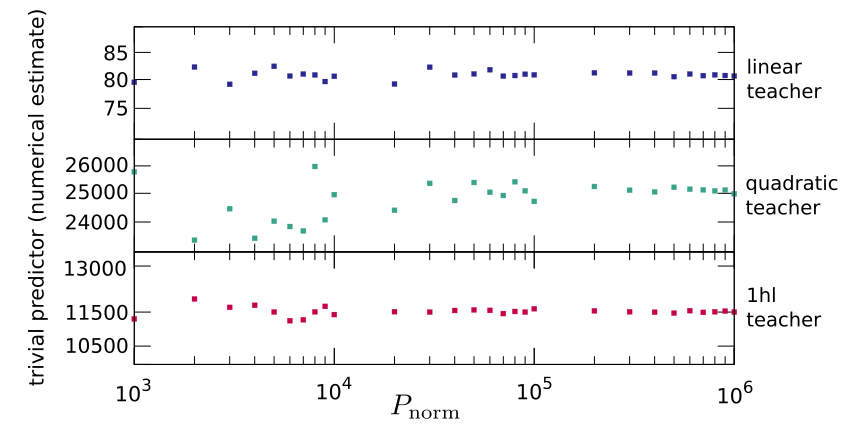}
    \caption{Numerical estimate of the trivial predictor $T$ as a function of the number of points $P_{\text{norm}}$ used in the approximation. These values describe a single realization of a random teacher function (linear, quadratic, 1hl). $T$ converges to a fixed value around $P_{\text{norm}} = 10^{6}$, which is the value chosen for all our numerical experiments.}
    \label{fig2:2AB-11}
\end{figure}
\clearpage
}

\section{Discussion and prespective}

In this chapter, I have introduced a novel asymptotic bound for the generalization gap of Deep Neural Networks (DNNs), derived through mean-field analysis. Although this analysis does not possess the complete rigor of formal theorems, it successfully establishes a significantly tighter bound for the generalization gap compared to those derived within the strict framework of statistical learning theory (SLT). 
In particular, SLT typically predicts an upper bound that is approximately proportional to $\sqrt{N_{\textrm{tot}}/P}$, where $N_{\textrm{tot}}$ represents the total number of network parameters \cite{bartlett2019nearly} \footnote{More accurately, $N_{\textrm{tot}}$ should be substituted with the Vapnik-Chervonenkis pseudo-dimension of the DNN \cite{pollard2012convergence}, or by the Rademacher complexity \cite{mohri2008rademacher}.}. 
Interestingly, our proposed upper bound is influenced by the number of network parameters solely through the width of the last layer. 
This observation is further supported by the fact that when considering DNNs capable of learning the rules, the generalization gap reaches a constant value (i.e., our bound) if we increase the total number of parameters while keeping $N_{\text{out}}$ fixed, as shown in Fig. \ref{fig2:2AB-12}.

Based on this, we can hypothesize that in deep architectures, the parameters of the initial layers and those of the last layer play distinct roles in addressing overfitting issues. In particular, the width of the last layer may have a unique and critical role in managing all the information extracted in the preceding layers.

\afterpage{
\begin{figure}[t]
    \centering
    \includegraphics[width = 0.95\textwidth]{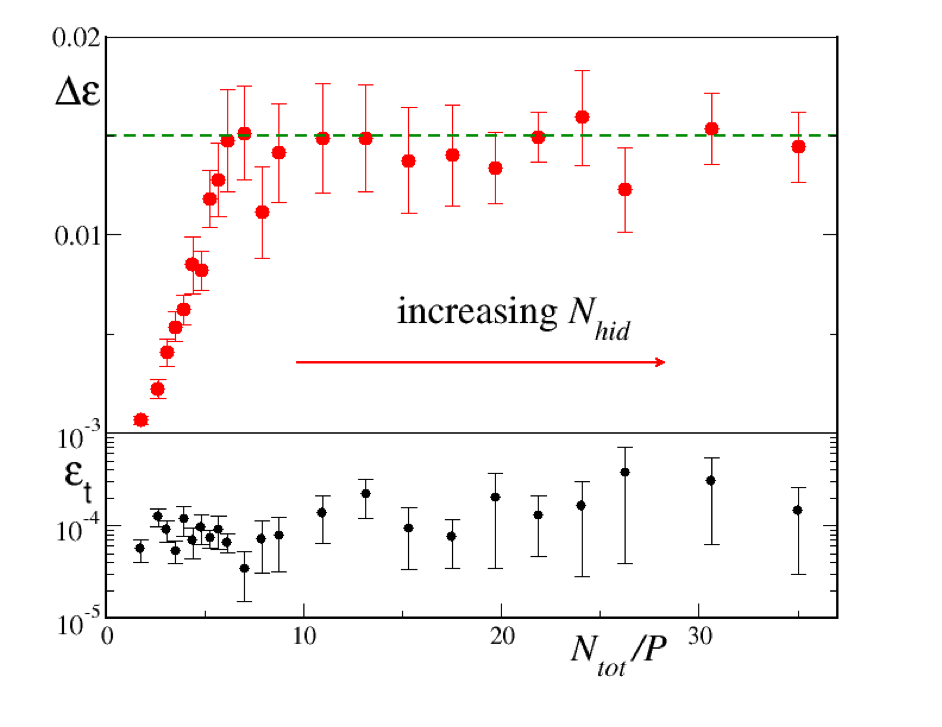}
    \caption{The behavior of the generalization gap in two-hidden-layer deep neural networks (DNNs) as a function of the ratio between the total number of parameters ($N_{\text{tot}}$) and the dataset size ($P$), where $N_{\text{tot}} = (D + N_{\text{out}})$. In the student setup, the input dimension ($D$) is set at 50, and the number of neurons in the output layer ($N_{\text{out}}$) is maintained at 300. The teacher model is a one-hidden-layer DNN, also with $N_{\text{out}} = 100$. The size of the training set is fixed at $P = 40,000$. The dashed green line represents the mean field upper bound $\delta\epsilon = 2 \cdot N_{\text{out}} / P = 0.015$. The red data points are averaged over 10 different realizations of the teacher model and the input. It is observed that the generalization gap increases with the $N_{\text{tot}}/P$ ratio until it reaches the upper bound (at $N_{\text{hid}} \approx 700$), after which it stabilizes. The accompanying graph below evaluates the training performance of our DNNs, confirming the student's ability to learn the rule set by the teacher model.}
    \label{fig2:2AB-12}
\end{figure}
\clearpage
}

The improved upper bound on the generalization gap that we have achieved can be attributed to incorporating the optimization process (i.e. learning), at least in the last layer of the network. Traditionally, SLT seeks to establish generalization gap results that are applicable to every function within a given model class. In contrast, our approach narrows the focus to specific, yet significant, elements within that class.

To be more specific, consider a Deep Neural Network (DNN) with parameters denoted as $\vartheta = \{ \mathbf{v}, \mathcal{W}\}$, where $\{\mathbf{v}\}$ are the output layer weights and $\mathcal{W}$ the remaining ones, which define the feature maps. Classical SLT bounds are applicable to every possible realization of $\vartheta$. However, our approach differs by assuming that, given a fixed $\mathcal{W}$, the weights $\mathbf{v}$ in the last layer are optimized with respect to the training set. This assumption of optimization in the context of a DNN significantly refines our approach and contributes to the establishment of a more stringent upper bound on the generalization gap.

Despite the remarkable result, this mean-field bound also has limitations:

\begin{enumerate}
    \item \textbf{Limitations Regarding Dataset and Architecture Size:} Traditional SLT bounds are applicable to datasets and architectures of any size, though their predictive power become useless when the architecture is larger than the dataset. In contrast, our mean field bound has been derived in the asymptotic limit of large $P$ and $N_{\textrm{out}}$ and does not provide finite size corrections. However, it remains predictive in overparametrized regime, where the classical bounds are "underpredictive".

    \item \textbf{Threshold for Bound Applicability:} As shown in Fig.~\ref{fig2:2ab-7}, our bound becomes relevant only beyond a certain value for the training set size, in our case approximately $10^4$ to $10^5$. This threshold may result from both finite P corrections and the difficulty in finding the optimal readout weights.  Our theory does not predict such a threshold value. However, our numerical experiments (see Fig.~\ref{fig2:2AB-9}) suggest that this threshold is relatively low in many empirically significant cases.

    \item \textbf{Gaussian Approximation in Analysis:} A Gaussian approximation was employed at the level of the replicated partition function. While similar approaches have been effective in studying the statistical physics of kernels and random feature models, with good numerical agreement (\cite{PhysRevLett.82.2975, canatar2021}), the quantitative accuracy of this approximation cannot be guaranteed in all scenarios.

    \item \textbf{Scope Limited to Regression:} Our results are currently limited to regression problems. Extending these non-rigorous tools to classification problems, as explored in \cite{loureiro2021learning}, is an intriguing direction for future research.
\end{enumerate}

Moreover, our investigation, in line with Statistical Learning Theory, primarily focuses on the generalization gap. It would be insightful to determine whether our findings can be interpreted through the lens of the classic bias-variance decomposition of the generalization error, as explored in studies such as Ref.~\cite{pmlr-v119-d-ascoli20a}. 

Some of the limitations identified in our study could potentially be addressed through a more rigorous approach. For instance, random matrix theory may be used to circumvent the need for replicas and the Gaussian approximation employed in our calculation.
Lifting the Gaussian approximation requirement to take the large D limit, could particularly illuminate the dependence of the generalization gap on the input dimension $D$, which we have numerically observed in the lazy training regime to approach the mean field bound exponentially.
For instance, the D dependence is explicitly treated in kernel learning studies \cite{pmlr-v119-bordelon20a, NEURIPS2020_b367e525, goldt2020gaussian, hu2020universality}. See also the next chapter for an analysis of the input dimension dependence in a different context.


Finally, an intriguing implication of our findings is that the mean field bound suggests a linear or sub-linear degradation in generalization performance with an increase in the size of the last layer, $N_{\textrm{out}}$. This leads to a speculative yet interesting hypothesis: to enhance generalization performance, it might be beneficial to design architectures with smaller last layer. A more systematic investigation of state-of-the-art architectures is necessary to determine whether this insight could contribute to the development of more performant deep neural networks in the future.

\clearpage
\printbibliography[heading=subbibnumbered, title=Chapter bibliography]
\end{refsection}

\chapter{Data-Dependent approach: Beyond the Infinite width limit}
\begin{refsection}
\label{ch.3BIW}
The remarkable success of deep learning has spurred significant interest in developing a theoretical framework capable of accurately predicting generalization performance. As we have seen in the previous chapters, the prevalent approach, inspired by the traditional approach to disordered systems, typically involves averaging over the training dataset distribution, thereby neglecting the impact of the the specific finite dataset's structure. However, there is a notable scenario where it's possible to construct a framework that doesn't require averaging over the dataset and instead makes predictions based on the training dataset itself. This scenario is known as the infinite width (IW) limit, characterized by the number of units $N_\ell$ in each hidden layer ($\ell=1,\dots, L$, $L$ denoting the network's depth) vastly outnumbering the number $P$ of training examples.
In this IW limit it is possible to show that a DNN behaves as a Gaussian Process. One can thus  rewrite the learning problem as a kernel problem, where the kernel is the covariance matrix of the Gaussian Process itself. 
While this is a notable result, it has been obtained in a regime that diverges from the typical scenarios in deep learning applications. In reality, especially in real-world applications, the number of neurons per layer is often on par with the training set size, presenting a stark contrast with the idealized infinite width limit.
In this Chapter, we aim to show how a somehow less rigorous approach can generate a similar scenario in a more realistic limit, where both $N_\ell$ and $P$ go to infinite but with a fixed ratio $\alpha = P/N_\ell$. This new framework entails several very considerable effects, as discussed in \cite{ariosto2023}. 

In the following, we begin with a simple introduction to the Bayesian approach to DNNs and a review of the known results obtained in the infinite width limit.
Our novel approach is then presented in the second part. 

\newpage
\vspace{4.5cm}
\minitoc
\vfill
\newpage

\section{Bayesian Learning for Neural Networks}
\subsection{Bayesian Learning}
Consider a series of quantities $\{y^{(1)}, y^{(2)}, \dots\}$, each generated by a Markovian process where every $y^\mu$ is independently subject to random variation. We can model this randomness using a probabilistic framework, where a set of unknown parameters, denoted as $\vartheta$, determines the probability distributions of the $y_i$'s. These probabilities, or probability densities, are represented as $\mathcal{P}(y_\mu \mid \vartheta)$, that is as the conditional probability of observing $y^{(\mu)}$ given parameters $\theta$.

To learn about $\vartheta$, we need to observe some values from this series, say $\{x^{(1)}, \dots, x^{(P)}\}$. The influence of these observations is encapsulated in the \textit{likelihood function}, $L(\vartheta) = L(\vartheta \mid x^{(1)}, \dots, x^{(P)})$, which expresses the probability of observing the data given the model parameters. Mathematically, it's defined as:

\[
L(\vartheta) = L(\vartheta \mid y^{(1)}, \dots, y^{(P)}) \;=\; \prod_{\mu=1}^{P} \mathcal{P}(y_\mu \mid \vartheta).
\]

The most straightforward approach to estimate $\vartheta$ is through \textit{maximum likelihood}, which seeks the value $\hat{\vartheta}$ that maximizes the likelihood $L$. This yields a {\it frequentist} estimate. However, Bayesian learning takes a different route. Rather than pinpointing a single estimate, Bayesian methods aim to determine a probability distribution for the parameters $\vartheta$, matching our beliefs about their likely values with the empirical observation of the finite dataset $\{y^{(1)}, \ldots y^{P} \}$.

To implement Bayesian learning, we start with a \textit{prior distribution} $\mathcal{P}(\vartheta)$, representing our initial beliefs about the parameters before considering the data. Upon observing the data $\{y^{(1)}, \dots, y^{(P)}\}$, we update this prior to a \textit{posterior distribution} $\mathcal{P}(\vartheta \mid y^{(1)}, \dots, y^{(P)})$, the conditional probability of the parameters being theta given the empirical observation $(y^{(1)}, \ldots y^{(P)})$, using Bayes' Rule:

\begin{equation}
    \mathcal{P}(\vartheta \mid y^{(1)}, \dots, y^{(P)}) = \frac{\mathcal{P}(y^{(1)}, \dots, y^{(P)} \mid \vartheta) \mathcal{P}(\vartheta)}{\mathcal{P}(y^{(1)}, \dots, y^{(P)})} = \frac{L(\vartheta \mid y^{(1)}, \dots, y^{(P)}) \mathcal{P}(\vartheta)}{\mathcal{P}(y^{(1)}, \dots, y^{(P)})}
    \label{3BIW:eq_posterior}
\end{equation}
where $\mathcal{P}(y^{(1)}, \dots, y^{(P)} \mid \vartheta)$ is the conditional probability of observe the sequence $\{x^{\mu}\}_{\mu=1}^P$ given $\vartheta$, and $\mathcal{P}(y^{(1)}, \dots, y^{(P)})$ is its marginal (typically called \textit{marginal likelihood}), that is $\mathcal{P}(y^{(1)}, \dots, y^{(P)} \mid \vartheta)$ integrated over the measure of $\vartheta$.

In this Bayesian framework, the posterior distribution integrates our initial assumptions (the prior) with the evidence from the data (the likelihood), offering a comprehensive view of what the parameters might be after considering both prior beliefs and observed evidence.

The selection of the prior distribution, $\mathcal{P}(\vartheta)$, is a crucial step in Bayesian inference and often a focal point for criticism. Critics argue that the choice of a prior can be somewhat arbitrary, potentially introducing subjectivity into the analysis. This concern is particularly pronounced when there is limited empirical data or we lack clear theoretical guidance on the form of the prior. The criticism can be summarized as follows:

\begin{itemize}
    \item \textbf{Subjectivity:} The prior is meant to encapsulate prior beliefs or knowledge about the parameters before observing the data. However, in many cases, these beliefs are subjective and can vary significantly among different analysts.
    \item \textbf{Influence on Results:} In situations where data is sparse, the choice of prior can have a substantial impact on the posterior distribution, potentially leading to different conclusions drawn by different analysts using different priors.
    \item \textbf{Arbitrariness:} Without strong empirical or theoretical grounds for choosing a specific prior, the process can seem arbitrary, undermining the objectivity that is often sought in statistical analysis.
\end{itemize}

In the Bayesian approach, the posterior, obtained from the observation of the P datapoints $x^{(1)}, \dots, x^{(P)}$, can be used to predict the probability distribution of a new datapoint. This is the so-called \textit{predictive distribution} given by:

\begin{equation}
    \mathcal{P}(x_{P+1} \mid x^{(1)}, \dots, x^{(P)}) = \int d\vartheta\; \mathcal{P}(x^{(P+1)} \mid \vartheta) \;\mathcal{P}(\vartheta \mid x^{(1)}, \dots, x^{(P)}).
    \label{3BIW:eq_pred_dis}
\end{equation}

This \textit{predictive distribution} for $x_{P+1}$ given $\{x^{\mu}\}_\mu^P$ is the culmination of Bayesian inference. One of the key advantages of the Bayesian approach is its ability to produce a distribution, that incorporates tour (hopefully informed) beliefs about the model parameters. This results in a more nuanced and informative prediction, especially useful in scenarios where understanding the range of possible outcomes is as important as predicting a single most likely outcome.

\details{
    \begin{center}
        \textbf{Example: coin toss}
    \end{center}
    Consider the problem of predicting the outcome of a coin toss. Let's use a Bayesian approach with a uniform prior $\theta \in [0,1]$ for the probability of the coin landing head in a single toss. If we denote the number of heads and tails observed in the first $P$ tosses as $h$ and $t$ respectively, we have the likelihood:
    \begin{equation}
        \mathcal{L} (\text{$h$ heads out of $P$ tosses})=\theta^h (1-\theta)^{P-h} \frac{P !}{h! (P-h)!}
    \end{equation}
    After a brief calculation, one gets the Bayesian prediction for the result of toss $P+1$ given by:
    \[
    \mathcal{P}(x_{P+1} \mid x^{(1)}, \dots, x^{(P)}) = 
    \begin{cases} 
        \frac{h+1}{P+2} & \text{if } x_{P+1} = \text{heads}, \\
        \frac{t+1}{P+2} & \text{if } x_{P+1} = \text{tails}.
    \end{cases}
    \]
    This result is intuitive: if we have observed more heads than tails, the probability of the next toss being heads is higher, and vice versa. 
    For a fair coin $(\theta_{head}=1/2)$ and a finite dataset, it slightly improves the frequentist or maximum likelihood estimate ($\mathcal{P}(x_{P+1} \mid x^{(1)}, \dots, x^{(P)})=h/P$ for heads and $t/P$ for tails) for any evidence $h \neq P/2$.
    
    However, the Bayesian approach shows its strength in situations where the maximum likelihood estimate might be misleading. For instance, if we have tossed the coin twice and it landed heads both times, the maximum likelihood estimate would lead us to believe that the coin will certainly land heads on the next toss, as $\hat{\vartheta} = 1$. In contrast, the Bayesian prediction with a uniform prior gives a more cautious probability of $3/4$ for heads. The Bayesian procedure avoids jumping to conclusions by considering not just the value of $\vartheta$ that best explains the data, but also other values of $\vartheta$ that explain the data reasonably well. These values contribute to the integral in the predictive distribution, as shown in equation \ref{3BIW:eq_pred_dis}.
    Note however that $i)$ for $P \to \infty$ the Bayesian and the frequentist predictions do converge and that $ii)$ the advantage of the latter on the former for a finite dataset is due to the choice of the prior and does not carry on to the case of an unfair coin for arbitrary sequences. 
}

\subsection{Bayesian Neural Network}

Consider a one-hidden-layer fully-connected neural network. This network processes a set of real inputs, denoted as $x^\mu$, to compute a scalar (for simplicity) output value, $f_{\text{DNN}}(x^\mu)$. As we have seen, the computation involves an initial affine transformation followed by a non-linear operation. The output is calculated as follows:

\begin{equation}
    \begin{split}
        f_{\text{DNN}}(x^\mu;\vartheta) = f_{\vartheta}(x^\mu) &= \sum_i^{N_1} v_{i} h^\mu_i + b^{(2)}, \\
        h^\mu_i &= \sigma\left(\sum_{j}^{N_0} W_{i,j} x_j^\mu + b^{(1)}_i \right),
    \end{split}    
\end{equation}

where $f_{\vartheta}(x^\mu)$ represents the output of the neural network for the input $x^\mu$. The parameters $v_{i}$ and $b^{(2)}$ are the weights and biases of the output layer, respectively, while the weights and biases of the hidden layer are given by $W_{i,j}$ and $b^{(1)}_i$. The term $h^\mu_i$ denotes the activated values of the neurons in the hidden layer, with $\sigma$ representing the activation function. The summations extend over all neurons in the respective layers.
It has been shown \cite{cybenko1989approximation,funahashi1989approximate,hornik1989multilayer} that this kind of neural network is capable of approximating any function defined on a compact domain to an arbitrary degree of accuracy, provided a large enough number of hidden units are employed. 
These results, known as {\it Universal approximation theorems} exist also for DNNs of arbitrary depths and architectures

The weights and biases in neural networks are learned based on a set of training cases, $\mathcal{T} = \{(x^{\mu}, y^{\mu})\}$, which provide examples of inputs, \(x^{\mu}\), and associated target labels, \(y^{\mu}\) (both of which may have several scalar components). Standard neural network training procedures adjust the weights and biases in the network to minimize a measure of ``error'' on the training set. 
For example, the most common way to find the weights and biases that minimize the chosen error function is to perform a  gradient-based optimization method. 
This involves using derivatives of the error with respect to the weights and biases, which are calculated by backpropagation \cite{rumelhart1986learning,rumelhart1986parallel}. 
Although there are typically many local minima in this optimization landscape, good solutions are often found despite this.
However, these neural networks can be also effectively utilized to define a probabilistic models of the input to label relation. This is achieved by using the outputs of the network to specify the conditional distribution $\mathcal{P}(x \mid y)$ of the target labels \( y\) given an input vector, \( x \). It's important to note that these models do not typically focus on modeling the distribution of \( x \) itself; in some cases, modeling this distribution might not even be relevant, as in the common case when the input values are user-specified or externally determined. For example, in the case of regression with real-valued targets, this conditional distribution for the targets might be defined to be Gaussian (for a physicist, this is a rather natural choice for its intuitive connection with Statistical Mechanics). Here, \( y \) would have a mean of \( f_{\vartheta}(x) \) and a variance of \( \beta^{-1} \). Then, the model can be expressed as:
\begin{equation}
    \mathcal{P}_{\vartheta}(y \mid x) =  \sqrt{\frac{\beta}{2\pi}}\, e^{-\frac{\beta}{2}\left[f_{\vartheta}(x)-y\right]^2}\,,
    \label{3BIW:eq_likehood_gauss}
\end{equation}
where \(\vartheta\) represents the network parameters (weights and biases).

We have seen that in the Bayesian approach to neural network learning, the objective is to find the predictive distribution for the labels of new ``test'' case, given the input for that case, and the inputs and targets in the training cases. As the distribution of the inputs is not being modeled, the predictive distribution of equation \ref{3BIW:eq_pred_dis} is modified as follows (for easyness of notation we omit the label $P+1$ denoting a new example):
\begin{equation}
    \mathcal{P}_{\mathcal{T}}(y \mid x) = \int d\vartheta \, \mathcal{P}_{\vartheta}(y \mid x) \; \mathcal{P}(\vartheta \mid \mathcal{T})\,.
    \label{3BIW:eq_pred_dis_bayes}
\end{equation}
According to Bayes' Rule Eq. \ref{3BIW:eq_posterior}, the posterior density $\mathcal{P}(\vartheta \mid \mathcal{T})$ is proportional to the product of the prior distribution for the DNN parameters and the likelihood function:
\begin{equation}
    \mathcal{P}(\vartheta \mid \mathcal{T}) \, \propto\, L(\vartheta \mid \mathcal{T}) \mathcal{P}(\vartheta)\,,
\end{equation}
where the likelihood is given by:
\begin{equation}
    L(\vartheta \mid \mathcal{T}) = \prod_\mu^P \mathcal{P}(y^\mu \mid x^{\mu}\,;\, \vartheta)
\end{equation}

The distribution for the target values, \( y^{\mu} \), given the corresponding inputs \( x^{\mu} \) and the parameters of the network, is defined by the type of loss with which the network is being trained. For regression, this distribution is typically given by Eq. \ref{3BIW:eq_likehood_gauss}.

Now consider a regression problem. If we wish to make a prediction for a new label \( y^{P+1} \) using squared error loss, a typical strategy is to guess the mean of its predictive distribution.
One has:
\begin{equation}
    \hat{y}^{(P+1)} = \int\; y \; \mathcal{P}_{\mathcal{T}}(y \mid x^{(P+1)}) \, dy   
\end{equation}

The meaning of this approach can be clarified by making use of Eq. \ref{3BIW:eq_pred_dis_bayes} we obtain:
\begin{equation}
    \hat{y}^{(P+1)} = \int\; dy\; y \; \int \mathcal{P}_{\vartheta}(y \mid x^{(P+1)}) \; \mathcal{P}(\vartheta \mid \mathcal{T})  \, d\vartheta
\end{equation}

Further inserting the Gaussian ansatz of Eq. \ref{3BIW:eq_likehood_gauss}, and invoking a saddle point approximation, this approach simplifies to the following guess:
\begin{equation}
    \hat{y}^{(P+1)} = \int f_{\vartheta}(x^{(P+1)}) \, \mathcal{P}(\vartheta \mid \mathcal{T})\, d\vartheta
\end{equation}

Therefore, the central object in a Bayesian approach to DNNs training is the conditional probability $\mathcal{P}(\vartheta \mid \mathcal{T})$.
In practice, this approach does not consider a single solution of the optimization process, but rather a probability distribution of the network parameters, (exponentially) weighted by a proper loss function.  Such a distribution clearly depends on the training dataset but also on our prior for the network parameters, that is, the distribution from which we draw the DNN initial weights. In this sense, one does not consider a single training procedure, starting from a given initialization, but integrates over many training instances (with the same dataset), thus potentially better exploring the loss complex landscape.
On a more technical ground, note that the saddle point approximation we have used becomes exact for a vanishing variance $\beta^{-1}$, that is for a fully deterministic rule in Eq. \ref{3BIW:eq_likehood_gauss}.

Naively, this Bayesian framework may not appear to bear any advantage for use with neural networks. Bayesian inference, in fact, deduces the posterior $\mathcal{P}(\vartheta \mid \mathcal{T})$ from a prior for the model parameters, which is intended to embody our prior beliefs about the problem. In a one-hidden-layer fully connected neural network, these parameters are the weights and the biases. The relationship of these parameters to any \textit{a priori} knowledge we might have about the problem is typically rather obscure (indeed DNNs are often classified as non-parametric models due to the lack of parameter interpretability). Furthermore, a network with a small number of hidden units can represent only a limited set of functions, which will generally do not exactly include the target function. 
Consequently, the Bayesian approach faces an initial hurdle: the challenge of defining a suitable prior, without which the Bayesian inference process risks stalling at the outset.

However, in the absence of strong priors, Bayesan analysis may be carried on with non-informative priors. Moreover, typical practical initializations draw weights ot of Gaussian distribution, which is indeed rather noninformative (apart for a preference towards small values). 
Even given a Gaussian prior, the task of computing the posterior is an unsolved problem: the likelihood $\mathcal{L}(\vartheta \mid \mathcal{T})$ is not easily computable as in our trivial coin toss example. 
We can circumvent this problem when the infinite width limit is taken into account: it turns out that in this limit, we can deduce the priors over the DNN outputs from the Gaussian priors for the network parameters \cite{neal1994priors,neal1994bayesian,williams1997computing}. The advantage of this change of prior, is that now one can explicitly compute the associated posterior and its corresponding predictive distribution.
In the following subsections, we will illustrate in details these fundamental results.

\subsubsection{Priors for Neural Networks with One Hidden Layer in the Infinite Width Limit}
Let's consider a single output one-hidden-layer fully connected neural network without bias (for notational simplicity, though they can be easily included in this analysis): 
\begin{equation}
    f_{\text{DNN}}(x^\mu) = \sum_{i=1}^{N_{1}} v_i h_{i}(x^\mu), \quad h_{i}(x^\mu) = \sigma\left(\sum_{j=1}^{N_0} W_{ij} x_j^\mu\right).
\end{equation}
Suppose that the weights of this neural network are randomly initialized from a normal distribution, i.e., \( v_i \sim \mathcal{N}(0, \sigma_v^2/N_1) \) and \( W_{ij} \sim \mathcal{N}(0, \sigma_W^2/N_0) \). 
This procedure is common in deep learning for weight initialization but, from a statistical viewpoint, can be seen as the selection of a prior distribution over the weights. 
Our interest lies in understanding what kind of prior over the output of the function \( f_{\text{DNN}} \) is induced by this specific prior over the weights.

Let's start from the first step: consider the activations \( h^\mu_i \) of the hidden layer. Given that the weights \( W_{ij} \) are independently and identically distributed (i.i.d.), each activation \( h^\mu_i \) is a function of a distinct set of these weights. Since the weights are drawn independently from the same distribution, the activations \( h^\mu_i \) across different neurons in the same layer are also independent of each other. However, due to the nonlinear activation $\sigma$, they have not generically Gaussian.

Nevertheless, in the architecture, the next step involves a randomly weighted sum of these i.i.d. activations. Considering the case of the limit infinite width (IW) limit \( N_1 \to \infty \), we can apply the Central Limit Theorem (if we have a finite number of examples $\mu = 1, \dots, P$ ). Doing so we deduce that the output function \( f_{\text{DNN}} \) is Gaussian distributed. The mean of this distribution is given by 
\begin{equation}
    \mu(x^\mu) = \mathbb{E}[f_{\text{DNN}}(x^\mu)] = \mathbb{E}\left[\sum_i v_i h_i^\mu\right] = \sum_i \mathbb{E}[v_i] \mathbb{E}[h_i^\mu] = 0,
\end{equation}
because \( \mathbb{E}[v_i] = 0 \) by definition. Note that $\mathbb{E}$ denotes the expectation value w.r.t. to the Gaussian prior. Moreover, the covariance matrix is:
\begin{equation}
    \begin{split}
        K(x^\mu, x^\nu) &= \mathbb{E}[f_{\text{DNN}}(x^\mu) f_{\text{DNN}}(x^\nu)] =  \mathbb{E}\left[\sum_i v_i h_i^\mu \sum_j v_j h_j^\nu\right] =\\
        &= \sum_{i,j} \mathbb{E}[v_i v_j] \, \mathbb{E}[h_i^\mu h_j^\nu] = \sigma_v^2 \sum_{i,j} \delta_{i,j} \,\mathbb{E}[h_i^\mu h_j^\nu] = \sigma_v^2 \; C(x^\mu, x^\nu).
    \end{split}
    \label{3BIW:KFUNC}
\end{equation}

Distributions over functions of this sort, in which the joint distribution of the values of the function at any finite number of points is (possibly multivariate) Gaussian, are known as \textit{Gaussian processes} (GP). To be more formal:
\begin{definition}[Gaussian Process]
    A set of (possibly infinite) variables is a Gaussian Process $\mathcal{GP}(\mu,K)$, with mean function $\mu(z)$ and covariance function $K(z,z')$, if any finite subset $\{z_\mu\}_{\mu=1}^M$ follows $\mathcal{N}(\mu^*,K^*)$ with:
    \begin{equation}
        \mu^*=\left[\begin{matrix} \mu(z_1)\\ \vdots \\ \mu(z_M)\end{matrix}\right] \qquad \qquad
        K^* = \left[\begin{matrix}
            K(z_1,z_1) & \dots & K(z_1,z_M)\\
            \vdots & \ddots & \vdots\\
            K(z_M,z_1) & \dots & K(z_M,z_M)
        \end{matrix}\right]\,.
    \end{equation}
\end{definition}
Therefore, in the infinite width limit, we can conclude that:
\begin{equation}
    \{f_{\text{DNN}}(x_\mu)\}_\mu^P \sim \mathcal{GP}(\mu,K) .
\end{equation}

GPs are widely used in various fields due to their flexibility and robustness in modeling uncertainties and complex phenomena. In geostatistics, GPs are fundamental for spatial data analysis and kriging, as detailed in Ref \cite{gelfand2010handbook}. In environmental science, they are employed for modeling climate and ecological systems, as explored in Ref. \cite{diggle2007model}. GPs also find applications in the field of robotics for trajectory optimization, as discussed in Ref. \cite{deisenroth2015gaussian}. These examples underscore the versatility of GPs in providing sophisticated solutions to complex problems across various scientific and engineering disciplines.

This approach can be extended to neural networks with multiple (vectorial) outputs (used for instance in classification) without loss of generality. In fact, it is possible to show that 
\begin{equation}
    \mathbb{E}[f_{\text{DNN}}(x^\mu)_k \, f_{\text{DNN}}(x^\nu)_{k'}] = 0 \quad \text{if} \quad k \neq k',
\end{equation}
since the weights leading into different output units are independent under the prior. As zero covariance implies independence for Gaussian distributions, knowing the values of one output does not provide any information about the values of other outputs, at these or any other input points. In other words, when the number of hidden units is infinite, there is no difference in training one network with two outputs or using the same data to train two separate networks, each with one output.

The central limit theorem requires summing contributions with a finite variance. Therefore the results in this section hold for any hidden unit activation function that is bounded and reasonably well-behaved (a discussion of this fine point is beyond the scope of this manuscript). For unbounded activations, such as ReLU, one should additionaly request bounded input data. They are also applicable for any prior on the input-to-hidden weights (and hidden unit biases) where the weights and biases for different hidden units are independent and identically distributed. Furthermore, the results extend to cases where the prior for hidden-to-output weights is not Gaussian, as long as the prior has finite variance. 

Note that there is an implicit requirement for this approach to be informative: the number of neurons in the layer \( N_1 \) must be of the same order as the input dimension \( N_0 \). This is a reasonable requirement, as the network should be able to capture the complexity of the input data. In fact, the off-diagonal elements of the kernel function \( K(x^\mu, x^\nu) \) often scale as \( \frac{1}{\sqrt{N_0}} \) and \( P > N_0 \). Therefore, in order to have a CLT variance that does not wash out these \( \frac{1}{\sqrt{N_0}} \) terms, we need that \( N_1 \sim N_0 \).


\subsubsection{Priors for Neural Networks with Multiple Hidden Layers in the Infinite Width Limit} \label{3BIW:sec:multi_layer}

The principles discussed in the previous section can be generalized to neural networks with multiple hidden layers. This generalization is achieved through an inductive process, where we consider the widths of the hidden layers to become infinitely large in succession (i.e., $N_1 \to \infty$, $N_2 \to \infty$, and so on). This approach ensures that the input to each layer under consideration is governed by a Gaussian Process (GP). Here, we present a more discursive proof, while in the next \textit{details} section a more accurate proof will be provided. 

We can express the deep neural network recursively (assuming again a single output for simplicity) as follows:
\begin{equation}
    \begin{split}
        f_{\text{DNN}}(x^\mu) &= \sum_{i=1}^{N_{L-1}} v_i^{(L)} \sigma\left(h_{i}^{(L-1)}(x^\mu)\right), \\
        h^{(L-1)}_{i}(x^\mu) &= \sum_{j=1}^{N_{L-1}} W^{(L-1)}_{ij} \sigma\left(h^{(L-2)}_j(x^\mu)\right), \\
        h^{(\ell)}_{i}(x^\mu) &= \sum_{j=1}^{N_{\ell-1}} W^{(\ell)}_{ij} \sigma\left(h^{(\ell-1)}_j(x^\mu)\right), \\
        h^{(1)}_{i}(x^\mu) &= \sum_{j=1}^{N_0} W^{(1)}_{ij} x^\mu_j.
    \end{split}
\end{equation}

Assuming that \( h^{(\ell-1)} \) follows a Gaussian Process \( \mathcal{GP}(0,K^{(\ell-1)}) \), and applying similar reasoning to that in the previous section, we observe that \( \sum_{j=1}^{N_{\ell-1}} W_{ij}^{(\ell)} \sigma(h^{(\ell-1)}) \) constitutes a sum of i.i.d. elements. Therefore, as \( N_{\ell-1} \to \infty \) while \( P \) remains finite, \( h^{(\ell)} \) converges to a Gaussian Process \( \mathcal{GP}(0,K^\ell) \). The covariance of this Gaussian Process is given by:
\begin{equation}
    \begin{split}
        K^{\ell}(x^\mu,x^\nu) &= \mathbb{E}\left[h^\ell(x^\mu) h^\ell(x^\nu)\right] = \\
        &= \sigma_{w_\ell}^2 \mathbb{E}\left[\sigma(h^{\ell-1}(x^\mu))\sigma(h^{\ell-1}(x^\nu))\right]_{h^{\ell-1} \sim \mathcal{GP}(0,K^{(\ell-1)})}.
    \end{split}
\end{equation}
This last expected value over the GP governing \( h^{\ell-1} \) is equivalent to the integral over the joint distribution of \( h^{\ell-1}(x^\mu) \) and \( h^{\ell-1}(x^\nu) \). In fact, we can write:
\begin{equation}
    \lim_{N_\ell \to \infty } K^\ell(x^\mu,x^\nu) = \int dt\,dt'\, \sigma(t)\sigma(t') \, \mathcal{N}\left(0,\sigma_{w_\ell}^2\left[ \begin{matrix} K^{\ell}(x^\mu,x^\mu) & K^{\ell}(x^\mu,x^\nu)\\ K^{\ell}(x^\nu,x^\mu) & K^{\ell}(x^\nu,x^\nu) \end{matrix}\right]\,\right)_{t,t'}
\end{equation}

This Normal distribution is two-dimensional with zero mean and with the covariance matrix that has only three entries: \( K^{\ell-1}(x^\mu,x^\nu) \), \( K^{\ell-1}(x^\mu,x^\mu) \), and \( K^{\ell-1}(x^\nu,x^\nu) \). We can summarize this observation by writing:
\begin{equation}
    K^\ell(x^\mu,x^\nu) = \sigma_{w_{\ell}}^2 \, F_\sigma\left[K^{\ell-1}(x^\mu,x^\nu), K^{\ell-1}(x^\mu,x^\mu), K^{\ell-1}(x^\nu,x^\nu)\right],
\end{equation}
to emphasize the recursive relationship between \( K^{\ell} \) and \( K^{\ell-1} \) via the deterministic function \( F_\sigma \), whose form depends solely on the nonlinearity \( \sigma \). 

This provides an iterative series of computations that can be performed to obtain \( K^L \) for the GP describing the network's final output. Using this recursion rule, we can also derive:
\begin{equation}
    K^1(x^\mu,x^\nu) = \mathbb{E}[h_i^{(1)}(x^\mu) h_i^{(1)}(x^\nu)] = \sigma_{w_{1}}^2 \sum_{i=1}^{N_0} \frac{x^\mu_i x^\nu_i}{N_0}.
    \label{3BIW:K1}
\end{equation}
These recurrence relations have appeared in other contexts. They are exactly the relations derived in the mean-field theory of signal propagation in fully-connected random neural networks \cite{poole2016exponential,daniely2016toward} and also appear in the literature on compositional kernels \cite{cho2009kernel}. For certain activation functions, $F_\sigma$ can be analytically computed, as we will see in a while for ReLU.

\details{
    \begin{center}
        \textbf{A more formal proof}
    \end{center}

    In this section, an alternative derivation is presented, demonstrating the equivalence between infinitely wide deep neural networks and Gaussian Processes through the process of marginalizing over the intermediate layers. This derivation assumes that both weight and bias parameters are independently drawn from Gaussian distributions with zero mean and variances scaled appropriately.

    Our objective is to determine the distribution \( \mathcal{P}(f_{\text{DNN}} \mid x) \) for the network outputs \( f_{\text{DNN}} \in \mathbb{R}^{N_L \times P} \), given the network inputs \( x \in \mathbb{R}^{N_0 \times P} \). Here, \( N_0 \) represents the input dimensionality, \( N_L \) the output dimensionality, and \( P \) the size of the dataset. The widths of the intermediate layers are denoted as \( N_\ell \), with the activation of the \( \ell \)-th layer represented by \( h^{\ell} \in \mathbb{R}^{N_{\ell+1} \times P} \). The second-moment matrix of the post-nonlinearity for each layer \( \ell \) is defined as:

    \begin{equation}
        K_{a b}^\ell = 
        \begin{cases}
          \frac{1}{N_0} \sum_i^{N_0} x^\mu_i\, x^\nu_i  \qquad \qquad \qquad \qquad \qquad \ell = 0\\
          \frac{1}{N_\ell} \sum_i^{N_\ell} \sigma(h^{\ell-1}_i(x^\mu_i))\, \sigma(h^{\ell-1}_i(x^\nu_i)) \qquad \ell > 0
        \end{cases}
        \label{eq3:KL}
    \end{equation}

    This methodology, as outlined in Ref. \cite{LeeGaussian}, involves considering intermediate random variables that correspond to the second moments defined above. By definition, $K^\ell$ only depends on ${\ell-1}$. In turn, the pre-activations $h^\ell$ are described by a Gaussian process conditioned on the second moment matrix $K_\ell$:

    \begin{equation}
        \mathcal{P}(h^\ell \mid K^\ell) = \mathcal{GP}_{h^\ell}(0\,,\,G(K^\ell))
    \end{equation}
    where $G(K) = \sigma_{w_\ell}^2 K^\ell + \sigma_{b_\ell}^2 \mathbf{1}\mathbf{1}^T$, to take into account also the bias. This correspondence of each layer to a GP, conditioned on the layer's second moment matrix, is exact even for finite width $N_\ell$ because the parameters are drawn from a Gaussian.

    So, we can write the process from the input to the output as:
    \begin{equation}
        \begin{split}
            \mathcal{P}(f_{\text{DNN}} =  h^L \mid x) &= \int \mathcal{P}(h^L, K^0,\dots, K^L \mid x) \, dK^0\,\dots\,dK^L =\\
            &= \int \mathcal{P}(h^L \mid K^L) \prod_{\ell=1}^L\left(\mathcal{P}(K^\ell\mid K^{\ell-1})\right)\mathcal{P}(K^0\mid x) \, dK^0\,\dots\,dK^L
        \end{split}
    \end{equation}
    where the decomposition holds because  because $K^L$ is a function only of $h^{L-1}$, $h^{L-1}$ depends only on $K^{L-1}$, $K^{L-1}$ depends only on $h^{L-2}$, and so on. 
    
    The sum in Equation \ref{eq3:KL} for $\ell > 0$  is a sum over i.i.d. terms. As $N_l$ grows large, the Central Limit Theorem applies, and $\mathcal{P}(K_\ell \mid K_{\ell-1})$ converges to a Gaussian with a variance that shrinks as $1/N_\ell$.
    Further, in the infinite width limit, it converges to a Dirac's delta:
    \begin{equation}
        \begin{split}
            \lim_{N_\ell \to \inf} \mathcal{P}(K^\ell \mid K^{\ell-1}) &= \delta\left(K^\ell - (F_\sigma \circ G)(K^{\ell-1}) \right)
        \end{split}
    \end{equation}
    Similarly, the dependence in the initial layer can be written as:
    \begin{equation}
        \mathcal{P}(K^0 \mid x) = \delta \left(K^0 - \frac{1}{N^0}x^T\,x \right) 
    \end{equation}
    
    Substituting, we finally obtain:
    \begin{equation}
        \lim_{N_1,\dots,N_L \to \infty} \mathcal{P}(f_{\text{DNN}}(x) \mid x) = \mathcal{GP}_{f_{\text{DNN}}} \left(0, \left(G\circ(F_\sigma\circ G)^L \frac{x^T\,x}{N_0}\right)\right) 
    \end{equation}

}

\subsubsection{The NNGP Kernel for predictions}

In the preceding section, we calculated the prior probability distribution for functions under the standard initialization of a neural network in the infinite width (IW) limit. We observed that this prior is a Gaussian with a mean of zero and a covariance defined by a specific two-point function, \( K(x,x') \). This function is crucial and is known as the \textit{NNGP Kernel}. It characterizes the correlation between the outputs of a deep neural network (DNN) as a function of its inputs.

With this new prior probability distribution, we can now perform Bayesian inference on a DNN. For a more illustrative approach, consider a one-hidden-layer neural network (1HL NN) encountering a new, unseen example \( \{x^*,y^*\} \). In the IW limit, the NN approximates a Gaussian Process (\( \mathcal{GP} \)), simplifying the calculation of joint probability distributions:
\begin{equation}
    P(f^*, y|x^*,x)  \sim  \mathcal{N}\left(\left[\begin{matrix} 0 \\ 0 \end{matrix}\right],\left[\begin{matrix} 
        K(x^*,x^*) & K(x^*,x)  \\ 
        K(x,x^*) & K(x,x)  \\ 
    \end{matrix}\right]\right),
\end{equation}
where \( f^* = f_{\text{DNN}}(x^*) \) and \( \{x,y\} \in \mathcal{T} \).
Utilizing this probability, we can derive the predictive distribution as follows:

\begin{equation}
    \mathcal{P}(f_{\text{DNN}}^* | x^*, \mathcal{T}) = \int \mathcal{P}(f_{\text{DNN}}^*, y | x^*, x) \, \mathcal{P}(y | x) \,dx
\end{equation}
In the context of regression, \( \mathcal{P}(y | x) \) is specified by \ref{3BIW:eq_likehood_gauss}.
Ultimately, we obtain: 
\begin{equation}
    \mathcal{P}(f_{\text{DNN}}^* | x^*, \mathcal{T}) =  \mathcal{N}(\gamma, \Gamma),
    \label{3BIW:BAYPFT}
\end{equation}
where:
\begin{equation}
    \begin{split}
        \gamma &= K(x, x^*) \left[K(x,x) +\frac{1}{\beta}\mathbf{1}\right]^{-1} y \\
        \Gamma &=  K(x^*, x^*) - K(x, x^*) \left[K(x,x) +\frac{1}{\beta}\mathbf{1}\right]^{-1} K(x^*, x)
    \end{split}
\end{equation}
It is important to note that considering the IW limit renders the model non-parametric. The inference process derived from this limit does not depend on the network's weights, \( \vartheta \).

Extension to deeper network is straightforward, as one just needs to substitute in the above formula the one hidden layer kernel $K$ with the recursive $K^L$ obttained in Section \ref{3BIW:sec:multi_layer}.

\subsection{A physical approach}\label{3BIW:PhisAPP}

Until now, we have employed the formalism and rigor typical of Statistics. We now propose a more intuitive explanation for the emergence of the Gaussian Process (GP) in the infinite width limit, drawing from concepts in Statistical Mechanics. 

Consider the one-hidden-layer neural network (1HL NN) without biases. Let us first discuss the common initialization scheme for such a network. Typically, this involves using an NN without normalization during the forward pass but starting with scaled initialized weights. However, we will consider a slightly different scenario where the weights are not scaled, but normalization is explicitly included during the forward pass of the NN. In this setting, the NN is represented as:
\begin{equation}
    f_{\text{DNN}}(x;W,v) = \frac{1}{\sqrt{N_1}} \sum_{i}^{N_1} v_i \sigma \left(\sum_{j}^{N_0}\frac{W_{i,j}\,x_j}{\sqrt{N_0}}\right)
\end{equation}
where \( \mathcal{P}(W) = \mathcal{N}(0,\sigma_W^2) \) and \( \mathcal{P}(v) = \mathcal{N}(0,\sigma_v^2) \). In this kind of parametrization, known as the Neural Tangent Kernel (NTK) parametrization, the scaling behavior of the entire function becomes more apparent.

Now we can exploit this to find the probability distribution of the output given a dataset \( \mathbf{x} = \{x_\mu\}_{\mu=1}^P \):
\begin{equation}
    \mathcal{P}(\mathbf{f} | \mathbf{x}) = \int \mathcal{D}W\,\mathcal{D}v\,\prod_{\mu=1}^P \delta\left(f^\mu - \frac{1}{\sqrt{N_1}} \sum_{i=1}^{N_1} v_i \sigma \left(\sum_{j=1}^{N_0}\frac{W_{i,j}\,x^\mu_j}{\sqrt{N_0}}\right)\right),
\end{equation}
where \( \mathcal{D}W = \prod_{i,j}\, \mathcal{P}(W_{i,j}) \,dW_{i,j} \) and \( \mathcal{D}v = \prod_{i}\, \mathcal{P}(v_{i}) \,dv_{i} \). 
These integrals are challenging to perform because the integral variables are coupled by the nonlinear activation function. To address this problem, we can introduce a set of Dirac's deltas to decouple each preactivation. We use the following identity:
\begin{equation}
    1 = \int \prod_{\mu=1}^P\prod_{i=1}^{N_1} dh_{i}^{\mu}\, \delta\left(h_i^\mu - \sum_{j=1}^{N_0}\frac{W_{i,j}\,x^\mu_j}{\sqrt{N_0}}\right)\,,
\end{equation}
and employ the Fourier representation of the Dirac's delta:
\begin{equation}
    \delta\left(h_i^\mu - \frac{\sum_{j=1}^{N_0}W_{i,j}\,x^\mu_j}{\sqrt{N_0}}\right) = \int \frac{d\bar{h}_i^{\mu}}{2\pi} e^{i\bar{h}_i^{\mu}\left(h_i^\mu - \sum_{j=1}^{N_0}\frac{W_{i,j}\,x^\mu_j}{\sqrt{N_0}}\right)}\,.
\end{equation}
In the same way, we can also introduce a set of Dirac's deltas to decouple the output of the network:
\begin{equation}
    \delta\left(f^\mu-\frac{1}{\sqrt{N_1}} \sum_{i=1}^{N_1} v_i \sigma \left(h^\mu_i\right)\right) = \int \frac{d\bar{f}^\mu}{2\pi} e^{i\bar{f}^\mu\left(f^\mu-\frac{1}{\sqrt{N_1}} \sum_{i=1}^{N_1} v_i \sigma \left(h^\mu_i\right)\right)}\,.
\end{equation}
By doing so, we obtain: 
\begin{equation}
    \begin{split}
        \mathcal{P}(\mathbf{f} | \mathbf{x}) &= \int \prod_{\mu=1}^P d\bar{f}^\mu \prod_{i=1}^{N_1} dh_i^\mu \,d\bar{h}_i^\mu\,e^{i\sum_{\mu=1}^P f^\mu \bar{f}^\mu + i \sum_{i=1}^{N_1}\sum_{\mu=1}^P h^\mu_i \bar{h}^\mu_i}\\
        & \qquad \qquad \times \int \mathcal{D}W \mathcal{D}v \, e^{-i \sum_{\mu=1}^P \frac{\bar{f}^\mu}{\sqrt{N_1}} \sum_{i=1}^{N_1} v_i \sigma(h^\mu_i) } \, e^{i \sum_{\mu=1}^P \sum_{i=1}^{N_1} \frac{\bar{h}^\mu_i}{\sqrt{N_0}} \sum_{j=1}^{N_0}W_{i,j}\,x^\mu_j}.
    \end{split}
\end{equation}
In this form, the integrals over the network's weights are Gaussian and can be easily performed. 
\begin{equation}
    \begin{split}
        \mathcal{P}(\mathbf{f} | \mathbf{x}) &= \int \prod_{\mu=1}^P d\bar{f}^\mu e^{i\sum_{\mu=1}^P f^\mu \bar{f}^\mu } \times \\ 
        &\qquad\qquad \times \left[\int \prod_{i=1}^{N_1} dh^\mu \,d\bar{h}^\mu\, e^{i \sum_{\mu=1}^P h^\mu \bar{h}^\mu}\, e^{-\frac{1}{N_1} \left(\sum_{\mu=1}^P \bar{f}^\mu \, \sigma(h^\mu)\right)^2- \frac{1}{2} \sum_{\mu,\nu=1}^P \bar{h}^\mu C_{\mu,\nu}^{-1} \bar{h}^\nu }\right]^{N_1},
    \end{split}
\end{equation}
where we have used the fact that \( i \) is a dummy index, and \( C_{\mu,\nu} = K^{(1)} = \sigma^2_W \sum_{j=1}^{N_0} \frac{x^\mu_j\,x^\nu_j}{N_0} \) as seen in \ref{3BIW:K1}.
Again, the integral over \( \bar{h}^\mu \) is Gaussian and can be performed:
\begin{equation}
    \mathcal{P}(\mathbf{f} | \mathbf{x}) = \int  \prod_{\mu=1}^P d\bar{f}^\mu e^{i\sum_{\mu=1}^P f^\mu \bar{f}^\mu } \left[\int \prod_{i=1}^{N_1} \frac{dh^\mu}{\sqrt{\det C}} \, \, e^{-\frac{1}{N_1} \left(\sum_{\mu=1}^P \bar{f}^\mu \, \sigma(h^\mu)\right)^2- \frac{1}{2} \sum_{\mu,\nu=1}^P h^\mu C_{\mu,\nu} h^\nu }\right]^{N_1}
\end{equation}
Now we consider the fact that the number of weights in the hidden layers goes to infinity, while the dataset remains finite. In this limit, we can consider the Taylor expansion around 1:
\begin{equation}
    \begin{split}
        \mathcal{P}(\mathbf{f} | \mathbf{x}) = \lim_{N_1 \to \infty}\int \prod_{\mu=1}^P df^\mu e^{if^\mu\bar{f}^\mu}\left[1-\frac{1}{2 N_1}\sum_{\mu,\nu=1}^P \bar{f}^\mu K_{\mu,\nu}\bar{f}^\nu\right]^{N_1} \to \mathcal{N}(0,K),
    \end{split}
    \label{3BIW:PFX}
\end{equation}
where we have used:
\begin{equation}
    \int \prod_{\mu=1}^P dh^\mu \, \mathcal{N}(0,C)_{\{h\}} \,  \sigma(h^\mu) \sigma(h^\nu) = K_{\mu,\nu}\,
    \label{3BIW:KERNELK}
\end{equation} 
as we have seen in \ref{3BIW:KFUNC}.

In this way, we have derived the probability densities of the prior over the function, and we aim to use it to establish a well-defined problem of Statistical Mechanics, which typically begins with a Partition Function. Let us discuss the meaning of optimizing a Deep Neural Network (DNN). Generally, this problem can be expressed as the search for parameters that minimize the loss function, or equivalently, maximize the following log-likelihood:
\begin{equation}
    \vartheta^* = \argmax_{\vartheta} \log \left[\mathcal{P}( \vartheta | \{x^\mu,y^\mu\}_{\mu=1}^P ) \mathcal{P}(\vartheta )\right]
\end{equation}

In the standard Statistical Mechanics approach, we can rephrase this problem as an optimization of the partition function, which can be written as:
\begin{equation}
    \mathcal{Z} = \int d\vartheta \, \mathcal{P}(\vartheta)\mathcal{P}( \vartheta | \{x^\mu,y^\mu\}_{\mu=1}^P )\,,
\end{equation}
with \( \mathcal{P}(\vartheta | \{x,y\}_{\mu=1}^P)= e^{-\beta \mathcal{L}(\vartheta;\{x,y\}_{\mu=1}^P)} \) and \( \mathcal{P}(\vartheta) \) being the standard Gaussian prior over the weights.
Using the following transformation, we can rewrite the partition function as an integral over the function prior.
\begin{equation}
    \begin{split}
        \mathcal{Z} &= \int d\vartheta \, \mathcal{P}(\vartheta) \, \mathcal{P}( \vartheta | \{x^\mu,y^\mu\}_{\mu=1}^P ) = \\
        &= \int d\vartheta \, \mathcal{P}(\vartheta) \int d \tilde{\mathbf{f}}\, \mathcal{P}(\tilde{\mathbf{f}}| \mathbf{y} )\, \delta(\tilde{\mathbf{f}} - f(\vartheta,\mathbf{x})) \\
        &= \int  d\tilde{\mathbf{f}}\, \mathcal{P}(\tilde{\mathbf{f}}| \mathbf{y} ) \int d\vartheta \, \mathcal{P}(\vartheta)\, \delta(\tilde{\mathbf{f}} - f(\vartheta,\mathbf{x})) \\
        &= \int d\tilde{\mathbf{f}}\, \mathcal{P}(\tilde{\mathbf{f}}| \mathbf{y} ) \,\mathcal{P}(\tilde{\mathbf{f}} | \mathbf{x})\,.
    \end{split}
\end{equation}
The partition function can be exactly found in the Infinite Width (IW) limit, where \( \mathcal{P}(\tilde{\mathbf{f}} | \mathbf{x}) \) is given by \ref{3BIW:PFX}, and \( \mathcal{P}(\tilde{\mathbf{f}} | \mathbf{y} ) \) descends directly from the forms of \( \mathcal{P}( \vartheta | \{x^\mu,y^\mu\}_{\mu=1}^P ) \), and reads:
\begin{equation}
    \mathcal{P}(\tilde{\mathbf{f}}| \mathbf{y} ) = e^{-\beta \sum_{\mu=1}^P (y^\mu-\tilde{f}^\mu)^2}
\end{equation}

So, in the end, we obtain the following form of the partition function:
\begin{equation}
    \mathcal{Z} = \mathcal{N}_{\{y^\mu\}_{\mu=1}^P}\left(0,K+\frac{1}{\beta}\right)
\end{equation}
Having the explicit form for \(\mathcal{Z}\), we can also calculate the average value of an observable. In particular, we are interested in the average error on an unseen datapoint (i.e., the generalization error):
\begin{equation}
    \langle\epsilon_g(x^\ast,y^\ast)\rangle = \int d\tilde{\mathbf{f}} \, \left(y^\ast - \tilde{f}(x^\ast)\right)^2 \, \mathcal{P}(\tilde{\mathbf{f}}| \mathbf{y} ) \,\mathcal{P}(\tilde{\mathbf{f}} | \mathbf{x})  \to (y^\ast-\gamma)^2 + \Gamma\,,
\end{equation}
which is the same result that one would obtain using the standard Bayesian approach in Eq. \ref{3BIW:BAYPFT}.

As a final comment on the infinite width limit, we can note that the success of the Gaussian prior can be linked to the well-known fact that in infinite architectures the network parameters are slow-moving and only change infinitesimally under gradient descent training \cite{jacot2018neural}. Therefore, the Gaussian prior turns out to represent a good solution of the learning problem.

In the context of gradient descent training, the infinite width limit leads of course to the celebrated Neural Tangent Kernel \cite{JacotNTK}.

\section{Beyond the Infinite Width limit}

In the previous section, we explored the power of Bayesian inference within the context of neural networks under the infinite width (IW) limit. This theoretical framework provides valuable insights into neural network behavior, particularly regarding how they generalize from training data. However, it's crucial to recognize that this approximation, despite its mathematical elegance, does not fully capture the complexities of neural networks as they are used in real-world applications.

In practical scenarios, the size of the training set often matches or even exceeds the number of neurons in each layer of the network. This situation is in stark contrast to the assumptions of the IW limit, where the network's width (i.e., the number of neurons in each layer) is considered to be infinitely large compared to the size of the training set. The traditional statistical approach, which heavily relies on the Gaussian Process (GP) assumption in the IW limit, is inadequate in these more balanced scenarios. It fails to account for the complex behavior that arises when the size of the training set and the network width are comparable.

This observation leads us to consider a new scenario, known as the \textit{proportional regime}, in which both the number of training examples and the number of neurons in each layer grow to infinity, but maintain a fixed ratio denoted as $\alpha_\ell = P/N_\ell > 0$. The proportional regime aims to bridge the gap between results obtained in the infinite-width limit and the practical realities of neural network training. In this regime, the conventional statistical approach of Gaussian Processes (GPs) is not suitable, as it is well-defined only for a finite subset of examples. Instead, the approach of Statistical Mechanics, though perhaps less rigorous to an mathematical eye, leads to valuable insights in this context.

The proportional regime (a kind of thermodynamic limit in the language of statistical mechanics) presents the challenge of developing new analytical tools and frameworks capable of accommodating the simultaneous scaling of training data and network size. This approach, however, offers a more realistic understanding of neural network behavior, moving beyond the idealized assumptions of the infinite width (IW) limit to address the nuances and complexities of real-world applications. By employing the tools of Statistical Mechanics, we can gain deeper insights into the behavior and generalization properties of neural networks, thereby enhancing the comprension of the theoretical aspects of machine learning.

\subsection{Linear Networks in the Proportional Regime} 
Before talking about real Fully Connected Deep Neural Network, let's consider a simpler case, the Fully Connected Linear Neural Network (LNN), as presented by Lee and Sompolinsky in \cite{PhysRevX.11.031059}.
In this case the LNN reads as:
\begin{equation}
    f_{\text{DNN}}(x;\vartheta) = \prod_{\ell=0}^{L-1} \left(\frac{1}{\sqrt{N_{L-\ell}}} W^{(L-\ell)} \right) x\,
\end{equation}
or, separating layer by layer, we have:
\begin{equation}
    \begin{split}
        f_{\text{DNN}}(x;\vartheta) &= \frac{1}{\sqrt{N_L}} \sum_i^{N_L} W^{(L)}_i \, h^{(L-1)}_i(x; \vartheta^{(L-1)})\,,\\
        h^{\ell}_i(x;\vartheta^{(\ell-1)}) &= \frac{1}{\sqrt{N_\ell}} \sum_j^{N_\ell} W^{(\ell)}_{i,j} \, h^{(\ell-1)}_j(x;\vartheta^{(\ell-1)})\,,\\
        h^{1}_i(x; \vartheta^{(1)}) &= \frac{1}{\sqrt{N_1}} \sum_j^{N_1} W^{(1)}_{i,j} x_j\,,
    \end{split}
\end{equation}
where $\vartheta^{(\ell)} = \{W^{(k)}\}_{k=1}^{\ell}$. As usual, we can consider a regression problem with the following \textit{Loss Function}:
\begin{equation}
    \mathcal{L}(\vartheta \mid \mathcal{T}) = \frac{1}{2}\sum_\mu^{P} \left( f_{\text{DNN}}(x^\mu;\vartheta) - y^\mu\right)^2  - \frac{1}{2\beta} \sum_{\ell=1}^{L} \frac{\lVert W^{(\ell)} \rVert^2}{\sigma_\ell^2} \,.
\end{equation}
The first term represents the mean-squared deviation of the network outputs for a set of $P$ training inputs from their corresponding target labels, essentially constituting the standard Mean Squared Error (MSE). The second term, scaled by $\beta^{-1}$, serves as a regularization term that promotes smaller L2 norms for the weights. If we want to link this regularization to the Bayesian setting, we can imagine it as a Gaussian Prior over the weights.

The inclusion of the temperature parameter $\beta$ in this term implies that the L2 regularization behaves as an entropic element. Specifically, in the regime of primary interest where $\beta \to \infty$, the first term prioritizes the minimization of training errors. Concurrently, the L2 term influences the statistical distribution of the weight vectors that achieve this minimization, skewing it towards weights with smaller norms.

In this discussion, we focus on the equilibrium distribution of weights, described by the Gibbs Distribution:
\begin{equation}
\mathcal{P}(\vartheta; \mathcal{T}) = \frac{1}{\mathcal{Z}} e^{-\beta \mathcal{L}(\vartheta ; \mathcal{T})},
\end{equation}
where \(\mathcal{Z}\) is the partition function, defined as follows:
\begin{equation}
\mathcal{Z} = \int d\vartheta e^{-\beta \mathcal{L}(\vartheta ; \mathcal{T})}.
\end{equation}
It's important to note that the Gibbs distribution aligns with the posterior distribution of weights under a Gaussian prior. This equivalence creates a bridge between statistical mechanics and Bayesian statistical methods.

Calculating \(\mathcal{Z}\) directly is not feasible straightforward since the linear case loss is not simply quadratic in the weights. To address this, the authors develop a technique to compute the partition function \(\mathcal{Z}\) through successive integrations in order to separate the different weights layers to recover a simple integrable quadratic form. Each integration involves a single-layer weights matrix, beginning with the final layer \(W_L\).

Following this idea, we can start by the final layer, considering that 
\begin{equation}
Z_L(\vartheta^{(L-1)}) = \int dW^{(L)} e^{-\mathcal{L}(\vartheta)} = e^{-H_{L}(\vartheta^{(L-1)})}.
\end{equation}
where \(\vartheta^{(\ell)} = \{W^{(k)}\}_{k=1}^{\ell}\). 
The revised Hamiltonian is:
\begin{equation}
\begin{split}
    H_L(\vartheta^{(L-1)}) &= \frac{1}{2\sigma^2_L} \Tr[(\vartheta^{(L-1)})^T\vartheta^{(L-1)}] + \frac{1}{2} \mathbf{y}^T\left(K_L(\vartheta^{(L-1)})+ \frac{\mathbf{1}}{\beta} \right)^{-1} \mathbf{y} + \\ & \quad + \frac{1}{2} \log\det \left(K_L(\vartheta^{(L-1)})+ \frac{\mathbf{1}}{\beta} \right),
    \label{3BIW:sompo1}
\end{split}
\end{equation}
where \((K_L)_{\mu,\nu} = \frac{\sigma_L^2}{N_L} h^{(L-1)}(x^\mu)h^{(L-1)}(x^\nu)\) is a \(P \times P\) kernel matrix, and \(\mathbf{y}\) is the vector of labels. The kernel \(K_L\) depends on the weights of the previous layers \(\vartheta^{(L-1)}\) through \(h^{(L-1)}\). 
Thus, starting from the previous partial partition function, we can integrate over the next weights of the previous layer:
\begin{equation}
    Z_{L-1}(\vartheta^{(L-2)}) = \int dW^{(L-1)} \, Z_L (\vartheta^{(L-1)}) = e^{-H_{L-1}(\vartheta^{(L-2)})}\,.
\end{equation}
Assuming for simplicity \(\sigma_\ell = \sigma\) and \(N_\ell = N\) for all \(\ell = 1, \dots, L\) and considering the zero temperature limit \(\beta \to \infty\), we have:
\begin{equation}
    \begin{split}
        H_{L-1} (\vartheta^{(L-2)}, u_{L-1}) &= \frac{1}{2\sigma^2}  \Tr[(\vartheta^{(L-2)})^T\vartheta^{(L-2)}]  + \frac{1}{2u_{L-1}} \mathbf{y}^T K^{-1}_{L-1}\mathbf{y} +\\ &\,+\frac{1}{2} \log \det (K_{L-1}u_{L-1}) - \frac{N}{2}\log u_{L-1}+ \frac{1}{2\sigma^2}N\, u_{L-1}   
    \end{split}
    \label{3BIW:sompo2}
\end{equation}
Here, \(u_{L-1}\) initially appears as an auxiliary integration variable. In the thermodynamic limit, it becomes an order parameter, determined self-consistently by minimizing \(H_{L-1}\):
\begin{equation}
    1-\frac{u_{L-1}}{\sigma^2} = \alpha \left(1-\frac{\mathbf{y}^T K^{-1}_{L-1}\mathbf{y}}{P\, u_{l-1}}\right)
    \label{3BIW:scaleu}
\end{equation}
As we can see at a single glance, the forms of equations \ref{3BIW:sompo1} and \ref{3BIW:sompo2} are identical, with the only distinction being the factor \( u_{L-1} \) that rescales the kernel \( K_{(L-1)} \). This procedure can be iteratively applied layer by layer, backpropagating the integration and renormalizing the kernel at each step. This is why this method is referred to as Backpropagating Kernel Renormalization.

Thanks to this framework, it is possible to characterize the performance of a Linear Neural Network (LNN) even beyond the infinite width limit. In fact, utilizing this Statistical Mechanics technique, there's no need to consider a finite set of examples. Instead, we can consider a number of examples \( P \) that scales with \( N \), and study how the performance changes as a function of the ratio \( \alpha = P/N \). Indeed, this quantity prominently features in the Eq. \ref{3BIW:scaleu} of self-consistency for \( u \).

Unfortunately, as highlighted in the initial chapter of this thesis, nonlinearities are crucial in the context of real-world Neural Networks (NNs). The inclusion of nonlinearities is essential because they substantially increase the expressiveness and complexity of Deep Neural Networks (DNNs), enabling them to model a wider range of functions and solve more complex problems.

\subsection{One Hidden Layer Neural Networks in the Proportional Regime}
In this section, I will present a discursive exposition of our work on the Statistical Mechanics framework for Deep Neural Networks (DNNs) beyond infinite width, as introduced in \cite{ariosto2023}. 
Notably, this framework adeptly accommodates DNNs characterized by well-behaved nonlinearities.
The technical details omitted from this exposition will be elaborated upon in the \textit{detail} sections.

As a first case, let us consider a simple 1HL NN, which can be expressed as follows:
\begin{equation}
    \begin{split}
        f_{\text{DNN}}(x^\mu;\vartheta) &= \frac{1}{\sqrt{N}} \sum_{i=1}^{N} v_i \, h_i^\mu(W),\\
        h_i^\mu (W) &= \frac{1}{\sqrt{N_1}} \sum_{j=1}^{N_0} W_{i,j} x^\mu_j,
    \end{split}
\end{equation}
associated with the following \textit{loss function}:
\begin{equation}
    \mathcal{L}(\vartheta \, \mid \, \mathcal{T}) = \sum_{\mu=1}^P \left[y^\mu - \frac{1}{\sqrt{N}}\sum_{i=1}^N v_i \sigma\left(\sum_{j=1}^{N_0} \frac{W_{i,j}\, x^\mu_j}{\sqrt{N_0}}\right)\right] + \frac{\lambda_1}{2\beta} \sum_{i=1}^N v_i^2 + \frac{\lambda_0}{2\beta} \lVert W \rVert^2.
\end{equation}
Following the physical approach presented in \ref{3BIW:PhisAPP}, our main focus will be on the partition function:
\begin{equation}
    \mathcal{Z} = \int dW \, dv \, e^{-\beta \mathcal{L}(\vartheta \, \mid \, \mathcal{T})}.
\end{equation}
As before, the first step is to introduce the Dirac's delta to disentangle the weights of each layer:
\begin{equation}
    \begin{split}
        &1=\int \prod_{\mu}^P ds^\mu\; \delta\left[s^\mu - \frac{1}{\sqrt{N}} \sum_{i}^{N} v_{i} \sigma(h_{i}^\mu)\right], \label{eq:delta_s}\\
        &1=\int \prod_{\mu=1}^P \prod_{i}^{N} dh_{i}^\mu \; \delta\left(h_{i}^\mu - \frac{1}{\sqrt{N_0}} \sum_{j}^{N_0} w_{i,j} x_{j}^\mu \right).
    \end{split}
\end{equation}
and then perform the Gaussian integration (thanks to the Gaussian prior) over the weights, to obtain:
\begin{equation}
    \begin{split}
        &Z= \int \prod_{\mu}^P \frac{ds^\mu d \bar{s}^\mu}{2\pi} e^{-\frac{\beta}{2}\sum_{\mu}^P\left( y^\mu - s^\mu \right)^2 + i\sum_{\mu}^P s^\mu \bar{s}^\mu} \times \\
        & \qquad \times \left\{\int \prod_{\mu}^P \frac{dh^\mu d \bar{h}^\mu}{2\pi} e^{i\sum_{\mu}^P h^\mu \bar{h}^\mu -\frac{1}{2 \lambda_1 N_1} \left[ \sum_{\mu}^P \bar{s}^\mu \sigma(h^\mu) \right]^2-\frac{1}{2 \lambda_0 N_0 }\sum_{j}^{N_0}\left(\sum_{\mu}^P \bar{h}^\mu x^\mu_{j}\right)^2}\right\}^{N_1},
    \end{split}
    \label{eq:Z_intermediate}
\end{equation}
where we have utilized the fact that \(i\) was a dummy index to factorize the integrals over $h^\mu_i$ and $\bar{h}^\mu_i$. 
This last set of integral is Gaussian and can be performed:
\begin{equation}
    \int \prod_\mu^P \frac{d \Bar{h}^\mu}{2\pi} e^{i\sum_\mu^P h^\mu \Bar{h}^\mu-\frac{1}{2 \lambda_0 N_0 }\sum_{j}^{N_0}\left(\sum_\mu^P \Bar{h}^\mu x^\mu_{j} \right)^2}  =\frac{e^{-\frac{1}{2}\sum_{\mu,\nu}^P h^\mu C_{\mu,\nu}^{-1} h^{\mu}  }}{\sqrt{(2\pi)^P \det C}}=\mathcal{P}_1(\{h^\mu\})\,,
\end{equation}
where 
\begin{equation}
    C_{\mu\nu} = \frac{1}{\lambda_0 N_0} \sum_{j}^{N_0} x^\mu_{j} x^\nu_{j}\,.
    \label{eq:P1}
\end{equation}

\details{
    \begin{center}
        \textbf{Details: The Covariance Matrix \(C\)}
    \end{center}

    The object \(C\), which naturally emerges from our calculations, is the same covariance matrix of the dataset over the data points, denoted as \(K^1\), referenced in the Bayesian approach shown in Eq. \ref{3BIW:K1}. This matrix encapsulates all the essential information about the dataset required for our analysis. 

    Technically, the covariance matrix \(C\) must be invertible. It can be demonstrated that invertibility fails when \(P > N_0\). However, we can address this issue by introducing a small diagonal term to \(C\). This extra regularization term is designed to be minimally invasive, ensuring it does not significantly affect the calculation's final outcomes.\\
    \,
}

To deal with the integral over $h^\mu$ we can include a further Dirac delta identity for the random variable $q = 1/\sqrt{\lambda_1 N_1}\sum_\mu \Bar{s}^\mu \sigma\left(h^\mu\right)$. This leaves us with the problem of finding the probability density $\mathcal{P}(q)$:
\begin{equation}
    Z= \int \prod_{\mu}^P \frac{ds^\mu d \bar{s}^\mu}{2\pi} e^{-\frac{\beta}{2}\sum_{\mu}^P\left( y^\mu - s^\mu \right)^2 + i\sum_{\mu}^P s^\mu \bar{s}^\mu} \left\{\int dq\; e^{-\frac{1}{2}q^2} \mathcal{P}(q)\right\}^{N_1}
\end{equation}
with:
\begin{equation}
    \mathcal{P}(q) = \int \prod_\mu^P dh^\mu\, \mathcal{P}_1{\{h^\mu\}}  \, \delta\!\left(q- \frac{1}{\sqrt{\lambda_1 N_1}}\sum_\mu \Bar{s}^\mu \sigma\left(h^\mu\right) \right)
    \label{3BIW:PQQQ}
\end{equation}
In the limit defined where $\alpha \geq 0$, this is exactly the case where we can apply the Breuer-Major theorems~\cite{BM,NourdinQuantitative, bardet2013} as shown in the following detail section. As such, it is sufficient that both the (regularized) covariance $C$ and the activation function $\sigma$ satisfy the hypotheses of the theorem to guarantee that the probability distribution $P(q)$ converges in distribution to a Gaussian: 
\begin{equation}
        P(q) \to \mathcal{N}_q(0,Q)\,.
    \label{eq:q_is_gaussian}
\end{equation}
with variance
\begin{equation}
    \begin{aligned}
        Q(\bar{s},C) &= \frac{1}{\lambda_1 N_1} \sum_{\mu,\nu}^P \Bar{s}^\mu\left[\int d^P h\, P_1(\{h^\rho\}) \sigma(h^\mu)\sigma(h^\nu)\right] \Bar{s}^\nu \\
        &=\frac{1}{\lambda_1 N_1}\sum_{\mu,\nu}^P \Bar{s}^\mu K_{\mu \nu}(C)\Bar{s}^\nu\,.
    \end{aligned}
    \label{eq:K_a_BM}
\end{equation}

Notably, the kernel obtained through this Gaussian approxiation is the same \emph{neural network Gaussian process} (NNGP) kernel that we have encountered previously. 

While it can be argued that there exist particular configurations $\bar s$ within the integration domain where invoking a Gaussian equivalence is not permissible, in our derivation, we operate under the assumption that the influence of these special configurations on the effective action becomes negligible in the thermodynamic limit. This point is further elaborated in the next \hyperlink{3BIW:detBM}{\textit{detail section}}.

Furthermore, our analysis presumes that the variable $q$ has a zero mean. This condition holds true under the equation:
\begin{equation}
\int d^P h \, P_1(\{h^\mu\}) \sigma(h^\nu) = 0,
\label{methods:eq:zeromean}
\end{equation}
which is valid when $\sigma$ is a zero-mean function. For a broader analysis that encompasses finite-mean activation functions, such as the ReLU, refer to a subsequent \hyperlink{finitemean}{\textit{detail section}}.

\details{
    \begin{center}
        \hypertarget{3BIW:detBM}{\textbf{Details: Breuer-Major theorems}}
    \end{center}

    Let us consider the following sequence of random variables:
\begin{equation}
S_N = \frac{1}{\sqrt{N}} \sum_{i=1}^{N} c_i F(x_i), \quad N \geq 1\,.
\end{equation}
It is evident that if the distribution of the vector $\mathbf{x} = (x_1, \dots, x_N)$ is factorizable over its coordinates, i.e., $p(\mathbf{x}) = \prod_{i} p(x_i)$, and if $F(x) = x$, then the random variable $S = \lim_{N \to \infty} S_N$ follows a normal distribution under specific conditions. These include the mean of each component $x_i$ being zero, $\mathbb{E}(x_i) = 0$, each $x_i$ having a finite variance, $\mathbb{E}(x_i^2) < \infty $, and the coefficients $c_i$ satisfying Lindeberg's condition. This conclusion holds true for cases where $F$ represents a well-behaved non-linearity..

The Breuer-Major theorem essentially extends this result to generic Gaussian Processes ($\mathcal{GP}$s). This extension provides sufficient conditions on both the covariance matrix of the GP and the non-linear function $F$, ensuring the convergence of the sequence $S_N$ to the normal distribution. We present here the modern formulation of the theorem, as presented in Ref. \cite{NourdinQuantitative}. This theorem's relevance is underscored in various fields where Gaussian Processes play a central role.

The initial consideration in the context of the Breuer-Major theorem is a stationary unidimensional $\mathcal{GP}$ , denoted as $x = \{x_k\}_{k \in \mathbb{Z}}$. In this scenario,the stationarity of the covariance matrix, which is not a mandatory requirement but rather a starting point for the theorem, implies that the covariance of the process, denoted as $C_{ij} = \mathbb{E}(x_i x_j)$, depends solely on the difference between the indices $i$ and $j$, i.e. $C_{ij} = C(i-j)$. This stationarity condition, however, will later be  replaced by a weaker condition to accommodate a broader range of Gaussian Processes.

The other technical requirement in the Breuer-Major theorem concerns the non-linear function $F$, which must have a well-defined \emph{Hermite rank} $R$. The Hermite rank is defined as the smallest positive integer that appears in the expansion of $F$ in terms of Hermite polynomials. This can be mathematically expressed as:
\begin{equation}
F(x) = \sum_{k=R}^{\infty} f_k \He_k(x),
\end{equation}
where $\He_k(x)$ represents the $k$-th Hermite polynomial, and $f_k$ denotes the coefficient of the expansion in this polynomial series. It is notable that for many practical and commonly used activation functions $F$, the Hermite rank $R$ is equal to 1.

\begin{theorem}[Breuer and Major, 1983]
    Let  $x = (x_k)_{k \in \mathbb Z}$ be a stationary unidimensional GP with covariance $C(i-j)$. Let $\mathbb E\left[ F(x_1)\right] = 0$ and $\mathbb E \left[F^2(x_1)\right] < \infty$ and assume that the function $F$ has Hermite rank $R \geq 1$. Suppose that:
    \begin{equation}
    \sum_{j\in \mathbb Z} \lvert C_{1j} \rvert^R  < \infty\,.
    \end{equation}
    Then $\sigma^2 := \mathbb E\left[ F(x_1)^2\right] + 2 \sum_{j =1}^\infty  \mathbb E\left[ F(x_1) F(x_j)\right]$ is finite. Moreover, one has that the sequence of random variables
    \begin{equation}
    S_N = \frac{1}{\sqrt N} \sum_{i=1}^N F (x_i)\, \quad N \geq 1
    \label{vanillaBM}
    \end{equation}
    converges in distribution to $\mathcal N(0,\sigma^2)$, i.e. to a Gaussian distribution with zero mean and variance $\sigma^2$.
\end{theorem}

For the purposes of our analysis, a slightly more robust version of the Breuer-Major theorem is required, which involves two key modifications:
\begin{itemize}
    \item[(i)] The covariance in our calculations will not adhere to the stationarity condition. This means that the covariance $C_{ij} = \mathbb{E}(x_i x_j)$ will not necessarily be a function of the difference $i-j$, allowing for a more general and non-stationary covariance structure.
    \item[(ii)] We will consider a more general sequence of nonlinear functions $c_i F(x_i)$ in our analysis. Unlike the standard formulation, where each term in the sum is identical, here each term of the sum is uniquely weighted by a factor $c_i \neq 1$. This introduces an additional layer of complexity and generalization to the theorem, as represented in the modified sum:
    \begin{equation}
    S_N = \frac{1}{\sqrt{N}} \sum_{i=1}^{N} c_i F(x_i),
    \end{equation}
    where the $c_i$'s are not necessarily equal to 1, diverging from the simpler case described in the vanilla Breuer-Major theorem.
\end{itemize}
These modifications, while adding complexity, allow for a broader applicability of the theorem,  including the random variable $h$ defined by a DNN.

Already in the original work by Breuer and Major \cite{BM}, it was demonstrated that the hypothesis of stationarity in the theorem could be relaxed. Instead of strict stationarity, a condition of uniform convergence for the elements of the covariance matrix is introduced. This is expressed as follows:
\begin{equation}
\sum_{j \in \mathbb{Z}} |C_{ij}|^R < B_0 \quad \forall i \in \mathbb{Z},
\label{eq:BMnonstationary}
\end{equation}
where $B_0$ is a d positive finite constant. 

Extensions (i) and (ii) to the Breuer-Major theorem have been further developed by Bardet and Surgailis in \cite{bardet2013}. The advancements are outlined as follows. Consider $\mathbf{x}^N$ to be an $N$-dimensional Gaussian vector with the properties $\expect[x^N_i] = 0$ and $\expect[(x^N_i)^2] = 1$. The covariance for this vector is defined as $C^N_{ij} = \expect[x^N_i x^N_j]$. For a given integer $m \geq 1$, the following conditions are assumed:
\begin{align}
   & \sup_{N \geq 1} \max_{1 \leq j \leq N} \sum_{i=1}^N |C^N_{ij}|^m < \infty, \label{eq:bardetMax}\\
   & \sup_{N \geq 1} \frac{1}{N} \sum_{\substack{1 \leq i, j \leq N\\ |i - j| > K}} |C^N_{ij}|^m \underset{K \to \infty}{\longrightarrow} 0. \label{eq:bardetSum}
\end{align}
Furthermore, define $\mathbb{L}^2_0(x)$ as the space of functions satisfying $\expect f(x) = 0$ and $\expect f^2(x) < \infty$, where $x$ is a standard normal variable. Then, 

\begin{theorem}[Bardet and Surgailis~\cite{bardet2013}, 1.ii]
    Assume~\eqref{eq:bardetMax},~\eqref{eq:bardetSum}. Let $f^N_i \in \mathbb{L}^2_0 (x)$ ($N\ge 1$, $1\le i \le N$) be a sequence of functions all having Hermite rank $m$ at least one. Assume that there exist a $\mathbb{L}^2_0(x)$-valued continuous function $\phi_\tau$, $\tau \in [0,1]$, such that
    \begin{equation}
        \sup_{\tau \in (0,1]} \expect [f^N_{[\tau N]}(x) - \phi_\tau(x)]^2 \underset{N\to\infty}{\longrightarrow} 0\,. \label{eq:bardetPhi}
    \end{equation}
    Moreover, let
    \begin{equation}
        (\sigma^N)^2 = \expect \left[\frac{1}{\sqrt{N}} \sum_{i = 1}^N f^N_i(x^N_i) \right]^2  \underset{N\to\infty}{\longrightarrow} \sigma^2 \,,
    \end{equation}
    where $\sigma^2 >0$. Then
    \begin{equation}
        \frac{1}{\sqrt{N}} \sum_{i = 1}^N f^N_{i}(x^N_i) \underset{N\to\infty}{\overset{d}{\longrightarrow}} \mathcal{N}(0,\sigma^2)\,,
    \end{equation}
    where $\underset{N\to\infty}{\overset{d}{\longrightarrow}}$ denotes convergence in distribution.
    \end{theorem}

    The hypotheses of this theorem serve as crucial conditions for several key components: the activation function $\sigma$, the rescaled input covariance matrix $C_{\mu\nu}$, and the dominant configurations $\bar{s}$ in the Fourier integral, as indicated in equation~\eqref{eq:K_a_BM}. These conditions are fundamental to the justification of our Gaussian ansatz presented in equation~\eqref{eq:q_is_gaussian}.

}

\details{
    \begin{center}
        \textbf{Details: Input dimension}
    \end{center}
    In this section, we explore the additional constraints on thermodynamic scaling (with $P,N_1 \rightarrow \infty$ while keeping $\alpha_1 = P/N_1$ finite) arising from the hypotheses of the Breuer-Major theorem related to the covariance matrix $C$. The primary condition to consider is equation~\eqref{eq:BMnonstationary}, expressed as:
    \begin{equation}
    \sum_{\mu=1}^P |C_{\mu \nu}|^R < B_0 \qquad \forall \,\nu=1,\dots,P,
    \label{eq:BM_Cmunu}
        \end{equation}
where $B_0$ is a specified finite constant, and $R$ is the Hermite rank of the activation function $\sigma$. In scenarios where inputs $\mathbf{x}$ have i.i.d. standard Gaussian coordinates, $C_{\mu\nu}$ manifests as a Wishart random matrix. Its off-diagonal entries are on the order of $1/\sqrt{N_0}$ with random signs. After accounting for the absolute values, the sum in Eq~\eqref{eq:BM_Cmunu} becomes of order $P(N_0)^{-R/2}$. This analysis reveals an infinite class of activation functions (those with Hermite rank $R \geq 2$) that are viable within the constraints of finite $\alpha_0 = P/N_0$. However, for activation functions with Hermite rank $R = 1$ (like Erf or ReLU), such assurance cannot be derived solely from the hypotheses of the Breuer-Major theorem.

It is also important to note that for any given odd (or non-odd) activation function $\sigma(x)$ with Hermite rank $R=1$, one can construct a new, reasonable activation function with Hermite rank $R=3$ (or $R=2$) by substituting the original function with $\sigma_1(x) = \sigma(x) - g_1 x$. Here, $g_1 = \langle \sigma(x) \He_1(x) \rangle$ represents the coefficient, with the average taken over a normal distribution of zero mean and unit variance. This approach provides a method to engineer activation functions that align with the requirements of the Breuer-Major theorem, particularly in terms of the Hermite rank.

We note that there exists at least one instance of an activation function with Hermite rank $R=1$ where the derivation remains valid at finite $\alpha_0$: specifically, the linear function $\sigma(x) = x$. In this case, the results can be obtained with finite $P, N_1, N_0$, as also demonstrated in Ref. \cite{hanin2023bayesian}. 
In a following detail section, we delve into the case of a quadratic activation function $\sigma(x) = x + x^2$, which also possesses a Hermite rank of $R=1$. Here, we derive the final effective action without relying on the Breuer-Major (BM) theorem. Similar to the linear case, this derivation remains robust at finite $\alpha_0$.
This observation leads us to hypothesize that the scaling $P = O(\sqrt{N_0})$, which is often suggested for activation functions with Hermite rank $R=1$, might be an overly pessimistic estimate.
    }

    The elements of the kernel matrix $K_{\mu\nu}(C)$ in \ref{eq:K_a_BM} can be effectively simplified from a complex $P$-dimensional integral to a more manageable two-dimensional form. This reduction is represented as:
    \begin{align}
    K_{\mu\nu}(C) = \int \frac{dt_1 dt_2}{\sqrt{(2\pi)^2 \det \tilde C}} e^{-\frac{1}{2} \mathbf{t}^T \tilde C^{-1} \mathbf{t}}  \sigma(t_1) \sigma (t_2), 
    \label{methods:eq:kernel}
    \end{align}
    where $\mathbf{t} = (t_1, t_2)^T$ and
    \begin{equation}
    \tilde C = \begin{pmatrix} C_{\mu\mu} & C_{\mu\nu} \\ C_{\mu\nu} & C_{\nu\nu} \end{pmatrix}
    \end{equation}
    is the reduced $2 \times 2$ input covariance matrix.

    Now, by integrating over the variable $q$, we obtain:
\begin{equation}
    \left[\int \frac{dq \,e^{ -\frac{q^2}{2}-\frac{q^2}{2 Q(\bar{s},C)} }}{\sqrt{2\pi Q(\bar{s},C)}} \right]^{\frac{N_1}{2}} = \left[Q(\bar{s},C)+1\right]^{-\frac{N_1}{2}}.
    \label{eq:P(Q)}
\end{equation}

In cases where $\alpha_1 = P/N_1$ remains finite, our analysis is reduced to the integrals over $s^\mu$ and $\bar{s}^\mu$.
\begin{equation}
    Z= \int \prod_{\mu}^P \frac{ds^\mu d \bar{s}^\mu}{2\pi} e^{-\frac{\beta}{2}\sum_{\mu}^P\left( y^\mu - s^\mu \right)^2 + i\sum_{\mu}^P s^\mu \bar{s}^\mu} \left[Q(\bar{s},C)+1\right]^{-\frac{N_1}{2}}
\end{equation}

To facilitate solving these integrals, it proves useful to incorporate an additional Dirac delta identity:
\begin{equation}
  1 = \int dQ \,\delta\left[Q - \frac{1}{\lambda_1 N_1} \sum_{\mu,\nu} \bar{s}^\mu K(C)_{\mu\nu} \bar{s}^\nu\right],
\label{EQ:Q}
\end{equation}
where $Q \geq -1$ is now treated as an integration variable independent of $\bar{s}$, thus removing the explicit dependence on $\sqrt{Q(\bar{s},C)+1}$ in the partition function. After that we can render the integrals over $s^\mu$ and $\bar{s}^\mu$ Gaussian, which is achieved by inserting another integral representation of the delta function with a conjugate variable $\bar Q$.
\begin{equation}
    Z= \int dQ d\bar{Q} e^{iQ\bar{Q}}\int \prod_{\mu}^P \frac{ds^\mu d \bar{s}^\mu}{2\pi} e^{-\frac{\beta}{2}\sum_{\mu}^P\left( y^\mu - s^\mu \right)^2 + i\sum_{\mu}^P s^\mu \bar{s}^\mu -i \frac{1}{\lambda_1 N_1} \sum_{\mu,\nu} \bar{s}^\mu K(C)_{\mu\nu} \bar{s}^\nu}
\end{equation}

Upon executing the last two Gaussian integrals, we arrive at a novel representation of the partition function, which is now dependent on just an integral over two order parameters, rather than an extensive number of weights:
\begin{equation}
    \mathcal{Z} = \int dQ \, d\bar{Q} e^{-\frac{N_1}{2} S(Q,\bar{Q})},
    \label{3BIW:finalZ}
\end{equation}
Here, the effective action $S$ is defined as:
\begin{equation}
    S = -Q\bar{Q} + \log(1+Q) + \frac{\alpha}{P}\Tr\log \left[ \beta\left( \frac{\mathbf{1}}{\beta} +\frac{\bar{Q}K}{\lambda_1}\right)\right] + \frac{\alpha_1}{P} \mathbf{y}^\top \left[ \frac{\mathbf{1}}{\beta} +\frac{\bar{Q}K}{\lambda_1}\right]^{-1} \mathbf{y},
    \label{effS}
\end{equation}
where $\alpha = P/N_1$.

\details{
    \begin{center}
        \textbf{Details: Multiple outputs}
    \end{center}
    The calculations performed in this section can be readily generalized to the case of a NN with multidimensional output. The starting point remains the same \textit{Loss functions}, but we now need to consider the fact that the DNN has $\kappa$ different outputs:
\begin{align}
    \mathcal{L} &= \frac{1}{2 \kappa} \sum_{\mu=1}^P \sum_{a=1}^{\kappa} \left[ y^\mu_{a} - (f_{\textrm{DNN}}(\mathbf{x}^\mu))_{a} \right]^2 + \mathcal{L}_{\textrm{reg}},\\
    \mathcal{L}_{\textrm{reg}} &= \frac{\lambda_1}{2\beta} \lVert v \rVert^2 + \frac{\lambda_{0}}{2\beta} \lVert W\rVert^2.
\label{lossA1}
\end{align}
The associated partition function is given by:
\begin{equation}
    \begin{split}
        Z = \int &\prod_{a,i_1}^{\kappa,N_1} dv_{a,i_1} \prod_{i_1,i_0}^{N_1,N_0} dw_{i_1 i_0} \exp\Biggl\{-\frac{\lambda_1}{2}  \lVert v \rVert^2  -\frac{\lambda_0}{2} \lVert W \rVert^2 \Biggr\}\\
        &\times \exp\Biggl\{-\frac{\beta}{2 \kappa} \sum_{\mu=1}^P\sum_{a=1}^{\kappa} \left[ y^\mu_{a} -\frac{1}{\sqrt{N_1}} \sum_{i_1}^{N_1}  v_{a,i_1} \sigma\left( \sum_{i_0}^{N_0}\frac{w_{i_1,i_0}x_{i_0}^\mu}{\sqrt{N_0}} \right) \right]^2\Biggr\}.
    \end{split}
\end{equation}

We decouple the layers in the loss through the addition of Dirac deltas, noting that there will now be an additional index $a$ for the outputs $s^\mu_a$. After integrating over the weights $w_{i_1,i_0}$ and $v_{a,i_1}$, which are Gaussian, the integrals in $h^\mu_i$ and $\bar{h}^\mu_i$ can be factorized over the index $i$. Performing the integrals over the variables $\bar{h}^{\mu}$, we find that the $h^{\mu}$ are Gaussian-distributed with zero mean and covariance matrix $C$, similar to the single-output case.

The crucial step is now to consider the joint probability distribution of the random variables:
\begin{equation}
    q_{a} = \frac{1}{\sqrt{\lambda_1 N_1}}\sum_\mu^P \bar{s}_{a}^\mu \sigma(h^\mu).
\end{equation}
In the asymptotic proportional limit $P/N_1 \sim O(1)$, we can hypothesize a Gaussian equivalence, assuming that the BM theorem can be extended to the multivariate case. Therefore, $P(\{q_a\}) \to \mathcal{N}(0,Q)$, where now $Q$ is the covariance matrix given by:
\begin{equation}
    Q(\bar{s},C)_{a,b} = \frac{1}{\lambda_1 N_1}\sum_{\mu,\nu}^P \Bar{s}^\mu_{a} K_{\mu \nu}(C)\Bar{s}^\nu_{b},
\end{equation}
with $K/\lambda_1$ being the NNGP kernel, as in the single-output case.

Following the same scheme as before, we obtain the multiple output effective action $S$: 
\begin{equation}
\begin{aligned}
S(\mathbf{Q},\Bar{\mathbf{Q}})=& -\Tr[\mathbf{Q}\Bar{\mathbf{Q}}^{\top}]+\Tr\log(\mathbf{1}_{\kappa}+\mathbf{Q})+\frac{\alpha_1}{P}\Tr\log \frac{\beta}{\kappa}\left[ \frac{\kappa}{\beta} \mathbf{1}_{\kappa}\otimes\mathbf{1}_P +\frac{\bar{\mathbf{Q}}\otimes K}{\lambda_1}\right]\\
&+\frac{\alpha_1}{P} \mathbf{y}^\top \left[ \frac{\kappa}{\beta} \mathbf{1}_{\kappa}\otimes\mathbf{1}_P +\frac{\bar{\mathbf{Q}}\otimes K}{\lambda_1}\right]^{-1} \mathbf{y}.
\end{aligned}
\end{equation}

}

\details{
    \begin{center}
        \textbf{Details: Linear activation function}
    \end{center}

    The case where the activation function \( \sigma \) is the identity  \( \sigma = \id \) has been thoroughly explored in existing literature, such as in~\cite{PhysRevX.11.031059, hanin2023bayesian}. We reference it here both for comparison and to demonstrate that our theory aligns with established results. Specifically, when the activation function is linear, the average over \( h \) in Eq.~\eqref{3BIW:PQQQ} after the Fourier transformation can be accurately computed at finite values of \( P \), \( N_1 \), yielding
    \begin{equation}
        \psi_{\text{lin}}(t) = \expect_h\left\{ e^{\frac{it}{\sqrt{\lambda N_1}} \sum_\mu \bar{s}^\mu \sigma(h^\mu)} \right\} =  e^{ -\frac{t^2}{2 \lambda N_1} \sum_{\mu,\nu} \bar{s}^\mu C_{\mu\nu} \bar{s}^\nu },
    \end{equation}
    where \( C \) is the kernel's value for \( \sigma = \id \). This holds true as long as \( C \) does not have zero eigenvalues, which is generally the case for \( N_0 > P \). To circumvent any potential issues with zero eigenvalues, a small regularization proportional to the identity matrix can be added to \( C \).
    
    This result is a direct consequence of the fact that the sum of jointly Gaussian variables remains Gaussian. This property holds for general Gram matrices \( C \) and any values of \( P \), \( N_1 \), even outside the asymptotic limit where \( P \sim N_1 \) are large. However, to evaluate the remaining integrals over the order parameters at the saddle point of the effective action, this limit is still necessary, as indeed performed in~\cite{PhysRevX.11.031059}. For finite \( P \), \( N_1 \), the partition function can be exactly expressed in terms of Meijer G-functions, as detailed in~\cite{hanin2023bayesian}.
    
}

\details{
    \begin{center}
        \textbf{Details: Quadratic activation function}
    \end{center}
    As we have suggested in the main text, there is a special case where the partition function can be computed without using the BM theorem. 
    Let us take now $C_{\mu\mu} = 1$ (normalized data) and quadratic (zero-mean) activation function:
\begin{equation}
    \sigma(x) = x + a(x^2-1)\,.
\end{equation}
The kernel is given by
\begin{equation}
    K_{\mu\nu} = \expect_h[\sigma(h^\mu)\sigma(h^\nu)] =  C_{\mu\nu}  +2 a^2  (C_{\mu\nu})^2\,.
\end{equation}
Also in this case the characteristic function in~\eqref{3BIW:PQQQ} (after the Fourier transformation) can be evaluated exactly:
\begin{equation}
\begin{aligned}
    \psi_{\text{quad}}(t) = \frac{\exp\left\{-\frac{t^2}{2\lambda N_1}  \bar{s}^\top C \left[\mathbf{1}_P - \frac{2i a t }{\sqrt{\lambda N_1}}\diag(\bar{s}) C \right]^{-1} \bar{s}\right\}}{\det[(\mathbf{1}_P - \frac{2 i a t }{\sqrt{\lambda N_1}}\diag(\bar{s}) C)]^{1/2}} \exp\left(- \frac{i a t}{\sqrt{\lambda N_1}} \sum_\mu 
    \bar{s}^\mu \right)\,.
    \label{supp:eq:quad_psi_exact}
\end{aligned}
\end{equation}
We can express the non-trivial matrices appearing in this expression as Neumann series:
\begin{align}
     \left[\mathbf{1}_P - \frac{2i a t }{\sqrt{\lambda N_1}}\diag(\bar{s}) C \right]^{-1} &=  \sum_{n=0}^{+\infty} \left(\frac{2i a t }{\sqrt{ \lambda N_1}}\right)^n [\diag(\bar{s}) C]^n \,,\label{supp:eq:quad_firstSeries}\\
     -\frac{1}{2} \Tr \log\left[\mathbf{1}_P - \frac{2i a t }{\sqrt{\lambda N_1}}\diag(\bar{s}) C \right] &= -\sum_{n=1}^{+\infty} \frac{1}{n} \left(\frac{2i a  }{\sqrt{\lambda N_1}}\right)^n \Tr\{[\diag(\bar{s}) C]^n\}\,.
     \label{supp:eq:quad_secondSeries}
\end{align}
To have Gaussianity, we must require the following asymptotic behaviors:
\begin{align}
    \frac{1}{N_1^{1+n/2}} \bar{s}^\top C  [\diag(\bar{s}) C]^n \bar{s} &= O(P/N_1^{1+n/2})\,, \label{supp:eq:quad_firstCondition} \\
    \frac{1}{N_1^{n/2}} \Tr\{[\diag(\bar{s}) C]^n\}  & = O(P/N_1^{n/2})\,, \label{supp:eq:quad_secondCondition}
\end{align}
so that in the regime where $\alpha_1 = P/N_1$ is finite only the $n=0$ term counts for~\eqref{supp:eq:quad_firstSeries} and the $n=1,2$ terms for~\eqref{supp:eq:quad_secondSeries}. 
In this case we can use the following approximation: 
\begin{equation}
    -\frac{1}{2} \Tr \log \left[\mathbf{1}_P - \frac{2i a t }{\sqrt{\lambda N_1}}\diag(\bar{s}) C \right] \approx \frac{i a t }{\sqrt{\lambda N_1}}\sum_\mu \bar{s}^\mu - \frac{a^2 t^2 }{\lambda N_1} \sum_{\mu,\nu} \bar{s}_\mu  (C_{\mu\nu})^2 \bar{s}_\nu\,,
 \end{equation}
to obtain our Gaussian behaviour:
 \begin{equation}
      \psi_{\text{quad}}(t) \sim  \exp\left[-\frac{t^2}{2 \lambda N_1} \sum_{\mu,\nu} \bar{s}^\mu K_{\mu\nu} \bar{s}^\nu \right]\,.
 \end{equation}
 The conditions in Equations \eqref{supp:eq:quad_firstCondition} and \eqref{supp:eq:quad_secondCondition} can be seen as prerequisites regarding the matrix \( C \) and the realization of the vector \( \bar{s} \) to ensure the property of Gaussianity. Let's consider the simplest case where the input data are independent and identically distributed (i.i.d.) standard normal variables, and \( \bar{s}^\top = (1, \ldots, 1) \).  Then, $C$ is a Wishart matrix with a finite spectrum in the regime $P\sim N_0$~\cite{marchenko1967}, and
\begin{equation}
   \frac{1}{P}\Tr(C^n) = O(1)\,,\qquad\frac{1}{P}\sum_{\mu,\nu} (C^{n})_{\mu\nu} = O(1) \,.
   \label{supp:eq:quad_hypothesis}
\end{equation}
The first behaviour follows from the fact that the eigenvalues are $O(1)$.
To prove the the second we can use the fact that the eigenvectors of a Wishart matrix (as $C$) are random and uniformly distributed on the sphere~\cite{forresterBook}, so that
\begin{equation}
    \frac{1}{P}\sum_{\mu,\nu} (C^{n})_{\mu\nu} = \frac{1}{P}\sum_{\rho} \lambda^{n}_\rho \sum_\mu U_{\mu\rho} \sum_\nu U^{-1}_{\rho \nu} = O(1)\,,
\end{equation}
where $\lambda_\rho$ is the $\rho$-th eigenvalue of $C$ and $U$ the matrix whose $\rho$-th column is the corresponding eigenvector.
Given this, properties \eqref{supp:eq:quad_firstCondition} and \eqref{supp:eq:quad_secondCondition} are satisfied, and \( q \) is Gaussian. It is important to note that there exist specific configurations of \( \bar{s} \), such as \( \bar{s}^\top = (1,0,\ldots,0) \), where this reasoning does not apply. Therefore, we assume that the contribution of these special configurations to the effective action is negligible in the thermodynamic limit.

In principle, Gaussianity can be also proven via diagrammatic techniques, as shown in \cite{ariosto2023}.

}

\details{
    \begin{center}
        \hypertarget{finitemean}{\textbf{Details: Finite mean activation function}}
    \end{center}

    In this detailed section, we demonstrate how the theory can be generalized in the case of finite-mean activation functions. Previously, our derivation assumed that the integral of the activation function over a centered Gaussian is zero, implying that the activation function is zero-mean. This section aims to show that removing this hypothesis modifies the effective action in the asymptotic limit. The key difference, compared to the case studied in the main text, is that if the activation function is not zero-mean, the random variable
\begin{equation}
q = \frac{1}{\sqrt {N_1 \lambda_1}} \sum_{\mu=1}^P \bar s^\mu \sigma(h^\mu) 
\end{equation}
now has a finite mean. Specifically:
\begin{align}
\braket{q}_{P(q)} = \frac{1}{\sqrt {N_1 \lambda_1}} \sum_{\mu=1}^P \bar s^\mu m^{\mu}, \qquad
m^{\mu} = \int \frac{dt}{\sqrt{2\pi C_{\mu\mu}}} e^{-\frac{t^2}{2C_{\mu\mu}}} \sigma(t) \,.
\end{align}
A straightforward calculation shows that the result for finite-mean activation is obtained by performing the replacement:
\begin{align}
\frac{\bar Q}{\lambda_1} K \to  K^{(R)}(Q,\bar Q) =  \frac{\bar Q}{\lambda_1} K - \frac{\left(\bar Q - \frac{1}{1+Q} \right)}{\lambda_1} K^{(1)}\,,
\end{align}
in the effective action in Eq. \eqref{effS} of the main text, with $K^{(1)}_{\mu\nu} = m^{\mu} m^{\nu}$. It is noteworthy that while in the zero-mean case there was a simple relationship between \( Q \) and \( \bar Q \) at any temperature, this property is lost when the mean is not zero, and the saddle-point equations become unsolvable exactly, even in the zero-temperature limit. However, in the infinite-width limit, we recover the previous result \( \bar Q = 1 \), \( Q = 0 \), and the rank-one matrix \( K^{(1)} \) does not contribute to the generalization error, as it always appears in combination with the scalar \( \bar Q - \frac{1}{1+Q} \) that vanishes in the infinite-width limit.
}

\subsubsection{Saddle point methods}

Using the saddle-point method, appropriate as $N_1 \to \infty$, we find the solutions $Q^*$, $\bar Q^*$ by solving the system of equations $\partial_Q S = 0$, $\partial_{\bar Q} S = 0$. In the zero-temperature limit, the analytical solution of the saddle-point equations, given the positive eigenvalues of the kernel $K$ in the asymptotic regime with $\alpha_1$, $\alpha_0$ finite, is:

\begin{align}
\bar Q = \frac{1}{1+Q}, \qquad Q = \frac{\alpha_1}{\bar Q} -\frac{\alpha_1}{\bar Q^2 }\frac{1}{P} \mathbf{y}^\top\left(\frac{K}{\lambda_1}\right)^{-1}\mathbf{y}.
\end{align} 
Considering the condition $Q \geq -1$, the unique exact positive solution for $\bar Q$ is:
\begin{equation}
\bar Q^* = \frac{\sqrt{(\alpha_1-1)^2 + 4 \alpha_1 \frac{1}{P} \mathbf{y}^\top\left(\frac{K}{\lambda_1}\right)^{-1}\mathbf{y} } - (\alpha_1 -1)}{2}.
\end{equation}
The infinite-width limit, re-obtained as $\alpha \to 0$, corresponds to the particular solution $Q^* = 0$, $\bar Q^* = 1$. 

\details{
\begin{center}
    \textbf{Details: Kernel Associated with the Most Commonly Used Non-linearities}
\end{center}
The NNGP is a well-known object, and for most of the common linear functions, its associated kernel has been analytically computed. For instance, for the \(\text{Erf}\) function, the kernel is known \cite{PANG2019270}:
\begin{equation}
K^{\textrm{Erf}}_{\mu\nu} (C) = \frac{2}{\pi} \arcsin\left(\frac{2 C_{\mu\nu}}{\sqrt{\left(1+ 2 C_{\mu\mu}\right)\left(1+ 2 C_{\nu\nu}\right)}}\right)\,.
\end{equation}
In the case of the \(\sigma = \text{ReLU}\) activation function, the kernel is given by \cite{cho2009kernel}:
\begin{align}
K^{\textrm{ReLU}}_{\mu\nu} (C) &= \sqrt{C_{\mu \mu} C_{\nu \nu}} \, \kappa \left( \frac{C_{\mu\nu}}{\sqrt{C_{\mu \mu} C_{\nu \nu}}} \right)\,, \\
\kappa(x) &= \frac{1}{2\pi} \left[x(\pi - \arccos(x))+ \sqrt{1-x^2}\right]\,.
\end{align}
}

\subsubsection{Generalization Error}
Utilizing the optimal values of the order parameters $Q^*$ and $\bar{Q}^*$, we can determine the average value of the generalization error. The calculation follows a similar approach to the previous one, with the addition of a new pair of variables $s^0$ and $h_i^0$, representing the output and the preactivation of a new test example, respectively. The average generalization error is given by:
\begin{equation}
\begin{split}
\braket{\epsilon_\text{g}(\mathbf{x}^0,y^0)} &= \frac{1}{Z} \int \frac{ds^0 d\bar{s}^0}{2\pi} \int \prod_{\mu=1}^P \frac{ds^\mu d\bar{s}^\mu}{2\pi} \, (y^0-s^0)^2 \\
&\quad \times e^{-\frac{\beta}{2}\sum_{\mu=1}^P (y^\mu-s^\mu)^2 + i\sum_{\mu=1}^P s^\mu \bar{s}^\mu +is^0\bar{s}^0} \\
&\quad \times \left[1+\frac{1}{\lambda_1 N_1}\left(\sum_{\mu,\nu=1}^P \bar{s}^\mu K_{\mu\nu} \bar{s}^\nu + 2 \bar{s}^0\sum_{\mu=1}^P \bar{s}^\mu \kappa_{\mu}(\mathbf x^0) + (\bar{s}^0)^2 \kappa_{0}(\mathbf x^0) \right)\right]^{-\frac{N_1}{2}},
\end{split}
\label{eq:gen_err0}
\end{equation}
where $\kappa_\mu$ and $\kappa_0$ are the train-test and test-test kernel integrals, respectively, defined by:
\begin{align}
&\kappa_\mu(\mathbf x^0) = \int \frac{dt_1 dt_2}{\sqrt{(2\pi)^2 \det \tilde C_\mu}} e^{-\frac{1}{2} \mathbf t^T \tilde C_\mu^{-1} \mathbf t}  \sigma(t_1) \sigma (t_2), \label{kmu} \\
&\kappa_0(\mathbf x^0) = \int \frac{dt}{\sqrt{2\pi C_{00}}} e^{-\frac{t^2}{2 C_{00}}}  \sigma(t)^2, \label{k0}
\end{align}
with
\begin{align}
&\begin{gathered}
\tilde C_\mu = \begin{pmatrix} C_{\mu\mu} & C_{\mu0}\\ C_{\mu0} & C_{00}\end{pmatrix}, \quad C_{\mu 0} =  \frac{1}{\lambda_0 N_0} \sum_{i_0}^{N_0} x^\mu_{i_0} x^0_{i_0},\\
C_{00} =  \frac{1}{\lambda_0 N_0} \sum_{i_0}^{N_0} (x^0_{i_0})^2.
\end{gathered}
\end{align}

Repeating the same procedure used in the calculation of the partition function, we arrive at an expression for the average generalization error:
\begin{align}
    &\braket{\epsilon_\text{g}(\mathbf{x}^0,y^0)} = \frac{1}{Z} \int \frac{dQ d\bar{Q}}{2\pi} e^{-\frac{N_1}{2}S(Q,\bar{Q})} \int \frac{d s^0 (y^0-s^0)^2}{\sqrt{2\pi\sigma^2}} e^{-\frac{(s^0+\Gamma_1)^2}{2\sigma_1^2}},
\end{align}
where
\begin{align}
&\Gamma_1 = \frac{\bar Q}{\lambda_1}\sum_{\mu\nu} \kappa_\mu(\mathbf x^0) \left(\frac{\mathbf{1}}{\beta}+\frac{\bar{Q}}{\lambda_1}K\right)^{-1}_{\mu\nu} y_\nu,\\
&\sigma_1^2 = \frac{\bar{Q}}{\lambda_1}\left[\kappa_0(\mathbf x^0) - \frac{\bar{Q}}{\lambda_1} \sum_{\mu\nu} \kappa_\mu(\mathbf x^0) \left(\frac{\mathbf{1}}{\beta}+\frac{\bar{Q}}{\lambda_1}K\right)^{-1}_{\mu\nu} \kappa_\nu (\mathbf x^0)\right].
\label{eq:generc_bias_var}
\end{align}

Completing the final integral and applying the saddle point method yields an analytical formula for the generalization error, which takes the familiar form of the bias-variance decomposition:
\begin{equation}
    \braket{\epsilon_\text{g}(\mathbf{x}^0,y^0)} = \left(y^0 - \Gamma_1\right)^2 + \sigma_1^2.
    \label{Eps_g}
\end{equation}

\subsubsection{Empirical test}

We can directly utilize Equation~\eqref{Eps_g} to derive testable predictions for the generalization error of finite-width single-hidden-layer (1HL) architectures within a Bayesian learning framework. The results are showcased in Fig.~\ref{fig3:3biw-1}, where we analyze two specific regression tasks based on the CIFAR10 and MNIST datasets. Detailed descriptions of these numerical experiments will be presented in the subsequent \textit{details} sections.

Our theoretical framework, initially developed for DNNs in the thermodynamic limit, exhibits a strong capability to predict the behavior of real-world DNNs when they are trained using a Bayesian procedure (i.e they are trained using Langevin dynamics \cite{welling2011bayesian}).
This observation suggests that our theory, despite its foundational assumptions of an infinite number of examples and weights, is adept at providing a more accurate description of finite DNNs than those models based on the Infinite Width limit. By accurately reflecting the traits of finite-sized networks, our model effectively bridges a critical gap in understanding how theoretical principles are manifested in practical deep learning applications. This aspect notably extends the relevance of our findings beyond the realm of idealized models to more tangible, real-world scenarios.

Moreover, it is important to notice that the generalisation curves for the two regression tasks are monotonically increasing (decreasing) as a function of $N_1$ depending on the fact that the observable $y^\top \left(K/\lambda_1\right)^{-1} y/P$ is smaller (larger) than one. The importance of this quantity in controlling the generalisation performance has been already noted in linear networks \cite{PhysRevX.11.031059, zavatone2022contrasting} as well as in direct perturbation theory at finite $P$ for non-linear networks \cite{zavatone2021asymptotics,zavatone2022asymptoticsJStatMech} and will be discussed in a following section of this Chapter.

We also highlight two semi-quantitative predictions regarding the general behavior of the generalization error, derived from examining how Equation \eqref{Eps_g} depends on the size of the hidden layer \(N_1\) and the Gaussian prior of the last layer \(\lambda_1\). At \(T=0\), the bias remains constant as a function of both \(N_1\) and \(\lambda_1\), a phenomenon also explicitly observed in the linear case in Ref.~\cite{PhysRevX.11.031059}. In contrast, the variance \ref{eq:generc_bias_var} is influenced by \(N_1\) and diminishes as \(1/\sqrt{\lambda_1}\) in the large-\(\lambda_1\) limit. These insights lead to two testable predictions: (i) increasing the magnitude of the Gaussian prior \(\lambda_1\) should systematically enhance the generalization performance for any \(N_1\); (ii) in the regime of large \(\lambda_1\), the dependence of the generalization error on \(N_1\) should diminish, as also evidenced by the numerical experiments presented in Fig.~\ref{fig3:3biw-2}.

\afterpage{
\begin{figure}
        \centering
        \includegraphics[width=0.90\textwidth]{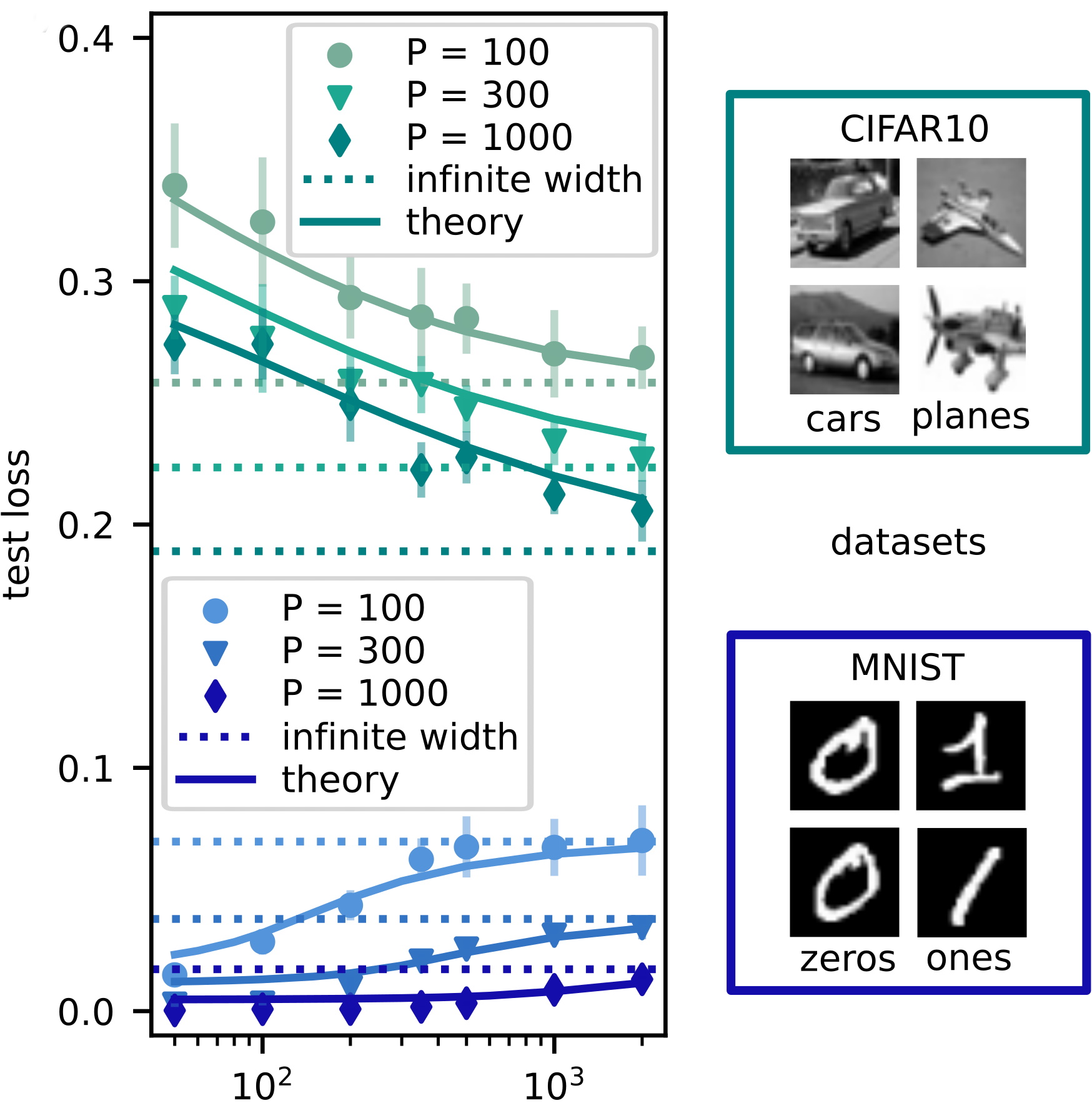}
        \caption{\textbf{Infinity Width vs Proportional regime (1).} Learning curves of 1HL architectures with Erf activation (trained with a discretised Langevin dynamics) as a function of the hidden layer size $N_1$ for two regression tasks on the CIFAR10 (above) and MNIST (below) datasets. Zero/one labels have been chosen in both cases and the images of the CIFAR10 dataset have been gray-scaled and down-scaled to $N_0 = 28 \times 28$. The experimental test loss at different values of the trainset size $P$ (points with error bars indicating one standard deviation) are compared with the theory computed from equation \eqref{Eps_g} (solid lines). The bar centres are computed as the average over an ensemble of $S = 450$ equilibrium configurations. Samples are taken every $10^4$ Langevin steps (after thermalisation of the dynamics). The error bar represents one standard deviations from the average.}
        \label{fig3:3biw-1}
    \end{figure}
\clearpage

\vspace*{5cm}
\begin{figure}[H]
        \centering
        \includegraphics[width=0.95\textwidth]{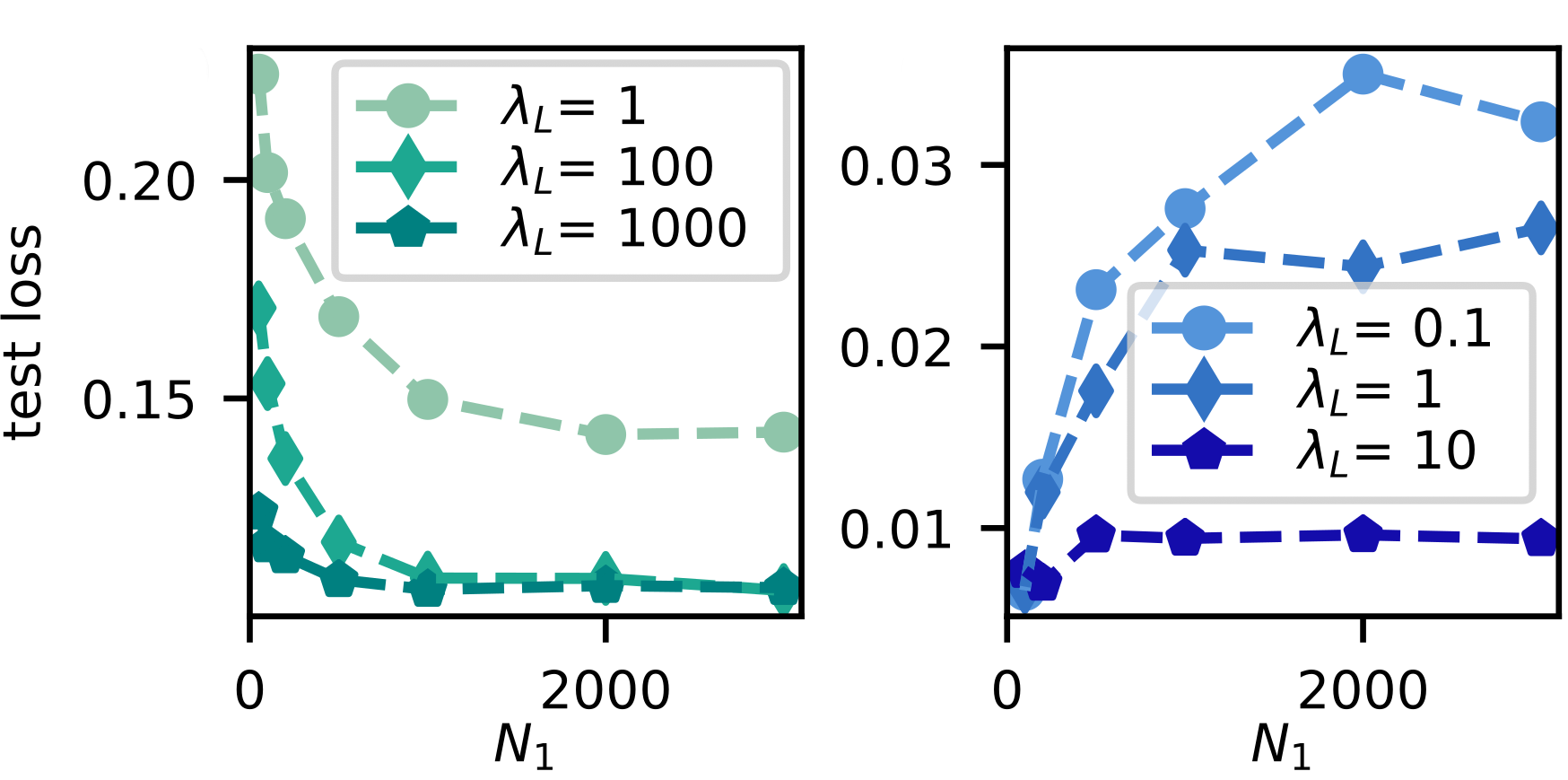}
        \caption{\textbf{Infinity Width vs Proportional Regime (2).} Experimental learning curves are presented as a function of \(N_1\) for increasing values of the Gaussian prior of the last layer \(\lambda_1\). Error bars are included within the points. Dashed lines connecting the points are provided to aid visual interpretation. The networks are trained on \(P = 3000\) examples from the CIFAR10 dataset in panel (b) and \(P = 500\) examples from MNIST in panel (c). Two key theoretical predictions at zero temperature are tested: (i) the generalization loss should decrease for any \(N_1\) as \(\lambda_1\) increases; (ii) the dependence of the learning curves on \(N_1\) vanishes in the large-\(\lambda_1\) limit, consistent with a constant bias (refer to the main text for additional details).}
        \label{fig3:3biw-2}
    \end{figure}
\clearpage

\vspace*{3cm}
\begin{figure}[H]
        \centering
        \includegraphics[width=0.70\textwidth]{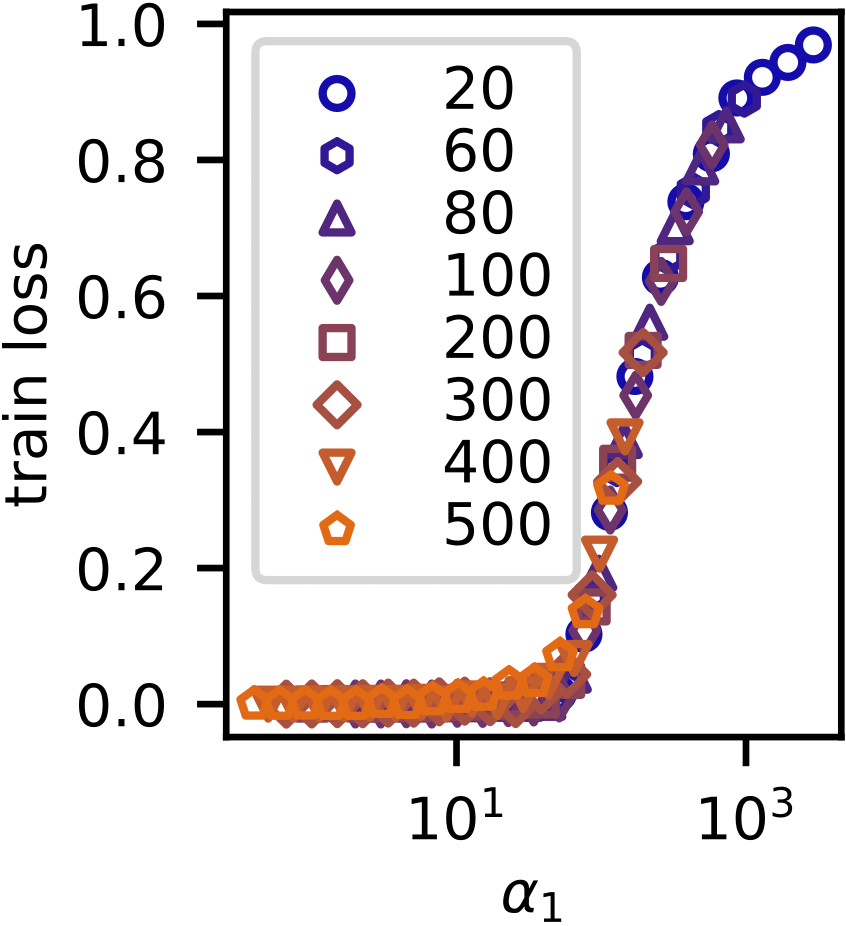}
        \caption{\textbf{Infinity Width vs Proportional Regime (3).} Training loss of various one-hidden layer architectures trained on a completely random task (where the inputs \(\mathbf{x} \in \mathbb{R}^{N_0}\) with \(N_0 = 50\) and the scalar outputs \(y\) are i.i.d. random variables drawn from a normal distribution with zero mean and unit variance) as a function of \(\alpha_1\).  The data suggest that DNNs trained with random labels exhibit a universal behavior. However, our current theoretical framework, which focuses on the overparametrised limit where the training error is precisely zero, does not account for this universal phenomenon.}
        \label{fig3:3biw-51}
    \end{figure}
\clearpage
}

\details{
    \begin{center}
        \textbf{Details: Numerical Experiments}
    \end{center}

    We conduct numerical experiments with deep fully-connected architectures focused on two regression tasks in computer vision. Specifically, we utilize the classes labeled ``0'' and ``1'' from the MNIST dataset, while we consider ``cars'' and ``planes'' in CIFAR10. CIFAR10 examples are coarse-grained to \(N_0 = 28 \times 28\) pixels and converted to grayscale.

    To ensure convergence of the posterior weights distribution to the Gibbs ensemble, we train our networks using discretized Langevin dynamics, as done in \cite{PhysRevX.11.031059,seroussi2023natcomm}. During each training step \(t\), the parameters \(\theta = \lbrace W^\ell, v\rbrace \) are updated according to:
    \begin{equation}
        \theta(t+1) =  \theta(t) - \eta \nabla_\theta \mathcal{L}(\theta(t)) +\sqrt{2T\eta}\epsilon(t)
    \end{equation}
    where \(T=1/\beta\) is the temperature, \(\eta\) is the learning rate, \(\epsilon(t)\) is a white Gaussian noise vector with standard normal distribution entries, and the loss function is defined in equation \eqref{lossA1}. We set \(T = \eta =  10^{-3}\) in all experiments, approximating the \(T=0\) dynamics in our regime. This dynamics requires \(10^5/10^6\) steps for thermalization, depending on the dataset and network sizes. We assess the generalization loss in a single run: after training error minimization and test loss thermalization, we average test loss values every \(10^3/10^4\) epochs, varying with \(P\), \(N_\ell\). For reference, the best test accuracy achieved on both datasets by 1HL architectures is 0.86 on CIFAR10 (with \(P = 3000\) and \(\lambda_1 = 1000\)) and 0.999 on MNIST (with the same Gaussian prior and \(P = 1000\)). The training accuracy consistently reaches 1.
    
    We finally note that, achieving perfect agreement between theory and simulations, especially when sampling from a Bayesian posterior at zero temperature, is a challenging task due to several technical numerical issues:
    
    \begin{itemize}

        \item[1.] Finite-size effects certainly play a role in explaining the small mismatch between theory and experiment. To address this point, we are currently performing high-precision numerical simulations with fixed $\alpha = P/N_1$ and increasing values of $N_1$ and $P$.
        \item[2.] The $T \to 0$ limit, which corresponds to perfect interpolation of the dataset and is the only case in which the saddle point equations can be solved analytically, was the most logic to address for starting, but turns out to be very hard to simulate. This is clear from some preliminary work we are doing, where we numerically solve the saddle point equations at generic $T$ for the saddle point variables $Q, \bar{Q} = f(Q)$. We find that the function $Q(T)$ changes rapidly for small temperatures.
        \item[3.] At $T = 0.001$, the autocorrelation time of the simulation is already very large, taking as little as $5 \cdot 10^6$ epochs to thermalize. As the temperature is decreased, the autocorrelation time increases, and we need hundreds of thousands of epochs to gain satisfactory statistics.
        \item[4.] The effect of a finite learning rate $\eta$ has to be taken into account as well. From our preliminary results, we empirically observe that finite-$\eta$ effects are larger at higher temperature. The standard way to take into account finite-$\eta$ effects is to perform the extrapolation to $\eta \to 0$ simulating different learning rates.
        \item[5.] Computing the theory in the case of $L > 1$ networks requires to numerically minimize a complex nested saddlepoint functional of the variables $\bar{Q}_\ell$. We are currently working on a numerical routine to efficiently perform this task.
    \end{itemize}

    The code used to perform experiments, compute theory predictions and analyze data is available at: \url{https://github.com/rpacelli/FC\_deep\_bayesian\_networks}.

}

\subsection{Deep neural networks in the proportional regime}

The theory developed for the case of a 1HL DNN can be generalized to a DNN with $L$ hidden layers. 
In a similar vein to the 1HL calculation, we introduce \(L\) sets of auxiliary variables \(h_{i_\ell}^{\mu}\) (where \(i_\ell=1,\dots, N_\ell\)) corresponding to the pre-activations at each layer. The strategy for performing the calculation is to demonstrate that the probability distribution of the preactivations at each layer, \(P_{\ell}(\{h_{i_\ell}^\mu\})\), can be computed recursively, starting from the input layer. This approach differs conceptually from the backpropagating kernel renormalization group introduced in Ref.~\cite{PhysRevX.11.031059}. It is a forward-propagating kernel renormalization group and represents a generalization of the kernel recurrence found in NNGPs~\cite{LeeGaussian}. Practically, our approach entails a systematic, layer-by-layer description of the pre-activation statistics, as demonstrated in the 1HL case. This is a quantitative correction to the standard Gaussian statistics recovered in the infinite width limit. Currently, we cannot re-derive the same result using the backpropagating method described in~\cite{PhysRevX.11.031059}.

We begin by integrating the weights of the first layer, which defines a probability distribution over the pre-activations of the first layer as follows:
\begin{equation}
\begin{aligned}
P_1(\{h_{i_1}^\mu\}) &= \int \mathcal D W^{(1)} \prod_{i_1, \mu} \delta\left(h_{i_1}^{\mu} -\frac{1}{\sqrt {N_{0}}} \sum_{i_0=1}^{N_{0}} W^{(1)}_{i_1 i_0} x_{i_0}^\mu  \right)\\
&= \prod_{i_1=1}^{N_1} \frac{e^{-\frac{1}{2} \sum_{\mu \nu} h_{i_1}^{\mu} C^{-1}_{\mu\nu}  h_{i_1}^{\nu} }}{\sqrt{(2\pi)^P \det C}}\,.
\end{aligned}
\end{equation}
where \(C\) is defined in~\eqref{eq:P1}. This result is straightforward and valid for any \(N_0\), \(P\), and \(N_1\), as the prior for the weights is Gaussian. At the second layer, we have:
\begin{align}
P_2(\{h_{i_2}^\mu\}) 
&= \int \mathcal D W^{(2)} \mathcal D h_1 P_1 (\{h_{i_1}^{\mu}\}) \notag\\
&\times \prod_{i_2, \mu} \delta\left(h_{i_2}^{\mu} -\frac{1}{\sqrt {N_{1}}} \sum_{i_1=1}^{N_{1}} W^{(2)}_{i_2 i_1} \sigma(h_{i_1}^\mu)  \right) 
\end{align}
We introduce conjugate variables \(\bar h^{\mu}_{i_2}\) to the activation of the second layer and the calculation proceeds similarly to the 1HL case. To make analytical progress, we need to make two fundamental approximations: (i) assuming that the set of random variables \(q_{i_2} = 1/(\sqrt{N_1} \lambda_1) \sum_\mu \bar h^{\mu}_{i_2} \sigma(h^\mu)\), where \(\mathbf h \sim \mathcal N(0, C)\), is Gaussian-distributed; (ii) neglecting correlations between different pre-activations of the second hidden layer. We ultimately obtain:
\begin{equation}
    \begin{split}
        P_2(\{h_{i_2}^\mu\}) =& \int dQ_1 d\bar Q_1 e^{-\frac{N_1}{2} (-Q_1 \bar Q_1 + \log (1+Q_1))}\prod_{i_2=1}^{N_2} \frac{e^{-\frac{1}{2} \sum_{\mu\nu}h_{i_2}^{\mu}\left(\bar Q_1 K(C)/\lambda_1\right)^{-1}_{\mu \nu} h_{i_2}^{\nu}}}{\sqrt{(2\pi)^P \det (\bar Q_1  K(C)/\lambda_1)}}\,,
        \label{preact12}
        \end{split}
\end{equation}
where \(K(C)\) is defined by equation \eqref{3BIW:K1}. Note that, except for the integration over the two variables \(Q_1\) and \(\bar Q_1\), this is analogous to the probability distribution of the 1HL system \eqref{eq:P1} if we replace \(C\) with \(\bar Q_1  K(C)/\lambda_1\). This reasoning can be iterated across layers, yielding:
\begin{equation}
    \begin{split}
        P_L (\{h_{i_L}^\mu\}) = \!\!\int& \prod_{\ell =1}^{L-1} \!dQ_\ell d\bar Q_\ell e^{-\sum_{\ell=1}^{L-1} \frac{N_\ell}{2}\left[ -Q_\ell \bar Q_{\ell} + \log(1+Q_\ell)\right]}\\ &
\times \prod_{i_L=1}^{N_L} \frac{e^{-\frac{1}{2} \sum_{\mu\nu}h_{i_L}^{\mu}\left( K^{(R)}_{L-1}(\{\bar Q_\ell\} )\right)^{-1}_{\mu \nu} h_{i_L}^{\nu}}}{\sqrt{(2\pi)^P \det (K^{(R)}_{L-1}(\{\bar Q_\ell\}))}}\,,
    \end{split}
\end{equation}
where \(K^{(R)}_{\ell}(\{\bar Q_\ell\})\) is a renormalized kernel that satisfies the recurrence relation:
\begin{equation}
    K^{(R)}_{\ell}(\{\bar Q_\ell\}) = \bar Q_{\ell}/\lambda_{\ell} K \circ\left[ K^{(R)}_{\ell-1}(\{\bar Q_\ell\}) \right]\,,\;\;\; K^{(R)}_0 = C\,,
    \label{K_LQ}
\end{equation}
where \(C\) is the covariance matrix of the inputs, as defined previously. It is important to emphasize that each \(K^{(R)}_{\ell}\) depends only on the variables \(\bar Q_1, \dots, \bar Q_{\ell-1}\). For the sake of completeness, we note that the recurrence relation for the infinite-width kernel \(K_L\) is given by equation \eqref{K_LQ} with \(\bar Q_\ell = 1\) for all \(\ell = 1, \dots, L\).
In the end, we obtain that the partition function can be expressed in terms of a $2L$-dimensional integrals:
\begin{equation}
Z_{\textrm{DNN}} = \int \prod_{\ell =1}^{L} dQ_\ell d\bar Q_\ell e^{-\frac{N_L}{2} S_{\textrm{DNN}}(\{Q_\ell ,\bar Q_\ell\})}\,,
\end{equation}
where the effective action is given by:
\begin{align}
S_{\textrm{DNN}}&= \sum_{\ell=1}^{L} \frac{\alpha_L}{\alpha_\ell}\left[ -Q_\ell \bar Q_{\ell} + \log(1+Q_\ell)\right] \nonumber+\frac{\alpha_L}{P}\text{Tr}\log \beta \left(  \frac{\mathbf{1}}{\beta} +K^{(R)}_L( \{\bar Q_\ell\})\right) \notag\\
&+\frac{\alpha_L}{P} y^T \left( \frac{\mathbf{1}}{\beta} + K^{(R)}_L( \{\bar Q_\ell\})\right)^{-1} y\,
\end{align}

The computation of the generalization error over a new example \((\mathbf x^0, y^0)\) yields:
\begin{equation}
\begin{split}
\braket{\epsilon_\text{g}(\mathbf{x}^0,y^0)} &= (y^0-\Gamma_L)^2+\sigma^2_L
\end{split}
\label{methods:eq:gen_err0deep}
\end{equation}
where \(\Gamma_L\) and \(\sigma^2_L\) are defined respectively by:
\begin{align}
    \Gamma_L &= \sum_{\mu\nu}\kappa^{(R)}_{L\mu} \left(\frac{\mathbf{1}}{\beta} + K^{(R)}_L( \{\bar Q_\ell\})\right)^{-1}_{\mu\nu} y_\nu, \label{GL}\\
    \sigma^2_L &= \kappa^{(R)}_{L0} -\sum_{\mu\nu}\kappa^{(R)}_{L\mu} \left(\frac{\mathbf{1}}{\beta} + K^{(R)}_L( \{\bar Q_\ell\})\right)^{-1}_{\mu\nu}\kappa^{(R)}_{L\nu}\label{sigmaL}
    \end{align}
and $\kappa^{(R)}_{L\mu}$, $\kappa^{(R)}_{L0}$ are recursive kernels hat generalize the train-test and test-test kernels (\ref{kmu})-(\ref{k0}) computed from the recurrence given in equation \eqref{K_LQ} using the input $\mathbf x^0$ in the initial conditions.

\details{
    \begin{center}
        \textbf{Details: scale independent kernels \texorpdfstring{$K(\alpha C) = K(C)$ ($\alpha>0$)}{K(alpha C) = K(C)}}
    \end{center}
    In the case of the sign activation function, it is straightforward to demonstrate that the behavior under scalar multiplication of the kernel \( K^{(R)}_L(C) \) follows from the property \(\textrm{sign} (\alpha x) = \textrm{sign} (x) \). This particular case significantly simplifies the effective action for deep learning, as the non-linear dependence of \( K^{(R)}_L \) on the variables \( \{\bar Q_\ell\}_{\ell\neq L} \) is eliminated. Consequently, this allows for the exact solution of the saddle-point equations. Specifically:
    \begin{equation}
    Q_\ell^* = 0\,, \quad \bar Q_\ell^* = 1\, \quad \forall \, \ell=1,\dots,L-1\,,
    \end{equation}
    while the functional form of the solution for \( \bar Q_L \) remains the same as in the one-hidden layer case (in the zero temperature limit):
    \begin{equation}
    \bar Q_L^* = \frac{\sqrt{(\alpha_L-1)^2 + 4 \alpha_L \frac{1}{P} y^T K_L^{-1}y}-(\alpha_L -1)}{2}\,.
    \end{equation}
    Practically, this solution indicates that deep architectures with sign activation (which are challenging to implement in practice due to difficulties in backpropagating derivatives) essentially behave like one-hidden layer neural networks in the proportional limit. The only indication of the depth \( L \) is encapsulated in the infinite-width kernel \( K_L \).
    
}

\details{
    \begin{center}
        \textbf{Details: piecewise linear kernels of the form \texorpdfstring{$K(\alpha C) = \alpha K(C)$}{K(alpha C) = alpha K(C)}}
    \end{center}
    The linear behavior of the kernel under scalar multiplication is evident for ReLU and leaky ReLU activation functions, as evidenced by the property \(\textrm{ReLU} (\alpha x) = \alpha\, \textrm{ReLU} (x)\). In this case, the effective action is given by:
    \begin{align}
    S_{\textrm{DNN}}(\{Q_\ell, \bar Q_\ell\}) &= \sum_{\ell=1}^{L} \frac{\alpha_L}{\alpha_\ell}\left[-Q_\ell \bar Q_{\ell} + \log(1+Q_\ell)\right]  + \\
    &  +\frac{\alpha_L}{P}\text{Tr}\log \beta \left[  \frac{\mathbf{1}}{\beta} +\left(\prod_{\ell=1}^L\bar{Q}_\ell\right) K_L(C)\right] \notag\\
    &+\frac{\alpha_L}{P} y^T \left[ \frac{\mathbf{1}}{\beta} +\left(\prod_{\ell=1}^L\bar{Q}_\ell\right) K_L(C)\right]^{-1} y\,.
    \end{align}
    
    Just like in the one-hidden layer case, the saddle-point equations simplify in the zero temperature limit, assuming that the $L$-hidden layers kernel $K_L$ has only positive eigenvalues:
    \begin{equation}
    Q_\ell \bar Q_\ell - \alpha_\ell +\frac{\alpha_\ell}{\left(\prod_{\ell_1} \bar Q_{\ell_1}\right)}\frac{1}{P} y^T K_L^{-1} y = 0
    \label{saddleReLU}
    \end{equation}
    for all \(\ell=1,\dots,L\).
    
    If \(\alpha_\ell = \alpha\) for all \(\ell=1, \dots, L\), it is straightforward to show that the only solution must satisfy \(Q_\ell^* = Q^*\) for all \(\ell\), which aligns with the heuristic mean field theory proposed in Ref. \cite{PhysRevX.11.031059}. This equivalence arises because the authors of \cite{PhysRevX.11.031059} developed the heuristic mean-field theory for ReLU activation by substituting the linear kernel with the corresponding NNGP kernel, recognizing that the ReLU kernel transforms like the linear one under scalar multiplication. Our derivation indicates that this substitution is not universally applicable for all activation functions (as illustrated by the case of sign activation previously discussed), but it is valid in this particular instance.
    
    For completeness, we also revisit the self-consistent equations proposed by Li and Sompolinsky \cite{PhysRevX.11.031059} in the linear case. The effective action for this scenario is:
    \begin{align}
    S_{\textrm{DNN}}(\{Q_\ell, \bar Q_\ell\}) &= \sum_{\ell=1}^{L} \frac{\alpha_L}{\alpha_\ell}\left[-Q_\ell \bar Q_{\ell} + \log(1+Q_\ell)\right]  \notag\\
    &+\frac{\alpha_L}{P}\text{Tr}\log \beta \left[  \frac{\mathbf{1}}{\beta} +\left(\prod_{\ell=1}^L\bar{Q}_\ell\right) C_L\right] + \\
    & + \frac{\alpha_L}{P} y^T \left[ \frac{\mathbf{1}}{\beta} +\left(\prod_{\ell=1}^L\bar{Q}_\ell\right) C_L\right]^{-1} y\,,
    \end{align}
    where \(C_L = C/(\prod_{\ell=1}^L \lambda_\ell)\) and \(C_{\mu\nu} = \mathbf x^\mu\cdot \mathbf{x}^{\nu}/(\lambda_0 N_0)\). Considering isotropic aspect ratios \(\alpha_\ell = \alpha\), \(\forall \ell=1,\dots, L\) and identical Gaussian priors at each layer \(\lambda_\ell = \lambda\), \(\forall \ell=0,\dots,L\), the saddle-point equations for \(\bar Q_\ell\) become: 
    \begin{equation}
        1-\bar Q_\ell = \alpha\left(1-\frac{\lambda^L}{\left(\prod_{\ell_1} \bar Q_{\ell_1}\right)}\frac{1}{P} y^T C^{-1} y\right)\,.
    \end{equation}
    This leads to the recovery of the equation for the renormalization parameter \(u_0\) in \cite{PhysRevX.11.031059}, with the solution for this system being \(\bar Q_\ell = \bar Q^*\) and the identification \(u_0 = \bar Q^*/\lambda\).
    
}

Even though, in principle, it is feasible to directly test the predictions for cases with more than one hidden layer, in practice, this proves to be particularly challenging due to the technical issues outlined in a previous \textit{details} section. To circumvent this drawback, we have conducted an alternative kind of analysis to verify the soundness of our theory.

Firstly, we observe that in these multi-layer cases, we can still perform a similar scaling analysis on the dependence of the generalization error on the Gaussian prior in the last layer \(\lambda_L\) (in the zero temperature limit). It emerges that the bias is independent of \(\lambda_L\), while the variance \(\sigma_L^2\) diminishes as \(1/\sqrt{\lambda_L}\) when \(\lambda_L\) approaches infinity. This suggests that training with large values of the Gaussian prior of the last layer should improve generalization for any network aspect ratio, even when the depth \(L > 1\). This general observation is supported by numerical experiments, as shown in panels (c) of Fig. \ref{fig3:3biw-3} and \ref{fig3:3biw-4}. However, unlike the 1HL case, we observe that the bias depends on the aspect ratio even in the zero-temperature limit, indicating that the dependence on the aspect ratios of the networks \(\alpha_\ell\) does not vanish in the \(\lambda_L \to \infty\) limit.

Another prediction of our theory for networks with \(L\) layers, which corroborates previous results on linear networks and perturbative calculations for non-linear networks \cite{PhysRevX.11.031059,zavatone2022contrasting, zavatone2021asymptotics,zavatone2022asymptoticsJStatMech}, can be obtained by considering the effective action for ReLU activation. A Taylor expansion around the infinite-width limit \(\alpha_\ell = \alpha = 0\) for all \(\ell=1,\dots,L\) reveals that the first correction to the test loss \(\Delta \epsilon_{\textrm g}\) is proportional to:
\begin{equation}
\Delta \epsilon_{\textrm g} \,\propto\, \alpha \left(\frac{1}{P}y^T K_L^{-1} y -1\right)\,.
\label{correction}
\end{equation}
Here, \(K_L\) is the solution of the recurrence in equation \eqref{K_LQ} for \(\bar Q_\ell = 1\) for all \(\ell=1,\dots, L\) and ReLU activation. This indicates the existence of a simple scalar observable that determines whether a finite-width deep neural network will outperform its infinite-width counterpart, generalizing the finding from the 1HL case:
\begin{equation}
s_L = \frac{1}{P}y^T K_L^{-1} y\,.
\end{equation} 
We anticipate that the finite-width network will surpass its infinite-width analogue whenever \(s_L < 1\). In panel (a) of Fig. \ref{fig3:3biw-3} and \ref{fig3:3biw-4}, this prediction is validated for deep architectures with ReLU activation on the same regression tasks as used in the 1HL scenario. Notably, \(s_L\) rapidly diverges to infinity as the number of hidden layers \(L\) increases. This is primarily due to the fact that the ReLU NNGP kernel \(K_L\) develops at least one zero eigenvalue as \(L \to \infty\). This phenomenon occurs because each element of the matrix \(K_L\) converges to zero as \(L\) increases (see panel (c) of Fig. \ref{fig3:3biw-3} and \ref{fig3:3biw-4}), as can be verified by examining the explicit recurrence relation for the NNGP ReLU kernel \cite{cho2009kernel, LeeGaussian}. We also observe that this singularity can be interpreted as the fixed point of the discrete dynamical map defined by the recurrence relation for the NNGP kernel, suggesting a potential link between the generalization performance in our asymptotic limit and the research on the edge of chaos in random neural networks \cite{ganguli2016chaos, yang2017residual}.

It is important to note that our theory predicts that a kernel limit should manifest in the asymptotic regime as the depth \(L\) approaches infinity. This prediction can be tested, for instance, in the isotropic limit where \(\alpha_{\ell} = \alpha\) for all \(\ell\), with ReLU activation. In this scenario, one can numerically solve the saddle-point equation for \(\bar Q\) at large \(L\) and verify that \(\bar Q \to 1\) for all values of \(\alpha\), as demonstrated in panel (b) of Fig.~\ref{fig3:3biw-52}. Consequently, even in this limit, we anticipate an equivalence with a kernel theory where the kernel is given by \(K_{\infty} (C)\). It is important to note that our framework explicitly assumes that the depth \(L\) approaches infinity only after the limits of \(P,N\). Therefore, we are not addressing the more complex simultaneous limit \(L,N\to \infty\) at fixed \(L/N\), which has been investigated in  \cite{hanin2023bayesian, hanin2023random,yaida2020nonGauss,roberts2022book}.

\afterpage{
\begin{figure}
        \centering
        \includegraphics[width=0.85\textwidth]{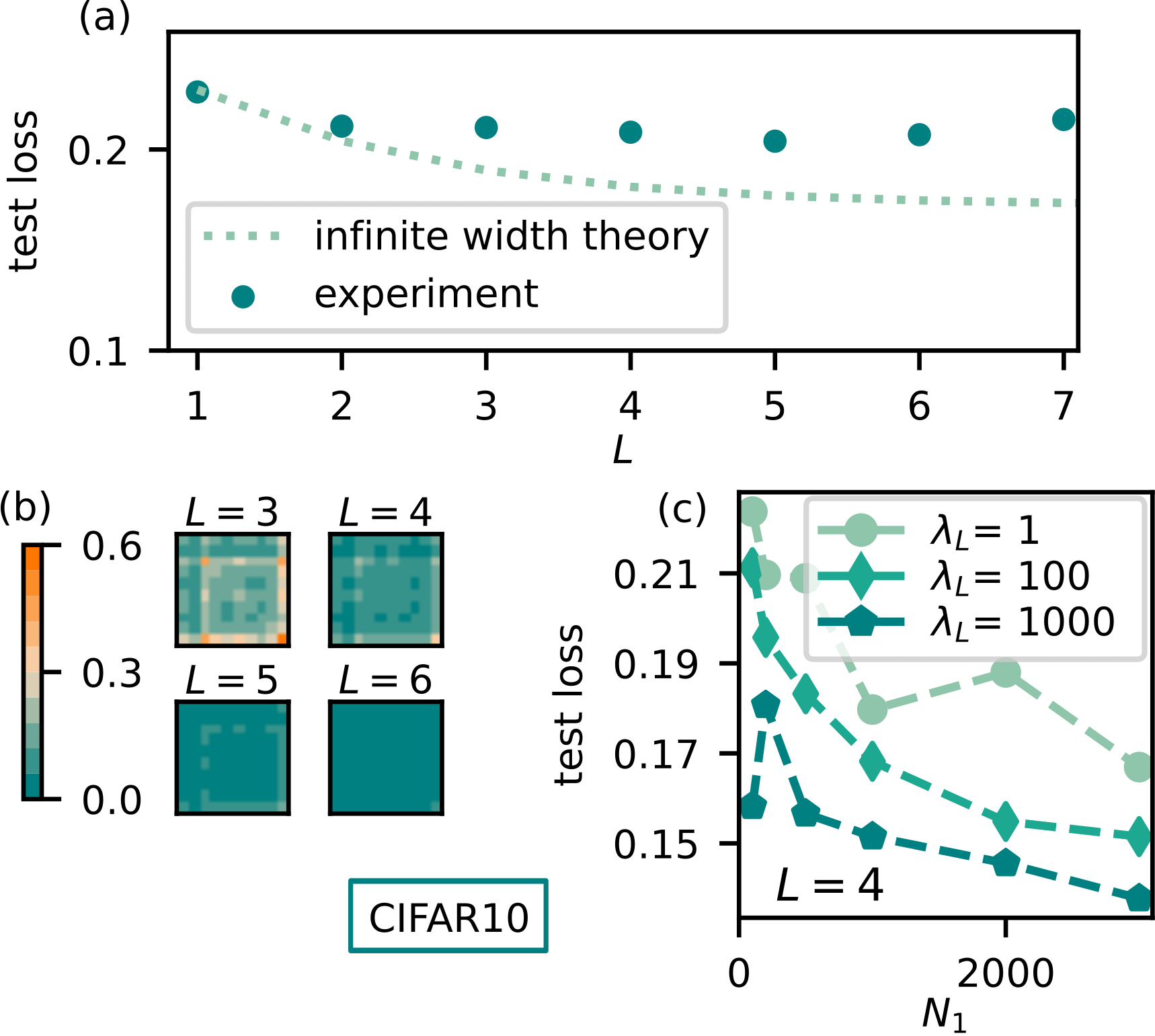}
        \caption{\textbf{Multilayer Neural Network in the Proportional Regime (1).}  (a) Test loss of an \(L\)-HL neural network with ReLU activation trained on the CIFAR10 dataset, as a function of depth \(L\), for \(P = 100\). The network is trained on a regression task in the small \(\alpha\) regime (\(\alpha = 0.1\)), close to the infinite-width limit. (b) Visualization of the entries of the infinite-width NNGP kernel at different network layers. The ReLU NNGP kernel tends to zero after repeated iterations, leading to almost vanishing eigenvalues and eventually making \(s_L\) always larger than one. (c) Test loss of a 4-HL network trained on \(P= 1000\) examples with different regularization strengths (with \(N_\ell = N = 1000\)). While increasing the magnitude of the Gaussian prior of the last layer still improves generalization for all \(N\), the curve at large \(\lambda_L\) is no longer constant as a function of \(N\), unlike the 1HL case. Dashed lines are included to guide the eye. In all panels, error bars are within the points.}
        \label{fig3:3biw-3}
    \end{figure}
\clearpage
}

\afterpage{
\begin{figure}
        \centering
        \includegraphics[width=0.85\textwidth]{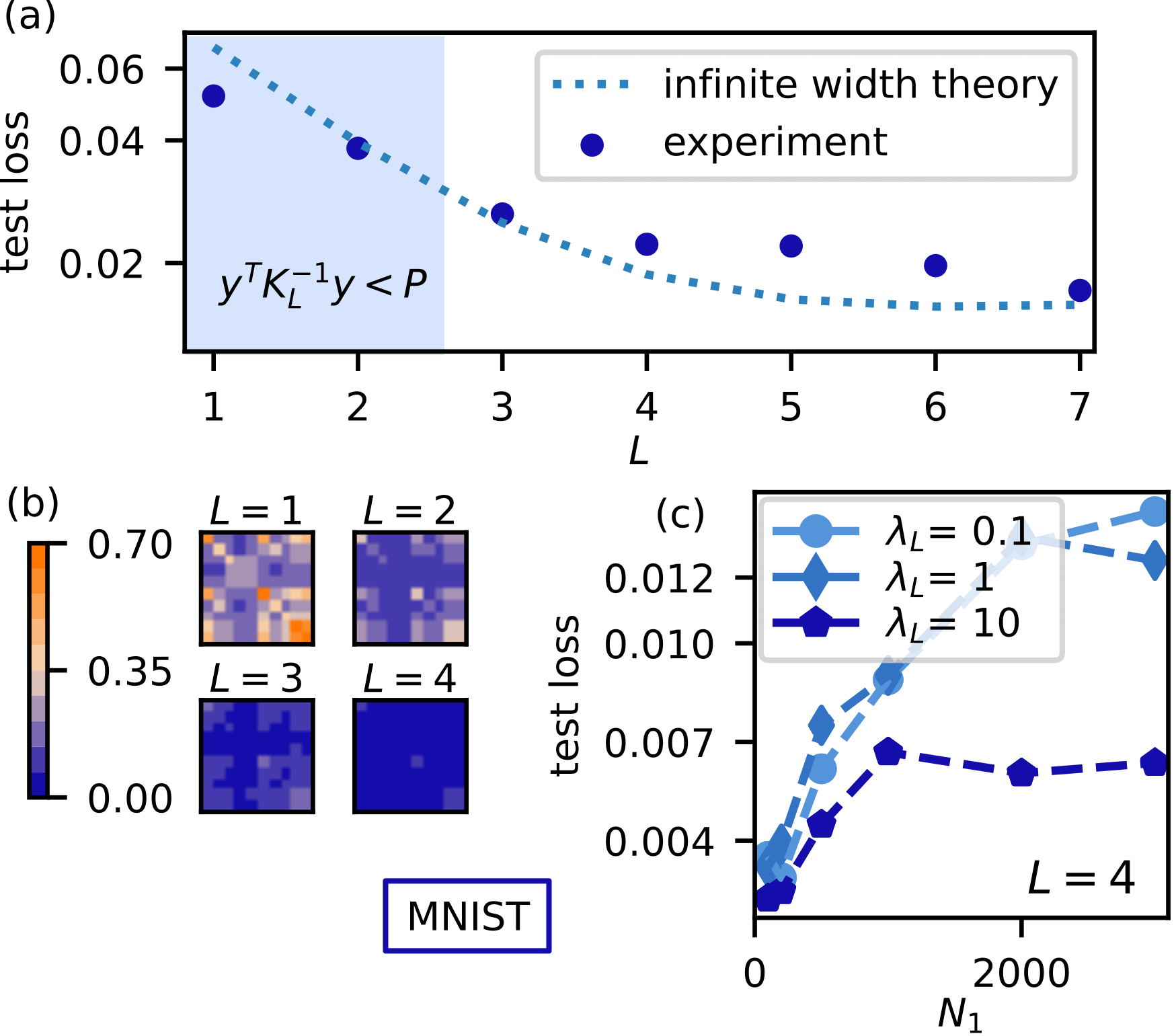}
        \caption{\textbf{Multilayer Neural Network in  Proportional Regime (2).} (a) Test loss of an \(L\)-HL neural network with ReLU activation trained on the MNIST dataset, as a function of depth \(L\), for \(P = 100\). The network is trained on a regression task in the small \(\alpha\) regime (\(\alpha = 0.1\)), close to the infinite-width limit. The finite-width network outperforms the infinite-width prediction only when $\mathbf{y}^\top K_L^{-1} \mathbf{y} < P$ (shaded area), i.e., only for depth \(L<3\). (b) Visualization of the entries of the infinite-width NNGP kernel at different layers. The ReLU NNGP kernel converges to zero after repeated iterations, leading to almost vanishing eigenvalues and making \(s_L\) eventually always larger than one. (c) Test loss of a 4-HL network trained on \(P= 1000\) examples with different regularization strengths (with \(N_\ell = N = 1000\)). Similar to panel (a), increasing the magnitude of the Gaussian prior of the last layer improves generalization for all \(N\), but the curve at large \(\lambda_L\) does not remain constant as a function of \(N\). The dashed line is included to guide the eye. In all panels, error bars are within the points.}
        \label{fig3:3biw-4}
    \end{figure}
\clearpage
}

\afterpage{
\begin{figure}
        \centering
        \includegraphics[width=0.70\textwidth]{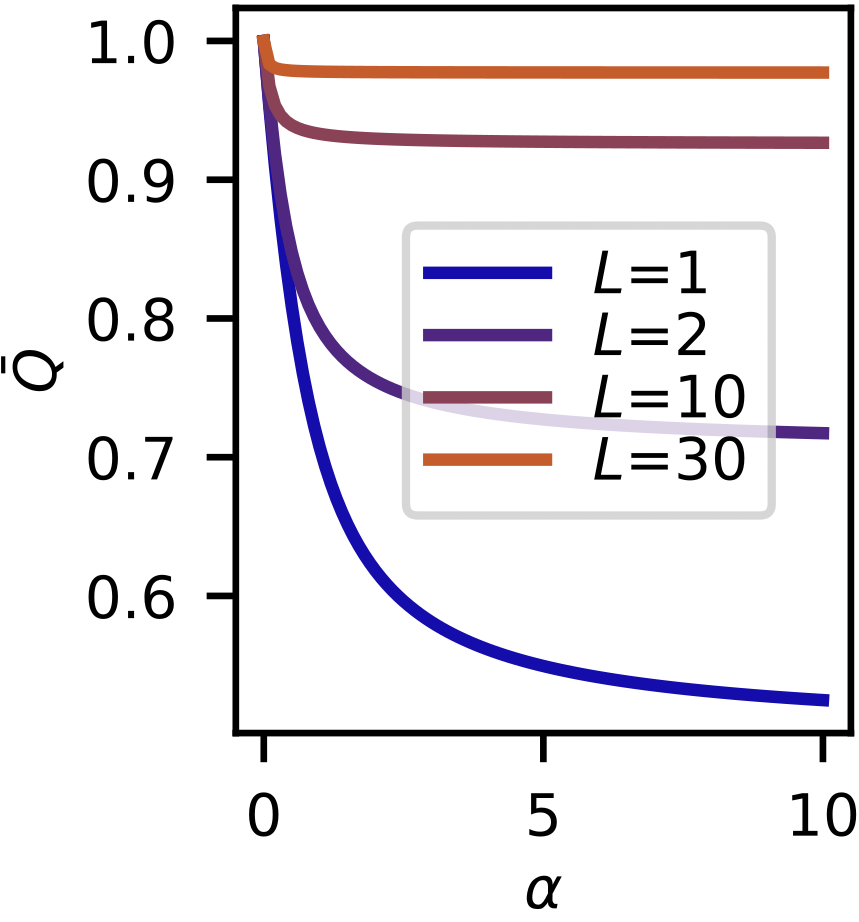}
        \caption{\textbf{Multilayer Neural Network in  Proportional Regime (3).} Numerical evaluation of the solution \(\bar{Q}^*\) is depicted for a ReLU activation function and isotropic network \(\alpha_{\ell} = \alpha\) for all \(\ell\), across different depths \(L\). As \(L\) increases (around \(L \sim 30\)), the parameter \(\bar{Q}^*\) rapidly converges to \(1\) for all values of \(\alpha\). This suggests that also DNNs in the asymptotic regime tend to a kernel limit in the sequential limit, where the depth \(L\) is taken to infinity after $P$ and $N$.}
        \label{fig3:3biw-52}
    \end{figure}
\clearpage
}

\subsection{Student's t-process as a DNN in the proportional regime}

From a more philosophical point of view, our theory can be formulated as a statement on the probability distribution of the output variables
\begin{equation}
\begin{gathered}
    s^\mu \equiv \frac{1}{\sqrt{N_1}} \sum_{i_1=1}^{N_1} v_{i_1} \sigma(h_{i_1}^\mu)\,,\\
\end{gathered}
\label{eq:smu}
\end{equation}
where $h\sim\mathcal{N}(0,C\otimes \mathbf{1}_{N_1}) $, $v \sim \mathcal{N}(0,\lambda_1^{-1} \mathbf{1}_{N_1})$. In fact, during the derivation of the partition function, we have seen that the probability density function of these variables can be written as a re-weighted Fourier transformation:
\begin{equation}
    P(s|\mathcal{T}_P) = \frac{e^{-\frac{\beta}{2}\sum_{\mu} (y^\mu - s^\mu)^2}}{Z} \int \prod_\mu\frac{d\bar{s}^\mu}{2\pi}\,  e^{i \bar{s}^\top s} \Xi(\bar{ s}) \,,
    \label{eq:P(s|data)}
\end{equation}
of the function:
\begin{equation}
	\Xi(\bar{ s}) = \left(1+ \frac{1}{\lambda_1 N_1}\sum_{\mu,\nu}^P \Bar{s}^\mu K_{\mu \nu}(C)\Bar{s}^\nu \right)^{-\frac{N_1}{2}}\,.
 \label{eq:Xi}
\end{equation}

It is evident that as \(N_1 \to \infty\) and \(N_1 \gg P\), the dependence on \(N_1\) vanishes, leading to
\begin{equation}
\Xi(\bar{s}) \to e^{-\frac{1}{2\lambda_1} \sum_{\mu,\nu}^P \Bar{s}^\mu K_{\mu \nu}(C)\Bar{s}^\nu}\,.
	\label{eq:P(Q)largeN}
\end{equation}
This result aligns with the NNGP literature, where for large \(N_1\) and finite \(P\), the variables~\eqref{eq:smu} are jointly multivariate Gaussian distributed as per the central limit theorem, as discussed in \cite{LeeGaussian}. This limit corresponds to the right-hand side of equation~\eqref{eq:P(Q)largeN} and forms the basis of mapping an infinite-width Bayesian neural network to a Gaussian Process (GP), as we have seen in Sec. \ref*{3BIW:PhisAPP}.

However, this is not the case when \(P\) is comparable to \(N_1\). Equation~\eqref{eq:Xi}, derived using the Gaussian equivalence principle based on the BM theorem in the proportional asymptotic limit \(P/N_1 \sim O(1)\), suggests that the variables \(\bar{s}^\mu\) follow a multivariate Student's \(t\)-distribution~\cite{Shah2014,Coolen_2020,Coolen_pre,Uchiyama2021}. Practically, our approach provides a layer-by-layer description of the pre-activation statistics by the Student's \(t\) distribution, a quantitative correction to the standard Gaussian statistics which is recovered in the infinite width limit.

The necessity of considering Student's \(t\)-processes as a generalization of NNGPs has been recognized in cases with different priors on the last layer's weights distribution \cite{Lee2022scale}. Similar non-Gaussian posterior forms, as in Eq. \eqref{eq:Xi}, have also been reported in \cite{aitchison2020bigger,zavatone2021depth, yang2023theory}. The emergence of this kind of process can be heuristically understood: when \(N_1\) and \(P\) are of the same order, we cannot take the limit \(N_1\to \infty\) before \(P\to \infty\), necessitating the use of the empirical covariance of the output variables \(s^\mu\) instead of their true covariance in estimating their probability distribution.

This observation is further supported by additional arguments. Referring to Fig.~\ref{fig3:3biw-1}, we observed that the monotonically increasing or decreasing behavior of the generalization curves as a function of \(N_1\) depends on whether the observable \(s_L = \mathbf{y}^\top (K/\lambda_1)^{-1} \mathbf{y}\) is greater or smaller than \(P\). A similar pattern of behavior is noted in \(L\)-hidden layer DNNs, as detailed in Eq.~\eqref{correction}, but with the corresponding kernel \(K_L\).
However, this criterion offers more than an empirical observation; it establishes an analytical link to Student's \(t\) inference. Specifically, Tracey and Wolpert \cite{tracey2018student} identified this criterion in their study of Bayesian optimization with Student's \(t\)-processes. They demonstrated that the value of \(s_L\) determines whether the Student's \(t\)-process under consideration exhibits a larger or smaller variance compared to a corresponding Gaussian Process (GP) with the same kernel. In a similar vein, our analysis employs this criterion to determine when an infinite-width neural network (analogous to a GP) outperforms a neural network in the proportional regime, which appears to be related to the Student's \(t\) process.

\section{Discussion and perspectives}

In this chapter we have described a strategy to investigate the statistical mechanics of deep neural networks beyond the infinite-width limit, that is in the finite asymptotic regime $P, N_\ell \to \infty$ with $\alpha_\ell = P/N_\ell >0$ as opposed to the simpler infinite-width limit $\alpha_\ell =0$, often considered in the literature. In the 1HL case, we conjecture that our evaluation is exact in the above thermodynamic limit. As such, we do not expect any additional corrections to the result, at least in the asymptotic regime. In particular, we have found a closed expression for the generalisation error that in principle provides a Bayesian estimator of the generalisation capabilities of fully-connected architectures for any given empirical dataset, provided that the chosen architecture is capable of perfectly fitting the trainset.

For the case of finite depth $L>1$ networks our results are only approximated, but it should be possible, at least in principle, to take systematically into account non-Gaussian corrections to the saddle-point action to check whether these are relevant or not for the theory at finite width, since the assumptions we made in deriving the results are clear {\cite{roberts2022book}}.

From the mathematical perspective, we find the link with Student's $t$-processes very promising. A more precise characterization of these neural network Student's \(t\)-processes (NNTPs) and the conditions under which they arise are intriguing topics for future research.

Our theory should work even in the case of non-zero training loss. Interestingly, numerical simulations performed with 1HL architectures of varying widths and random training labels show that the training loss follows a universal behavior \cite{gerace2022gaussian, zavatone2022contrasting} with respect to $\alpha_1$ (see Fig. \ref{fig3:3biw-51}), even in the regime where the DNN is not capable of perfectly fitting the data. It would be desirable to study this phase further, but our theory must be extended to the case of $\alpha_1 \gg 1$ to fully address this computational result.

It is important to emphasize that our theory describes only the equilibrium regime (at any temperature, although we provide the optimal order parameters only in the zero temperature limit). However, it would be interesting to understand to what extent this mean-field static analysis can be extended beyond equilibrium to assess the full training dynamics. In fact, such a theory would make it possible to investigate the performance of the many (often heuristic) learning algorithms used to train deep neural networks.

We conclude by pointing out that it would be interesting to compare our theory at fixed data with the data-averaged cases studied in \cite{zavatone2022contrasting, cui2023optimal}.

Moreover, similarly to what has been done for the infinite width limit \cite{novak2019bayesian}, there results are  extendable to the case of  convolutional networks \cite{aiudi2023local}.

\clearpage
\printbibliography[heading=subbibnumbered, title=Chapter bibliography]
\end{refsection}

\chapter{Task-explicit approach: retrieve the structure}
\begin{refsection}
\label{ch.4RTS}
The most remarkable feature of state-of-the-art Deep Neural Networks (DNNs) is their exceptional ability to generalize well. Unfortunately, it is often not straightforward to determine whether a network has achieved the ``perfect'' weights configuration, that is, the absolute minimum of the population loss. However, there exists some specific scenarios in which the DNN can perfectly mimic the target function (this would corresponds to strictly zero generalization error) and this optimal set of configurations can be analytically determined. 

In this chapter, I will present some novel (preliminary) numerical results regarding a simple case in which the teacher function is strictly contained in the student hypothesis space. Additionally, we propose an argument demonstrating how the capacity of DNNs to reach this special configuration is influenced by the number of constraints, namely the size of the dataset.

\vspace{0.5cm}
\minitoc
\vfill
\newpage

In this final chapter, we aim to leverage and complement the insights gained in the previous chapters to address a new facet of the DNNs behavior. As suggested by the bound we put forward in Chapter \ref*{ch.2AB}, different layers within a neural network play different roles in the learning process. Moreover, it has been conjectured that the initial layers are responsible for extracting relevant features from the dataset, while the last layer is tasked with executing the actual functions, such as regression or classification.

Simultaneously, we know that in the scenario of an infinitely wide network, the network is capable of learning the dataset. However, in this configuration, it behaves akin to a Gaussian Process, seemingly failing to extract any meaningful feature from the dataset. In Chapter \ref{ch.3BIW} we explored the statistical mechanics of DNNs beyond this infinite width limit.

In this chapter, our objective is to discern under what circumstances a finite network learns a training dataset by truly discovering the underlying rules of the dataset, as opposed to merely memorizing the finite dataset.

\section{By Memorizing or By Learning?}

One of the most remarkable aspects of modern neural networks is their ability to generalize from training datasets to new, unseen data, a capability that is crucial for practical applications where networks must operate effectively on diverse and real-world data. 
This generalization ability is influenced by several factors including the complexity of the network, the volume and variety of training data, and the regularization techniques employed during training. 
A well-trained network, which avoids overfitting on training data, can recognize and apply learned patterns to new data, thus demonstrating strong generalization capabilities. 
While in classical machine learning algorithms achieving this balance is challenging, as it requires the algorithm to adapt to training data without overfitting, thus losing the ability to apply these learnings to new and varied situations, modern deep neural networks do not seem to suffer this backfire. 
At the base of this phenomenon is the concept of \textit{feature learning} \cite{6472238, yu2013feature}, that is the automatic identification of relevant features from data for specific tasks. 
Unlike traditional methods where features need to be manually selected and engineered, neural networks can learn these features directly from data through hidden layers that progressively transform the input into more abstract and useful representations for the task at hand. 
In deep learning, this transformation occurs across multiple layers, allowing the network to construct a hierarchy of features from simple to complex. Note that this could happen only in the case of standard deep networks. In different scenarios, such as Infinite Width networks that we will discuss later, the network cannot learn the features. However, they can still achieve good generalization performance.

The rise of this feature is very known in DNNs, especially in the case of convolutional neural networks (CNNs). In this case, the networks can exploit the symmetries of the data to efficiently identify the relevant features of the training images, making them excellent tools for tasks such as image recognition and computer vision \cite{46832}. A similar phenomenon was also observed in simpler cases of Fully Connected DNNs (FCNs) \cite{ingrosso2022data}.

As we have seen in the previous Chapter \ref{ch.3BIW}, it is well known that in the \textit{infinite-width regime} the stochastic process that describes the inner information flow in the deep neural network is a Gaussian process, which is completely determined by a non-linear kernel, as we have shown for the NNGP kernel in the Bayesian setting. An alternative approach, based on the gradient optimizations, leads to a different kind of kernel, the so-called \textit{NTK} kernel.

A key insight from this observation is that learning in fully connected (or better, not {\it sparsely} connected, see below) networks of infinite width with the NTK initialization (where the weight variance scales as the inverse of the layer size) is equal to kernel learning, where the kernel is static and defined by the initial weight statistics \cite{canatar2021spectral,cortes1995support,dietrich1999statistical,bordelon2020spectrum}. This static nature of the kernel during training implies that the network weights vary very little during training.
In other words, this regime ensures that the learning dynamic behaves similarly to its linearized form around the initialization point.

From these teoretical results, we can find out a very important empirical implication: in the infinite-width limit a typical DNNs is not able to learn the feature of the dataset, as highlighted in \cite{vyas2022limitations}. 
However, it should be noted that this result does not hold in the sparse connectivity regime \cite{petrini2023learning}. This can be imposed a-priori on the network topology, or it could be obtained by a different weight initialization, the so-called mean field limit \cite{MEIMONTA}, with a small initialization scale such as the weight (standard deviation) scaling as the inverse of network width. 
Sparse connectivity can be also obtained by weight regularization during training \cite{de2020sparsity,bach2017breaking}

Conversely, finite DNNs are expected to be able to explore the solution space beyond the linearized regime and to learn dataset features. This is for instance the case of the convolutional neural networks (CNNs), which are known to be able to extract and learn features from the data thanks to their geometrical structure. This is a crucial aspect of their success in image recognition and computer vision tasks. As a consequence, their infinite-width limit typically perform worst than their finite version, as shown in \cite{aiudi2023local,lee2020finite,cagnetta2023can, favero2021locality}. Efforts to understand the effect of feature learning in convolutional networks have generated very interesting insights, as seen in Refs. \cite*{seroussi2023separation,naveh2021self}. 

The situation is less clear for fully connected DNNs. Empirical work \cite{lee2020finite,atanasov2022onset} suggest that in many situations infinite-width DNNs (kernel methods) can indeed outperform their finite width counterparts, but this trend is not universal and it is possible to find counterexamples, see for instance our own result in the previous chapter. 
Moreover, the connection between a good generalization performance and the ability to learn an internal representation of the dataset features (in the following: {\it feature learning}) are not fully understood. In Ref. \cite{petrini2023learning}, for instance, it is argued that feature learning in FCNs may lead to sparse weight matrices, and that this configuration could perform badly on certain ``smooth'' tasks such as image recognition.

In general, the problem of the presence (or the lack) of feature learning and its impact on generalization capabilities is central in the nowadays research on DNNs, as shown by the large numbers of manuscripts about it \cite{aiudi2023local,aitchison2020bigger,yaida2020non,antognini2019finite,hanin2022random,roberts2022principles,zavatone2021exact,zavatone2021asymptotics,zavatone2022contrasting}. 

To put it in simpler terms, we can think of the opposite situations where DNNs learn either \textit{``by memorizing''} the training set or effectively \textit{``by learning the underlying rule''}, where the latter being associated with feature learning.

Although these two scenarios - memorization and rule learning - are distinctly different, it is not always clear when one or the other is more likely to occur in practical DNNs (with standard initialization and scaling). Furthermore, if there are two phases, what do we observe around the phase transition point? Do we see a mix of both behaviors in competition, or do we encounter a completely different scenario?

These questions are of fundamental importance for understanding the generalization capabilities of DNNs, as well as for interpreting the networks, which is crucial in real-world applications of DNNs. A clear solution to these problems would provide definitive answers about when and why infinite width networks can outperform finite width ones.

To shed light on these questions, we have examined a simplified teacher-student setting where the teacher is fully contained in the student hypothesis space. By doing so, we can identify feature learning with the student weights reaching the optimal configuration perfectly reproducing the teacher. In this setting, feature learning is always optimal, and cannot be outperformed by lazy training as in \cite{petrini2023learning}.   Moreover these ``optimal'' weights, those corresponding to a fully learned rule, can be analytically determined. Our focus is on understanding how the hyperparameters (i.e the size of the dataset, the number of neuron, etc etc) influence the network's ability to achieve this specific, optimal weight configuration.

It is crucial to emphasize that the findings discussed in this chapter are in a preliminary stage and are subject to ongoing development and refinement. Our goal is to empirically investigate the specific conditions under which a neural network develops an internal representation of a dataset's features. In essence, from our simple example, we seek to provide insights into the circumstances that enable real-world Deep Neural Networks (DNNs) to transition from simple memorization to actual rule learning, and to understand how their architecture influences this process.
This inquiry is of significant importance as it contributes to our understanding of \textit{interpretability} in DNNs. Gaining a deeper understanding of this aspect seems to us a required step in order to address the fundamental and ethical considerations of AI safety, which is a critical concern in the application of DNNs in real-world scenarios.

\section{An exact solution for a simple DNNs}
 
In this section, we explore simplest possible multilayer setup we can think of that allows for the identification of an optimal solution. We consider a fully connected neural network with a single hidden layer. This network comprises $D$ input neurons, $N$ hidden neurons, and a single output neuron. This simplified structure serves as a fundamental model for understanding how neural networks process and learn from data, providing a basis for more complex architectures and applications.
\begin{equation}
    s = \sum_{i=1}^N v_i \cdot \sigma\left(\sum_{j=1}^D W_{i,j} x_j\right)\,,
\end{equation}
where $\sigma$ is the \textit{ReLU} activation function, $W_{i,j}$ are the weights connecting the input to the hidden layer randomly extracted from $\mathcal{N}(0,\sigma_w)$, and $v_i$ are the weights connecting the hidden layer to the output randomly extracted from $\mathcal{N}(0,\sigma_v)$. In a more compact form form we can write
\begin{equation}
    s = v \cdot \sigma (W x)
\end{equation}
where $\sigma$ acts element-wise, $x$ and $v$ are vectors in a $D$ dimensional space and $W$ is a $N\times D$ matrix.
The training dataset is a set of $P$ input-output couples. The inputs are $D$-dimensional vectors $x$, while corresponding output $y^\mu$ is given by a simple linear rule:
\begin{equation}
    y^\mu = \sum_{i=1}^D T_i x^\mu_i\,,
    \label{4:teacher}
\end{equation}
where the vector $\mathbf{T}$ is the teacher vector randomly extracted from $\mathcal{N}(0,1)$.

In this special scenario, where the teacher model is entirely encompassed within the student model, the critical question is whether the network is genuinely learning and internalizing the underlying principles taught by the teacher, or if it is merely memorizing the training set. In the case of memorization, the network lacks the ability to generalize, resulting in a non-zero generalization error. Conversely, if the network is indeed learning the underlying rules, it will be able to generalize effectively, leading to a zero generalization error.

To begin with, let us focus on the optimal configuration. In particular, we want to find the weights that not only minimize the loss function but perfectly fit the teacher rule. In other words, we want to find the weights by which the DNNs fully recover the linear teacher in Eq. \ref{4:teacher}.
In the case of a linear DNNs this problem has been extensively discussed in \cite{ziyin2022exact}. Here we generalise their results to the case of a ReLU nonlinearity. It is easy to convince ourselves that there are two families of such weights:

\begin{enumerate}

\item The first family of optimal weights directly encodes the teacher vector $T$ in the rows $W_i$ of $W$. $N_+ >0$ of them will be parallel to $T$, $W_i \sim T$, while $N_->0$ will be antiparallel, $W_i \sim -T$. Up to irrelevant permutations one can write:
\begin{equation}
    \begin{split}
        W_{i} = + \alpha_i T \,, &\qquad \text{for } i = 1,\ldots,N_+ \,,\\
        W_{i} = - \alpha_i T \,, &\qquad \text{for } i = N_++1,\ldots,N_++1N_- \,.
    \end{split}
    \label{4:optimal1}
\end{equation}
Here, $\alpha_i>0$ represents neuron-specific rescaling factors, while the corresponding readouts satisfy the condition 
\begin{equation}
    \sum_{i=1}^{N_+} v_i \alpha_i = 1 \quad \text{and} \quad \sum_{i=1+N_+}^{N_+ + N_-} v_i \alpha_i = - 1\,.
\end{equation}

The logic behind this solution is straightforward: the weight of the first layer align (or anti-align) itself with the teacher vector, while the second layer performs a linear combination of the outputs. Due to the ReLU nonlinear activation function it is not sufficient to have weights rows $w_i$ that only align with the teacher, as the ReLU function would prevent the network from learning inputs such that $T \cdot x<0$. Conversely, anti-aligned $W_i$ could not learn the cases $T \cdot x>0$. Therefore, we need at least one weight that is aligned and another that is anti-aligned with the teacher.
Note that, it is not necessary to enforce the condition $N_+ + N_- = N$, but it's crucial that the other readout neurons with $i>N_-+N_+$ have $v_i=0$ (the associated $W_i$ values are irrelevant in this case).

This choice of weights is obviously capable of fully reconstructing the teacher. In this case, the output of the hidden layer can be described as follows:

\begin{equation}
    \begin{split}
        \text{If }\,\, T\cdot x > 0\,\, &\Rightarrow\,\, s = \sum_{i=1}^{N_+} v_i\cdot \sigma(\alpha_i\cdot T\cdot x)=\sum_{i=1}^{N_+} v_i \cdot\alpha_i\cdot T\cdot x = T\cdot x\,,\\
        \text{If }\,\, T\cdot x < 0 \,\,&\Rightarrow\,\,s= \sum_{i=1+N_+}^{N_+ + N_-} v_i\cdot \sigma(- \alpha_i T\cdot x)= - \sum_{i=1+N_+}^{N_+ + N_-} v_i\cdot \alpha_i\cdot T\cdot x = T\cdot x\,.
    \end{split}
\end{equation}

In practice, with this optimal solution, the weight matrix W directly encodes the teacher vector T, which is the only relevant dataset feature (the input $x^\mu$ themselves are featureless, being drawn out of a multivariate Gaussian). Trained DNNs approaching this solutions thus will exibit a simple form of {\it feature learning}.

\item It should be noted, however, that a second family of exact solutions does exist. It is compsed by $M$ rows $W_i$ that are not parallel or anti-parallel to $T$, but whose linear combinations (via the readout weights $v_i$) equal T. However, due to the ReLU nonlinearity, these linear combinations always require a dual set of other $M$ rows with opposite orientation:
\begin{equation}
    \begin{split}
        \sum_{i=1}^M v_i \cdot W_i &= +  T \,,\\
        \sum_{i=M+1}^{2M} v_i \cdot W_i &= +  T \,,\\
        v_{i+M} = - v_i \quad&\quad \text{for} i = 1,\ldots,M  \,,\\
        W_{i+M} = - W_i \quad&\quad \text{for} i = 1,\ldots,M  \,.
    \end{split}    
    \label{4:optimal2}
\end{equation}

In simpler terms, this family of solutions involves a set of weights $W_i$ which, when appropriately linear combined by $v_i$, collectively sum up to the teacher vector. To ensure that the network can learn the entire rule of the teacher, another set of weights $W_{i+M}$ is required. These weights, weighted by $v_{i+M}$, also sum up to the Teacher. However, it's crucial that the weights $W_{i+M}$ and $v_{i+M}$ are the exact opposites of the weights $W_i$ and $v_i$.

\end{enumerate}

Note that these two families of solutions are not mutually-exclusive. In fact, it is possible to have some weights that belong to the first family and some that belong to the second family. In this case is important to normalize properly the output weights so that the total network output equals $T \cdot x$. In any case, the second family still captures the teacher vector, but in a more subtle way (by the proper linear combination of rows $W_i$) so that its interpretability is less transparent than the first. However, we will show that the occurrence of solutions close to this second family is strongly suppressed in our trained DNNs, so that in the following we will focus exclusively on the first family of optimal solutions.

\section{Learning curves and explicit feature visualization}

Now that we have identified the optimal weights, we can ask ourselves if and when the network is able to reach this configuration. In particular, we want to understand how the size of the dataset $P$, the dataset size $D$, and the number of neurons $N$ in the hidden layer influence the network's ability to achieve these specific, optimal weight configurations. To obtain this information, we have performed extensive simulation  training the networks using the Adam optimization scheme \cite{kingma2014adam} with a learning rate  $lr = 0.01$ and weight decay of $\lambda_{wd} = 10^{-4}$, over $60,000$ epochs (except when specified otherwise), keeping fixed to zero the biases.
The network is trained using the mean squared error (MSE) loss function with weight decay, which is given by:
\begin{equation}
    \mathcal{L} = \frac{1}{2P}\sum_{\mu=1}^P \left(y^\mu - s^\mu\right)^2+\frac{\lambda_{wd}}{2} \sum_{i,j} w_{i,j}^2 +\frac{\lambda_{wd}}{2} \sum_{i} v_{i}^2  \,.
\end{equation}
First of all, lets discuss the learning curves (training and generalization errors) as a function of the training dataset size and of the network size. The learning curves are  defined as the average of the loss function respectively over the training dataset $\mathcal{T}$, and over the underlying distribution of the dataset:
\begin{equation}
    \begin{split}
        \epsilon_t &= \frac{1}{P}\sum_{\mu=1}^P \left(T\cdot x^\mu - s(x^\mu)\right)^2\,,\\
        \epsilon_g &=  \int \mathcal{D}x\,\left(T\cdot x - s(x)\right)^2\,.
    \end{split}
\end{equation}

We can see in Fig. \ref{4RTS:fig0} that the students are always (for any dataset size $P$) able to obtain vanishing training errors. This means that the network can always find a set of weights that nearly satisfy the constraints imposed by the finite training set. 
However this is not true for the generalization error, that evaluates the trained network on the full dataset distribution. Generalization seems to be finite for DNNs trained on small datasets,  and start to match the vanishiong training error only for large dataset.

\afterpage{
\begin{figure}
    \centering
    \includegraphics[width=0.9\textwidth]{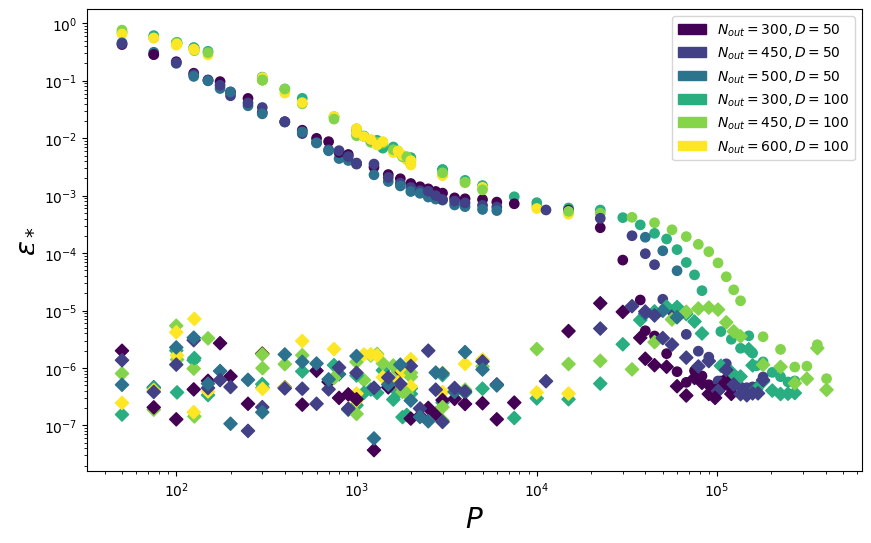}
    \caption{\textbf{Learning Curves as a function of the Dataset Size $P$.}
    Generalization (circle) and training (diamond) curves as a function of the dataset size $P$, for different choices of the number of neurons in the hidden layer and input dimensions. The single performance point is obtained averaging over 10 distinct runs at the end of the training procedure ($epochs = 10,000$).} 
    \label{4RTS:fig0}
\end{figure}
\clearpage
}

Interestingly, when the dataset size \( P \) is appropriately scaled by the total number of weights in the network \( N_{tot} = N(D+1) \), the learning curves exhibit a collapsing behavior, as illustrated in Fig. \ref{4RTS:fig01}. This phenomenon is consistent across various choices of \( N \) and \( D \). This observation leads to the identification of two distinct regimes, each characterized by unique properties.

The first regime, where \( P/N_{tot} \ll 1 \) (or better $P/N_{tot} < 10^{-2}$), is the \textit{overparametrized regime}. In this regime, the DNN has enough capacity to fit the training data perfectly (achieving zero training errors), yet it struggles to deal with new examples, leading to high (i.e. non zero) generalization errors. The second regime, where \( P/N_{tot} \gg 1 \) (or better $P/N_{tot} > 5$), is the \textit{underparametrized regime}. Here, the DNN not only fits the training data well but also shows remarkable generalization performance, achieving nearly zero generalization errors.

These two regimes are separeted by the so-called \textit{interpolation threshold}, in particular at \( P = N_{tot} \) \cite{singh2022phenomenology, schaeffer2023divergence}. Here, the \textit{double descent} phenomenon occurs, where the generalization error initially increases with the dataset size, before decreasing again. This is not evident in our data due to early stopping issues. Indeed, increasing the number of training epochs (see Fig. \ref{4RTS:fig10}) reveals this phenomenon also in our case. A more careful investigation of this \textit{intermediate regime} is left for future work. However, it is fair to say that the choice of the regime ranges in $P$ is arbitrary, and we do not have formal proof of these numbers. We have chosen them based on the observation of the data, and the expectation of different behaviors in the cases of $P/N_{1} \to 0$ (that implies $P/N_{tot}\to 0$) and $P \,\propto\, N_{tot} $.

\afterpage{
\begin{figure}
    \centering
    \includegraphics[width=0.9\textwidth]{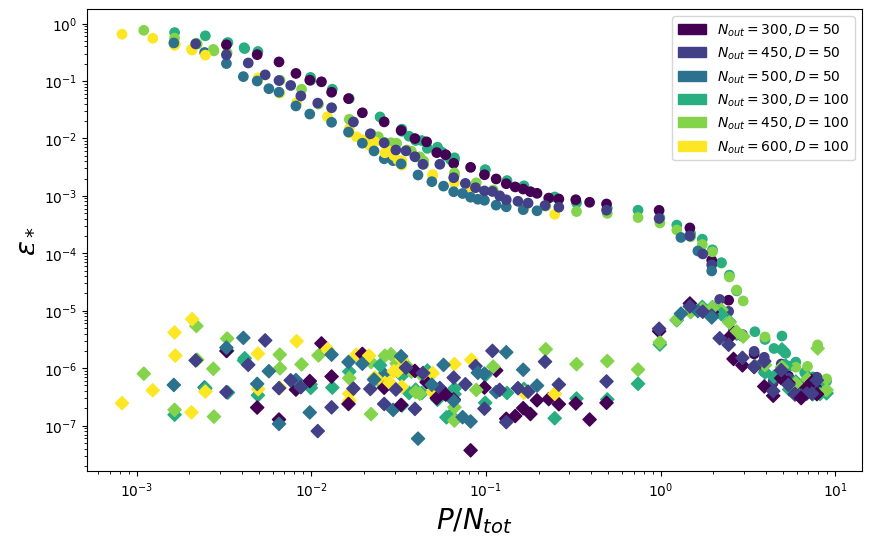}
    \caption{\textbf{Learning Curves as a function of the Dataset Size $P$.}
    Generalization (circle) and training (diamond) curves as a function of the rescaled dataset size $P/N_{tot}$, for different choices of the number of neurons in the hidden layer and input dimensions. The single performance point is obtained averaging over 10 distinct runs at the end of the training procedure. We can see how this rescaling shows the common behaviour of the learning curves for different choice of $N$ and $D$. Any single points is obtained averaging over 10 distinct runs at the end of the trainig procedure ($epochs = 10,000$).} 
    \label{4RTS:fig01}
\end{figure}
\clearpage
}


In the following, we are interested in assessing the trained network weights structure across the above two regime, or more in general, as a function of the rescaled hyperparameter $\gamma = P/N_{tot}$.

To visualize the weights distribution of the network, we can use the following approach. We consider the angles between the teacher and the post-optimization input weight vectors $W_i$. 
In particular, we can consider the following quantity:
\begin{equation}
    \theta_i = \arccos\left(\frac{T\cdot W_i}{||T||||W_i||}\right)\,.
\end{equation}

If the network is approaching the first family of exact solutions, the angles $\theta_i$ should be approach either 0 (for $W_i \sim T$) or $\pi$ (for $W_i \sim -T$). Moreover, near zero angles $\theta_i$ should be associated with a positive readout weight $v_i$, while one expects negative readout weights $v_i$ for $\theta_i$ approaching $\pi$. Finally, non aligned (or antialigned) input weights row should be associated with zero readout weights, $v_i=0$.


In Figure \ref{4RTS:fig3}, we present the weight distributions for DNN with $D=100$, $N=600$ and  three distinct dataset sizes: \( P=80 \), \( P=60000 \), and \( P=200000 \), each representative of a different regime. In the first scenario (\( P=80 \)), corresponding to the overparametrized regime with \( P/N_{tot} \approx 0.001 \), the weights do not align with the teacher and appear randomly distributed. However, this distribution cannot be entirely random, as the network still achieves zero training errors. Within this simple framework, we can infer the nature of this solutions.

Consider an input \( x \in \mathcal{T} \) from the training dataset. We hypothesize the existence of a subset of weights \( \tilde{\mathbf{W}} \) such that \( \sum_{i \in \tilde{\mathbf{W}}} v_i \cdot \sigma\left( W_i \cdot x \right) = T \cdot x \), while the other weights \( \tilde{\mathbf{W}}' = \mathbf{W} \setminus \tilde{\mathbf{W}} \) satisfy \( \sum_{i \in \tilde{\mathbf{W}}'} v_i \cdot \sigma\left( W_i \cdot x \right) = 0 \). These relationship is expected to hold for every datapoint \( x \in \mathcal{T} \) in the training dataset. 
In practice there exist a subset of input weight rows vectors and readout weights that reproduce the teacher $T$ on the finite trainig set $\mathcal{T}$.
We expect such solutions to be feasible only in a strongly overparametrized network, leading inevitably to the non-zero generalization error. This is because the weights are distributed in a manner that holds predictive power solely for the training dataset and not for the entire input space. We interpret this scenario as a typical example of \textit{memorization}, that is the ability to reproduce the training data without learning any of its essential features. To be fair, we must highlight that in this case, the \(L2\) regularization on the weights makes the weight distribution different from the purely random case (where we would have a bivariate Gaussian distribution centered at \((\pi/2, 0)\)). In fact, it seems that the regularized network dynamics prefer to set to zero all the read-out weights \(v_i\) that are useless for learning (i.e. the ones in $\tilde{\mathbf{W}}'$). The same analysis without regularization will be performed in the future.

In the opposite regime, the underparametrized one, where the dataset size is \( P = 200000 \) (and \( P/N_{tot} \approx 3.3 \)), the weights nearly align with the teacher vector. The angles \( \theta_i \) are close to \( 0 \) or \( \pi \), suggesting that the weights belong to the first family of solutions. This alignment is indicative of \textit{feature learning}.

The behavior of the weights in these two extreme regimes — overparametrized and underparametrized — will be further elucidated in the next section. There, we will utilize an analytical approach to demonstrate how the hyperparameters scaling significantly influences the volume of the solution space for the weights. This analysis aims to provide deeper insights into the mechanism that generate the distinct characteristics of each regime.

In the intermediate regime, characterized by a dataset size of \( P = 60000 \), the weights do not align with the teacher. Instead, they exhibit a distinct behavior. This phenomenon is more pronounced in Figs. \ref{4RTS:fig6} and \ref{4RTS:fig11}, where the weights \( v \cdot W \) appear equidistant from the teacher. These configuration still exibit a non zero generalization gap, showing that they do not belong to the exact solution families. Although this regime is not the main focus of our study, it presents intriguing characteristics that merit further investigation. The dataset size \( P = 60000 \) is specifically chosen as it equates to \( P/N_{tot} = 1 \), placing us at the \textit{interpolation threshold}. This concept is elaborated in \cite{singh2022phenomenology, schaeffer2023divergence}.

Let me highlight the fact that in this learning setting the chances to find solutions from the second family are pretty low. This is due to the implementation of \textit{Weight Decay}, a regularization technique that force the systems to prefer weights with smaller norms. For instance, in the case with $M=2$ and assuming $||v_i||=1$ for simplicity, the minimal optimal weights of the first family would have norm  $||W_1||+||W_2||=2||T||$. In contrast, the minimal norm for the weights of the second family is $||W_1||+||W_2||+||W_3||+||W_4||=2||W_1||+2||W_2||$. Given the triangular inequality, $2||T|| \leq 2||W_1||+2||W_2||$, it follows that solutions from the second family are less probable in our experimental setup.


\afterpage{
\begin{figure}
    \centering
    \includegraphics[width=0.9\textwidth]{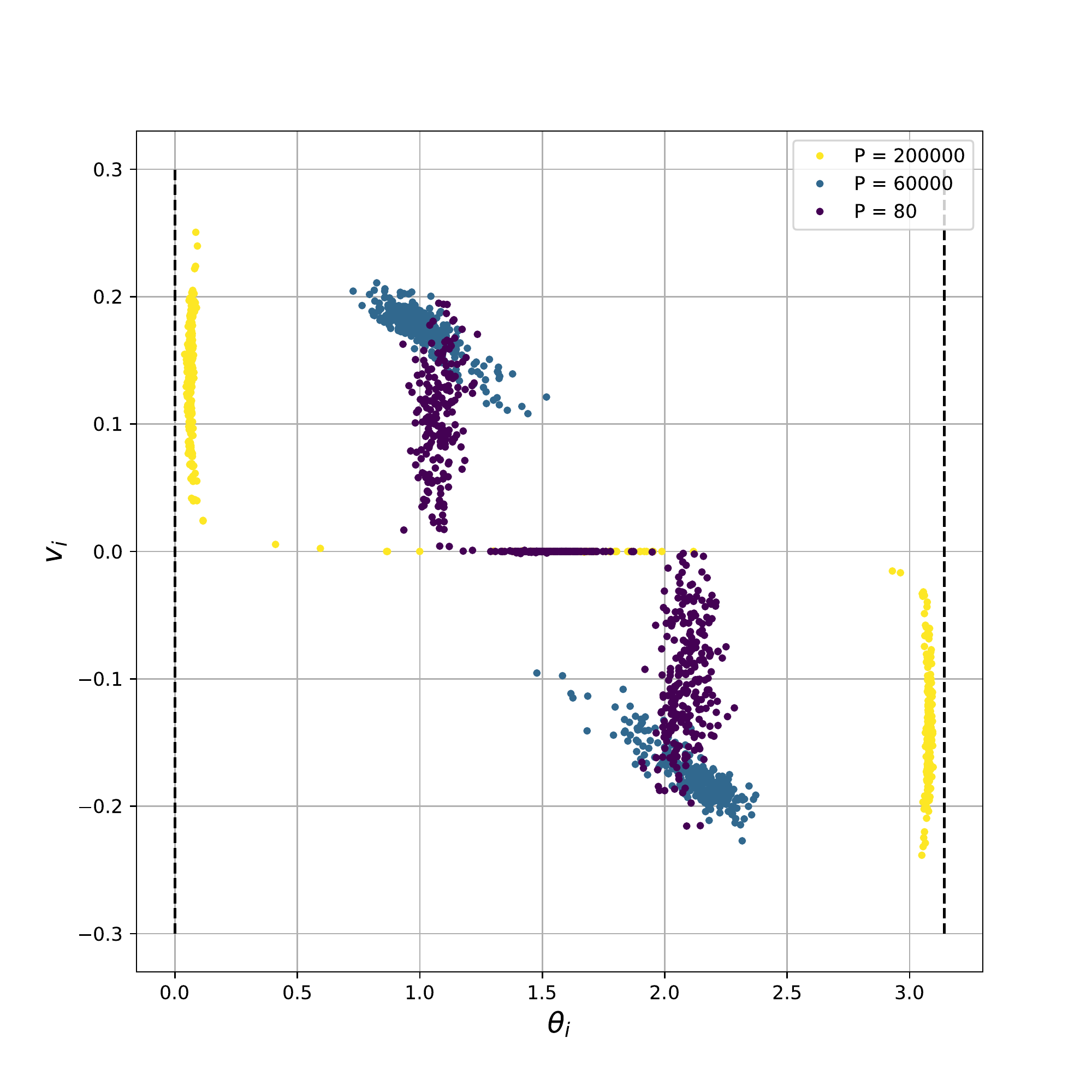}
    \caption{\textbf{Final Weight Distribution Across Different Dataset Sizes $P$.} This graph illustrates the weight distributions for three different dataset sizes $P$. On the x-axis, we have $\text{Arccos}(W_i\cdot T)$, representing the angle between the weights and the teacher, while the y-axis indicates the magnitude of the corresponding readout weights $v_i$. In the first scenario (depicted in purple), the dataset size is $P=80$. In the second scenario (in blue), it is $P=60,000$, and in the third (in yellow), the size is $P=200,000$. This visualization highlights how, with larger datasets, the weights $W$ tend to align more closely with the teacher, whereas in smaller datasets, the alignment disappears.
    The input size $D$ is set at $100$, and the hidden layer contains $N=600$ neurons. The weights are initialized using a standard normal distribution and updated using the Adam optimization scheme with a learning rate of $0.01$ and a weight decay of $10^{-4}$, over a span of $60,000$ epochs. The black dashed line represents the optimal alignment between the weights and the teacher.}
    \label{4RTS:fig3}
\end{figure}
\clearpage
}

\afterpage{
\begin{figure}
    \centering
    \includegraphics[width=0.9\textwidth]{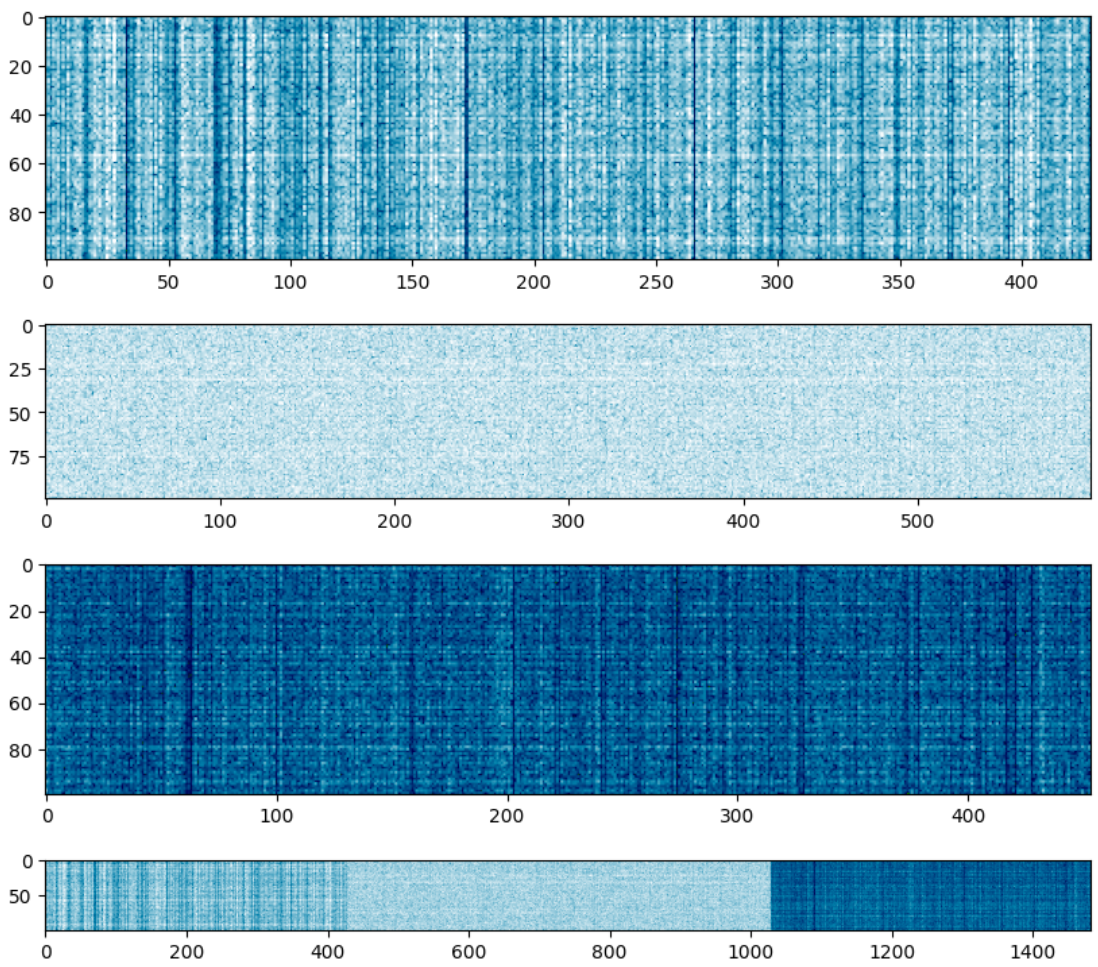}
    \caption{\textbf{Alignment between the Weights and the Teacher in Three Different Scenarios.} This graph depicts the alignment between the weights and the teacher in three distinct scenarios. The first row corresponds to a dataset size of $P=80$, the second row to $P=60,000$, and the third row to $P=230,000$. The final row provides a comparative overview of all three cases. In each row, we plot the matrix constructed as $v_i \cdot W_i - \tilde{T}$, where $\tilde{T}$ is the normalized teacher vector with the same norm as $v_i \cdot W_i$. The matrix size is $N \times D$. Darker pixels indicate matrix entries closer to zero. The first row reveals random distribution of weights, the second row shows that all weights are pretty equidistant from the teacher, and the third row demonstrates alignment of the weights with the teacher. Note that in the first and in the third row, we have removed the weights associated to a null redout weights $v_i= 0 $.}
    \label{4RTS:fig6}
\end{figure}
\clearpage
}

\afterpage{
\begin{figure}
    \centering
    \includegraphics[width=0.9\textwidth]{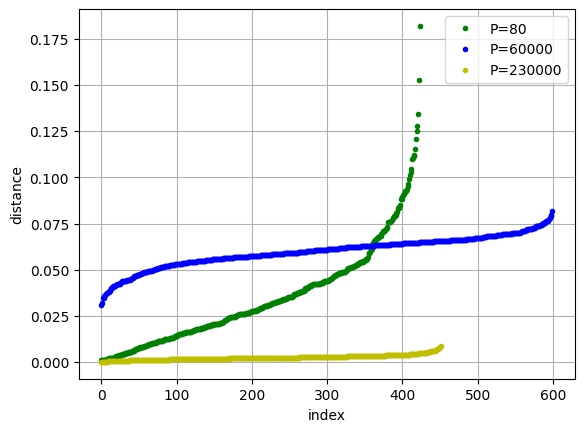}
    \caption{\textbf{Distance between the Weights and the Teacher in Three Different Scenarios.}  This graph depicts the Euclidean distance between the weights $v\cdot W$ and the rescaled teacher $\tilde{T}$ in three distinct scenarios: $P=80$ in green, $P=60,000$ in blue,and $P=230,000$ in yellow. We can see how in the overparametrized regime the vector are randomly distribuited, while in the  underparametrized the typical distance is very small. In the intermediate regime (P=60000) the weights seems to be equidistant from the teacher.}
    \label{4RTS:fig11}
\end{figure}
\clearpage
}


We have observed that the distribution of weights in a neural network is significantly influenced by the size of the dataset \( P \), particularly in the two extreme regimes. To more effectively visualize the behavior of the weights as a function of \( P \), a different approach is required. 

Consider the construction of the matrix \( J \) as follows:
\begin{equation}
    J = \left[ v_1 \cdot W_1, \dots, v_N \cdot W_N \right] = 
    \begin{bmatrix}
        v_1 \cdot W_{1,1} & v_2 \cdot W_{1,2} & \dots & v_N \cdot W_{1,N} \\
        v_1 \cdot W_{2,1} & v_2 \cdot W_{2,2} & \dots & v_N \cdot W_{2,N} \\
        \vdots & \vdots & \ddots & \vdots \\
        v_1 \cdot W_{D,1} & v_2 \cdot W_{D,2} & \dots & v_N \cdot W_{D,N}
    \end{bmatrix}.
\end{equation}

The matrix \( J \) is an \( N \times D \) matrix where each column \( J_i \) represents the product of the readout weight \( v_i \) and the associated input weight \( W_i \). This matrix \( J \) effectively encapsulates all the information about the distribution of weights. Ideally, eigenanalysis would be suitable for extracting information about the alignment between the weights and the teacher if \( J \) were a square matrix. However, given that \( J \) is not always square, eigenanalysis is not applicable. Instead, we can consider the singular value decomposition (SVD) of \( J \):
\begin{equation}
    J = U \Sigma V^\top,
\end{equation}
where \( U \) is an \( N \times N \) orthogonal matrix, \( \Sigma \) is an \( N \times D \) diagonal matrix containing the singular values, and \( V \) is a \( D \times D \) orthogonal matrix. The entries along the principal diagonal of \( \Sigma \), denoted as \( \lambda_i \), are the singular values that provide valuable insights into the distribution of the weights.

Specifically, in the context of \textit{feature learning}, where all vectors in matrix \( J \) align with the teacher (i.e., \( v_i \cdot W_i \parallel T \)), this results in only the first singular value being nonzero and all others being zero. Conversely, in the case of \textit{memorization}, the singular values should display a random distribution.
In order to compare the singular values fairly, they should be rescaled as follows:
\begin{equation}
    \sigma_i = \frac{\lambda_i}{\sum_i \lambda_i}.
\end{equation}

In Fig. \ref{4RTS:fig12}, we display the rescaled singular values of the matrix \( J \) for various dataset sizes \( P \). In the context of \textit{feature learning}, we observe that all singular values except the first are close to zero, while the first is much bigger, thus showing a clear large gap. Conversely, in the case of \textit{memorization}, the singular values are more uniformly close to zero. 
This observation is consistent with the fact that the network's weights align with the teacher when the dataset size is large, demonstrating feature learning. In contrast, when the dataset size is small, the weights appear to be randomly distributed, indicative of memorization.

\afterpage{
\begin{figure}
    \centering
    \includegraphics[width=0.99\textwidth]{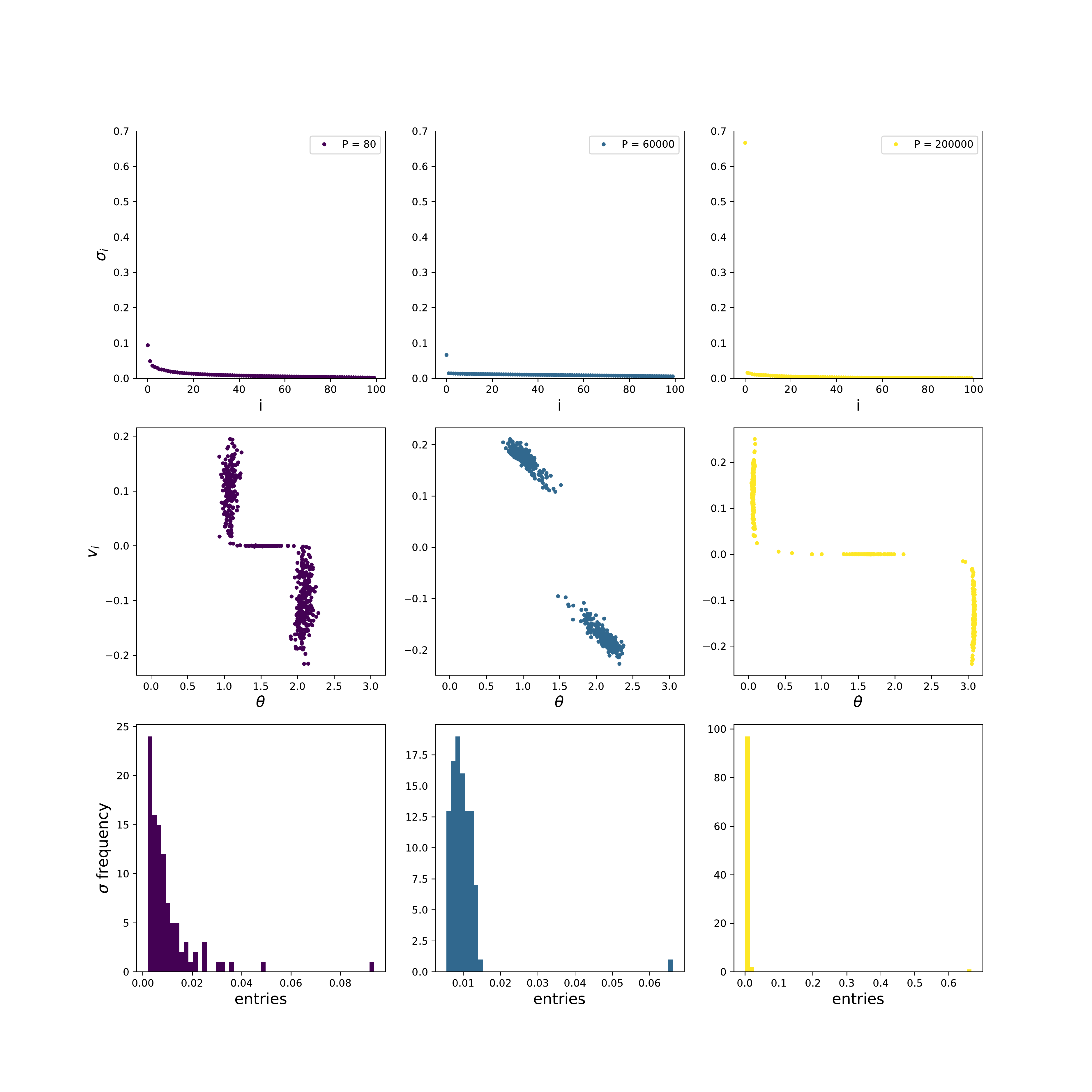}
    \caption{\textbf{Rescaled Singular Values of Matrix \( J \) for Different Dataset Sizes \( P \)}. In the first row, we plot the rescaled singular values of matrix \( J \) for different dataset sizes \( P \). The second row features plots of the weight distributions, as a reminder, in the same way that we have used Fig. \ref{4RTS:fig3}. The last row presents the rescaled singular value distributions. In the context of \textit{feature learning} (on the righthand side, in yellow, $P=200,000$), we observe that all singular values, except the first, are close to zero, while the first singular value is significantly larger. Conversely, in the case of \textit{memorization} (on the lefthand side, in purple, $P=80$), the singular values are uniformly close to zero. This pattern is consistent with the network's weight alignment with the teacher in cases of large dataset sizes, while, for smaller dataset sizes, the weights appear to be randomly distributed. The behaviour in the \textit{intermediate regime} (in the middle, in blue, $P=60,000$) is more complex, and it is an interesting feature that we will investigate in the future.}
    \label{4RTS:fig12}
\end{figure}
\clearpage
}

We would like to condense the information from these singular spectrum into a single metric. One promising approach is to consider the \textit{entropy} of the rescaled singular values. We define it as follows:
\begin{equation}
    S = -\sum_{i=1}^{P} \sigma_i \log(\sigma_i),
\end{equation}
where \( \sigma_i \) represents the normalized singular values of the matrix \( J \). In scenarios of \textit{feature learning}, the entropy is expected to be zero, indicating a dominant singular value. Conversely, in cases of \textit{memorization}, the entropy should be greater than zero, reflecting a more uniform distribution of singular values.

In Fig. \ref{4RTS:fig4}, we show the entropy of singular values of the matrix \( J \) as a function of the dataset size \( P \). The entropy value obtained at large $\gamma=P/N_{tot}$ is consistent with a good alignment between the weights and the teacher, while the results obtained at small $\gamma$ suggest a random alignment between the weights and the teacher. Even if these results are preliminary, they are consistent with the idea that the network is capable of recovering the teacher's rule only in the (severely) underparametrized regime, where the dataset size is larger than the total number of parameters of the DNN. 

\afterpage{
\begin{figure}
    \centering
    \includegraphics[width=0.9\textwidth]{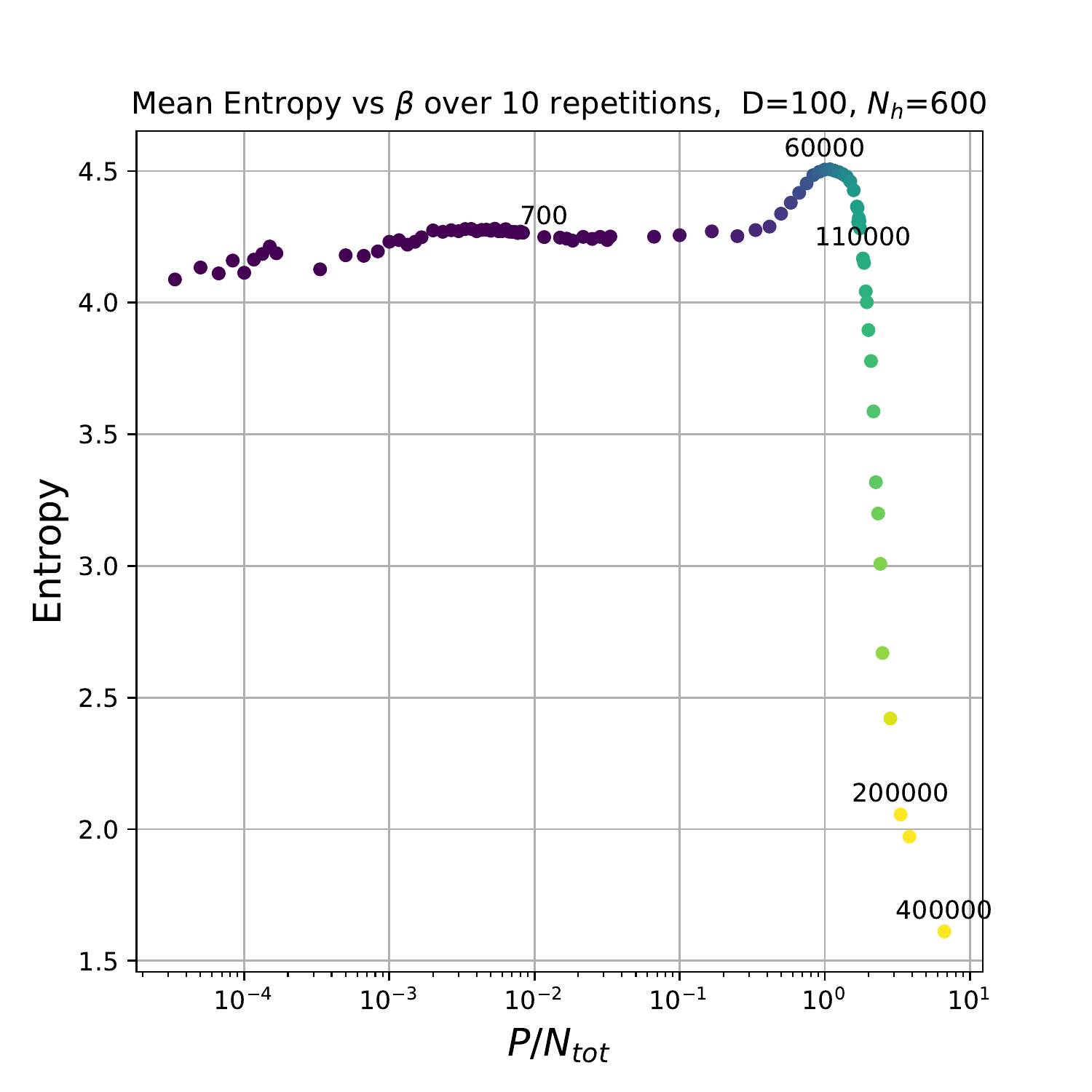}
    \caption{\textbf{Entropy of Singular Values of Matrix \( J \) as a Function of Dataset Size \( P \).} This graph illustrates the entropy of the singular values of matrix \( J \) as a function of the dataset size \( P \). The entropy value obtained at large dataset sizes is consistent with a good alignment between the weights and the teacher, indicating very small errors, while the results obtained at small dataset sizes align with a random alignment between the weights and the teacher. The peak at the interpolation threshold \(( P = 60000) \) shows an entropy larger than the random case, and it is an interesting feature that we will investigate in the future.}
    \label{4RTS:fig4}
\end{figure}
\clearpage
}

From a fully optimal perspective, in the case of perfect alignment between the weights and the teacher, the associated entropy \( S \) should be zero. However, in our simulations, we observe that the entropy does not exactly reach zero, but instead stops to a value of \( S \approx 2 \). 
This result is not in contradiction with our framework; rather, it is consistent with the fact that the weights are not exactly, but closely aligned with the teacher. In fact, in Fig. \ref{4RTS:fig5}, we demonstrate how an alignment error on the order of \( 1 \times 10^{-3} \) can generate an entropy of approximately $S=2$. This slight misalignment can be justified by the fact that achieving perfect alignment is challenging in a high-dimensional problem such as this one.

\afterpage{
\begin{figure}
    \centering
    \includegraphics[width=0.9\textwidth]{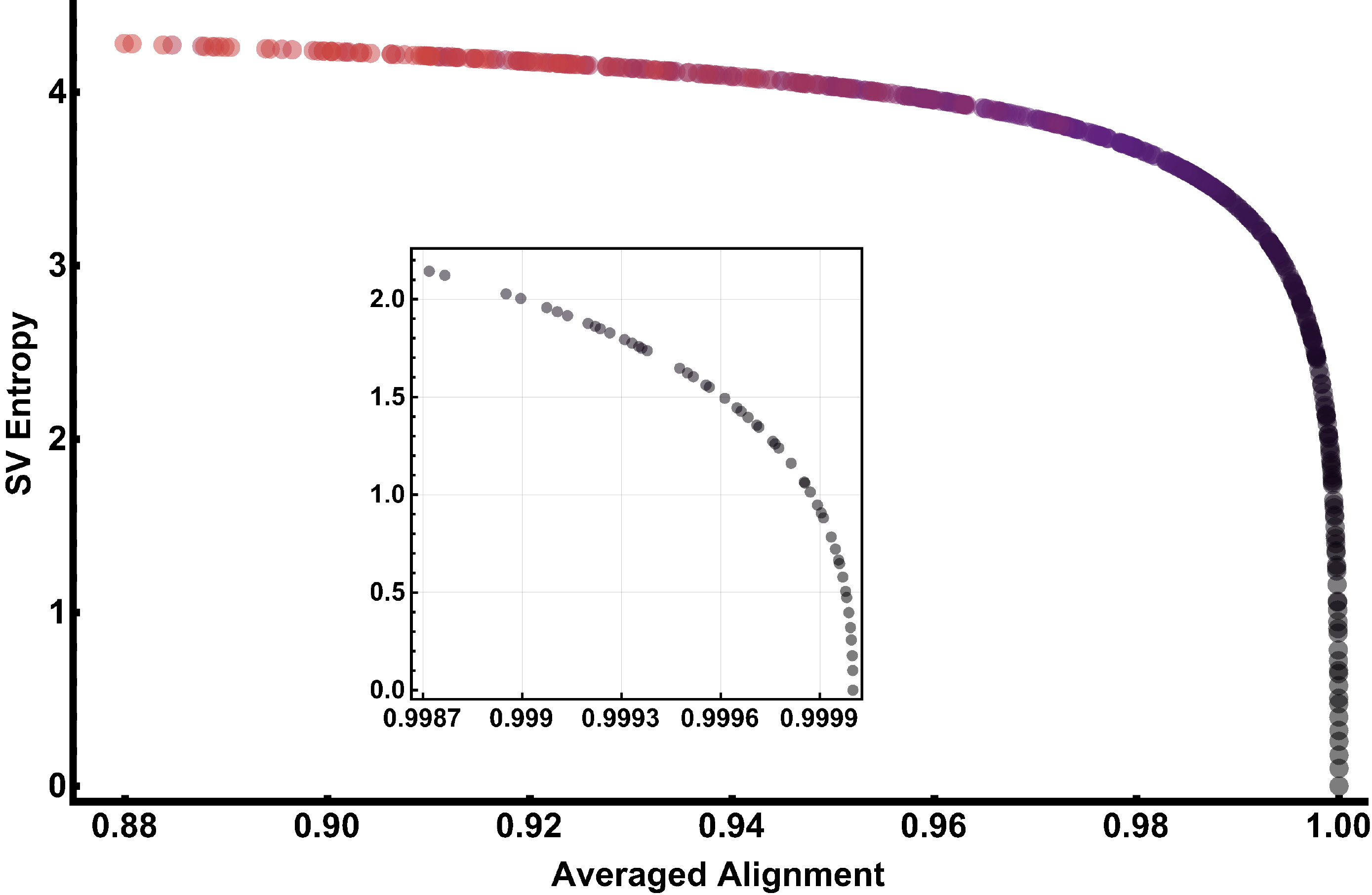}
    \caption{\textbf{Entropy of Singular Values of Matrix \( J \) as a Function of Alignment Degree.} Starting from a teacher vector, we create a hypothetical matrix \( J \), taking 600 copies of the vector \( T \) to which we add Gaussian noise with varying strengths. This graph plots the entropy of the singular values of matrix \( J \) as a function of the average alignment, calculated as \(\text{Arccos}(W \cdot T)\). We observe that even a small misalignment of about 0.1\% generates an entropy value of \( S = 2 \). In the inset, we focus on cases with small misalignments.}
    \label{fig:entropy_alignment}
    
    \label{4RTS:fig5}
\end{figure}
\clearpage
}

In this section, we have observed that, when the dataset size $P\gg N_{\text{tot}}$, the network is capable of recovering the teacher's rule and aligning its weights accordingly. 
Conversely, when $P\ll N_{\text{tot}}$ the network is not able to recover the teacher's rule, and the weights are distributed in a way that generates zero training errors, but without recovering the teacher. In this latter case, consistently with the lazy training paradigm, the trained weights distribution still seems to be random, as as indicated by the entropy values converging to those of random vectors, $S \approx 4$, a phenomenon observable in Fig. \ref{4RTS:fig5}. 
This non-optimal weights distribution determines the bad generalization performance of the network, which is greater than zero. 
In the intermediate case, where $D < P < N_{\text{tot}}$, the weights show a very complex behavior an intriguing behavior which at present time we do not fully understand. See for instance, the peak in entropy observed in this regime that is intriguing and not easily interpreted. We leave these interesting issues for future research.

\section{Volume of the solution space}

In this section, we introduce a theoretical argument in order to justify the different behavior observed the underparametrized and overparametrized regimes. We will attempt to estimate the volume of the solution space, which is the space of all the possible weights configurations that exactly satisfy the constraints imposed by the request of a zero loss on the training set. Armed with this result, we will examine how the volume of the solution space changes as a function of the dataset size \( P \), the dataset size \( D \), and the number of neurons \( N \) in the hidden layer. 
In the following we will consider $D$ large to make use of a central limit theorem and we will control the overparametrized and underparametrized limits by the ratio $\alpha=P/N$, in particular using $\alpha \to 0$ in the former and $\alpha \to \infty$ in the latter.

We can define the volume of the solutions in the following way:

\begin{equation}
    V = \int \mathcal{D}W\,\mathcal{D}v \, \prod^P_\mu \delta \left(y^\mu -\frac{1}{\sqrt{N}}\sum_i^N v_i\,\sigma\left(\sum_j^D\frac{W_{i,j}x_j^\mu}{\sqrt{D}}\right)\right)\,.
    \label{V_init}
\end{equation}

Note that in Eq. \ref{V_init} the measure $\mathcal{D}v \sim \mathcal{N}(0,\sigma_v)$ and $\mathcal{D}W \sim \mathcal{N}(0,\sigma_w)$, i.e. are assumed to be Gaussian distributed, as is typical both in the infinite-width limit analytic scheme and in the real world DNNs. The delta functions in Eq. \ref{V_init} imposes that the output of the DNNs of a given input $x^\mu$ must be equal to the associated label $y^\mu$, for anyone of the $P$ elements in the training set. The volume of the exact  solution space is thus the integral over all possible weights configurations that satisfy these constraints.

The first step is to introduce a set of auxiliary variables $\Delta^\mu$ to rewrite the Dirac's Delta functions in Eq. \ref{V_init} by its integral representation as follows:

\begin{equation}
    V = \int \mathcal{D}W\,\mathcal{D}v\,\int \left(\prod_\mu^P \mathcal{D}\Delta^\mu\right) \, e^{-i\sum_\mu^P\Delta^\mu\left(y^\mu -\frac{1}{\sqrt{N}}\sum_i^N v_i\,\sigma\left(\sum_j^D\frac{W_{i,j}x_j^\mu}{\sqrt{D}}\right)\right)}\,.
\end{equation}

To proceed we make use of the technique introduced in the previous chapter: via Dirac's delta and their integral representation, we introduce a set of auxiliary variables \( h_i^\mu \) and $s^\mu$ (together with their conjugated counterparts) to rewrite the DNN function in a more easy to handle form. After this step, we obtain:

\begin{equation}
\begin{split}
    V = &\int \frac{d\Delta^\mu}{\sqrt{2\pi}}  \int \frac{ds^\mu \, d\bar{s}^\mu}{2\pi} e^{i\sum_\mu^P \Delta^\mu(y^\mu-s^\mu)} e^{is^\mu\bar{s}^\mu} \int \frac{dh_i^\mu\, d\bar{h}_i^\mu}{2\pi} e^{ih_i^\mu\bar{h}_i^\mu}  \\ 
    &\int \mathcal{D}W\, e^{-i\sum_\mu\sum_i \bar{h}_i^\mu \sum_j \frac{W_{i,j}x_j^\mu}{\sqrt{D}}} \int \mathcal{D}v\, e^{-i \sum_\mu \bar{s}^\mu \sum_i \sigma(h^\mu_i)}
\end{split}
\end{equation}

We can see that the integral over the \( s^\mu \) and \( h_i^\mu \) variables are analogous to the one encountered in the calculation of Chapter \ref{ch.3BIW}. These integral can be performed exactly, with the same technique introducing the same $Q$ and $\bar Q$ order parameters and proceeding accordingly by the same Gaussian-like assumptions for $\bar{s}$. In the end, after performing a Wick's rotation in the variable $\bar{Q}$, we obtain the following expression for the volume of the solution space:

\begin{equation}
    \begin{split}
        V &= \,\frac{N}{2i}\int \frac{dQ\, d\bar{Q}}{2\pi} \,\left(1+Q\right)^{-\frac{N}{2}}\,e^{\frac{N}{2}Q\bar{Q}}  \times \\
        &\times \int \prod_\mu \frac{d\Delta^\mu}{\sqrt{2\pi}}  \frac{1}{\sqrt{\det \frac{\Bar{Q}K}{\sigma_v}}} \frac{1}{\sqrt{\det \left(\frac{\Bar{Q}K}{\sigma_v}\right)^{-1}}} e^{-\frac{1}{2} \sum_{\mu,\nu} \Delta^\mu\frac{\Bar{Q}K_{\mu,\nu}}{\sigma_v} \Delta^\nu +i\sum_\mu \Delta^\mu y^\mu}\\
        &= \,\frac{N}{2i}\int \frac{dQ\, d\bar{Q}}{2\pi} \,\left(1+Q\right)^{-\frac{N}{2}}\,e^{\frac{N}{2}Q\bar{Q}} \frac{1}{\sqrt{\det \frac{\Bar{Q}K}{\sigma_v}}} e^{-\frac{1}{2} \sum_{\mu,\nu} y^\mu\left(\frac{\Bar{Q}K_{\mu,\nu}}{\sigma_v}\right)^{-1} y^\nu }=\\
        &= \frac{N}{2i} \int \frac{dQ\, d\bar{Q}}{2\pi} \; e^{\frac{N}{2}Q\bar{Q} - \frac{N}{2} \log(1+Q) - \frac{1}{2}\text{Tr} \log \left(\frac{\Bar{Q}K}{\sigma_v}\right)-\frac{1}{2} \sum_{\mu,\nu} y^\mu\left(\frac{\Bar{Q}K_{\mu,\nu}}{\sigma_v}\right)^{-1} y^\nu }=\\
        &= \frac{N}{2i} \int \frac{dQ\, d\bar{Q}}{2\pi} \; e^{-\frac{N}{2} S(Q,\bar{Q}) }\,,
    \end{split}
\end{equation}
where the action \( S(Q,\bar{Q}) \) is given by:
\begin{equation}
    S(Q,\bar{Q}) = -Q\bar{Q} + \log(1+Q) + \frac{1}{N}\text{Tr} \log \left(\frac{\Bar{Q}K}{\sigma_v}\right)+\frac{\alpha}{P} \sum_{\mu,\nu} y^\mu\left(\frac{\Bar{Q}K_{\mu,\nu}}{\sigma_v}\right)^{-1} y^\nu\,,
\end{equation}
with $\alpha = P/N$ and $K$ being the NNGP kernel, defined as:
\begin{equation}
    \int \prod_{\mu=1}^P dh^\mu \, \mathcal{N}(0,C)_{\{h\}} \,  \sigma(h^\mu) \sigma(h^\nu) = K_{\mu,\nu}\,
\end{equation} 
where \( C \) is the covariance matrix of the input data \( x^\mu \) and \( \mathcal{N}(0,C)_{\{h\}} \) is the Gaussian measure over the space of the hidden layer activations \( h^\mu \) (for more datails, see Chapter \ref{ch.3BIW}, and in particular Eq. \ref{3BIW:KERNELK}).

From a physical point of view, such a calculation is essentially analogous to the one presented in Chapter \ref{ch.3BIW}, with the differnce that here we consider the microcanonical ensamble rather that the canonical one. 
In this way one realizes that the volume of the exact solution space is given by the integral over the action \( S(Q,\bar{Q}) \), which is a function of the auxiliary variables \( Q \) and \( \bar{Q} \).

We can now consider the behavior of the volume of the solution space as a function of the dataset size \( P \), the dataset size \( D \), and the number of neurons \( N \) in the hidden layer.
To do so, we need to study the behavior of the action \( S(Q,\bar{Q}) \) as a function of these parameters. As usual, to perform this last integral we can make use of a saddle point approximation.

Following once again the same procedure of the previous chapter, we can obtain the following results:
\begin{equation}
    \begin{split}
        \bar Q^{*} &= \frac{1}{1+Q^{*}}\qquad\text{or}\qquad Q^{*} = \frac{1-\bar Q^{*}}{\bar Q^{*}}\,,\\
        \bar{Q}^{*} &= \frac{(1-\alpha) + \sqrt{(1-\alpha)^2 + 4\alpha\Gamma}}{2}\,,
    \end{split}
\end{equation}
where $\Gamma$ is defined as:
\begin{equation}
     \Gamma = \frac{1}{P} y^T \left(\frac{K}{\sigma_v}\right)^{-1} y\,.
\end{equation}

This quantity plays a crucial role in calculating the volume of the exact solution space. Specifically, its scaling could significantly affect the volume behavior in both over- and under-parametrized regimes. Numerical evidence, as shown in Fig. \ref{4RTS:figh1} and Fig. \ref{4RTS:figh2}, indicates that $\Gamma < c\cdot D$, where $c$ is a positive constant which, in our case, appears to be approximately $c \sim 2$. This finding aligns with the fact that $\Gamma$ is constructed from a double summation over P and another double summation over $D$ of random objects. According to the central limit theorem, this summation is expected to scale as $P\cdot D$. Considering the prefactor $1/P$ in the definition of $\Gamma$, we retrieve the observed scaling $\Gamma \propto D$. However, the situation differs when $P > D$ because in this case the matrix $K$ is not full rank, making the inverse of $\Gamma$ a not well defined object. In such cases, we have numerically observed that $\Gamma$ (regularized by a negligible amount of noise) scales as $D^2/P$.

In the following, we will also need to take into account the Hessian of the action \( S \), which is given by:
\begin{equation}
    \begin{split}
        \partial_{QQ} S &= - \left(\Bar{Q}^{*}\right)^2\\
        \partial_{Q\bar Q} S &= -1\\
        \partial_{\bar Q \bar  Q} S &= -\frac{\alpha}{(\bar Q^{*}) ^2 } + \frac{2\alpha\Gamma}{(\bar Q^{*})^3}\,,
    \end{split}
\end{equation}

We are interested in the behaviour of our volume in two different regimes: the overparametrized regime, where \( P \ll N_{tot}=(N+1)\cdot D \), and the underparametrized regime, where \( P \gg N_{tot} = (N+1)\cdot D \). 

\begin{itemize}
\item \textbf{Overparametrized regime}: To study the overparametrized regime, we must consider that $\gamma \equiv P/(N\cdot(D+1)) \ll 1$. Unfortunately, from the saddle point, we don't have direct access to $\gamma$. However, we can circumvent this limitation by considering a different condition, namely $\alpha \Gamma \ll 1$. As previously discussed, $\Gamma$ is expected to scale at most with $D$. We have that $\alpha \Gamma \to 0$ is a stronger request than $P/(N\cdot D) \to 0$, because in the first case we have to impose $\alpha \cdot D \to 0$ (considering the worst case scenario where $\Gamma \,\propto \, D$), while in the second it is enough to impose that $\alpha/D \to 0$. 
With this assumption, we can expand $\bar Q^{*}$ in the limit \( \alpha \Gamma \to 0 \) to obtain:   
\begin{equation}
    \begin{split}
        \bar{Q}^{*} &\approx \frac{1-\alpha}{2}\left[1+1+\frac{1}{2}\frac{4\alpha\Gamma}{(1-\alpha)^2}\right] \approx 1 + \alpha\Gamma
    \end{split}
\end{equation}
Substituting this result into the action, we obtain:
\begin{equation}
    \begin{split}
        S \approx S_{0} &= 1+\alpha\Gamma-1 -\log(1+\alpha\Gamma) +\alpha\log(1+\alpha\Gamma) +\frac{\alpha}{P} \text{Tr}\log\left(\frac{K}{\sigma_v}\right) + \frac{\alpha\Gamma}{1+\alpha\Gamma}\\
        &\approx \alpha\left(\Gamma + \frac{1}{P}\text{Tr}\log\left(\frac{K}{\sigma_v}\right)  \right)\,.
    \end{split}
    \end{equation}
    In order to correctly compute the exact solution volume, one need to compute the full saddle point solution taking into account the Hessian $H_0$ of the action \( S \). In the limit we are taking into account, we obtain:
    \begin{equation}
        \begin{split}
            \det H_{0} &\approx -1 +\alpha -\alpha\Gamma
        \end{split}
    \end{equation}
   The final saddle point result for the overparametrized limit $\alpha \to 0$ gives:
   \begin{equation}
    \begin{split}
            V_0 &\simeq \frac{N}{2i}\cdot\left(N^2\det(H_0)\right)^{-\frac{1}{2}}\cdot e^{-\frac{N}{2} S_0}\\
            &\simeq \frac{N}{2i}\cdot\left(-N^2\right)^{-\frac{1}{2}}\cdot e^{-\frac{N}{2} \alpha\left(\Gamma + \frac{1}{P}\text{Tr}\log\left(\frac{K}{\sigma_v}\right)  \right)}\\
            & \simeq \frac{1}{2} e^{-\frac{N}{2}\alpha\Gamma  - \text{Tr}\log\left(\frac{K}{\sigma_v}\right)}\,.
    \end{split}
    \end{equation}

    \item \textbf{Underparametrized regime}: 
    To study this regime, we must consider that $\gamma = P/(N\cdot(D+1)) \gg 1$. To address this regime, we can considerthe limit $\Gamma/\alpha \to 0$ (with $\alpha \to \infty$). As previously discussed, $\Gamma$ is expected to scale at most with $D$. Thus, we can obtain that $\Gamma/\alpha < (D\cdot N)/P \ll 1$, which effectively represents the definition of the underparametrized regime. 
    In this regime, we can expand $\bar Q^{*}$ in the limit \( \Gamma/\alpha \to 0 \) (with $\alpha \to \infty$) to obtain:
    \begin{equation}
        \begin{split}
            \bar{Q}^{*} &\approx \frac{1-\alpha}{2}\left[1-1-\frac{2\alpha\Gamma}{\alpha^2}\right] \to \Gamma
        \end{split}
    \end{equation}
    As before, we can calculate the action $S_\infty$ and the determinant of the Hessian of the action:
    \begin{equation}
        \begin{split}
            S \approx S_{\infty} &= \beta - 1 -\log(\Gamma) +\alpha\log(\Gamma) +\frac{\alpha}{P} \text{Tr}\log\left(\frac{K}{\sigma_v}\right) + \frac{\alpha\Gamma}{\Gamma}\,,\\
            &\to \alpha\left(1+\log(\Gamma) +\frac{1}{P} \text{Tr}\log\left(\frac{K}{\sigma_v}\right) \right) - \Gamma -1 -\log(\Gamma)\\
            \det H_{\infty} &\approx -\alpha+1 \,.
        \end{split}
    \end{equation}
    The final saddle point result in the underparametrized regime reads as follows:
    \begin{equation}
        \begin{split}
                V_\infty &\simeq \frac{N}{2i}\cdot(N^2\det(H_\infty))^{-\frac{1}{2}}\cdot e^{-\frac{N}{2}S_\infty}\\
                &\simeq \frac{N}{2i}\cdot(-N^2\alpha)^{-\frac{1}{2}}\cdot e^{-\frac{N}{2}\alpha\left( 1+\log \Gamma +  \frac{1}{P}\text{Tr}\log\left(\frac{K}{\sigma_v}\right) \right)}\\
                &\simeq \frac{1}{\sqrt{\alpha}}\cdot e^{-\frac{N}{2} \alpha \left( 1+\log \Gamma\right) - \text{Tr}\log\left(\frac{K}{\sigma_v}\right) }\,.
        \end{split}
        \end{equation}
\end{itemize}

Before commenting on these two results, it is crucial to acknowledge the technical limitations of these estiamtes:

\begin{itemize}
    \item The calculations are conducted for activation functions with zero mean, while the ReLU activation functions has finite mean. This approximation is chosen because, for zero mean activation functions, the scaling behavior can be analytically determined. In contrast, ReLU activation functions present a more complex scenario that necessitates reliance on numerical simulations, as done in previous Chapter for the canonical counterpart of this calculation. 
    In the canonical case, non-zero mean activation only introduce a small correction; therefore we conjecture that the volume qualitative scaling will not be influenced by this corrections. Future work will test this assumption to verify its validity.
    
    \item In principle, the Breuer-Major teorem introduced to justify the Gaussian statistics of the auxiliary variable $\bar{s}$, originally introduced in \cite{ariosto2023}, only holds for non-diverging $\alpha$. Our extension to the underparamterized $\alpha \to \infty$ case thus relies on a somehow less controlled Gaussian approximation. However our focus is on understanding how the solution space volume scales with  large \( \alpha \), rather than on the volume in the strict limit. This lead us to believe that this procedure should remain informative for our analysis.
    \item In both limits considered, we assume that the scaling of \( \Gamma \) is dominated by the one of $\alpha$, so that either $\alpha \Gamma \to 0$ (overparametrized) or $\Gamma/\alpha \to 0$ (underparametrized). Furtermore, in the underparametrized limit we also need  \( \Gamma > e^{-1} \). This assumption has been numerically verified (as shown in Fig.\ref{4RTS:figh1} and Fig.\ref{4RTS:figh2}), and it consistently holds in our case. However, in the future it could be important to deepen our understanding of the behavior of this parameter, especially since our simulations suggest that \( \Gamma \) scales with the input size \( D \) and the dataset size \( P \). A different scaling behavior of \( \Gamma \) could significantly impact the asymptotic behavior of the solution space volume.
\end{itemize}

\afterpage{
\begin{figure}
    \centering
    \includegraphics[width=0.9\textwidth]{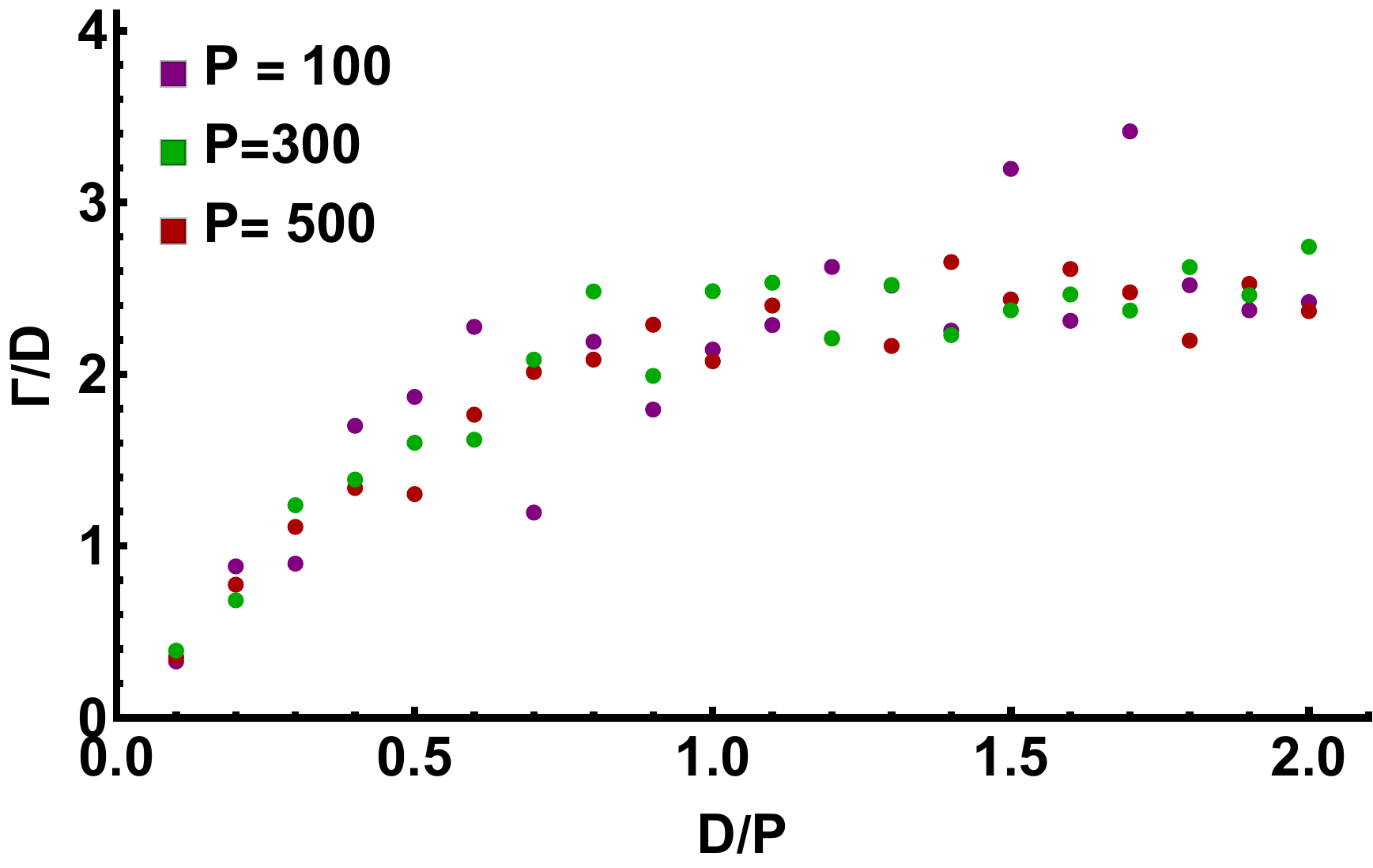}
    \caption{\textbf{Behavior of \( \Gamma \) as a function of the dataset size and the input size.} This graph illustrates the behavior of  \( \Gamma \), rescaled by $D$, as a function of the ratio between the input size \( D \) and the dataset size \( P \). We can see how, for small value of $D/P$ the behaviour of $\Gamma$ is non linear, as it scales as $\Gamma\, \propto\, D^2/P$. For large value of $D/P$, instead, the behaviour is linear, as it scales as $\Gamma\, \propto\, D$.}
    \label{4RTS:figh1}
\end{figure}
\clearpage
}
\afterpage{
\begin{figure}
    \centering
    \includegraphics[width=0.9\textwidth]{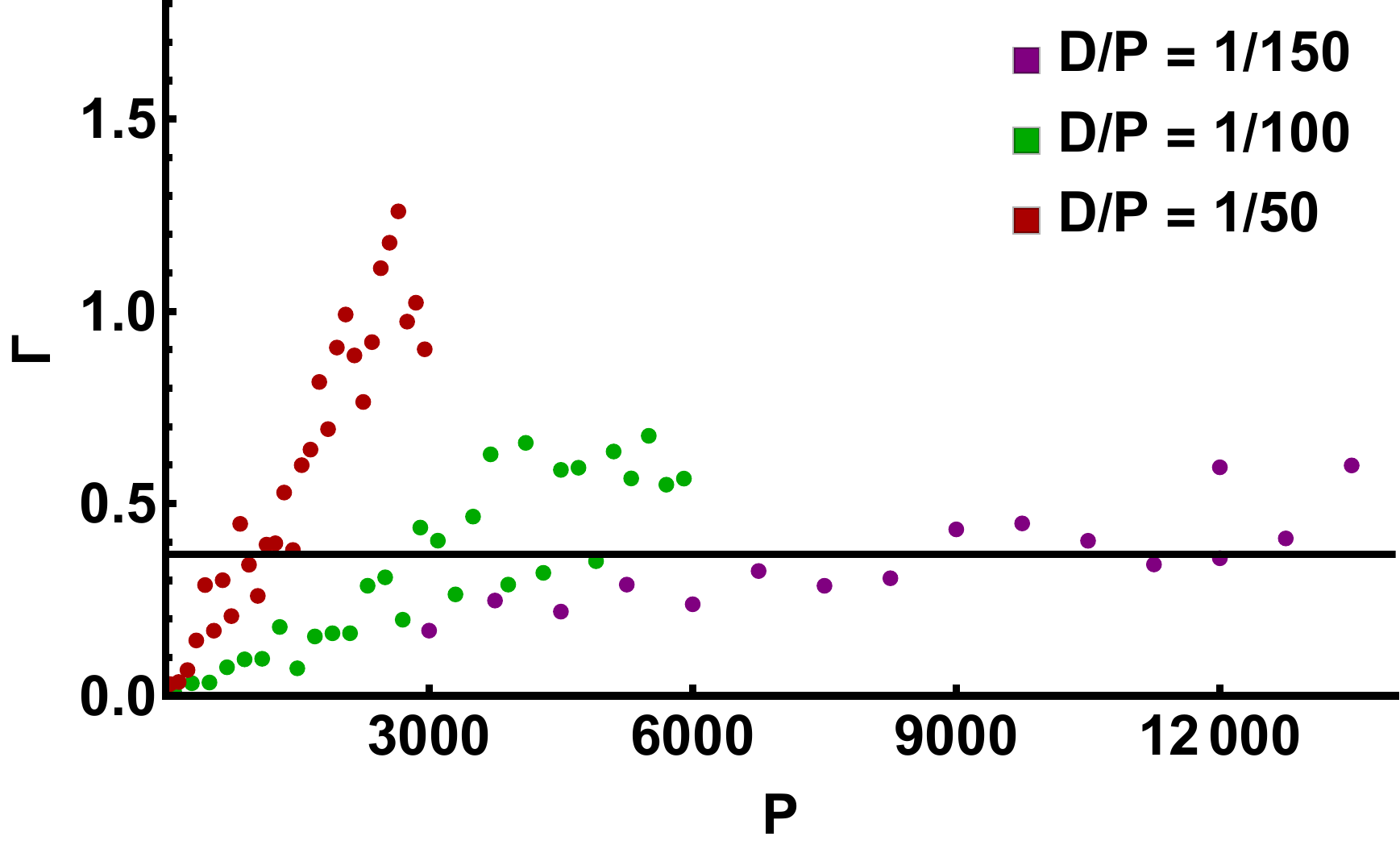}
    \caption{\textbf{The subleading behaviour of \( \Gamma \) in P guarantee that $\Gamma > e^{-1}$, at least for large value of $P$. } This graph illustrates the behavior of  \( \Gamma \), as a function of the dataset size \( P \), for fixed value of $D/P$. We can see how, for large value of $P$, $\Gamma$ is always greater than $e^{-1}$.}
    \label{4RTS:figh2}
\end{figure}
\clearpage
}

Now that we have discussed the technical limitations, we can analyze the results obtained in the two different regimes.
First we have to highlight that the term $\frac{1}{P} \text{Tr} \log \left(\frac{K}{\sigma_v}\right)$ is a at most a finite values, as it is the trace of $P\times P$ matrix, rescaled by P. However, we have to take into accout the fact that if $P > D$, the matrix $K$ is not full rank, and it scaling could be different. This is a technical point left for future investigations.

In the overparametrized regime, the asymptotic volume of the solution space is given by:
\begin{equation}
    \begin{split}
             V_0 &\simeq  e^{-\frac{N}{2}\alpha\left(\Gamma- \text{Tr}\log\left(\frac{K}{\sigma_v} \right) \right)}\,,
    \end{split}
\end{equation}

In this (strongly) overparametrized limit, we have assumed that the number of examples grows more slowly than the number of neurons $N$ in the hidden layer. Including this fact into the analysis of the volume of the solution space reveals that, in this scenario, the volume is exponentially suppressed. However, the rate of this suppression is slower than $N$. Note that we have also assumed that $\Gamma > \frac{1}{P} \text{Tr} \log \left(\frac{K}{\sigma_v}\right) $. This assumption, yet uncontrolled at the present state, seems to hold due to the fact that for  $P<D$, $\Gamma$ scales as $D$, while the right hand side of the inequality is of order one.

In contrast, in the case of the underparametrized regime, the asymptotic volume of the solution space is given by:
\begin{equation}
    \begin{split}
            V_\infty &\simeq \frac{1}{\sqrt{\alpha}}\cdot e^{-\frac{N}{2} \alpha \left( 1+\log \Gamma - \text{Tr}\log\left(\frac{K}{\sigma_v} \right)  \right)}\,.
    \end{split}
\end{equation}
Here, we have that the volume of the solution space is exponentially suppressed with an exponent that is greater that $N$.
Note that we have assumed that $\Gamma > e^{-1}$ and $1+\log\Gamma > \frac{1}{P} \text{Tr} \log \left(\frac{K}{\sigma_v}\right)$. While the first assumpion is justified by the behavior observed in Fig. \ref{4RTS:figh2}, the second one, yet partially uncontrolled at the present state, can be roughly justified by the fact that in the case of $P>D$, the matrix $K$ is not full rank, and, as a consequence the second term is suppressed by $1/P$.

This argument suggests that increasing the number of constraints in the training set (i.e., increasing the dataset size $P$) leads to an exponential decrease in the volume of the solution space. This suggests that, when $P$ is smaller than $N_{tot}$ (i.e. in the overparametrized regime), the solution space contains many spurious solutions (those that do not belong to the optimal set), so that the network is likely to find one of these suboptimal solutions, failing to recover the exact teacher's rule. Conversely, if $P$ is larger than $N_{tot}$, the volume of the solution space is exponentially reduced, leaving predominantly those solutions that belong to the first optimal family. This is indeed consistent with the empirical results discussed in the previous section.

A future extension of this argument should involve a more detailed characterization of the behavior of the terms \( \Gamma \) and $\frac{1}{P} \text{Tr} \log \left(\frac{K}{\sigma_v}\right)$, in order to better understand how they scale with the dataset size \( P \) and the input dimension \( D \). This will also enhance our comprehension of the volume of the solution space in the intermediate regime, where \( P \) is larger than \( D \) but smaller than \( N_{\text{tot}} \).
This extension requires careful consideration but appears feasible, thanks to this controlled environment where we have an explicit expression for the teacher.
Additionally, it is crucial to apply this argument to activation functions with non-zero mean, such as the ReLU, to confirm our earlier conjecture and accurately determine the real scaling of the solution space volume.

Another possible analytical approach could be consider the volume fraction occupied by the optimal solutions (from both the first and the second families) within the full solution space for a finite training set. This analysis should reveal how, in the overparametrized regime, the volume of optimal solutions constitutes only a small fraction of the total volume. Conversely, in the underparametrized regime, we aspect the volume of optimal solutions to be the dominant component. This alternative analytical perspective could also be instrumental in understanding the network's behavior in the intermediate regime and will be  subject of future work. In particular, it will be insightful to study the behavior of networks at the interpolation threshold through this analytical lens. Specifically, efforts should be directed towards understanding the existence of the peak in entropy observed at this point. 

Let us conclude by highlighting the strong agreement between our analytical analysis and empirical observations: when the number of examples in the training set is sufficiently large, the volume of the solution space is exponentially suppressed, focusing on the optimal solutions. Conversely, when the number of examples is small, the solution space is considerably larger, containing both optimal and numerous spurious solutions. Given their abundance, the network is more likely to converge to these suboptimal solutions. This outcome aligns consistently with the empirically observed behavior of the weights and the entropy of the singular values of the matrix \( J \) as discussed in the previous sections.

\section{Gaussianity Breaking}

In order to conclude this chapter, we aim to check whether finite deep neural networks (DNNs) behave akin to Gaussian Processes as their inifinite width counterparts or if and when deviation may appear. Indeed, we know that in the infinite width limit, the DNN weights tend to evolve ``slowly'', making the learning process in this regime similar to a problem of kernel learning.
Moreover, the slowly moving weights are fully compatible with the idea that, in this regime, the DNNs converge to a spurious solution, as we have seen in the previous section. In fact, we argued that this spurious solutions are denser in the solution space, so it is likely that it always exist at least one of this solutions near to the network random initial conditions.

From this perspective, it becomes intriguing to investigate whether a finite width network needs to deviate from its Gaussian behavior to effectively recover the teacher's rule. In fact, we argued that the optimal solutions are rare in the solution space, so it reasonable to conjecture that the process to align the weights to the teacher requires a deviation from the Gaussian behavior. Moreover, in the previous chapter we have already seen that that in the case of finite $\alpha$  the GP arising in the infinite-width limit of Bayesian neural networks  should be generalized to a Student's t stochastic process.

To investigate this question, we consider the statistics of error made the DNNs:
\begin{equation}
    \Delta(x) = s(x) - T\cdot x\,.
\end{equation}
If the DNNs behave as a Gaussian Process, we would expect the kurtosis \( \kappa \) of the error, defined as:
\begin{equation}
    \kappa = \frac{\langle \Delta(x)^4 \rangle}{\langle \Delta(x)^2 \rangle^2} - 3\,,
\end{equation}
to vanish. Here, the average is taken over full input distribution.

As an initial experiment, we have measured the kurtosis behaviour during the training, for different dataset sizes \( P \) and for a fixed DNN size ($D=100$ and $N=600$). It is convenient to visualize the Kurtosis not directly as a function of the training epochs but rather as a function the corresponding generalization error. In Fig. \ref{4RTS:figboh}, we can see how, to improve their generalization performance above a certain threshold, the DNNs (or at least the ones that can do that) need to dynamically break their Gaussianity.
In particular, we can see how, as expected, the kurtosis is always zero in the overparametrized regime (small values of P in dark purple symbols),explodes in the undeparametrized regime (large values of P, green to yellow) when the generalization error falls below a given threshold ($\epsilon_g \simeq 10^{-4}$). 
Interestingly the Kurtosis curves seem to collapse when plotted as a function of the generalization error rather than of the training epoch. This indeed suggests that the in order to reach optimal generalization capabilities in the underparametrized feature learning regime investigated in the previous sections the DNNs has to break its Gaussianity (at least in its output statistics).



\afterpage{
\begin{figure}
    \centering
    \includegraphics[width=0.9\textwidth]{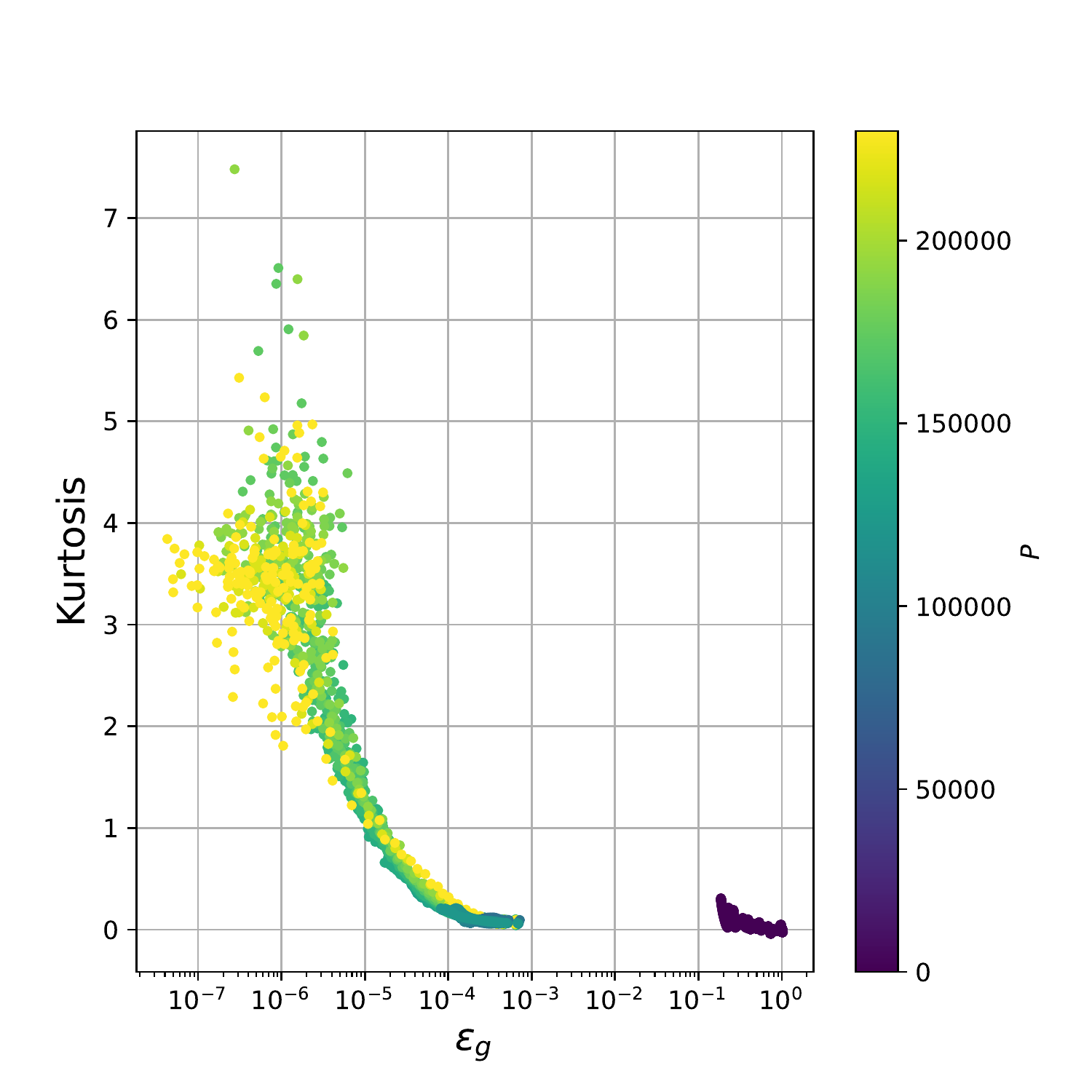}
    \caption{\textbf{Kurtosis of the Output Error as a Function of the Generalization Error, for different values of \( P \).} This graph illustrates the behavior of the kurtosis of the error as a function of the generalization error during the trining, for dataset size \( P\). In particular, we focus on the underparametrized regime ($P > 70000 $, in yellow and green), and the overparametrized regime  ($P<200$, in purple), for fixed values of $D = 100$ and $N_{out} = 600$.} 
    \label{4RTS:figboh}
\end{figure}
\clearpage
}

In Fig. \ref{4RTS:fig7}, we present the post-training behavior of kurtosis of the outputs as a function of the rescaled dataset size \( P/N_{\text{tot}} \), for fixed values of \( D = 100 \) and \( N_{\text{out}} = 600 \). As hinted at by the dynamic approach, the kurtosis explodes after the interpolation threshold (in the underparametrized regime), while it remains zero for small values of \( P \) (i.e., in the overparametrized regime). Interestingly, in the intermediate regime, the kurtosis exhibits a more complex pattern.

\afterpage{
\begin{figure}
    \centering
    \includegraphics[width=0.9\textwidth]{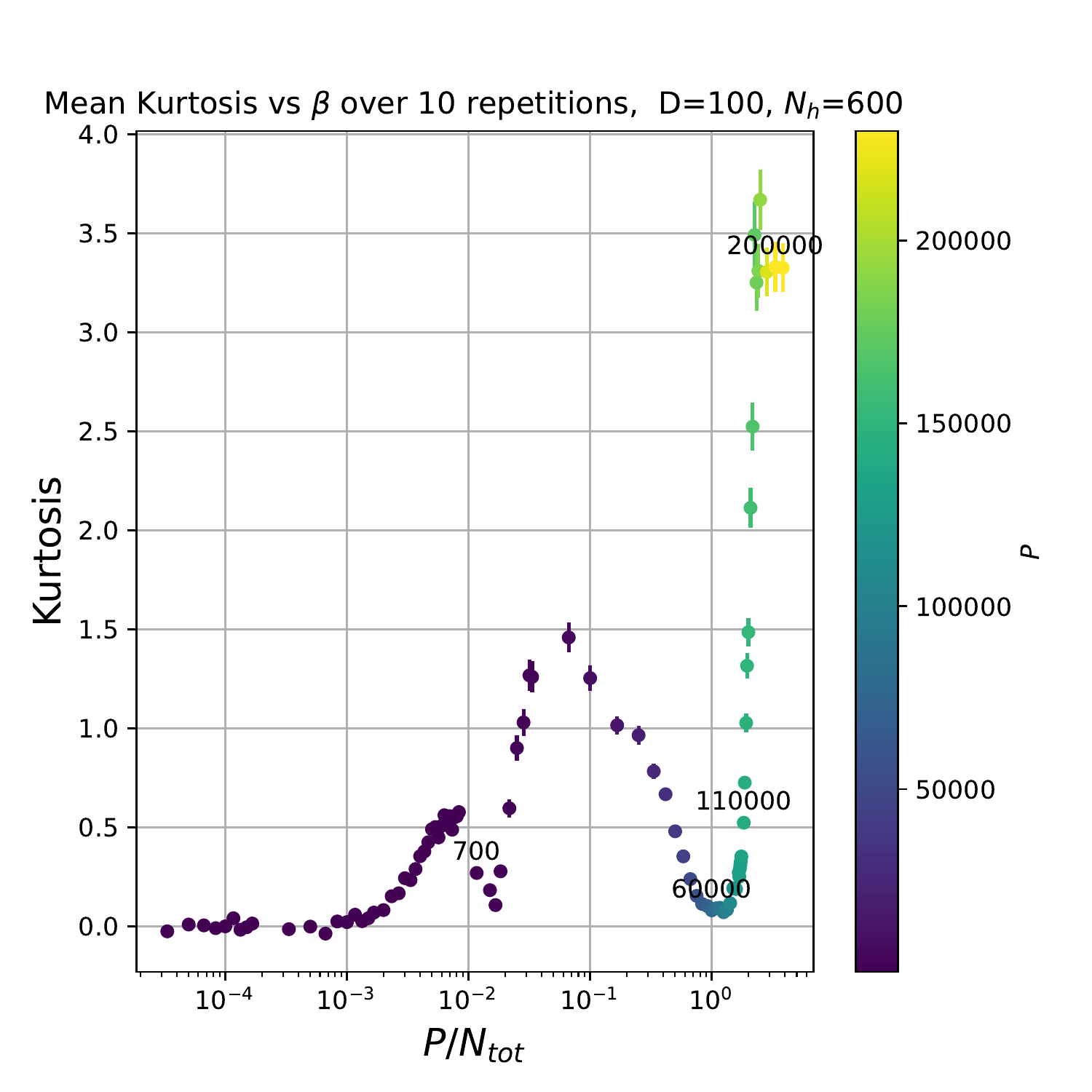}
    \caption{\textbf{Kurtosis of the Output Error as a Function of the Rescaled Dataset Size \( P/N_{\text{tot}} \).} This graph illustrates the behavior of the kurtosis of the output as a function of the rescaled dataset size \( P/N_{\text{tot}} \), for fixed values of \( D = 100 \) and \( N_{\text{out}} = 600 \). For greater clarity, the in-graph labels marks selected values of P.}
    \label{4RTS:fig7}
\end{figure}
\clearpage
}

This result suggests that the output of deep neural networks, trained using gradient-based methods, is not always consistent with a Gaussian Process. Specifically, in the underparametrized, feature learning regime, it appears that the network, in order to reach the optimal solution, must develop internal correlations. These correlations lead to a leptokurtic distribution of the output.

We have observed, in trained DNNs, a close correlation between the behavior of kurtosis and the behavior of the weights, especially in the two extremal under- and over-parametrized regimes. This correlation can be directly confirmed by utilizing the entropy measure defined earlier. In fact, in Fig. \ref{4RTS:fig9}, we note an increase in kurtosis concurrent with a decrease in entropy. This confirms that for the network to successfully recover the teacher's rule, it needs to deviate from its initial Gaussian distribution and develop a positive kurtosis. Conversely, within the overparametrized regime, kurtosis remains at zero, and entropy stays fixed at its initial value, characteristic of random vectors rows in the input weight matrix $W$.



\afterpage{
\begin{figure}
    \centering
    \includegraphics[width=0.9\textwidth]{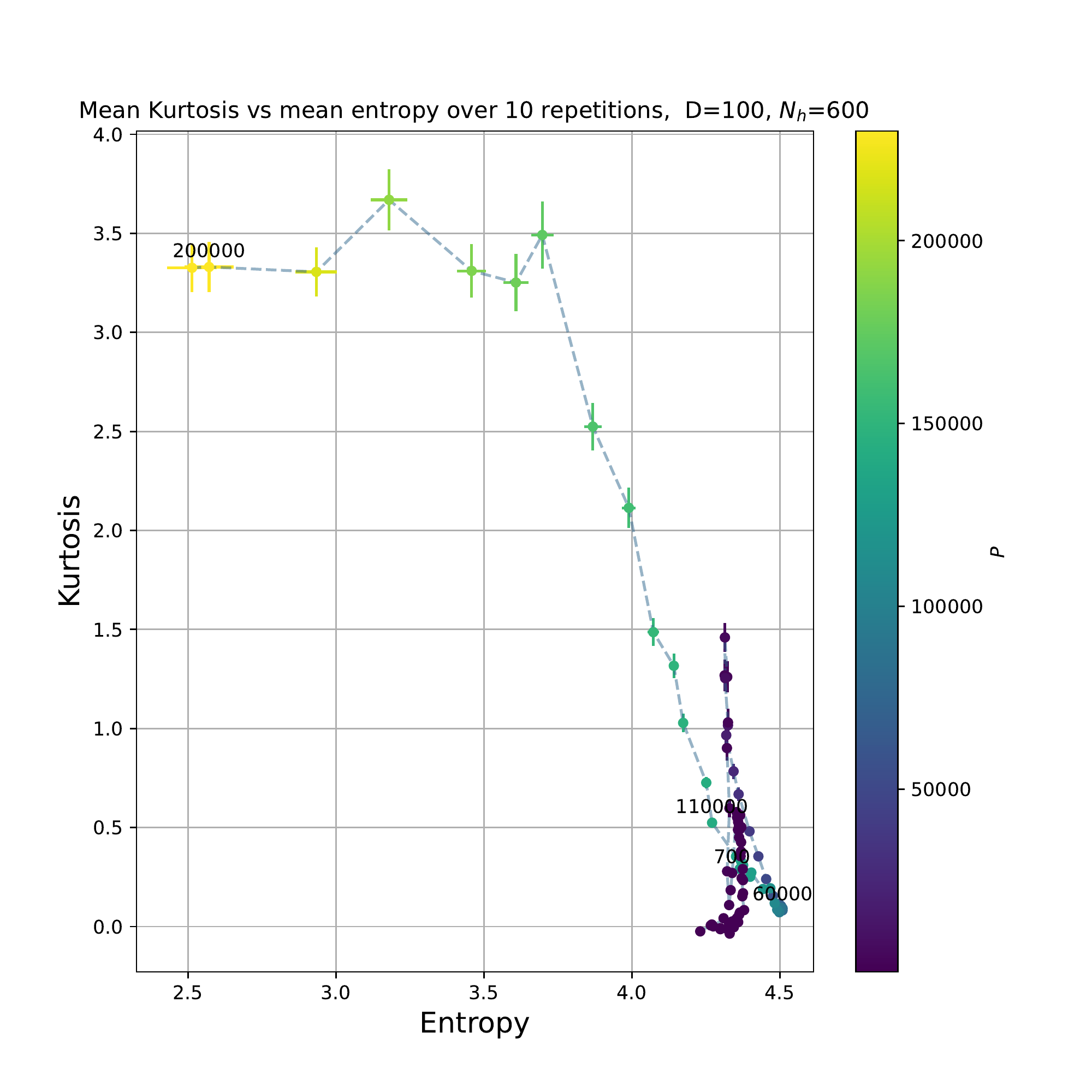}
    \caption{\textbf{Kurtosis of the Output Error as a Function of the Entropy of the Singular Values of Matrix \( J \).} This graph illustrates the behavior of the kurtosis of the output as a function of the entropy of the singular values of matrix \( J \), for different choice of $P$, and fixed values of \( D = 100 \) and \( N_{\text{out}} = 600 \). For greater clarity, the in-graph labels marks selected values of P. The kurtosis increases when the entropy starts to decrease, indicating the network's ability to recover the teacher's rule. Conversely, in the overparametrized regime, the kurtosis remains zero, and the entropy is fixed to its value in the case of random vectors.}
    \label{4RTS:fig9}
\end{figure}
\clearpage
}

While the behavior in the two extremal regimes in compatible with our expectations, the intermediate regime presents a more complex pattern. We observe that, the generalization error (as a function of $P/N_{tot}$) go through a series of changes in concavity. 
In fact, iniitially, just beyond the memorization regime, the generalization error exhibits a convex behavior. After an unidentified threshold, the generalization error demonstrates a concave behavior. Another change in concavity is observed near the interpolation threshold, where the curve reverts to convexity. Finally, after the interpolation threshold, the DNNs enter the underparametrized regime with yet another change in concavity. Intriguingly, these changes in concavity correlate with changes in kurtosis. This phenomenon is depicted in Fig. \ref{4RTS:fig10}, where the change in convexity is emphasized.

It is conceivable that changes in kurtosis (and the accompanying shift in the convexity of $\epsilon_g$) are linked to a transition in the network's learning focus. Intriguingly, one can thus be tempted to interpret the generalization error as a thermodynamic function of interest. However, these preliminary findings necessitate more detailed and focused simulations to thoroughly understand the behavior of DNNs in this complex intermediate regime.

To further investigate this phenomenon, simulations in more complex setups, such as a two-hidden-layer (2HL) student network learning from a one-hidden-layer (1HL) teacher, seem necessaryto verify if this behavior is consistent across different architectures and dataset. This will be the subject of future work.



\afterpage{
\begin{figure}
    \centering
    \includegraphics[width=0.9\textwidth]{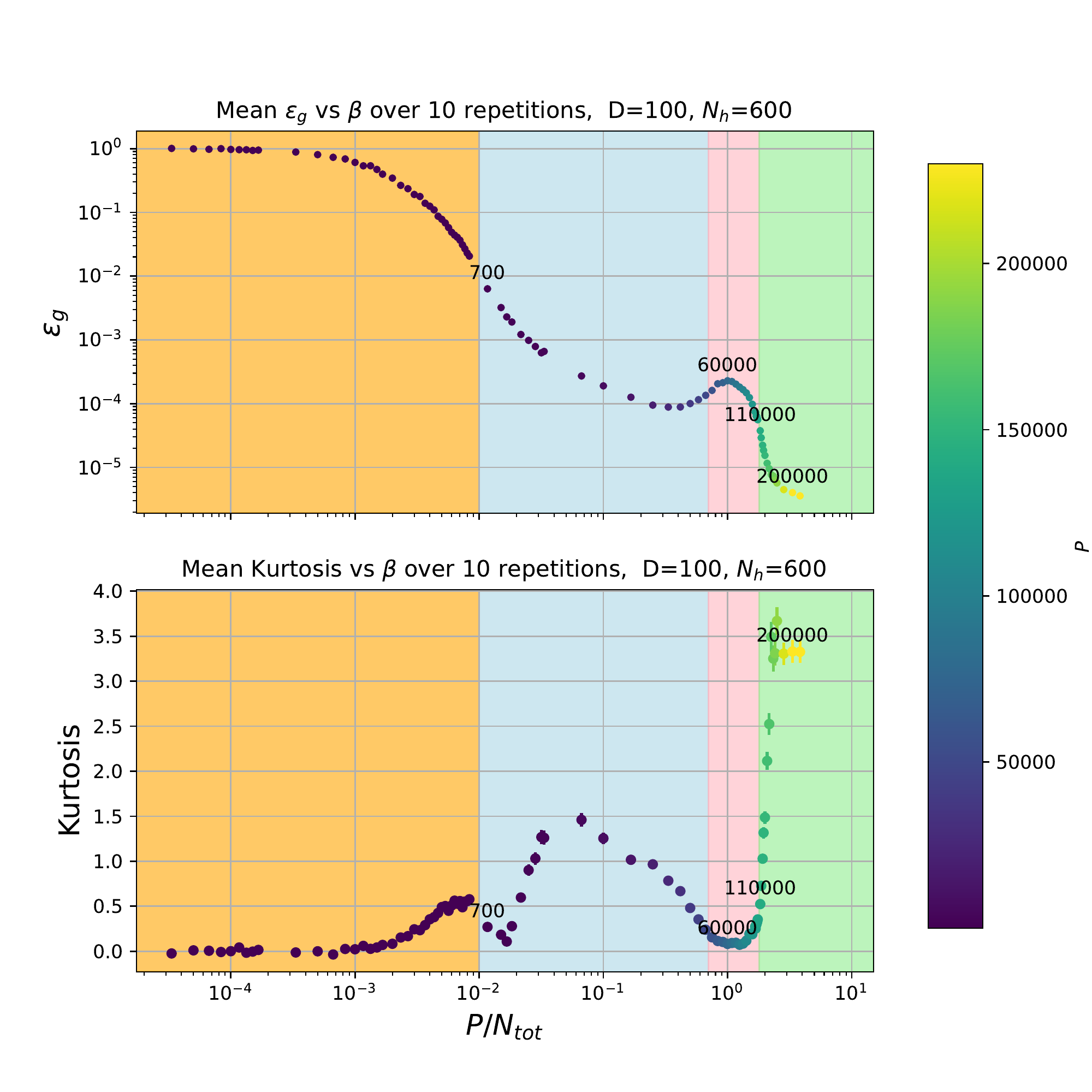}
    \caption{\textbf{Concavity Changes in Generalization Errors Corresponding to Peaks and Valleys in Kurtosis.} The generalization error (top) and Kurtosis (down) as a function of $\gamma=P/N_{tot}$. The various colored regions highlight changes in the concavity of the generalization error. In-graph labels highlight selected values of P.
    Archtecture size is fixed (N=600 and D=100), while the dataset size $P$ is varied.}
    \label{4RTS:fig10}
\end{figure}
\clearpage
}

\section{Conclusion and perspective}

In this chapter, we have presented an initial analysis of the behaviour of a DNNs weights in a controlled setting, where the teacher is fully contained in the student, and the optimal solutions are known.
We have shown that the weights of the student network reach the optimal solutions family (i.e they align with the teacher) in the underparametrized regime, while they remain misaligned with the teacher in the overparametrized case. We have also put forward a preliminary a theoretical argument to justify these results, based on the volume of the solution space. This argument suggests that the volume of the solution space is exponentially suppressed as the dataset size increases, leading from a dense set of sporiuous solution to the predominance of rare optimal solutions, and it is consistent with the observed behavior of the weights. 
Even if these results are preliminary, they show how the network, when the dataset is large enough, is capable of feature learning, that is to generate itself the relevant feature of the teacher. Otherwise, if the number of pattern studied is not enough, the network is only able to memorize the training set, in a lazy training regime.
Let me highlight that this results are observed in simulations performed in a vanilla (i.e. fully connected DNN with standard scaling inizialization and standard learning procedures) realistic setting, considering the usuals scaling in the initialization of the weights.

Additionally, in the second part, we have investigated the behavior of the kurtosis of DNNs' outputs to glean insights about their (lack of) Gaussianity. We observed that the kurtosis is vanishing for small values of $P$ in the overparametrized regime, while it significantly increases past a certain threshold for large values of $P$ in the underparametrized regime. 
This finding shows that the output of DNNs is not always Gaussian distributed and suggests that the breaking of this Gaussianity is linked to the network's ability to recover the teacher's rule in the feature learning regime.

This intriguing result calls for further investigation to understand the origins of the internal correlations that the networks develop during training in the underparametrized regime (as indicated by the leptokurtic distribution) and how these correlations contribute to achieving zero generalization error. In particular, we limited our numerical analysis to the DNN output statistics, but it could be interesting to asses the statistical behavior of the hidden layers activations.  

Finally, we observed that changes in kurtosis correlate with shifts in the sign of the concavity of the generalization error as a function of the dataset size. This phenomenon is particularly interesting and requires more detailed and focused simulations for a comprehensive understanding.

It is necessary to highlight that these findings and results are preliminary and need to be confirmed by more detailed and focused simulations and analyses. Analytically, to better characterize the scaling of the solution space volume, understanding the behavior of the parameter \( \Gamma \) and how it scales with the dataset size \( P \) and the input dimension \( D \) is crucial. Furthermore, extending our argument to activation functions with non-zero mean, such as ReLU, could confirm our conjecture on the volume scalings.
From an empirical standpoint, extending the kurtosis analysis to different input sizes \( D \) and numbers of hidden layers \( N \) would be beneficial to verify the generality of the Gaussianity breaking scenario. The results presented in this chapter are mostly obtained for a DNN with \( D = 100 \) and \( N = 600 \), and while our initial analysis of different architectures suggest generality, it is prudent to extend all our observations to different network dimensions to strengthen our results. Beyond this, it would be very interesting to extend the analysis to more complex setups, such as a two-hidden-layer student learning from a one-hidden-layer teacher, to verify if the observed behaviors in simpler cases persist.

\clearpage
\printbibliography[heading=subbibnumbered, title=Chapter bibliography]
\end{refsection}

\chapter{Outlook}
\begin{refsection}
\label{ch.outlook}
Deep neural networks (DNNs) are reshaping our world through their remarkable capabilities, which often equal or even surpass human abilities in various tasks. Despite their wide deployment, in complex inner workings of these algorithms is largely shrouded in mystery, so much so that they deserve the epithet  `black boxes'. Although it is possible to empirically measure and improve their performance, fully understanding their complex internal dynamics remains an elusive goal.

In this thesis, I have tried to show how the application of physics-based methodologies can enrich our understanding of these algorithms.

In my first work, I explored how the use of the Statistical Mechanics framework could unearth new insights on the overfitting problem. Building on a typical Statistical Physics approach we realized that an effective upper bound for the generalization  can be derived from known estimates of the network performance.
This insight has allowed us to put forward a mean-field like upper bound for the so-called generalization gap that is much more stringent that classical results of Statistical learning theory and can be meaningfully applied in the overparametrized regime.
In particular, our calculation is able to reveal how the generalization performance of a network can be bounded by the ratio of two architectural hyperparameters: the number of neurons in the readout layer and the dataset size. This theory is tested against both simplistic toy models and advanced state-of-the art DNNs, showing how our new asymptotic bound holds in practical scenarios.
These findings lead us to a key insight suggesting a differentiated role for the network's layers. It seems that initial layers are specified in the extraction of feature from the dataset, while the final layer focuses on processing these features. 

In the third chapter, we introduce a novel framework designed to describe the inner workings of DNNs in the `proportional regime', a regime akin to the celebrated thermodynamic limit in Statistical Mechanics, where both the number of neurons in the hidden layers $N_i$ and the dataset size $P$ diverge, while keeping a constant ratio. 
Moreover, we are able to adopt a data dependent perspective and do not necessarily take an average over the training dataset measure.
We are thus able to derive an approximate partition function for fully connected deep neural architectures, which encodes information on the trained models. This advance allows us to obtain: (1) a closed formula for the generalization error associated with a regression task in a one-hidden layer network with finite width; (2) an approximate expression of the partition function for deep architectures (via an effective action that depends on a finite number of order parameters); and (3) a closed, data-dependent formula capable of comparing the generalization capabilities of finite and infinite-width networks on a given dataset.
Furthermore, a fascinating connection emerges between DNNs in the proportional limit and a Student's t Process, contrary to the classical infinite width regime results,  where there is a well established a link between the DNNs behaviour and Gaussian processes.

Building on the insights gained in the previous work, in a final (and still unpublished) research project, we shift our focus to analyzing the DNNs' capability for feature learning as opposed to simple pattern memorization. 
In this work, we consider a simplified deep-learning setting that allows us to identify the optimal weight distribution, by which a student DNNs can exactly reproduce the teacher function.
These optimal weights potentially indicate genuine learning of underlying rules, as opposed to mere memorization.
Empirical observations in this simple setting, show that DNNs are able to achieve the optimal distribution only in the underparametrized regime. Conversely, in the overparametrized regime, the weight distributions of DNNs appear to be random. 
An analytical argument for the onset of these two different learning regime, based on the scaling properties of the solution space's volume, is put forward.
Finally, we attempt to link the occurrence of these two learning regimes to the trained network statistics. In particular, we consider the behavior of the kurtosis of the output in trained DNNs. Notably, the kurtosis deviates from zero in the underparametrized regime, indicating that severe deviations from the Gaussian behavior are associated to the network capability to extrapolate features. On the other hand, in the overparametrized regime, the kurtosis keeps vanishing values confirming Gaussianity in the lazy training regime. Additionally, we observe that in the intermediate regime, the kurtosis exhibits a complex, but still unexplained, behavior that can be linked to the changes in the concavity of the generalization curves. Although these results are fascinating, it is important to highlight their preliminary nature, necessitating further simulations and analysis for a complete understanding and validation.

In conclusion, this thesis explored the potential of physics-based approaches to demystify the inner workings of Deep Neural Networks (DNNs). Attempting to apply principles from Statistical Mechanics, beyond the most common pathways, we hope to have gained a new unique and insightful perspective, enabling a deeper understanding of DNN behavior. 

Hopefully, this is just the beginning of a long, fruitful, and exciting journey. Much remains to be discovered.



\clearpage
\end{refsection}

\nocite{*}
\printbibliography[heading=bibnumbered, title=Bibliography]



\end{document}